\documentclass[justified,nobib]{tufte-book}
\hypersetup{colorlinks}
\usepackage[T1]{fontenc}    
\usepackage{hyperref}       
\usepackage{url}            
\usepackage{booktabs}       
\usepackage{amsfonts}       
\usepackage{nicefrac}       
\usepackage{microtype}   
\usepackage{graphicx}
\setkeys{Gin}{width=\linewidth,totalheight=\textheight,keepaspectratio}
\graphicspath{{graphics/}}

\newcommand{\etal}{\textit{et al}. }
\newcommand{\dd}{\mathrm{d}}
\usepackage{amsmath}

\usepackage{booktabs}

\usepackage{fancyvrb}
\fvset{fontsize=\normalsize}

\usepackage[titletoc]{appendix}

\usepackage{xspace}

\newcommand{\monthyear}{%
  \ifcase\month\or January\or February\or March\or April\or May\or June\or
  July\or August\or September\or October\or November\or
  December\fi\space\number\year
}

\usepackage{makeidx}
\makeindex

\title{Waves, patterns and bifurcations:\\ a tutorial review on the vertebrate \\segmentation clock}

\author{%
Paul Fran\c cois and  Victoria Mochulska
}

\newcommand{\chap}[1]{\chapter{#1}}
\newcommand{\chapsec}[1]{\section{#1}}
\newcommand{\chapsubsec}[1]{\subsection{#1}}

\newcommand{\appsec}[1]{\chapter{#1}}
\newcommand{\appsubsec}[1]{\section{#1}}
\newcommand{\bookpub}[1]{\publisher{#1}}
\newcommand{\ba}{\cleardoublepage\chapter{Abstract}}
\newcommand{\ea}{}
\newcommand{\bmarf}{\begin{marginfigure}}
\newcommand{\emarf}{\end{marginfigure}}
\newcommand{\sizefig}{\textwidth}
\newcommand{\ssizefig}{\textwidth}
\newcommand{\bnormf}{\begin{figure}}
\newcommand{\enormf}{\end{figure}}
\newcommand{\bwt}{}
\newcommand{\ewt}{}

\newcommand*{\myref}[1]{\hyperref[#1]{\textit{\nameref*{#1}}}}
\newcommand{\startappendix}{\begin{appendices}
\setcounter{secnumdepth}{0}
{\Huge \itshape Appendix}

}
\newcommand{\terminateappendix}{\end{appendices}}

\bookpub{Department of Biochemistry and Molecular Medicine \\
Universit\'e de Montr\'eal\\
 MILA Qu\'ebec\\
 \\
Department of Physics \\
McGill University\\
\\
Montr\'eal, Qu\'ebec\\
\\
\texttt{paul.francois@umontreal.ca}
}

\begin{document}

\maketitle
\ba
 Proper vertebrae formation relies on a tissue-wide oscillator called the segmentation clock. Individual cellular oscillators in the presomitic mesoderm are modulated by intercellular coupling and external signals, leading to the propagation of oscillatory waves of genetic expression eventually stabilizing into a static pattern. Here, we review 4 decades of biophysical models of this process, starting from the pioneering Clock and Wavefront model by Cooke and Zeeman, and the reaction--diffusion model by Meinhardt. We discuss how modern descriptions followed advances in molecular description and visualization of the process, reviewing phase models, delayed models, systems-level, and finally geometric models. We connect models to high-level aspects of embryonic development from embryonic scaling to wave propagation, up to reconstructed stem cell systems. We provide new analytical calculations and insights into classical and recent models, leading us to propose a geometric description of somitogenesis organized along two primary waves of differentiation.
 \ea
\newpage
\tableofcontents
\listoffigures

\newpage

\chap{Verterbrate segmentation for theorists: why?  }

The French naturalist Geoffroy Saint Hillaire noticed in the XIXth century a universal feature of the body of many common animals \cite{Pourquie2022}: they are primarily built on the repeat of metameric units along their anteroposterior axis.  Canonical examples include segments in arthropods, or our vertebrae.  This organization is  so fundamental that entire phylogenetic groups have been named in reference to units of their body plan, e.g. \textit{annelids} or \textit{vertebrates}. Fossil records suggest that this segmental organization is an extreme form of organ metamerism, that possibly accompanied the Cambrian explosion 600 million years ago \cite{Couso2009}. As such, metamerism can be considered a major evolutionary innovation leading to modern animal life. The segmental organization is generally assumed to provide multiple evolutionary advantages, for instance having multiple connected body parts allows for versatile body movements, and division between units allows for subsequent evolutionary specializations of individual segments \cite{Bodo1995}.

Vertebrae precursors in embryos are called somites, and the process of somite formation is called "somitogenesis". Somites first appear as pairs of epithelial spheres on both left and right sides of the neural tube, and sequentially form from anterior to posterior during axis elongation, Fig. \ref{fig:Malpighi}. Multiple tissues derive from somites so a proper understanding and control of somite formation might potentially lead to both fundamental advances and practical application in regenerative medicine \cite{Chal2015}. Somitogenesis is particularly appealing to physicists for multiple reasons. As we will describe below, it is now established that somitogenesis is tied to the presence of a global genetic oscillator, called the segmentation clock \cite{Palmeirim1997}, which is associated with multiple waves propagating in embryonic tissues \cite{Aulehla2008, Delaune2012,Oates2012,Masamizu2006}. The periodicity of this process further allows for multiple observations within one single experiment,  making it an ideal system for developmental biophysics. Examples of experimental perturbations include recovery of oscillation following perturbations \cite{Dubrulle2004, Riedel-Kruse2007} and entrainment \cite{Sanchez2021}. Individual cells can oscillate when dissociated \cite{Rohde2021}, and it is now clear that the segmentation clock at the tissue level is an emergent, self-organized process \cite{Tsiairis2016}. Somites are \textit{in fine} well-defined physical units, so somitogenesis also presents a nice example of interaction between genetic expression, signaling, and biomechanical processes leading to morphogenesis. Lastly, it should be pointed out that the existence of an oscillator controlling somite formation has been predicted theoretically using advanced mathematical concepts (catastrophe theory) \cite{Cooke1976} 21 years before its definitive experimental proof \cite{Palmeirim1997}. So vertebrate segmentation is a good example of "Figure 1" scientific endeavor \cite{Phillips2015}, where theoretical predictions suggest experiments, and where a fruitful back and forth between experimental biology and theoretical modeling has occurred.

 Important recent advances include more controlled experimental setups such as explants \cite{Lauschke2013,Sonnen2018}, stem cell systems \cite{Hubaud2016, Diaz-Cuadros2018} and even synthetic developmental biology assays \cite{matsuda2020, yamanaka2023, McNamara2022} such as somitoids/segmentoids \cite{Budjan2022,Miao2022}. Feynman's famous quote   "What I cannot create, I do not understand" is often invoked (see e.g. \cite{Singer2016,Thompson2022}) to motivate such \textit{in vitro} reconstruction of biological systems. Indeed, great insights can be drawn by creating and manipulating minimal \textit{experimental} models. It is however important to stress that this quote, found on Feynman's blackboard upon his death  \cite{CaltechAr}, likely reflects the mindset of a theoretical physicist, further known for his pedagogical insights  \footnote{Schwinger even qualified Feynman diagrams as "pedagogy, not physics'' \cite{Gleick1994}}. While experiments are of course necessary,  "creation'' in Feynman's mind might also refer to the building of a predictive mathematical model, seen as the \textit{sine qua non} for understanding. This program is best described by Hopfield \cite{Hopfield2018} :
\begin{quote}
    "The central idea was that the world is understandable, that you should be able to take anything apart, understand the relationships between its constituents, do experiments, and on that basis be able to develop a \textbf{quantitative} understanding of its behavior. Physics was a point of view that the world around us is, with effort, ingenuity, and adequate resources, understandable in a predictive and reasonably quantitative fashion. Being a physicist is a dedication to a quest for this kind of understanding."
\end{quote}

This tutorial aims to introduce such a quantitative understanding of somitogenesis. Excellent reviews have been recently written on a more biological side, e.g. \cite{Negrete2020,Diaz-Cuadros2021, Pourquie2022}, or on developmental oscillations in general, \cite{DiTalia2022}, see also  \cite{Venzin2020} for a review of synchronization in the present context. We hope to provide here a modern mathematical introduction to the field of somitogenesis, allowing for conceptual discussions framed with non-linear models, in a language amenable to physicists. We are careful to relate models to experimental biology as much as we can.

 In the following,  we first briefly summarize the main biological concepts and molecular players. The field is still evolving and new aspects are still being discovered to this date (we write those words in 2023). This justifies a more theoretical and conceptual discussion. We then follow approximately a chronological approach, describing how the (theoretical) understanding of the field has progressed with time. Importantly,  classical models proposed before the molecular biological era have been crucial to suggest experiments and ideas, and our ambition is to describe them in detail because they are still relevant today, at the very least to frame the theoretical discussion. 
 
The field has then been strongly driven by the constant experimental progress in molecular biology, genetics, imaging, and more recently synthetic biology, allowing scientists to explore more complex and refined scenarios, that we will describe. Many of the most recent ideas described in the following also find their origin in the era of "systems biology", with a focus on the (emergent) properties of gene regulatory networks \cite{Alon2006}. For this reason,  there is a bias in both experimental and modeling works, towards the signaling aspects of the system, which we would loosely define as the dynamics of gene expression in time and space, described by non-linear models.  We discuss the experimental reasons why such an approach makes sense in retrospect, but also describe works exploring other aspects (e.g. mechanics). We eventually connect those models to current descriptions grounded in dynamical systems or catastrophe theory \cite{Rand2022}, with the hope to infer some general principles and scenarios  \cite{Bialek2023,Francois2023} (see e.g. summary Figure \ref{fig:WaveScenario}). In Appendix A, we put together a condensed discussion of classical results on non-linear oscillators and bifurcations, with examples relevant to the present context (phase oscillator, phase responses, and some introduction to relaxation oscillators and excitability).  Appendix B contains calculations associated with the main text. We also include multiple Jupyter notebooks to simulate the multiple models presented in this tutorial : \url{https://github.com/prfrancois/somitetutorial}

\bnormf
\includegraphics[width=\sizefig]{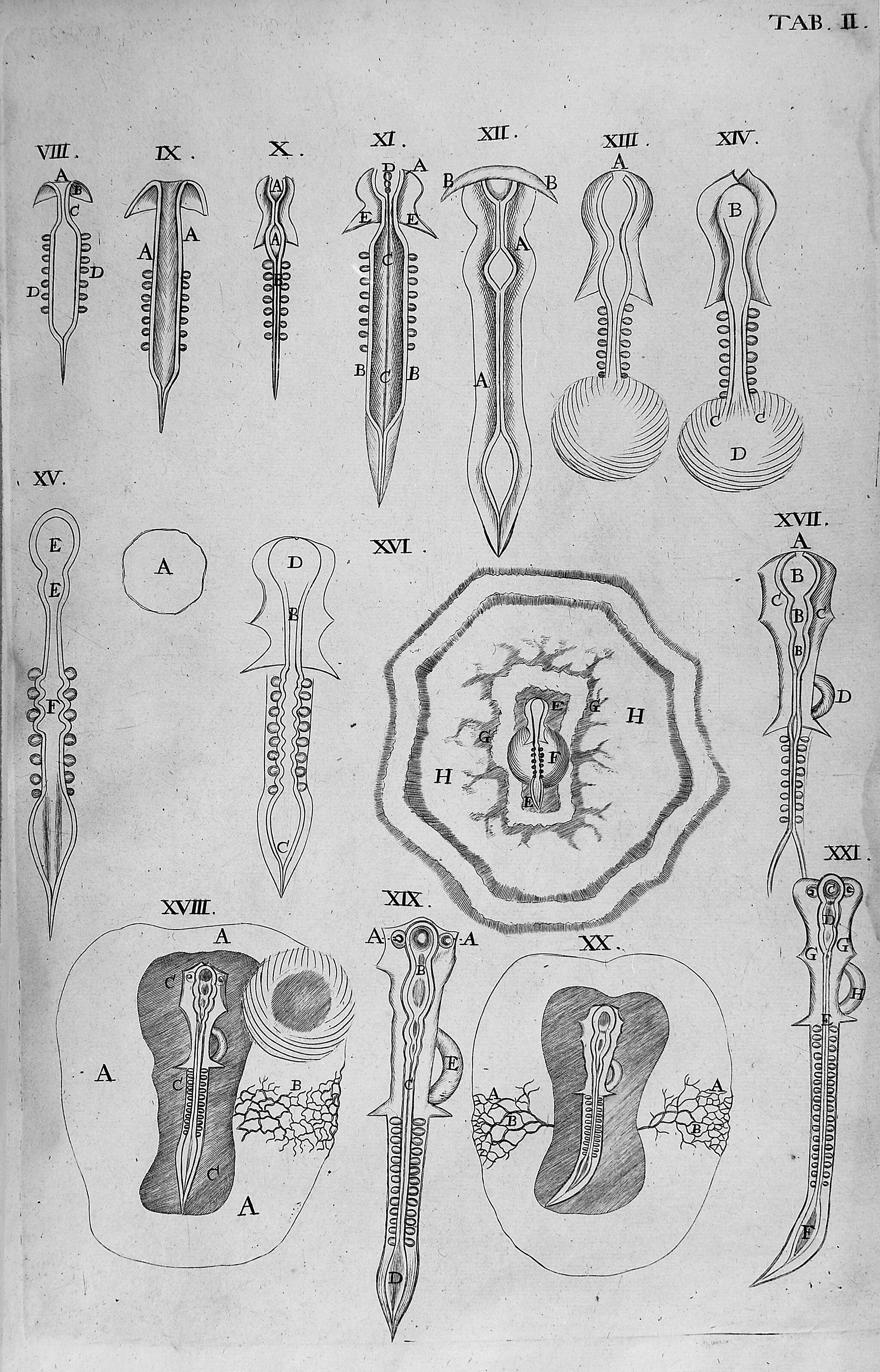}
   \caption[Sketches of Malpighi illustrating the process of somite formation in chick embryos]{Sketches of Malpighi illustrating the process of somite formation in chick embryos }\label{fig:Malpighi}
\enormf



 

 


 \newpage

\subsection*{Conventions and definitions used in tutorial}

In Table \ref{table_convention}, we summarize a couple of notations used throughout this tutorial

\begin{table}[h]
  \begin{center}
    \footnotesize
\begin{tabular}{ll}
\toprule
Symbol & Definition\\
\midrule
$\theta_i(t)$  & phase of a given oscillator at time $t$ and discrete position $i$. \\

$\theta(x,t)$  & phase of a given oscillator at time $t$ and position $x$ (continuous limit of $\theta_i$) \\
$\dot \theta$  & time derivative of variable $\theta$ \\

$\omega_i$   & frequency of an oscillator at position $i$ \\

$\omega(x)$ & continuous limit of $\omega_i$ \\

$\Omega$  & (global) frequency of the segment formation process \\

$T$ &  period of the segment formation process\\

$\phi$ & phase of an oscillator in a moving frame of reference \\

$\psi$ &  relative phase of an oscillator with respect to a reference oscillator (usually $\phi$) \\

$v$ & speed of propagation of the front \\

$S$ & size of somites \\

$L$ & size of the tissue (e.g. presomitic mesoderm) \\

$\tau$ & delay in the differential equations and/or the coupling \\
 
 $\lambda$ &  wave length of the pattern

 \end{tabular} 
   \end{center}
 \caption{Some notations used in this review}
 \label{table_convention}
\end{table}

When discussing biology, we follow a standard convention where specific gene names are italicized (e.g. \textit{Mesp2, Lfng}, names of pathways or gene families are kept in normal font (e.g. Notch pathway).

 For many theoretical works, we represent the spatio-temporal behaviour of the system by so-called  "kymographs", which are pictures showing the spatio-temporal values of a variable as different colour/gray level. We will follow the convention for representing kymographs from \cite{Lauschke2013} and other works: columns of the kymographs correspond to different times (with time increasing from left to right), and lines to different positions in the embryo (with most anterior cells on the top and tail bud cells on the bottom), see Fig. \ref{fig:kymo} below. For models including growth, instead of imposing some moving boundary condition, it is common to typically extend the tail of the embryos as a fictitious extended region in space with a homogeneous pattern of expression, as is represented at the bottom of Fig.\ref{fig:kymo}.

\bnormf
\includegraphics[width=\sizefig]{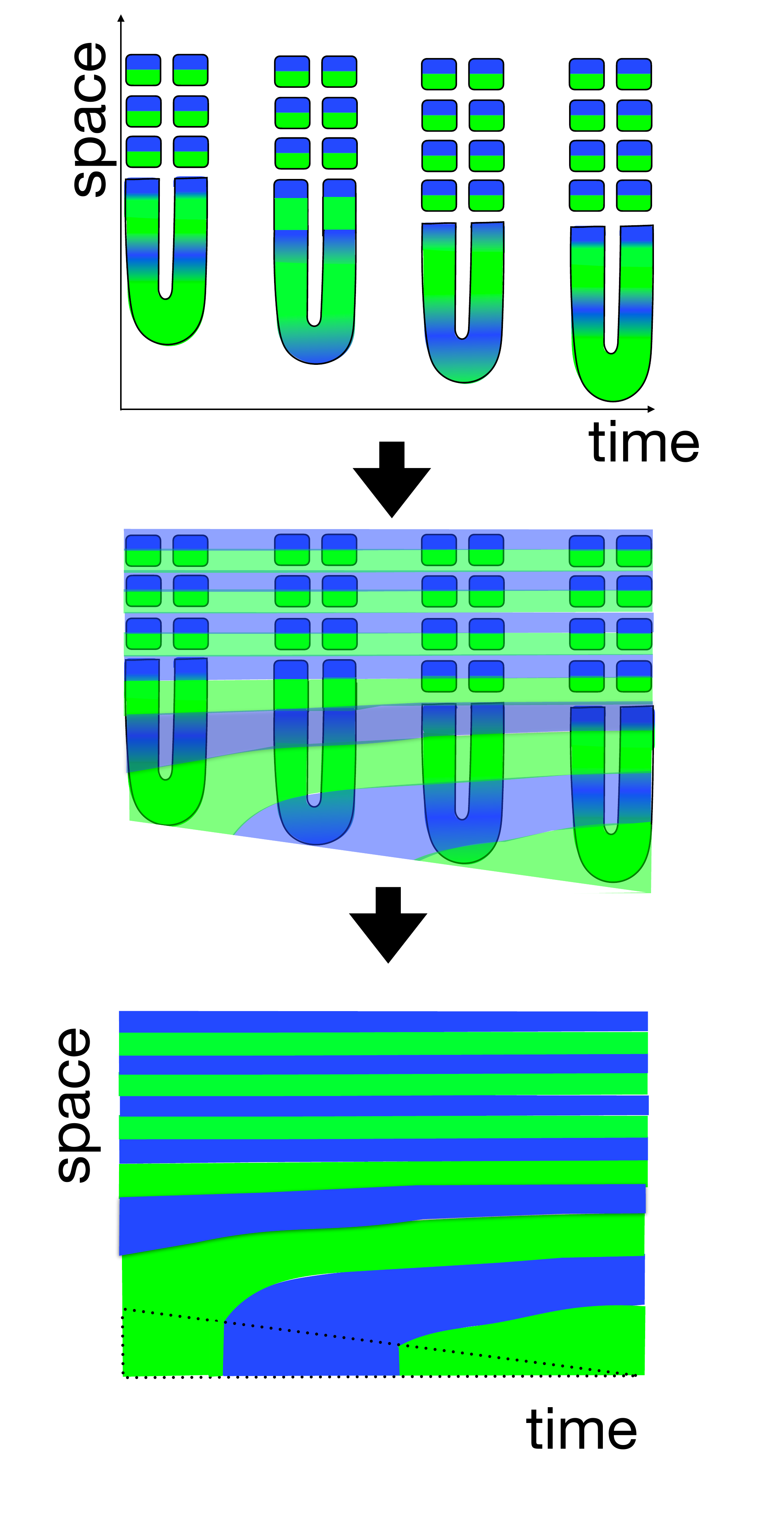}
   \caption[Definition of Kymographs]{Correspondance between the observed spatial-temporal pattern of genetic expression (in an embryo, top) and a theoretical Kymograph (bottom). When representing the behaviour of theoretical models, as a convention, we extend the tail expression pattern to a fictitious region posterior to tail (dotted triangle at the bottom)}\label{fig:kymo}
\enormf
 
 \newpage
\chap{Characterizing vertebrate segmentation : clock, waves, morphogens }

\chapsec{Early concepts}
\label{sec:early_concepts}

\chapsubsec{Vertebrate segmentation}
 One of the first recorded observations of somite formation is due to Marcello Malpighi, a medical doctor who pioneered the use of the microscope for scientific observation. In \textit{Opera Omnia} \cite{Malpighi1687}, published in 1687, Malpighi drew several stages of chick embryonic development,  Fig. \ref{fig:Malpighi} (reproduced from \cite{Malpighisource}). Somites were represented as balls of cells on both sides of the neural tube. For the first time, it was visible from these drawings that somite formation is a dynamic process, where somites sequentially form from anterior to posterior as the embryo is elongating.

 It took a few more centuries to get a more detailed view of embryogenesis and of somite dynamical formation. In 1850, Remak observed that future vertebrae arise from the fusion of the posterior part of a somite with the anterior part of the following one \cite{Remak1850, Pourquie2022}, suggesting that somites are not homogeneous and present functional anteroposterior (another biological term for this being 'rostral-caudal') polarity. Fast forward another century, a more precise description of somitogenesis (the "genesis" of somites) was made possible with progress in the manipulation of chicken and amphibian embryos, and was motivated in parts by theoretical questions. We refer to Pourqui\'e's recent review \cite{Pourquie2022} for a very detailed description and now proceed in describing key theoretical proposals of those pioneering times.


 \chapsubsec{Morphogens}

Turing's seminal work on "The Chemical basis of morphogenesis" \cite{Turing1952} represents a conceptual turning point in theoretical embryology, . Turing introduced several key ideas that have deeply shaped the entire field of developmental biology, up to this day. In particular :
\begin{itemize}
\item Turing suggested that some morphogenetic events find their origin in differences of concentrations of \textit{chemical} substances.  While he explicitly discussed in the introduction the role of mechanics in morphogenesis, he was the first to consider a model where the chemical and mechanical aspects can be separated.
\item Chemical substances driving development are called "morphogens", a term now widely used in biology.  Turing postulated that morphogens interact with each other via \textbf{reaction and diffusion}. This can give rise to patterns (now generally called "Turing patterns") at the origin of biological shapes. 
\end{itemize}

The typical Turing 'interaction network' is made of two morphogens : one 'activator' morphogen, that diffuses slowly (thus with short range activity), self-activating and activating a 'repressor' morphogen, that diffuses rapidly (thus with long range activity). A simulation of 1 D Turing mechanism with (almost) homogeneous initial condition indeed gives rise to a periodic pattern, where islands of the activator morphogens are limited by more broadly expressed repressors. Turing patterns thus  are a natural candidate for the formation of metameric units, similar to the ones observed in vertebrate segmentation. Alternation of stripes in a Turing model could either correspond to a somite/nonsomite pattern or the anterior/posterior parts of somites. 

Diffusion is crucial in the establishment and maintenance of Turing patterns, for instance if a physical barrier is put in place, the long-range repression effect is impinged, and new activating regions can emerge. Another key feature of Turing patterns is their intrinsic length scale, which is a function of the parameters such as diffusion constant. This led to a direct experimental test of a Turing-based model of somitogenesis by Waddington and Deuchar \cite{Waddington1953} (and later Cooke \cite{Cooke1975}), who generated amphibian embryos of different sizes by adding/removing tissue at the gastrula stage. They observed that somite size is scaling accordingly, i.e. bigger embryos have bigger somites in all dimensions. This excludes a process where the length scale is set by a simple reaction-diffusion process. Another difference can be found in the dynamical aspect of the process: as said above, the formation of somites is sequential, from anterior to posterior, while stripes or spots in a Turing system a priori form simultaneously.

 \chapsubsec{Positional information}
Further considerations of the scaling of structures in embryos of different sizes led to many conceptual discussions on how genetically identical cells can take different fates, which are worth mentioning to better understand the current theoretical framework. In 1969, Lewis Wolpert introduced the notion of positional information in development \cite{Wolpert1969}.  Information here should be understood in the colloquial sense: positional information is more akin to a zip code or an address (rather than a physics-inspired definition of information in relation to entropy). Wolpert's underlying idea is that cells have ways to "know" (or to compute) their position within an embryo, and to differentiate accordingly. Then problems such as embryonic scaling boil down to the problem of specification of positional information (which should actively scale with, e.g., cell number).

\bnormf
\includegraphics[width=\textwidth]{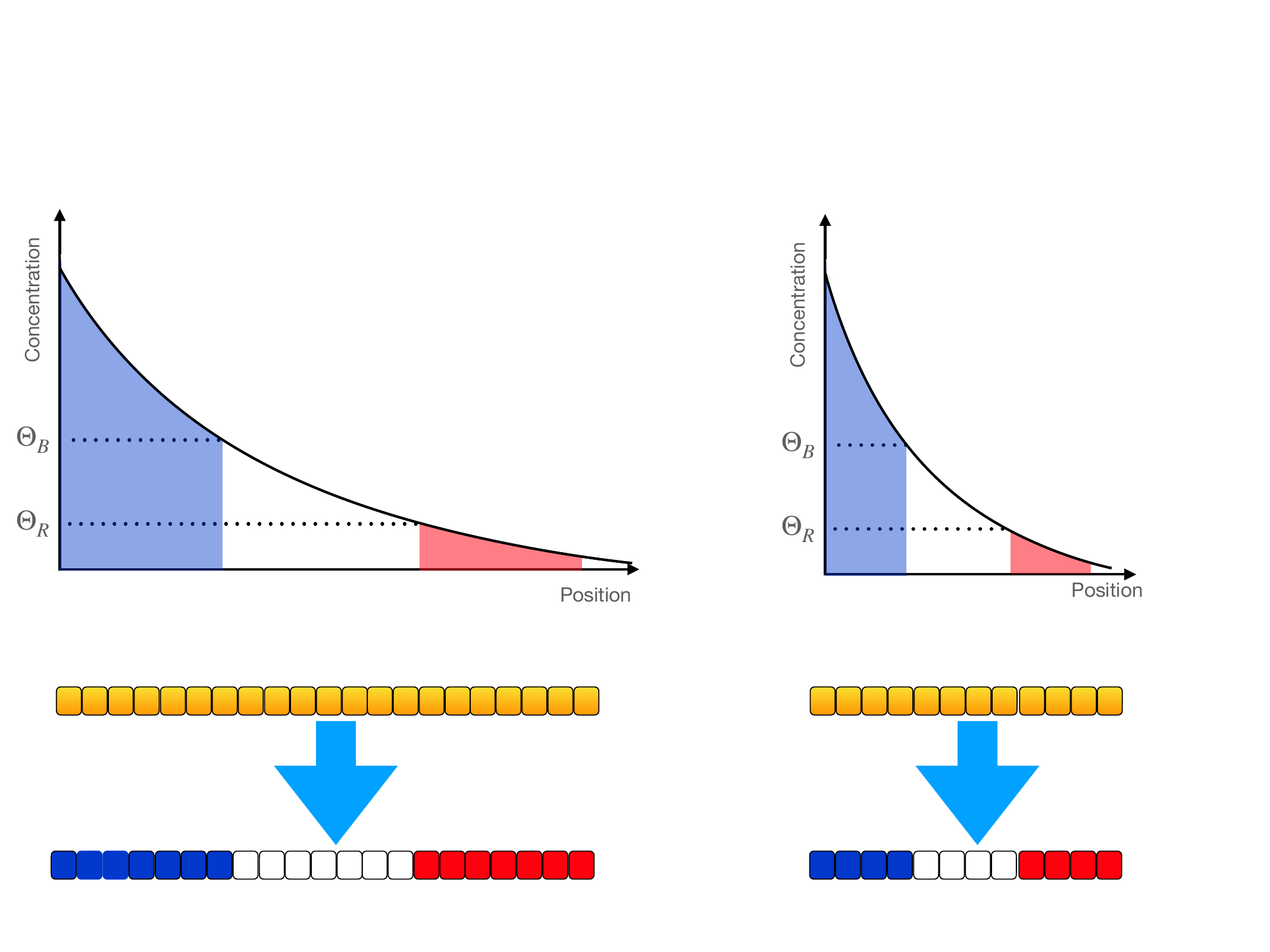}
   \caption[The French Flag Model]{The French Flag Model (Adapted from \cite{Wolpert2006}). A graded morphogen concentration is used as an input for cells to define three domains - via two thresholds of activity $\Theta_{B,R}$. Those three domains are depticted using the three colours of the French Flag.
   (left). If the embryo size is reduced but the morphogen gradient properly scales, this ensures a scaled pattern of cellular fates even if the number of cells itself is not conserved (right)}\label{fig:FFL}
\enormf

The paradigmatic example of positional information in biology is Wolpert's famous French Flag Model \cite{Wolpert2006}, Fig. \ref{fig:FFL}. Imagine an embryo as a line of cells (with the position of a given cell defined by its coordinate $x$), and imagine that there is a graded concentration of a morphogenetic protein (let us assume it is exponential of the form $C(x)=C_0e^{-x/L}$ where $L$ is the size of the tissue to pattern). Then, cells have access to local concentration $C(x)$ and can decide their fate based on this. For instance,  imagine that there are two thresholds respectively at $\Theta_B$ and $\Theta_R$, then cells observing concentration lower than $\Theta_R$ can develop into a "red" fate, cells with observing concentrations between $\Theta_R$ and $\Theta_B$ develop into a "white" fate and cells with concentrations higher than $\Theta_B$ can develop into a "blue" fate, giving rise to a paradigmatic French Flag picture Fig. \ref{fig:FFL}.  The French Flag paradigm provides a parsimonious explanation of embryonic scaling. If the number of cells is changing, one can possibly scale patterning within an embryo by scaling the morphogen gradient itself, which is arguably a much simpler problem to solve (both for biology and for theorists). For instance, Crick proposed in 1970 a "source-sink" model where a gradient of a diffusing protein is maintained at concentration $C_0$ at one extremity of the embryo and at $0$ at the other extremity \cite{Crick1970}. A solution of the 1D diffusion equation with those boundary conditions clearly is a linear, steady-state profile,  which thus naturally scales with the size of the diffusing field. To ensure scaling, one simply needs to impose boundary conditions, which is consistent with the existence of  embryonic regions such as organizers \cite{Wolpert2006}.  Such ideas led to multiple discussions on the theory/conceptual side. For instance, it is not clear if one can separate any informational content from the processing of this information. Some of those early debates are summarized by Cooke \cite{Cooke1981}, who observed that the proportional allocation of cells to different tissues in embryos of vastly different sizes can not be very easily explained with simple morphogen gradients or reaction-diffusion models . He suggested some coupling between protein production rates and the size of tissue might rather play a role as a 'proportion sensor'. It should be mentioned that our understanding of such scaling properties remains incomplete to this date.

Coming back to segmentation, a natural idea within the positional information framework would be to assume that different thresholds of one or several morphogens would define somite locations. The
problem is that many animals (snakes, centipedes) can have many segments (more than 200 vertebrae in snakes). In a French Flag/positional information picture, the potential number of thresholds needed to explain somite formation appears unlikely huge \cite{Cooke1976}. Another issue is that from one animal to the other,  there is some variability in the number of somites even within the same species, which implies a degree of versatility with respect to the overall body plan in the encoding of somite position \cite{Cooke1976}. Other explanations are thus needed both for the process of segmentation itself and the underlying scaling mechanisms.

\chapsec{Statics and dynamics of metazoan segmentation}

\chapsubsec{Establishment of the (fly) segmentation paradigm }
In parallel, starting in the early 1980s, molecular details of developmental processes in general have been established and refined with increasing progress in molecular biology, genetics, and, later on, imaging. The fruit fly (\textit{Drosophila}) model organism is the first organism for which the key principles of segmentation and associated genes  have been identified, starting in 1981 with a groundbreaking series of papers by Christiane N\"usslein-Volhard and Eric Wieschaus \cite{Nusslein-Volhard1980} (who were awarded the Medecine Nobel Prize for this work in 1995.)

In a nutshell, fly segmentation appears, maybe surprisingly, largely consistent with the  "French Flag model" view, \cite{Wolpert2006}, Fig. \ref{fig:phenomenology_fly}). Multiple morphogenetic gradients were discovered over the years: the \textit{bicoid} gradient defines identities in the anterior part of the embryo, while posterior gradients such as \textit{nanos} and \textit{caudal} define identities in the posterior part of the embryo \cite{Rivera1996, Lawrence92}. Those gradients are generally called ``maternal",  since they are initially defined by localization of RNA molecules in the egg by the mother (and subsequent cross-regulation, e.g. \textit{caudal} translation is repressed by \textit{bcd} ).

 In their original papers, Wieschaus and  N\"usslein-Volhard identify so-called "gap-like" phenotypes, in which mutants have parts of their body missing. Those gap phenotypes are due to the mutation of so-called gap genes, themselves normally expressed in the part of the body missing in the mutants. Gap genes's expressions are positioned and controlled by the maternal gradients, and consistent with this, cellular identities can be shifted anteriorly or posteriorly by changing the levels of the maternal gradients \cite{Jaeger2011}. 
 
 Downstream the gap genes, we find pair-rule genes, then segmentation genes \cite{Carroll1989, Lawrence92}, Fig.  \ref{fig:phenomenology_fly}. The pair-rule genes correspond to periodic structure every 2 segments, while segmentation genes are expressed in all segments. Those genes  are expressed in periodic  stripes corresponding to future segments and their sub-compartments. Such striped patterns naturally evoke reaction-diffusion mechanisms to physicists, but quite astonishingly, it turns out that those different stripes are encoded in the genetic sequence and regulated more or less independently from one another. As an example, an Eve2 stripe genetic module can be identified on the fly DNA, regulated by a subset of gap genes independently from all other stripe modules \cite{Stanojevic1991}, Fig.  \ref{fig:phenomenology_fly} B. Those discoveries thus suggested a very local and feedforward view of development and positional information, where concentrations of morphogens dictate local fates all the way to segmentation genes. Remarkably, it has been shown since then that the \textit{bicoid} gradients and the gap genes downstream of it contain exactly the right amount of information (in the physics sense) to encode identity with a single cell resolutions along the entire fly embryo \cite{Gregor2007, Dubuis2013,Petkova2019, Seyboldt2022}. 
 
 Those discoveries considerably shaped the subsequent discussions on segmentation in vertebrates as well. First, they firmly established the morphogen gradient paradigm, where different levels define different identities or properties. Second, they argue against models where reaction-diffusion processes are crucial for robust patterning. The view coming from fly segmentation is more local and modular: the definition of cellular fates is done through a given gap gene combination \cite{Petkova2019,Seyboldt2022}, which is specific to the cell location, independently from all other locations within the embryo. Consistent with this view, there is some variability in the pattern of gap genes' expression (and likely regulation) from one species to the other in "long germ band" insects (forming their segments like flies) \cite{Goltsev2004,Wotton2015}, see also \cite{Rothschild2016} for simulations of underlying network evolution.  
 

 That said, it rapidly turned out that flies are to some extent evolutionary exceptions. The almost paradigmatic morphogen,  \textit{bicoid}, does not exist outside of \textit{Drosophila}.  Long germ segmentation further appears highly derived evolutionary: it occurs in an egg of approximately fixed size, with segmentation genes expressed more or less simultaneously, while in most other metazoans, segmentation is sequentially coupled to embryonic growth  \cite{Peel2003} \footnote{it should be pointed out though that long germ segmentation still evolved many times independently, suggesting deep evolutionary forces are at stake to move towards such mode of segmentation } .  Gap phenotypes are also not observed in vertebrates, suggesting that segmentation is a more global, integrated process in opposition to a more local process where identities are defined by local morphogen concentrations. Finally, as said above, flies have a relatively small number of segments compared to some vertebrates such as snakes.

\bnormf
\includegraphics[width=\textwidth]{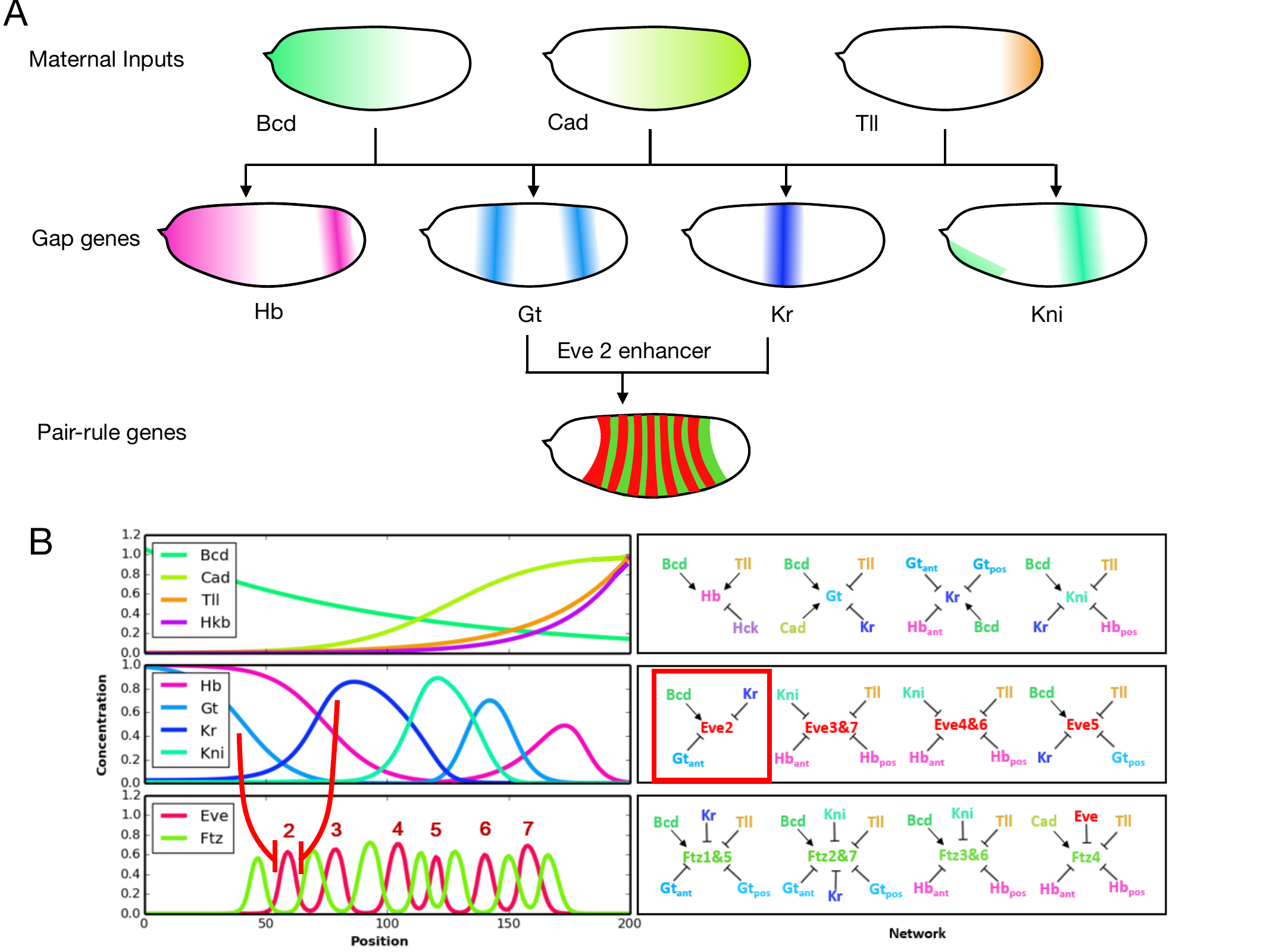}
  \caption[Fly segmentation]{ Summary of Fly segmentation (A) Schematic of the expression pattern of some of the main genes regulating segmentation in Drosophila. Maternal genes, control gap genes and gap genes in turn control the expression of pair-rule genes.  (B) A simplified, hierarchical feed-forward model for fly segmentation, reproduced from \cite{Rothschild2016}. The embryo is simulated as a one-dimensional field. The left panel shows the behaviour of the model, the maternal profiles are imposed and the network "computes" the concentration of downstream genes from top to bottom. The right panel shows the topology of the corresponding gene regulation network. Genes interaction are symbolized by arrows, regular arrows correspond to activation, T-shaped arrows to repression. For instance, one can see how individual Eve stripes are regulated differently. We highlight Eve2, which is  activated by Bcd and repressed by both Anterior Gt and Kr in this model, and as a consequence appears at the interface between those two genes. }\label{fig:phenomenology_fly}
\enormf



\chapsubsec{Discovery and phenomenology of the segmentation clock}

All animals are evolutionary related and, as a spectacular consequence, many of the lower level controls of the animal physiology are similar even in very different-looking animals  \cite{Couso2009}. This is especially striking for molecular controls of embryonic development : many developmental genes are highly conserved, and play the exact same role in many animals. A spectacular example are Hox genes, which prescribe anterior-posterior identities of cells in similar ways in all animals, to the point that Hox genes were proposed as a 'defining character of the kingdom Animalia' \cite{Duboule2022,slack1993}. Coming back to segmentation, given the crucial role of pair-rule genes in fly, several groups then proceeded to identify and study their homologs in vertebrates.

It quickly appeared that vertebrate proteins closely related to the fly \textit{hairy} genes presented patterns in developing vertebrate embryos somehow reminiscent of what happens in the fly. For instance,  \textit{her1} in zebrafish \footnote{\textit{her} stands for 'hairy-E(spl) related' which is the name of the broader family of these proteins} was first described to present patterns, with broad stripes in the presomitic mesoderm (PSM) and narrower stripes in somite primordia \cite{Muller1996}. In 1997, Palmeirim \etal  identified a homologous of \textit{hairy} in chick (called \textit{ c-hairy}), and carefully studied its behavior in a seminal work \cite{Palmeirim1997} redefining the entire field. 

Palmeirim \etal proceeded to study the pattern of genetic expression of \textit{c-hairy}.  Comparing embryos to embryos, they confirmed that \textit{c-hairy} presents two distinct patterns of expression. In the anterior part of the embryos, \textit{c-hairy is} expressed in the posterior half of formed somites. But the pattern of gene expression in the non-segmented pre-somitic mesoderm (i.e. posterior to formed somites) appears much more complex. Depending on the embryo, \textit{c-hairy} is expressed broadly in the posterior, or into increasingly narrower and more anterior stripes of genetic expression, not unlike what happens in zebrafish for  \textit{her1}  \cite{Muller1996}.

The "Eureka" aspect of this work was to realize that this pattern of gene expression in the posterior actually corresponds to snapshots of the dynamics of a propagating (and narrowing) wave of \textit{c-hairy} expression from posterior to anterior, which appears clearly when embryos are reordered as a function of a pseudo-time (see schematic in Fig. \ref{fig:phenomenology_somites} A,  \textit{c-hairy}  would correspond to the green colour). To unambiguously show that such a wave originates from a posterior oscillator, Palmeirim \etal used an ingenious trick of chick embryology. They cut the embryo into two pieces, fixed one side of the embryo, then waited before fixing the other side. Assuming the dynamics on either side of the embryo are independent of what happens on the other side, this allows the capture of two time-points of the same dynamical process (essentially a two-point kymograph), and from there to reconstruct the entire process using multiple embryos. This technique indeed shows that the variability of the \textit{c-hairy} pattern comes from a dynamical gene expression, since in the very same embryo one effectively sees a stripe of c-hairy gene expression move towards the anterior, similar to what is observed for the gene expressions in Fig. \ref{fig:phenomenology_somites} A). Furthermore, fixing the two halves of embryos with a time difference of 90 mins, one sees the same pattern of gene expression of \textit{c-hairy}, but with one extra somite on the right side vs the left (compare first and last time in Fig. \ref{fig:phenomenology_somites} A). This indicates that a periodic mechanism drives the waves of genetic expression and is indeed correlated to somite formation, as expected from the oscillatory models proposed previously (see Section \myref{sec:pioneer}). The very same technique of fixing one half of the embryo while keeping the other alive was later used to show the existence of a segmentation oscillator in \textit{Tribolium} \cite{Sarrazin2012}.

\bnormf
\includegraphics[width=\sizefig]{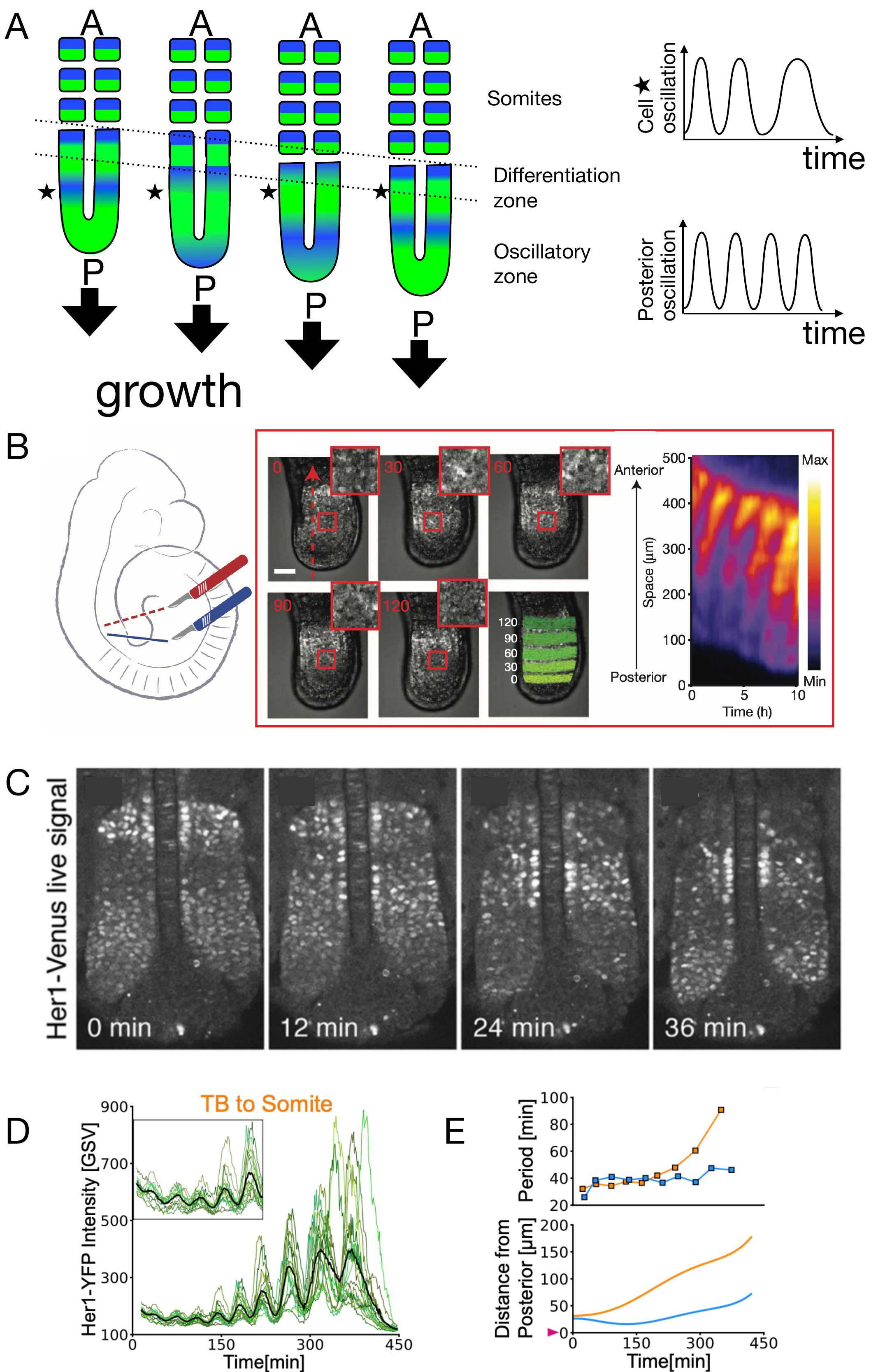}
  \caption[Vertebrate segmentation]{ Segmentation in vertebrate. (A) Phenomenology of segmentation, coupled to growth. Oscillatory gene expressions in the posterior give rise to kinematic waves in the embryo propagating from posterior to anterior. The wave refines and stabilizes in the anterior to structure future somites into anterior and posterior compartments. There is a differentiation zone corresponding to the region where cellular oscillations stop and physical boundaries form. On the right, we show schematically the oscillation phases for a cell staying in the posterior (bottom) and for a cell (star) reaching the differentiation zone at the end of the time window depicted. The oscillation is slowing down as cells get more anterior, giving rise to a period gradient within the oscillatory zone. (B) Experimental visualization of the segmentation clock in a mouse embryo, adapted from \cite{Lauschke2013}. The Middle panel shows snapshots of a movie with a dynamical \textit{ Lfng} reporter (see Section \myref{sec:molfor}), with recapitulation of the location of the last "stripe" for each wave. The right panel shows a corresponding Kymograph . (C)  Experimental visualization of the segmentation clock in zebrafish embryos, with a single cell resolution, adapted from \cite{Delaune2012}. A live reporter for  \textit{ Her1} is used. (D) \textit{Her-1} oscillation for single cells in a zebrafish embryo, reproduced from \cite{Rohde2021}. The cell is moving (relatively) from tail bud until a somite where it is integrated and where the oscillation stops. (E) Inferred period (top) for two single-cell oscillators in a zebrafish embryo as a function of time, reproduced  from \cite{Rohde2021}. The distance measured from the posterior (bottom)  allows to identify the relative positions of the cells within the presomitic mesoderm (PSM), and to correlate it to their respective period. The cell staying close to the tail bud oscillates with a constant period, while the period is increasing as a function of time for cell moving towards the anterior, indicative of the existence of a period gradient within the PSM.   }\label{fig:phenomenology_somites}
\enormf

 \chapsec{The segmentation clock paradigm}

\chapsubsec{Phenomenology of the segmentation clock}
It is now generally acknowledged that the work of  Palmeirim \etal showed the existence of what is now called the "segmentation clock".  In this review, by "segmentation clock", we mean the ensemble of periodic gene expressions, at the embryo level, which controls the periodic formation of somites.  
Before we focus on molecular details in the next section, we wish to point out four high-level components and properties underlying the segmentation clock, which will be central to the discussion in this review. The segmentation clock paradigm is summarized in Fig. \ref{fig:phenomenology_somites}  A, with experimental illustrations in subsequent panels (B-E).

Firstly, the segmentation clock emerges through cellular oscillators, clearly visible in Fig. \ref{fig:phenomenology_somites}  B-C. Cells in the presomitic mesoderm PSM display coordinated oscillations of multiple genes, thus defining a global oscillator at the PSM level.
Importantly, cellular oscillators are synchronized but not in phase:  waves of oscillations sweep the embryo from posterior to anterior, as first evidenced in the work of  Parlmeirim \textit{et al}, and can now be seen using real-time reporters Fig. \ref{fig:phenomenology_somites}  B-C. Those waves are related to the fact that, as cells get more anterior, the period of their internal oscillator is increasing (see e.g. the oscillation in the starred cell compared to posterior oscillation in the schematic in Fig. \ref{fig:phenomenology_somites} A, and see experimental measurements of the period in single cells in   Fig. \ref{fig:phenomenology_somites} D-E). There are thus parallel anterior-posterior period and phase gradients in the PSM. One of the key theoretical questions is to figure out how those gradients are related: do cells modify their intrinsic period (slow down) so that a phase gradient builds up, or is there a phase gradient building up (e.g. via cell-to-cell interactions) leading up to an apparent period slowing down?

As the waves of genetic expression move towards the anterior, and as the local period of the oscillators increases, the wavelength decreases, before stabilizing into a fixed pattern. Some genes, like \textit{c-hairy} discussed in the previous section, then form a stripe pattern of genetic expressions, localized in half a somite. The formation of those stripes appears to be tightly coupled to the formation of a somite boundary, Fig. \ref{fig:phenomenology_somites} B, middle panel. Somites eventually display an anterior-posterior  (or rostral-caudal) pattern of genetic expression, with some specific genes expressed in the anterior half of the somite, and some others in the posterior half of the somite, see Fig. \ref{fig:phenomenology_somites} A  --within the same somite, blue gene is rostral, and green gene is caudal. Notice that this pattern is to some extent reminiscent of pair-rule patterning in flies, compare Fig. \ref{fig:phenomenology_somites} A with Eve and Ftz in Fig. \ref{fig:phenomenology_fly} .
The region where cellular oscillations stop and where, subsequently, boundaries form between future somites, is labeled as 'differentiation zone' in Fig. \ref{fig:phenomenology_somites}, and is the second important component of the segmentation process. Specific genes are expressed in this region. Very often in the literature, this region is designated not as a zone, but phenomenologically reduced to a single front, often called 'wavefront', largely because of the initial Clock and Wavefront model that we describe in section \myref{sec:pioneer}. Notice that the slowing down of the cellular oscillations is tied to stable patterns in somites, following posterior to anterior waves of genes such \textit{c-hairy}. So clock and differentiation front might not be considered as independent processes. They seem at the very least coordinated, which raises the fundamental question of the nature of the front and its spatial extension,  a central question discussed in this review (see Fig. \ref{fig:WaveScenario} for a synthesis).

Thirdly, segmentation is tied to embryonic growth. Schematically, as the tail is growing,  cells move anteriorly relative to the growth zone (Fig. \ref{fig:phenomenology_somites} E bottom), so that, as said above, a phase gradient is accumulating and their period appears to increase (Fig. \ref{fig:phenomenology_somites} E top). They eventually differentiate and integrate into somites. It is well established that embryonic growth is connected to anterior-posterior gradients of various morphogens, and thus it is natural to think that those gradients likely regulate somite formation in some way, especially in line with the French Flag and the Fly paradigms where anterior to posterior gradients largely control segment position. Since somitogenesis is a much more dynamical process, there are two additional questions: how do gradients control cellular oscillators 
 themselves (e.g. their period and amplitude ?), and how do they control the location of the differentiation zone? Again those questions are not independent and we will comment on them in this review.

Fourthly, vertebrate segmentation is a tissue autonomous process: interruption of continuity of the presomitic mesoderm (PSM) - the undifferentiated tissue from which somites derive - does not impinge somite formation. Furthermore, local inversion of fragments within the PSM  leads to an "inversion" of the progression of somite formation. This suggests that once cells exit the tail bud, they are largely preprogrammed to oscillate and eventually differentiate in a precise way, and as we will see below it seems that indeed dissociated cells behave very similarly to cells within the embryo, suggesting that many processes are largely cell autonomous. From the theoretical standpoint, it is not clear how this large degreee of cell autonomy eventually gives rise to weill proportioned, multi-cellular somites.



 


To finish this section, it is important to point out that the existence of an oscillator (or clock) driving the formation of somites was first predicted and studied by Cooke/Zeeman and Meinhardt in two pioneering models, that we describe in details in section \myref{sec:pioneer}. This is a nice example in biology where theory was far ahead of experimental biology and inspired it.

\chapsubsec{The molecular forest }
\label{sec:molfor}

The phenomenology of the segmentation waves first described in  \cite{Palmeirim1997} and summarized in the previous section has been confirmed and generalized to other model organisms. Furthermore, it has been established in subsequent works that not only the phenomenon of oscillations and waves is broadly observed, but also that a plethora of genes is oscillating, forming multiple parallel waves of gene expression during vertebrate segmentation  \cite{Dequeant2006, Dequeant2008}. Listing here all phenotypes and interactions discovered would be both impossible and potentially confusing, but to understand the principles underlying current modeling, it is important to summarize some of the biological players, as well as some crucial biological mechanisms they have been suggested to regulate.  It should be pointed out that a major difficulty is that interactions are not conserved between different species \cite{Krol2011}, e.g. a gene oscillating in one species might not oscillate in another one. This renders the study of molecular segmentation clock very difficult, and to this date, no clear conserved molecular mechanism controlling the segmentation oscillator has been established, and in fact, segmentation waves likely work in slightly different ways in different organisms (see  section \myref{Species}). We summarize some important results in this section, with a special focus on mouse somitogenesis, but will also comment on some results on other animals.

\smallskip
\textit{Segmentation waves : 3 pathways}

Three major signaling pathways have been implicated in the segmentation waves: Notch, Wnt, and FGF  \cite{Dequeant2008}. The current consensus is that the core oscillator is related to the Notch signaling pathway, implicated in cellular communication \cite{Venzin2020}. Notch ligands (called \textit{delta}s) are produced and membrane-bound at the surface of cells, and interact with Notch receptors at the surface of neighboring cells, driving transcriptional response. Lunatic Fringe (\textit{Lfng}), a glycotransferase modifying  Notch activity, is at the heart of the chick segmentation clock \cite{Dale2003}. Misexpression of \textit{Lfng} disrupts somite formation and anteroposterior compartmentalization in chick \cite{Dale2003}, and similar phenotypes are observed in mouse \cite{Zhang1998, Evrard1998}.
\textit{Lfng} does not oscillate in zebrafish though, and studies in this organism have rather focused on other components of  Notch signaling pathway. Notch ligands (\textit{delta})  are implicated in many segmentation phenotypes. Perturbation of Notch signaling results in clear somite formation defects \cite{Riedel-Kruse2007}. Mutations of \textit{delta} ligands do not prevent segmentation but impact the coherence of segmentation waves, prompting the suggestion that the main role of Notch signaling is to synchronize cellular oscillators \cite{Jiang2000, Ozbudak2008}. Indeed, real-time monitoring  has since then confirmed that in delta mutants, individual cells oscillate but are desynchronized \cite{Delaune2012}.   \textit{Lfng} has  actually been shown to play a role in this synchronization as well in mouse by modulating \textit{delta} ligand activity and thus Notch signaling in neighboring cells \cite{Okubo2012}.
 The Hes/Her transcription factors, phylogenetically related to the fly \textit{hairy} gene mentioned above, appear to play a major role in the core part of the oscillator\cite{Oates2002, Holley2002, Giudicelli2007}. Interestingly, serum-induced oscillations of \textit{Hes1}  (a Notch effector) are observed in multiple types of cultured cells (myoblasts, fibroblasts, neuroblastoma) with a 2-hour period consistent with somitogenesis period in several organisms \cite{Hirata2002}, suggesting that it could be part of a more general core oscillator based on a negative feedback loop \cite{Lewis2003}.  \textit{Hes5} oscillations have also been implicated in neurogenesis \cite{biga2021dynamic}

Another major oscillating pathway is Wnt.  \textit{Axin2}, a negative regulator of the Wnt pathway oscillates in mouse, even when Notch signaling is impaired \cite{Aulehla2003}. Perturbation of Wnt signaling pathway results in segmentation phenotypes, e.g. \textit{Wnt3a} is required for oscillating Notch signaling activity. Importantly, a posterior to anterior gradient of $\beta$\textit{-catenin} (a key intracellular mediator of Wnt transcription) is also observed \cite{Aulehla2008}, and crucially, mutants with constitutive (i.e. highly expressed)  $\beta$\textit{-catenin} display non-stopping traveling waves of gene expression within the PSM, suggesting that Wnt plays a crucial role in the stopping of the segmentation waves. However, Wnt does not oscillate in zebrafish


The last major player is FGF. Many genes related to the FGF pathway oscillate  \cite{Dequeant2006}, but the major feature of FGF is that it appears to control the location and the size of somite. \textit{FGF8} presents a graded expression, from posterior to anterior \cite{Dubrulle2001, Dubrulle2004}. FGF8 overexpression disrupts segmentation by maintaining cells in a posterior-like state (characterized by the expression of many characteristic markers and associated posterior morphology). Dubrulle \etal used beads soaked with \textit{FGF8} to show that local overexpression of FGF leads to strong segmentation phenotype in chick (monitored by looking at the expressions of the Notch ligand \textit{c-delta}) \cite{Dubrulle2001, Dubrulle2004}. If the bead is initially placed in a posterior region, as elongation proceeds and the bead gets more anterior, major changes are observed, with several small somites anterior to the bead and one big somite posterior to the bead. If the bead is placed midway in the PSM, a similar phenotype is observed but only around the bead, up to a well-defined anterior boundary, 4 somites posterior to the first somite boundary. Grafts of FGF beads in this region yield no phenotype.

\smallskip
\textit{Anterior PSM : Stabilization and pattern formation}

Some genes are also (in)activated following an apparent front moving from anterior to posterior, likely controlling somite formation. For instance, in mouse, \textit{Tbx6}  is expressed only in the oscillating PSM region\cite{Saga2012}. Furthermore, in the most anterior section of the presomitic mesoderm, segmentation oscillators slow down, and genetic waves of expression either stabilize or simply disappear. In the region where the system leaves the oscillatory regimes, new genes are expressed, such as \textit{Mesp2}. \textit{Mesp2} is first expressed in a few broad stripes, possibly slightly bigger than a somite size, before restricting itself to the anterior part of the somite \cite{Saga2001, Saga2012}. \textit{Mesp2} activates \textit{Ripply2}, which then turns off  \textit{Tbx6}.

 Somites present Anterior-Posterior (or rostrocaudal) polarity. As said above, within a somite, \textit{Mesp2}  is eventually becoming anterior (A) within a somite. Other Notch  signaling pathway genes get stably expressed in the posterior part (P) of somites, such as \textit{Dll1} or \textit{Uncx4.1}. \cite{Saga2012}. Interestingly, the boundary formation between somites is clearly correlated to the Posterior-Anterior boundary between \textit{Notch} signaling in the posterior part of a future somite and \textit{Mesp2}  in the anterior part of the next one \cite{Oginuma2008, Oginuma2010}. 

One issue, first discussed by Meinhardt \cite{Meinhardt1982}  is the problem of the symmetry of AP vs PA boundary to define the somite boundary. This is visible on kymographs such as the one in Fig. \ref{fig:kymo} focusing only on the expression of oscillator genes :  the boundaries at steady state between the green and the blue region do not distinguish between internal or external somite boundaries. Meinhardt suggested that there might be a third state (X) to define the such boundary. Experiments in zebrafish possibly falsify the existence of such intermediate state: mutants for convergence extension \footnote{a process of cellular convergence towards an axis, so that, because of volume conservation, tissue is thinning perpendicular to the axis and extending in the direction of the axis } give rise to broad, large somites with well-defined boundaries, but only two-cell wide in the anteroposterior direction \cite{Henry2000}. So in such somites,  there can not be any cell corresponding to a hypothetical X state [A possible caveat is that those cells are polarized so that there could be \textit{subcellular} divisions allowing for the existence of the X state]. Coming back to mouse, in \cite{Oginuma2010}, a solution is suggested where the clock would in fact impose a rostrocaudal gradient of \textit{Mesp2}  inside the somite, imposing a natural polarity, where the PA border between somites is "sharper" than the AP border within somites, leading to a local "sawtooth" pattern. This exactly fits the pattern of downstream genes implicated in cellular adhesion  \cite{McMillen2016}.

\bnormf
\includegraphics[width=\textwidth]{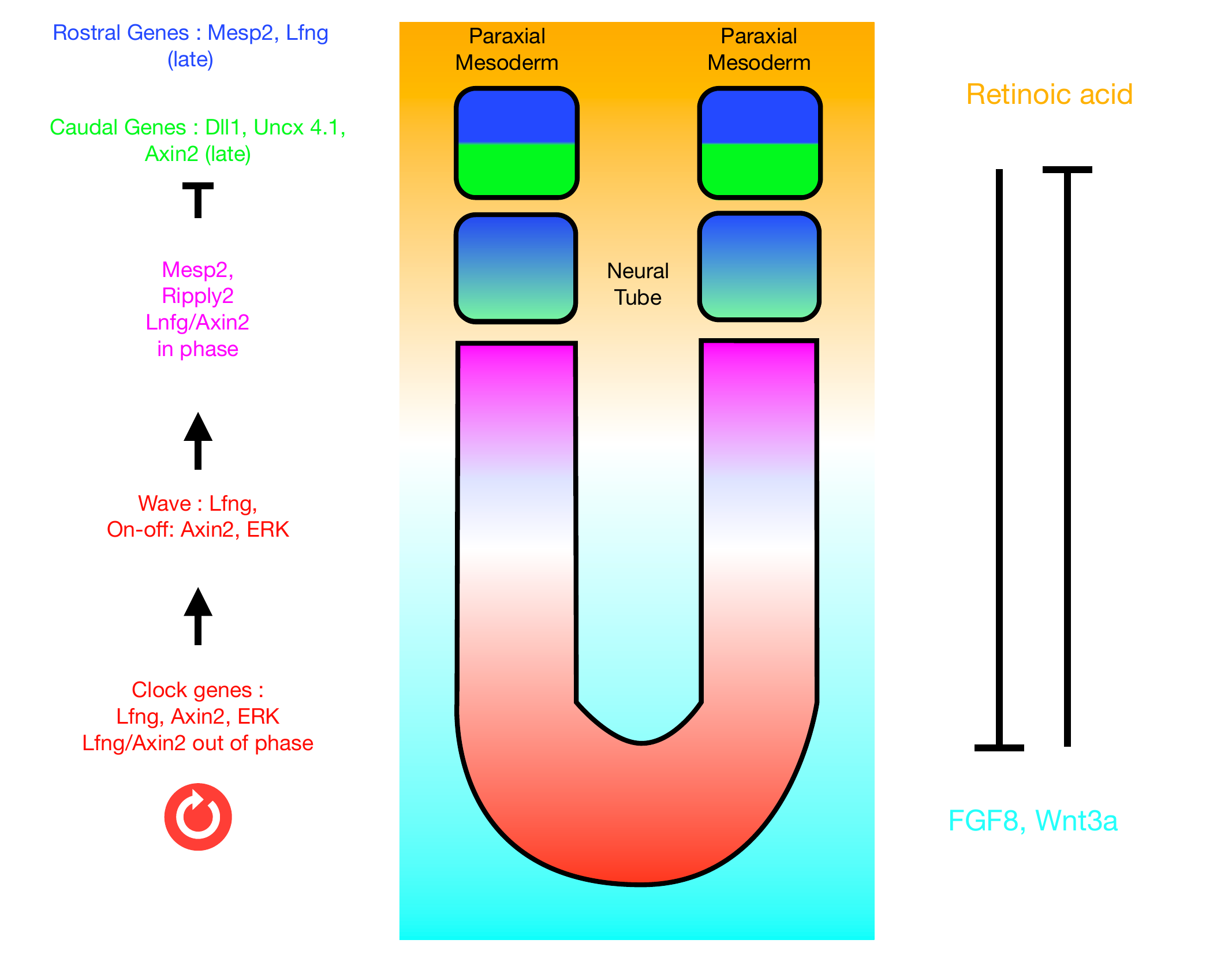}
   \caption[Sketches of Molecular Players in Somitogenesis]{Schematic of some key molecular players in somitogenesis, using mouse genes as examples. Anterior is on the top, posterior at the bottom. In mouse, there is only one wave (i.e. roughly a $2\pi$ phase shift) of genetic expression within the presomitic mesoderm. }\label{fig:MolPlay}
\enormf

It is worth mentioning at this stage a few other higher-order molecular controls modulating somitogenesis formation. Retinoic Acid (RA) is well-known to form an anteroposterior gradient opposite to FGF in metazoan embryos.  RA mutants display smaller somites \cite{DiezdelCorral2004}. So a natural question is the impact of RA mutation on FGF gradient and the segmentation clock,  \cite{Vermot2005,Vilhais-Neto2010}. Surprisingly, embryos deprived of retinoic acid form asymmetrical left-right somites. The associated phenotype is highly dynamic: for the first 7 somites, \textit{Lfng} and \textit{Hes7} waves are symmetrical, but afterward somites on the right side of the embryo form later than on the left side, with one to three cycle delay. The wave pattern is asymmetrical, and \textit{Mesp2} is more anterior on the right side. This somite asymmetry is a consequence of the general left-right, \textit{Nodal} induced asymmetry (driving in particular internal organs asymmetry) \cite{Tanaka2005, Vermot2005}, so that RA appears in fact to act as a buffer of this already present asymmetry.

There are also many interesting modulations on the formation of the somite boundaries. For instance, it is possible to induce separation between the rostral and caudal parts of a somite by modifying cadherin and \textit{cad11}  \cite{Horikawa1999}, thus reavealling a length scale half of somite size in mouse. Conversely, in zebrafish, disruption of \textit{her7} creates somites with alternating weak and strong boundaries, suggesting the system can also generate an intrinsic length scale twice the somite size \cite{Henry2002}.



\chapsubsec{Visualizing oscillations in embryos}

Recent years have seen the development of multiple fluorescent reporters, allowing for the real-time observations of some of the clock components. In mouse, the current toolbox includes reporters for Notch signaling pathway, such as a destabilized luciferase reporter for \textit{Hes1} \cite{Masamizu2006}, destabilized Venus reported for \textit{Lfng} (LuVeLu) \cite{Aulehla2008}. An \textit{Axin2} reporter associated with the Wnt signalling pathway is also available \cite{Sonnen2018} as well as \textit{Mesp2}  and FGF \textit{Erk} reporters \cite{Niwa2011}. In zebrafish, reporters for the Notch signaling pathway are available as well, mostly based on \textit{Her1} fluorescent fusion proteins, and a single cell resolution to visualize oscillations has been achieved \cite{Delaune2012, Soroldoni14, Shih2015, Webb2016}. It should be pointed out that it is not necessarily easy to combine reporters to visualize multiple components of the system in real-time, one reason being that some of them are based on similar fluorescent proteins and would not be easily distinguishable in the same cells \cite{Sonnen2018}.

Oscillations of Notch signaling pathway in single cells present a characteristic profile, where both the average and the amplitude of the oscillations increase as cells mature towards the anterior PSM. In zebrafish, the last peak-to-peak time difference is approximately twice the period in the tailbud  \cite{Shih2015}, consistent with the strong slowing down first inferred from in situs \cite{Giudicelli2007}. Waves of oscillations move from posterior to anterior to the very anterior PSM, so that the most anterior cells within a somite are the last ones to stop oscillating (as measured by the timing of the last peak of oscillation \cite{Shih2015}). This contrasts with the idea of a differentiation front moving continuously from anterior to posterior: there, within a future presumptive somite, anterior cells are expected to differentiate (and stop their clock) before posterior cells. Such a mechanism creates an asymmetry in the wavefront, with a $\pi$ phase shift within a future presumptive somite, giving a "sawtooth" pattern within the presumptive somite. This could define anterior and posterior somite compartments \cite{Shih2015}, and relate to the previous observation that the system can generate a length scale twice the normal somite-size \cite{Henry2002}.

It is also possible to monitor mitotic cells in embryos, providing a natural perturbation of the segmentation oscillator. Mitosis delays oscillation in cells, but divided cells eventually resynchronize with their neighbors after roughly one cycle \cite{Delaune2012}. Interestingly, sibling cells  are statistically more synchronized with one another than with their neighbors, which shows that single-cell oscillations are rather robust and only modulated by interactions. Lastly, there is a clear interaction between the cell cycle and the segmentation oscillator, since mitosis happens preferentially at a well-defined phase, when Notch activity is the lowest (which possibly provides a natural mechanism for noise robustness in presence of equal partitioning of proteins) \cite{Delaune2012}. In Notch pathway mutants, single cells still oscillate, but in a desynchronized way and with a longer period. The amplitude of Notch oscillations in mutants appears bigger than in WT, with possibly a modest increase towards the anterior, but there is no obvious increase in period length in those mutants.

\chapsubsec{Biomechanical aspects}

 When treated with \textit{Noggin} (an inhibitor of another signaling pathway called BMP), non-somite mesoderm spontaneously segregates into somite-like structures \cite{Dias2014}. Those have sizes similar to normal somites, and when grafted instead of normal somites, express normal somite markers. Contrary to normal somites, they form almost simultaneously without the need for a clock, and are not linearly organized but rather look like "a bunch of grapes". Importantly, they do not have well-defined rostrocaudal identities: rather, cells within those somite-like structures display patchy expressions of rostral and caudal markers. This suggests that normal anteroposterior patterning within somites might in fact be one of the main outputs of the clock \cite{ClaudioDStern2015}.

The biomechanical program responsible for somite segregation can thus be triggered independently of the segmentation clock. This suggests that there is a level of biomechanical self-organization in the system, with associated length scales, which raises the question of the multiple scaling effects at play and of downstream self-organization within a given somite \cite{Piatkowska2022}. Consistent with this, it has been recently shown in normal somitogenesis that tension forces allow for a correction of initial left-right asymmetries in somite size \cite{Naganathan2020}. This possibly suggests an overall view where slightly imprecise signaling mechanisms (clock, wavefront, somite anteroposterior polarity) are later canalized/corrected/adjusted by downstream biophysical processes, such as tissue mechanics \cite{Naganathan2020}.

\chapsubsec{Difference between species}

\label{Species}

While the phenomenology of somitogenesis is roughly conserved between species, it is also worth pointing out rather striking quantitative and qualitative differences.

The segmentation period varies widely between species: around 30 mins for zebrafish, 90 minutes for chicken, 2 hours for mice, and 5 hours for humans \cite{Kageyama2022}. More direct comparisons between mammalian cells, have been done using stem cell cultures differentiated into PSM cells (See the section \myref{Section:synthetic}  for  more details) \cite{Matsuda2020a, Lazaro2022}. Mouse and human cells were first compared \cite{Matsuda2020a}, and later on, a segmentation `zoo' was designed, including marmoset, rabbit, cattle, and white rhinoceros \cite{Lazaro2022}. The segmentation clock periods in this new zoo range from 150 mins in rabbit to 390 mins in marmoset, and are comparable to the ones in embryos. `Swap' cells where e.g. human sequences for the \textit{Hes7} gene is introduced in mouse cells show a period increase of 20 to 30 mins, so only a fraction of the 200 mins difference of periods between the two species. This suggests that internal cellular biochemistry (rather than specific coding sequences) plays a role in setting up the segmentation period.


Those scaling dependencies appear rather specific to the segmentation clock though: the authors estimate parameters for other genetic cascades and protein degradation rates in mice vs humans, and, while  degradation rates are slower in human cells than in mice cells, the typical differences are at most by a few tens of percents (while the segmentation period varies by more than two-fold), and for some important mesodermal proteins like \textit{Brachyury} (also called T) there is hardly any difference at all.  All in all, those experiments suggest that the biochemical reactions specifically implicated in the segmentation clocks are essentially scaled in one species vs another. Interestingly, this scaling could be rather global in the sense that the segmentation clock period scales with embryogenesis length (defined as the time from fertilization to the end of organogenesis). Of note, similar scaling of embryonic developmental steps is often observed, for instance, different fly species living under different climates (and thus different temperatures) present scaling developmental stages \cite{Kuntz2014}. See more discussions on scaling in the Appendix, section \myref{app:scaling}.

Beyond the time scales of the segmentation period and development,  it is worth pointing out that the wave pattern observed in the PSM widely varies between species, Fig. \ref{fig:waves}. In Mouse and Medaka, there is only one `wave' of genetic expression within the PSM (meaning that the oscillators close to the front are less than one cycle phase-shifted compared to the oscillators in the tail bud). In zebrafish, there are three waves, and in snake, there are 8 to 9 waves. This suggests that the relative clock period as a function of relative position within PSM varies widely between species. While in mice, the period close to the front is only slightly longer than the period in the tailbud, in other animals such as zebrafish and snakes, the relative period in the anterior appears to be at least 3 times longer, possibly more \cite{Giudicelli2004, Gomez2008,Gomez2009}. Interestingly the period profile as a function of relative position within the PSM is highly non-linear, almost diverging towards anterior PSM, and rather identical between zebrafish and snake, see \cite{Gomez2008} for a comparison. This could indicate some common mechanisms ensuring the coordination of the slowing down of individual cellular oscillators.

\bnormf
\includegraphics[width=\textwidth]{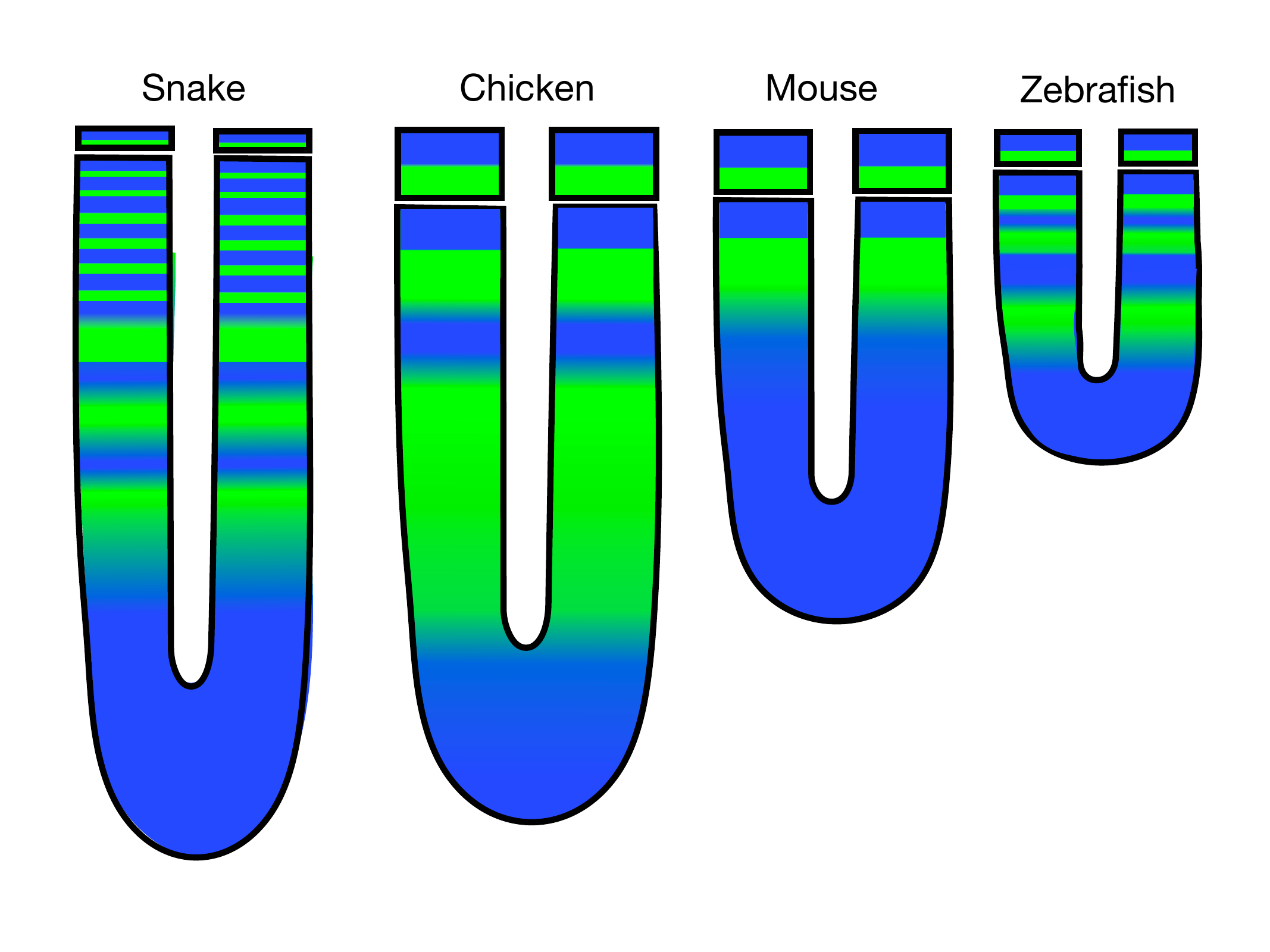}
   \caption[Different wave patterns in different species]{Schematic of the different wave patterns in different species. Adapted from \cite{Gomez2009,Oates2012,Pourquie2022}.}\label{fig:waves}
\enormf

It is proposed in \cite{Gomez2008} that the extensive number of segments in snake vs other animals is indeed due to a relatively slower overall growth rate compared to the segmentation clock. Imagine for instance a zebrafish growing at half the normal rate, but with a segmentation clock keeping the same pace, then it would naturally have twice as many segments. This scenario is supported by the following back-of-the-envelope calculation  :
\begin{itemize}
    \item assuming PSM growth is completely driven by the cell cycle, period $T_{cycle}$, the number of generation times for the PSM to fully grow is $n_g=T_{tot}/T_{cycle}$ where $T_{tot}$ is again the total developmental time
    \item the length of a somite approximately is $S=\alpha L T$, where $T$ is the period of the segmentation clock and $\alpha = \ln 2/ T_{cycle}$ is the growth rate of the PSM ($\ln 2$ factor converts into time via cell division)
    \item eliminating $T_{cycle}$ one gets $n_g=\frac{T_{tot}}{T} \frac{S}{L \ln 2}= n_s \frac{S}{L \ln 2}$ where $n_s$ is the number of somites (assuming a constant period of the segmentation clock).
\end{itemize}
Now $n_s$ is 315 in snake and 65 in mouse, but $\frac{S}{L \ln 2}$, the rescaled ratio of somite vs PMS is also 5 times lower in snake than in mouse, so that both effects compensate and the number of generation $n_g$ is the same independent of the organism. This suggests a picture where $n_g$ is constant across species for other reasons, and that inter-species variability in the number of stripes  indeed primarily comes from different values of $T/T_{tot}$ or similarly $T/T_{cycle}$. Notice that,  if the segmentation clock period gradient  within the PSM is (once rescaled) the same in all species irrespective of PSM size, then if  a cell spends relatively more time (in cell cycle units) to go from tail bud to the front compared to other species, it accumulates much more extensive phase gradient, which results in more waves within the PSM, consistent with what is seen in snake (see more detailed calculations in Appendix \myref{Appendix:growth}).

Going into more molecular details, it turns out that there is quite some variability/plasticity between species in the genes oscillating  \cite{Krol2011}. Microarrays \cite{Dequeant2006} identify 40 to 100 oscillating genes in the PSM, mostly involved in signaling and transcription. In mouse, genes in Notch, Wnt and FGF pathways oscillate, but in zebrafish it seems only Notch pathway clearly oscillates. Phase relations between pathways also appear to vary between species.  Interestingly, only Hes1 and Hes5 orthologs appear to oscillate in the three species considered in \cite{Krol2011} (mouse, zebrafish, and chick), meaning that there is likely "very limited conservation of the individual cycling genes observed", and consistent with the hypothesis that the Hes gene family includes the "core" oscillator. Needless to say, those differences might matter a lot when modeling the segmentation process. There could be big differences between segmentation processes in different species, and for this reason, it is all the more important to discuss, contrast and compare multiple models. Also, since individual cycling genes are likely, not conserved, this justifies more top-down approaches, focused on higher levels, that can eventually be related to actual gene expressions, rather than bottom-up approaches too closely tied to the molecular implementation in a given species.

\newpage

\chap{Early models}
\label{sec:pioneer}
We now review models of vertebrate segmentation spanning more than 40 years of theoretical work. We start with two pioneering models proposed \textit{before} the discovery of the segmentation clock : the Cooke/Zeeman clock and wavefront model, and the reaction-diffusion Meinhardt model. Those two models frame the conceptual discussion and still inspire experiments to this date, but they are also useful reference points for subsequent models. We also review in this section a cell-cycle model,  proposed shortly after the discovery of the segmentation clock, to some extent as an alternative explanation and also providing a slightly different viewpoint (see also review in \cite{Baker2003}).

\chapsec{The clock and wavefront framework}

\label{CWsection}

In 1976,  Cooke and Zeeman \cite{Cooke1976} proposed a "clock and wavefront" model for somite formation to recapitulate many  aspects known at that time. In a nutshell, the model argues that a simple way to build a \textit{spatially periodic} pattern (e.g. vertebrae) is to  imprint a spatial record of a \textit{time-periodic} signal (i.e. a clock).

\chapsubsec{Qualitative view : wavefront}

Such imprint is done with the help of a \textit{moving} variable coupling positional information to developmental time :

\begin{quote}
"There will thus be a rate gradient or timing gradient along these columns, and we shall assume a fixed monotonic relation (non necessary linear) between RATE of an intracellular evolution of development process, and local positional information value experienced by a cell at the time of setting that rate."
\end{quote}

It is not difficult to imagine such a variable in the context of embryonic development since in many metazoans, growth happens in the anterior to posterior  (AP) direction, with anterior cells laid down before posterior ones. This is represented in Fig. \ref{fig:CWV} A : here  we define it as the age of the embryo when the cell is born and positioned, counted from the beginning of embryonic growth (anterior cells have age $0$, posterior cells have higher age), so that the "positional information value" is linear in the position. While they were not known at the time of the Cooke and Zeeman publication, we know now that \textit{Hox} genes \cite{Duboule2022} encode a  similar discretized version of such coordinates, and are likely controlled by a more continuous variable \cite{wacker2004initiation,Francois2010}.
Notice that if, for some reason, the growth rate is twice as small, cells laid at a given distance from the head are twice 'older' compared to a reference embryo, so that the positional information value grows at a doubled rate in absolute unit in space. Thus positional information naturally scales with embryo length, Fig. \ref{fig:CWV} A right. This naturally solves the scaling problem mentioned in Section \myref{sec:early_concepts}.

Cooke and Zeeman propose that such positional information variable could then be used to set the time for future developmental transitions. A simple model would be that a developmental process is triggered after a time proportional to the positional information value defined in Fig. \ref{fig:CWV} A.  Phenomenologically, this results in what we would call today a timer \cite{Francois2010,Clark2021}, where the time at which the process happens at a given position is proportional to the relative position along the A-P axis.

In such a case, one would observe  a developmental wavefront, moving along the anterior-posterior axis. Thus in this picture developmental time (when a cell is positioned along the AP axis) defines positional information, later setting the stage for a kinematic wave of  developmental transition moving from anterior to posterior. [Importantly from a physics standpoint, the term wave does not refer to any oscillation here, but rather is,  to quote Zeeman, the "movement of a frontier separating two regions" \cite{Zeeman1974}, see \myref{sec:ZeemanPrimary} for the definition of primary and secondary waves]. Again, an important aspect of such proposal is that the kinematic wave would move at a speed  scaling with the embryo size since a temporal coordinate related to growth is properly positioned relatively to an embryo of any size, Fig. \ref{fig:CWV} A right, consistent with experiments where the number of cells is artificially reduced\cite{Cooke1975}.

\chapsubsec{Qualitative view : clock}

However, such a kinematic wave moves smoothly from anterior to posterior, while the aim is to define discrete units (somites). 
To induce such change, Cooke and Zeeman propose to introduce a periodic variable or "clock". A simple description of the mechanism is illustrated on Fig \ref{fig:CWV} B. Imagine there is a global oscillator in the embryo, or at the very least that there are synchronized oscillators so that 
\begin{quote}
[pre-somite cells] "are entrained and closely phase-organized (...) because of intercellular communication."
\end{quote}
 Now assume that the front is moving from head to tail with a speed $v$. The assumption is that as the front moves, it interacts with the clock to switch the local state  of a cell from undifferentiated (not somite) to differentiated (somite). Importantly, the timing of the transition depends on the phase of the clock when the front passes, to ensure a synchronous commitment.

To fix ideas, let us assume that a segmental boundary is formed if and only if the clock interacts with the wavefront at phase $\phi=\phi^*$ (Fig \ref{fig:CWV} B). Then starting from an initial segmental boundary where the front is present (phase $\phi=\phi_*$ at $x=x_1$), the clock goes on ticking  (period $T$) while the front is passing. No boundary is formed until the clock reaches again the phase $\phi=\phi_*+2 \pi$, i.e. after waiting for the period $T$. During that time, the front has moved from position $x=x_1$ to position $x_2=x_1+vT$, where the next segmental boundary is formed. This entire process is then:
\begin{quote}
"converting the course of the wavefront into a step function in time, in terms of the spread of recruitment of cells into post-catastrophe behavior."
\end{quote}

It is thus clear that segments of size $S=vT$ are sequentially formed. Importantly, this process recapitulates the minimum phenomenology of somite formation. Somites form periodically in time, and sequentially in space. Future somite boundaries are encoded in the tissue by the kinetics of the wavefront and the clock, so way before boundaries form.  Notice that as soon as we assume the existence of a clock with period $T$ and of a wavefront of speed $v$, the size of the pattern to be proportional to $S=vT$ by dimensional analysis, irrespective of the details of the model, so that if the clock period $T$ period is fixed, the size of the segment is proportional to $v$ (which should thus scale with embryonic size) .  See Appendix section \myref{app:scaling} for discussions of other possible scaling laws.

\bnormf
\includegraphics[width=\ssizefig]{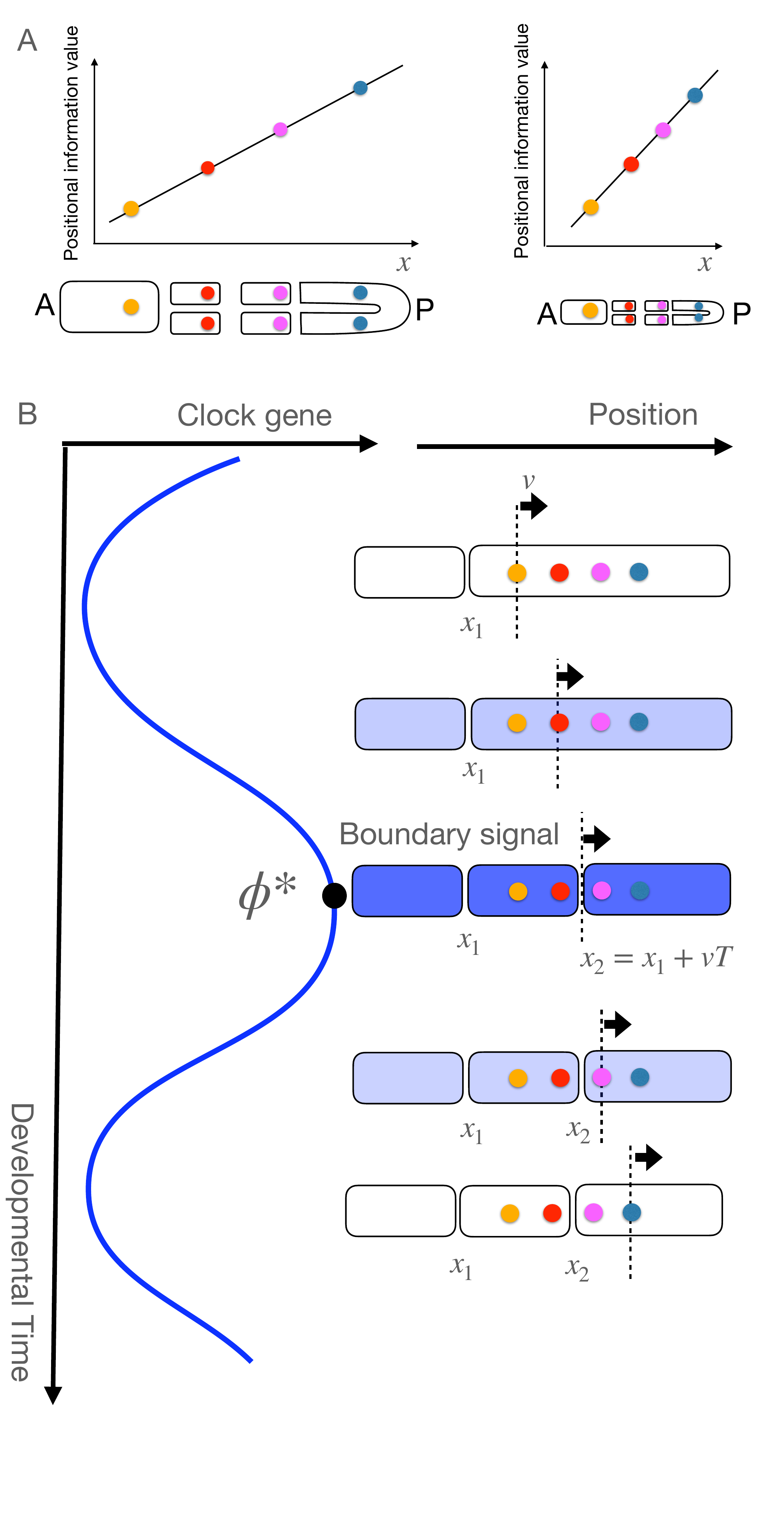}
   \caption[Qualitative view of the Clock and wavefront model. ]{Qualitative view of the Clock and wavefront model. (A) A temporal coordinate is imposed on the embryo, via some monotonic process (e.g. growth), defining positional information. Different colours indicate different values of the temporal coordinate, notice that cells along the same anterior posterior position have the same coordinate. Also if the embryo is smaller (right), the temporal coordinate should scale with the size of the embryo. (B) From top to bottom, one cycle of the segmentation clock (left), as the wavefront (vertical dashed line) progresses with speed $v$ along the temporal coordinate defined in A (right). Phase $\phi^*$  of the clock defines when the new somite boundary is formed, here at position $x_2$ }\label{fig:CWV}
   \enormf

\chapsubsec{Mathematical model :  Wavefront}
 \label{CW_maths_discussion}
Cooke and Zeeman's paper is also groundbreaking because it uses seminal mathematical notions to describe developmental transitions.
The model is  inspired by catastrophe theory, a branch of applied mathematics concerned with a systematic classification of qualitative changes in behaviors of dynamical systems.  There the state of a cell is defined as a vector in a multidimensional space, which generally localizes on a small number of attractor domains (defining different cell states). The idea is that cells move smoothly within each attractor domain, but developmental transitions occur when cells abruptly change their attractor domain (akin to a "catastrophe" \cite{Thom1969}, see also the work of Zeeman \cite{Zeeman1976,Rand2022}).   As pointed out in \cite{Baker2006}, there is no explicit equations provided for their model, but their exact reasoning can easily be put into equations, which we do in the following.

 Cooke and Zeeman graphically suggest in their Fig. 4 \cite{Zeeman1976} that somite formation is induced by a bistable/cusp catastrophe, and that space and time define the two parameters controlling the transition. Calling $t$ the time, $p$ the positional information (which should be related to the anteroposterior position in the embryo, higher $p$ being more posterior), and $z$ the variable representing the state of the cell, let us then define a potential :

\begin{equation}
    F(t,p,z)=z^4/4-p z^2/2+\mu tz \label{cusp}
\end{equation}
This functional form is identical to  the one generated by the so-called "Zeeman Catastrophe Machine" \cite{Rand2022} (see also section \myref{sec:ZeemanPrimary}). A cell at the local position and time $p,t$ has a state variable $z$, driven by the landscape defined by Eq. \ref{cusp} (Fig. \ref{fig:CW_Waddington}). All cells are independent and each cell has its own landscape and state variable $z$; it is implicitly assumed here that a positive value of $z$ corresponds to an undifferentiated state, while negative values correspond to a differentiated (somite) state. For simplicity, we put time and space in different monomials in Eq. \ref{cusp}, which is not generic, but we will comment on more general forms below.

\bnormf
\includegraphics[width=\textwidth]{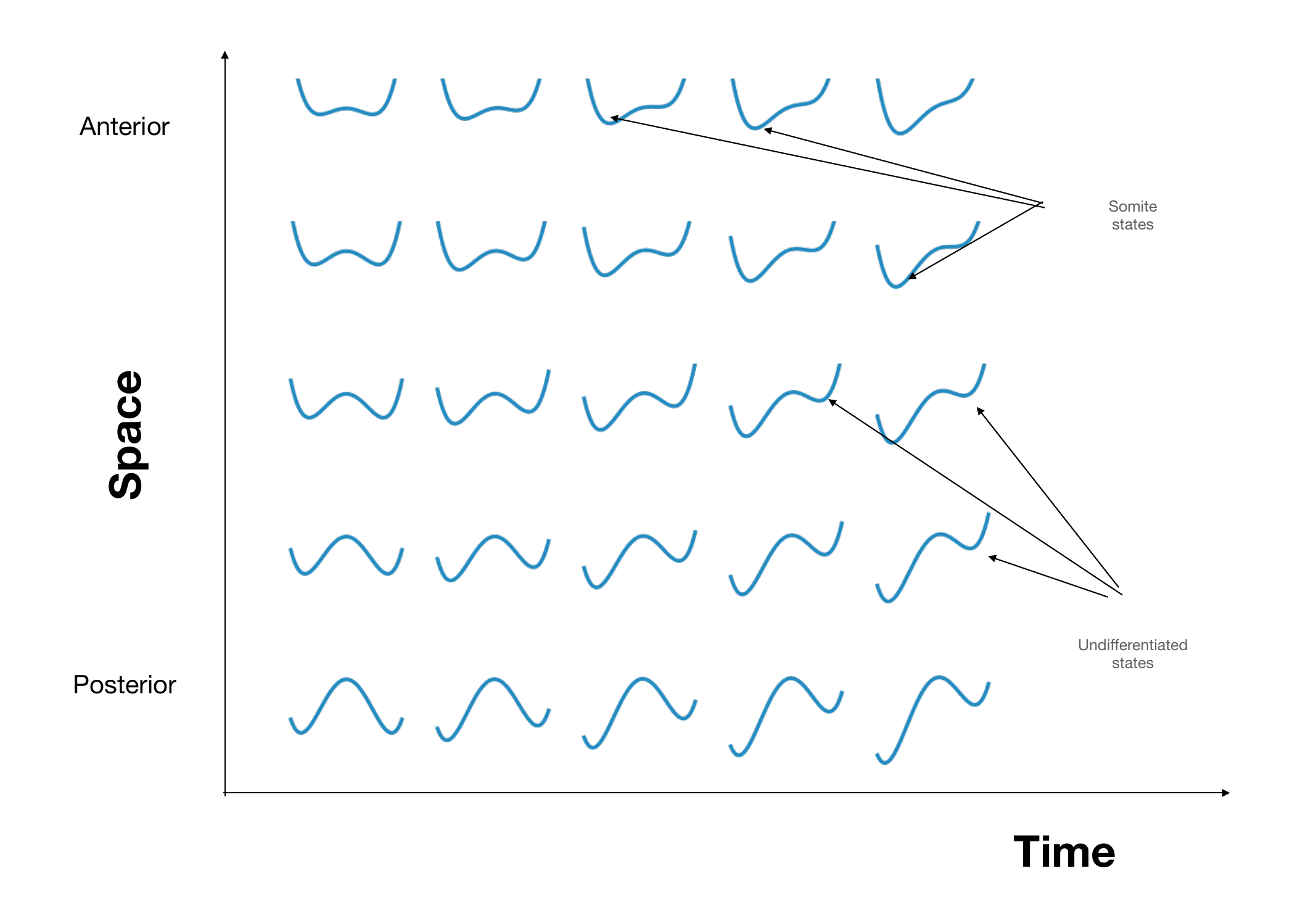}
   \caption[Two-well landscape for the clock and wavefront model]{Representation of the two-well landscape depending on time and space defined  by Eq. \ref{cusp}. The somite state corresponds to the left well, and the undifferentiated state to the right well}\label{fig:CW_Waddington}
\enormf

The equilibrium points are given by the solution of the third-order polynomial equation $\frac{\partial F}{\partial z}=z^3-pz+\mu t=0$.  Assuming the system is such that it rapidly stabilizes, we first see that for $t\rightarrow -\infty$, the system is in a "positive" $z \propto t^{1/3}$ state  (so corresponding to an undifferentiated state) while for $t\rightarrow \infty$, the system is in a "negative" $z \propto -t^{1/3}$ state (corresponding to a somite state). Using classical algebra, it is not difficult to show that for $p<0$, the system is monostable, i.e. $z$ can only take one stable positive value, so can not differentiate. The interesting behaviour occurs for $p>0$, for which there is a bistable region (i.e. $z$ can take two stable values), delimited by $p=\left(\frac{27}{4}\right)^{1/3}(|\mu t|)^{2/3}$. The most interesting behavior occurs along the line $p=\left(\frac{27}{4}\right)^{1/3}(\mu t)^{2/3}$, which corresponds to the saddle-node bifurcation where the high $z$ (i.e. undifferentiated) state disappears (this line corresponds to what Zeeman calls a "primary" developmental wave in \cite{Zeeman1974}, see section \myref{sec:ZeemanPrimary}). Inverting the expression, and assuming the system quickly relaxes to a steady state, at time $\mu t_c(p)=\left(\frac{4}{27}\right)^{1/2}p^{3/2}$, the system at position $p$ has no other choice than to suddenly jump from the positive to the negative state (Fig \ref{fig:CW_maths_1} A-B).  Notice this jump happens (much) later for higher $p$. 

In this view, there would be a kinematic differentiation front, continuously moving at higher $p$ values as a function of time, which is what Cooke and Zeeman refer to when they say the actual differentiation wavefront involves :

\begin{quote}
"a kinematic `wave' controlled, without ongoing cellular interaction, by a much earlier established timing gradient."
\end{quote}

Cooke and Zeeman point out that such variable $p$ could be easily set up by a smooth, anteroposterior (timing) gradient.


\chapsubsec{Mathematical model : Clock }
To make a somite, we shall not need a smooth wave propagation, but rather a simultaneous differentiation for a block of cells - for a range of different positions in the embryo $p$. To account for such "block" differentiation, one needs to introduce a clock. There are multiple ways to put that into equations, but to fix ideas, let us thus consider the following addition to the cusp catastrophe model :

\begin{equation}
    \dot z_p= -\frac{\partial F}{\partial z}- k\delta_T(t)=-\mu t+pz_p-z_p^3-k\delta_T(t) \label{CW_maths}
\end{equation}

\bnormf
\includegraphics[width=\sizefig]{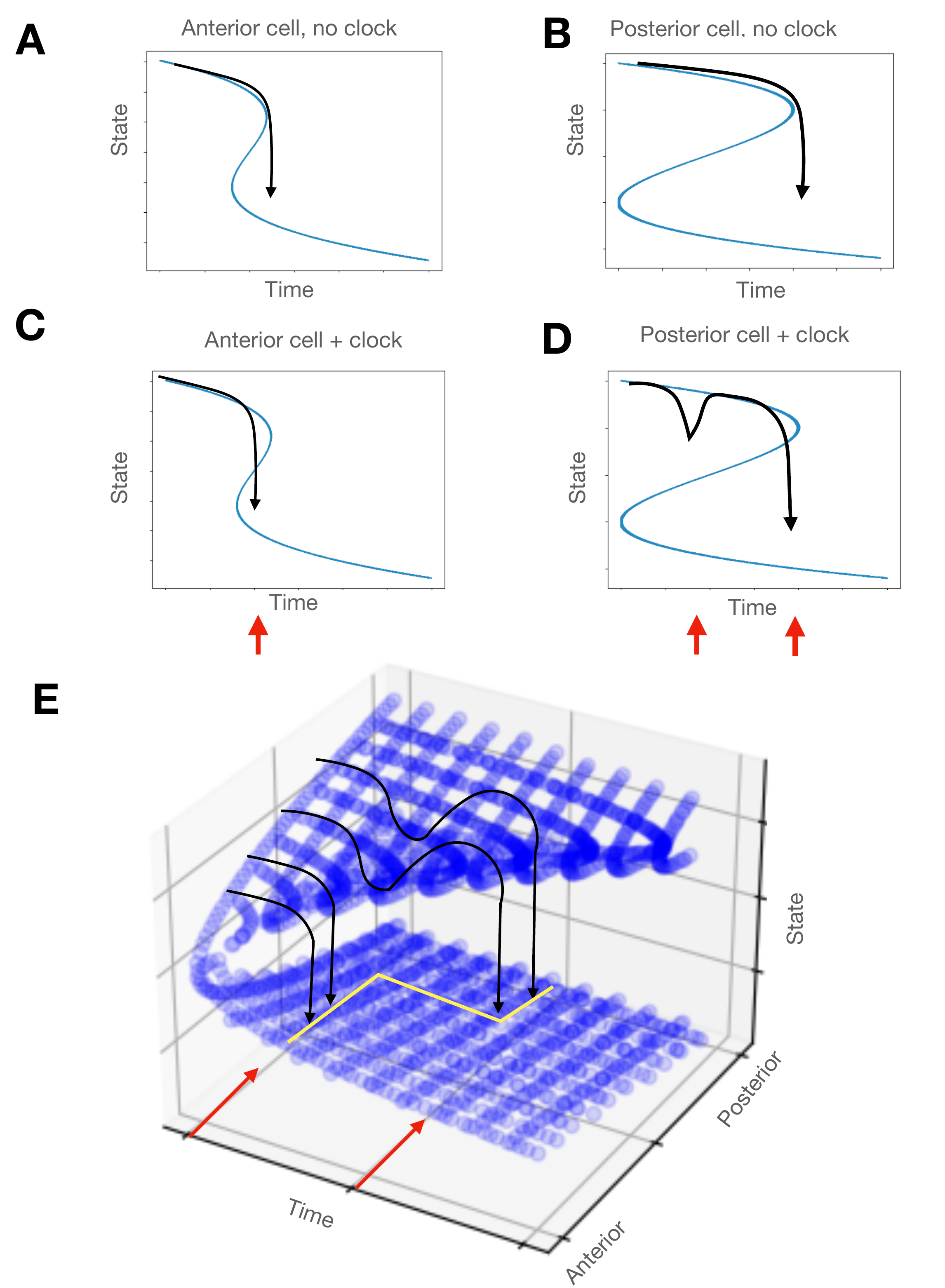}
   \caption[Mathematical formulation of the Clock and Wavefront model.]{Mathematical formulation of the original Clock and Wavefront model. (A-D) The blue curve indicates the possible steady-state values of the state variable $z$ from Eq. \ref{CW_maths} as a function of space and time. The actual dynamics of variable $z$ are sketched with an arrow. High $z$ corresponds to undifferentiated cells, and low $z$ to somites. When $t$ is high enough the system goes through a saddle-node bifurcation from a bistable to a monostable system, and $z$ suddenly jumps from high to low value. In the absence of a clock, this transition happens at a later time for more posterior cells (compare A and B). (C-D) The effect of the clock (red arrow) is to periodically lower $z$, so that cells close to the bifurcation will jump from the high to low state branch. (E) Ensemble of cells close enough to the bifurcation jump at the same time, thus defining discrete blocks. This is illustrated with a yellow line} \label{fig:CW_maths_1}
\enormf

\bnormf
\includegraphics[width=\textwidth]{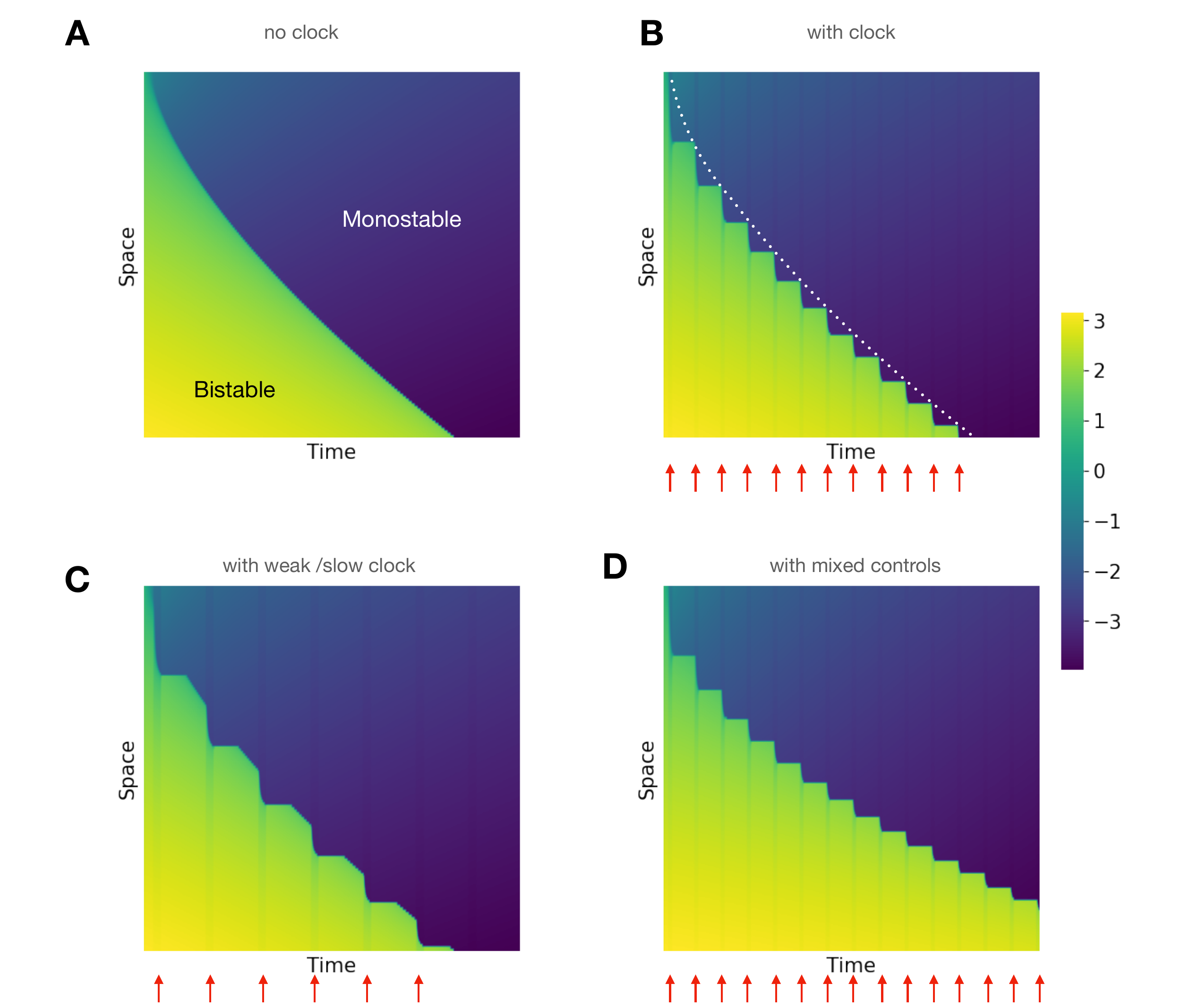}
   \caption[Simulations of the Clock and Wavefront model. ]{(A) Kymograph for $z$ in the absence of the clock, Bistable and Monostable zones are indicated for reference (B) In presence of the clock (red arrows), modeled as periodic kicks uniform in space, blocks of cells are simultaneously induced from high to low $z$ state, modeling somite commitment. Notice that somite commitment happens \textit{below} the bifurcation line of (A), which is indicated by a dotted while line, thus corresponding to the wavefront  (C) Effect of a slower clock. In this case, some cells reach the bifurcation line before the next pulse of the clock, so that the front follows the bifurcation line with the periodic commitment of smaller blocks (D) Changing time and space dependencies of the control parameters changes the shape of the bifurcation line and of the front.} \label{fig:CW_maths_2}
\enormf

where we consider the time evolution of the state $z_p(t)$ for a cell with positional information $p$. Here, $\delta_T(t)$ is a function periodically kicking the value of \textit{all} $z$ (magnitude $k$) towards a more negative value. In such a situation, for cells close enough to the jump (saddle-node bifurcation), the periodic kicking might induce differentiation \textit{earlier} than $t_c(p)$ (Fig \ref{fig:CW_maths_1} C-D). In particular, following a tick of the clock, we expect multiple cells close to bifurcation to jump simultaneously to the negative $z$ state, defining a somite in Cooke and Zeeman's view. More posterior cells with higher positional information $p$ initially stay in the high $z$ state, but as they get closer to the bifurcation they will eventually jump.  Notice that in physics terms, the differentiation timing exactly corresponds to the first passage time  from the right well to the left well in the time-evolving landscape of Fig. \ref{fig:CW_Waddington}, under the control of the clock periodically kicking towards the left.  A 3D plot in Fig. \ref{fig:CW_maths_1} E further summarizes the overall dynamics in the spirit of Fig. 4 of the initial Cooke and Zeeman paper \cite{Cooke1976}.

\chapsubsec{Simulated  Clock and Wavefront model}

Fig \ref{fig:CW_maths_2} displays actual simulations of Eq. \ref{CW_maths} under various conditions, see also attached Notebook. Fig. \ref{fig:CW_beads} also illustrates what happens within a landscape description (see also Supplementary Movie 1).  The bistable/monostable regions are illustrated in Fig \ref{fig:CW_maths_2} A by simulating the system without the clock. Fig \ref{fig:CW_maths_2} B shows what happens with the clock, where blocks of cells jump in a coordinated way as desired. Notice that the new somite boundary after each pulse is always below the bifurcation line, i.e. in the absence of the clock, cells would be committed later compared to a situation with the clock. Interestingly, there is a balance between the position of the bifurcation line and the period/strength of the signal induced by the clock, a situation not studied in \cite{Cooke1976}. For instance, if the clock is either weaker or slower enough, it can happen that some cells will reach the bifurcation line between two cycles of the clocks, leading to a "jagged" front, \ref{fig:CW_maths_2} C. The intuition for this result is simpler: in the limit of no clock, the cells only transition when they go through the bifurcation, so if the clock is both slow and weak, only cells very close to the bifurcation would periodically transition to the differentiated state.


Cooke and Zeeman further comment on an interesting geometrical feature of the wavefront: as can be clearly seen from Fig \ref{fig:CW_maths_2} , the front is not a straight line, which means that the speed of the wavefront is not constant in the coordinate defined by the positional information $p$. Here, the saddle-node bifurcation happens for $p \propto t^{2/3}$, so we expect the speed of the differentiation front (in units of positional information) to be proportional to $t^{-1/3}$ as well, i.e. going to $0$.  If positional information is directly proportional to the actual position, this means that  that boundary $i$ is located at a position scaling as $ (iT)^{2/3}$, and thus the size of a somite $i$ would then be $S_{i} \propto i^{-1/3}T^{2/3}$, so that the size of somites would go to $0$ as well. This could explain why somites can get smaller during development. This scaling law comes  from the fact that position and time are in separate coefficients in the polynomial of $z_p$ in Eq.\ref{CW_maths}, a choice we made here for simplicity. A more generic model would be to mix time and space dependency, e.g. we can add a temporal dependency in the linear term $z_p$ that modulates the front speed and shape, see e.g. Fig. \ref{fig:CW_maths_2} D : the speed front would then go to zero and  a stable boundary would form separating the monostable and the bistable region, thus leaving a permanently undifferentiated region.

\bnormf
\includegraphics[width=\textwidth]{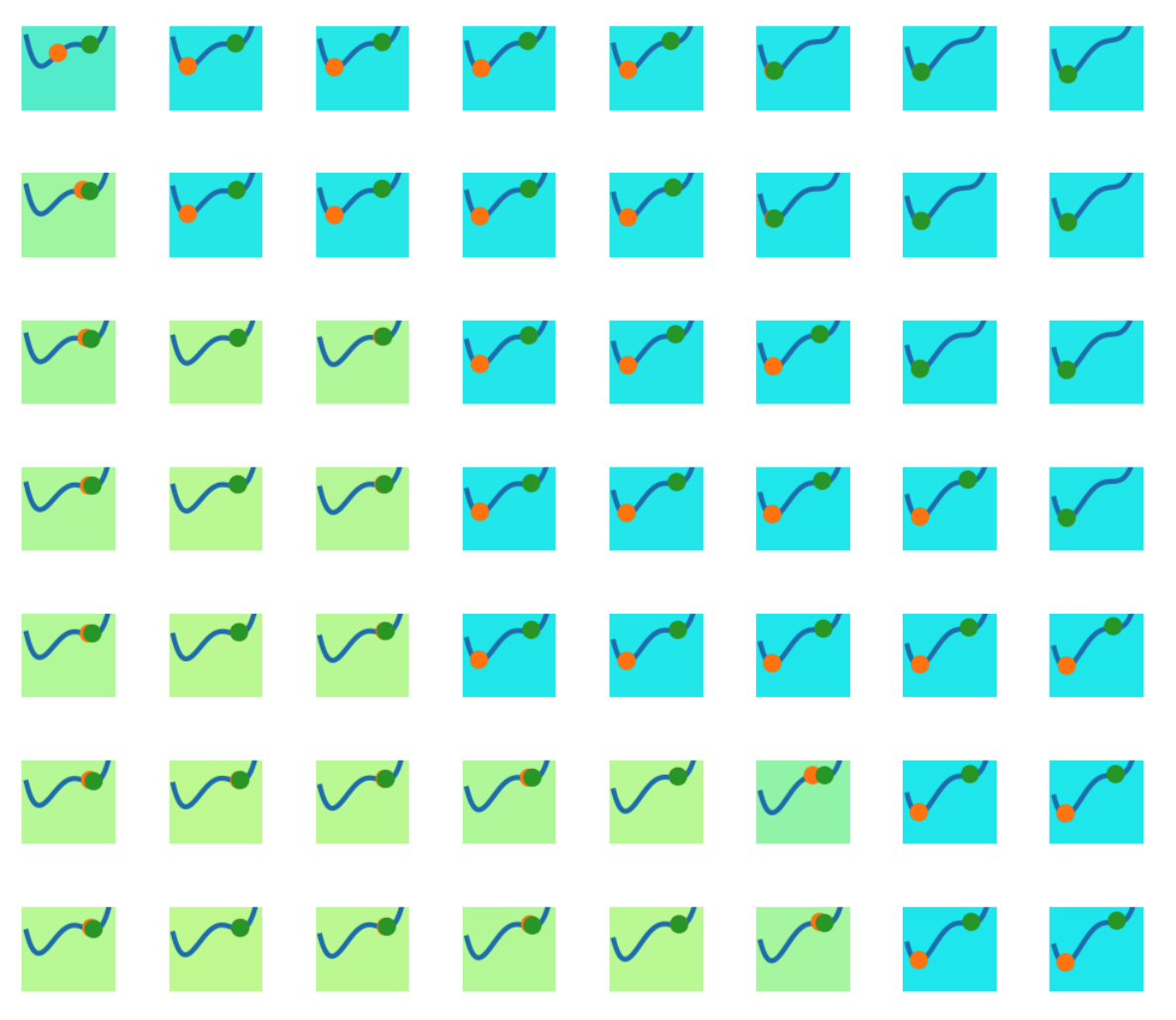}
   \caption[Landscape Dynamics in the Clock and Wavefront model]{Time space dependency of the cellular states in the Landscape defined by Eq. \ref{cusp}, with same conventions as Fig. \ref{fig:CW_Waddington}. Different lines correspond to different positions (top is more anterior), and different columns to different times. Green beads correspond to the time evolution of the system without the clock, and orange bead to time evolution with the clock as defined by Eq. \ref{CW_maths}. A cycle of the clock is completed every three columns. The background colour is a function of the state of the cell in the Clock and Wavefront model (light green: undifferentiated, light blue: differentiated) } \label{fig:CW_beads}
\enormf


Lastly, it is worth mentioning that in Cooke and Zeeman's view, the clock is an external pacemaker, essentially independent from the catastrophe controlling differentiation, and could go on oscillating with minimal impact, even in differentiated cells. Remarkably, the clock has an effect on the state of the cells only close to the primary wave defined by the saddle-node bifurcation. There are important experimental consequences of this observation: for instance, if one could find an external way to manipulate the variable $z_p$, one could induce somite formation without a clock, for all cells within the bistable zone. Conversely, one should be able to largely manipulate features of the clock (such as the period) without impacting the potential driving the dynamics of the variable $z_p$. The most direct way to test this would be to change the clock period, to see how this impacts the speed of the regression and the size of the somites. However, there could be  new features arising in a regime where the clock is very slow, or has only a weak influence on $z_p$: as illustrated in Fig \ref{fig:CW_maths_2} C, one can obtain a mixed system with both discrete and continuous jumps for weak or slow clocks.

This example illustrates one issue in defining the wavefront: depending on the parameters, the jump in $z_p(x,t)$  can be discrete within a block of cells, continuous, or both. Thus the actual wavefront of differentiation is an emergent feature of the interactions of the system, that might not be easily associated with some simple observable  (e.g. a given level of a morphogen). There is an even more general lesson here: processes that are independently regulated (here the clock on the one hand and the possible states of the cell $z_p$ on the other hand) might become more coupled close to a bifurcation (i.e. at criticality \cite{Mora2011}), with important phenotypical consequences. For this reason, it might be desirable that both the clock and the kinematic wave induced by the $z$ jump are in fact coordinated upstream in some way. For instance, one could imagine models where the `constant' term in the right-hand side of Eq. \ref{CW_maths} could also depend more explicitly on $p$ and the phase of the clock, or we could imagine that the strength of the clock increases with clock period to prevent a situation like Fig. \ref{fig:CW_maths_2} C. Conversely, a weaker clock might be desirable, for instance, the jagged line in \ref{fig:CW_maths_2} C could be used to define anteroposterior polarity within one somite, so again requiring some level of fine-tuning or coupling between the clock and the primary wave.

\chapsubsec{Generalization : Zeeman's primary and secondary waves}
\label{sec:ZeemanPrimary}

 The Clock and Wavefront model is related to an earlier proposal by Zeeman regarding the existence of "primary" and "secondary" waves for spatially extended dynamical systems \cite{Zeeman1974}.  Zeeman proposes a much more general theory, with illustrations from epidemiology,  ecology, and developmental biology.

 The general idea is to consider the propagation of a boundary separating two regions with different steady states.
 \begin{quote}
 "By a wave, we mean the movement of a frontier separating two regions. We call the wave primary if the mechanism causing the wave depends upon space and time."
 \end{quote}
 
 An example offered by Zeeman in the context of embryonic development is a field of cells, where initially cells are in a B state, but where cells can also exist in an A state because of bistability. A primary wave can then propagate from a region of A  cells into a region of B cells if cells lose their ability to be in the B state. This can happen for instance via a saddle-node bifurcation, say in response to a disappearing morphogen. For this reason, in this review,  we will associate primary waves with bifurcations and will be slightly more generic by including bifurcations associated with the disappearance of oscillating states.
Secondary waves are defined as such 
\begin{quote}
     "We call the wave secondary if it depends only upon time, in other words it is series of local events that occur at a fixed time delay after the passage of the primary wave."
\end{quote}

For instance, in a pandemic context, a primary wave would consist in the propagation of a disease in a population, while the secondary waves would consist of the delayed appearance of symptoms. This example illustrates in particular how the secondary wave might reveal the existence of a hidden primary wave. Similarly, in biology, the actual differentiation of cells might be a secondary wave following a primary wave directing cells to go to different fates depending on positional information depending on space, and time. 
 
 To fix ideas and be more quantitative, let us consider a slightly more general potential  than Eq. \ref{cusp}, similar to the example that Zeeman uses in Fig. 5 of \cite{Zeeman1974}
 
 \begin{equation}
    F_{\epsilon,\alpha}(t,p,z)=\epsilon(z^4/4-(p+\alpha t)z^2/2+\mu tz) \label{wave}
\end{equation}
 with the associated dynamics $\dot z=-F'(z) $, with various examples displayed in Fig. \ref{fig:Zeeman_Waves}, see also attached Notebook. Initially, all cells are in the same state (at $t\rightarrow\infty$), and then as bifurcation occurs cells end up in two different states, clearly visible in Fig \ref{fig:Zeeman_Waves}. The primary wave then coincides with the bifurcation line from bistability to monostability separating the two regions. Notice that the wavefront in the Cooke Zeeman model is such a primary wave and that the role of the clock is mainly to anticipate the "catastrophic jump" associated to such primary wave.
 
The case $\epsilon=1, \alpha=0$ gives the same example as  Fig \ref{fig:CW_maths_2} A. There, the primary and secondary wave essentially coincides because there is a very fast relaxation of $z$ following the jump from high to low $z$ values on the saddle-node bifurcation line. As pointed out above, this is a bit of a particular case because the polynomial coefficients should rather mix space and time, so that a more general case is displayed in  the middle panel of Fig. \ref{fig:Zeeman_Waves}, where $\epsilon=1, \alpha= 0.02$. In such a case, the bifurcation line does not move completely towards the posterior, so the primary wave "invades" a portion of the field before stabilizing, leading to the sharp and fast definition of two regions. For slow dynamics of $z$, e.g. $\epsilon=0.001$ in the right panel of Fig. \ref{fig:Zeeman_Waves}, the dynamics of domain separation is not sharp and there rather is a refinement process. The primary wave is identical to the middle panel of  Fig. \ref{fig:Zeeman_Waves}, but because of the smallness of $\epsilon$ the dynamics take a long time to relax to smaller values of $z$, leading to the slow propagation of a secondary differentiation wave. Noteworthy, the final steady state in the latter case is identical to the former one but will take a much longer time to reach, giving the feeling that some boundary sharpens, while it was in fact defined much earlier by the primary wave.

\bnormf
\includegraphics[width=\textwidth]{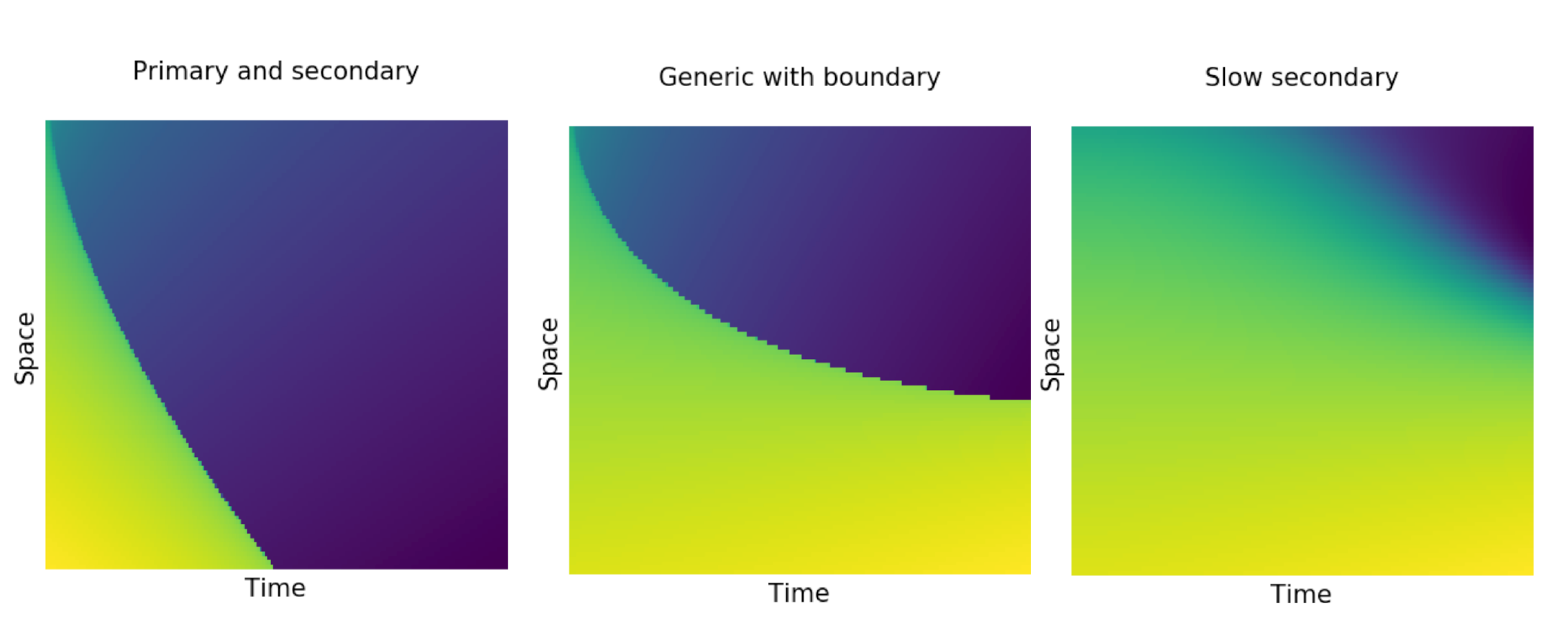}
\caption[Primary and Secondary waves]{Different dynamics of the primary and secondary waves described by Eq. \ref{wave}. On the left panel and middle panel, primary and secondary wave are essentially simultaneous ($\epsilon=1$), the right panel has same $\alpha$ as middle panel but with a much slower $\epsilon$, giving rise to an identical (hidden) primary wave and a much  later secondary wave}\label{fig:Zeeman_Waves}
  \enormf

\chapsec{Meinhardt's model}

In a series of papers in the 70s, an alternative view was defended by Gierer and Meinhardt, who proposed that reaction-diffusion processes combining activator and inhibitors were at the origin of segment formation in metazoans \cite{Gierer1972}. In 1977 Meinhardt applies them to fly, proposing the following model \cite{Meinhardt1977, Meinhardt1986} :

\begin{eqnarray}
\dot A &=& c A^2/H-\mu A + D_a \Delta A +\rho_0 \label{fly1}\\ 
\dot H &=& c A^2-\nu H + D_h \Delta H +\rho_1 \label{fly2}\
\end{eqnarray}
where $\Delta=\frac{\partial^2}{\partial x ^2}$ is the one-dimensional diffusion operator. This model is  `Turing-like', with an activator $A$ that self-activates and activates a repressor $H$, both diffusing. 
Later, in 1982, Meinhardt argued that the addition of a segment from a growth zone, with subcompartmentalization, required new mechanisms to produce an alternation of Anterior and Posterior states within one segment. In particular, it is very natural to assume there is an oscillator generating such alternation, that can further be coupled to an external morphogen. Meinhardt calls this the "pendulum-escapement model" :

\begin{quote}
    "Imagine a grandfather clock. The weights are at a certain level (corresponding to the local morphogen concentration). They bring a pendulum into movement, which alternates between two extreme positions. The escapement mechanism allows the pointer to advance one unit after each change from one extreme to the other. As the clock runs down, the number of left-right alternations of the pendulum and hence the final position of the pointer is a measure of the original level of the weights (level of morphogen concentration)."
\end{quote}

The "extreme" positions of the pendulum correspond to the anterior-posterior segment states, both being generated by an oscillator and modulated by the presence of an explicit morphogen to control the pattern (e.g. the number of segments). So while Meinhardt proposes the existence of a clock his work differs from the Cooke and Zeeman model in a subtle but crucial way. In the Cooke and Zeeman model, the oscillator defines blocks of cells corresponding to somites. In Meinhardt's model, the oscillator defines alternating fates of genetic expression, in modern terms corresponding to somite compartments (anterior and posterior).

To model such alternation, Meinhardt essentially combines his fly segmentation model reproduced above with its own negative mirror image, to include another alternating fate. Remarkably, the addition of this fate allows for the natural emergence of oscillations. 
More precisely, Meinhardt assumes that two variables are present, called $A$ and $P$ (that correspond respectively to anterior/posterior markers of somites). $A$ and $P$ also activate fast diffusing variables $S_A$ and $S_P$, respectively limiting extension of $A$ and $P$, so that the pairs $(A,S_A)$ and $(P,S_P)$ define two (so far independent) Turing systems. Meinhardt then adds mutual exclusion between the two Turing systems, via a repressor $R$ which is activated similarly to both $A$ and $P$, Fig. \ref{fig:Mein_Model} .

\bnormf
\includegraphics[width=\textwidth]{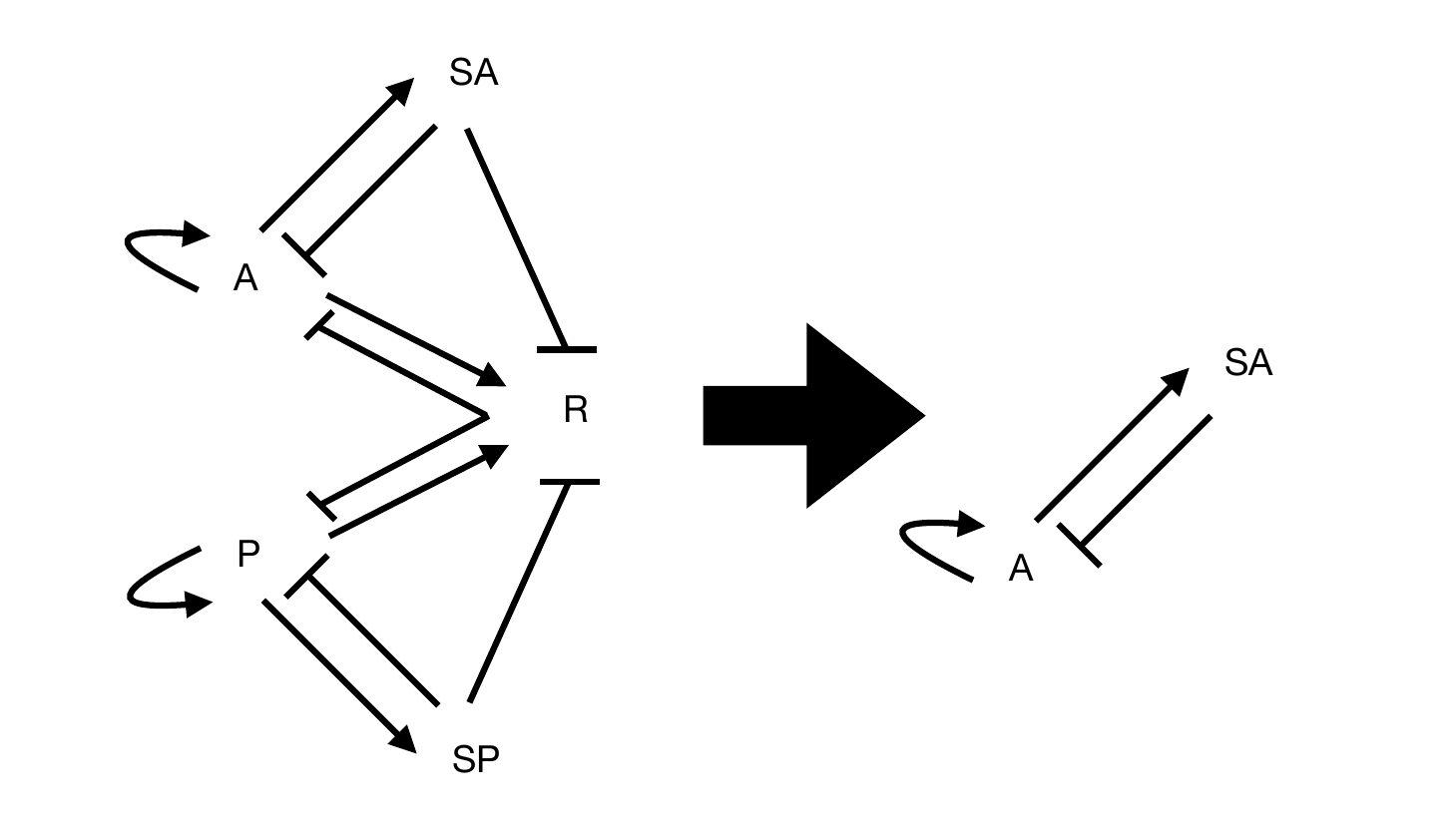}
   \caption[Meinhardt model and its reduction]{ Meinhardt model and its reduction to the Meinhardt-VanDerPol Model. A and P are anterior and posterior genes within the same segment, they are mutually exclusive via the interaction with an extra variable R. SA and SP are diffusing genes limiting the expansion of respective genes A,P. See the main text for detailed equations.} \label{fig:Mein_Model}
\enormf

\chapsubsec{Mathematical formulation}

We could not find an explicit mathematical description of this model from Meinhardt, but it can be reconstructed both from Meinhardt's other similar models and from the BASIC code used to generate his figures, found in appendix of \cite{Meinhardt1982}, Fig. \ref{fig:Mein_Model}, left.
Meinhardt's model  can thus be described with 5 variables :

\begin{eqnarray}
\frac{dA}{dt}&=& \rho_0-d_A A +\frac{cA^2}{RS_A}\\
\frac{dP}{dt}&=& \rho_0-d_P P +\frac{cP^2}{RS_P}\\
\frac{dR}{dt} &=& \frac{cA^2}{S_A}+\frac{cP^2}{S_P}-\beta R\\
\frac{d S_A}{dt}  &=& \gamma_A(A-S_A)+D_A \Delta S_A \\
\frac{d S_P}{dt}  &=& \gamma_P(P-S_P)+D_P \Delta S_P 
\end{eqnarray}

Because of the presence of $R$, in the absence of diffusion, the whole system oscillates, while in the presence of diffusion a stabilizing wavefront propagates, converting the temporal oscillation into a spatial one  \cite{Meinhardt1982}.

The initial Meinhardt model requires 5 variables, so is rather complicated to analyze. But we can use its natural symmetries to simplify it and extract the core working mechanism.

To make a better sense of what happens, let us take $d_A=d_P=d$, $\gamma_A=\gamma_P=\gamma$, and $D_A=D_P=D$. In the following we also assume that $\rho_0$ is small. We start with a quasi-equilibrium assumption on $R$ so that 

\begin{equation}
\beta R=\frac{cA^2}{S_A}+\frac{cP^2}{S_P}
\end{equation}

This gives  
\begin{equation}
\frac{d (A+P)}{dt}=2\rho_0-d(A+P)+\beta
\end{equation}
This suggests performing a new quasi-static assumption
\begin{equation}
A+P=\frac{\beta+2\rho_0}{d}=C_0
\end{equation}
Notice then that $A$ and $P$ are inversely correlated, corresponding to the intuition that they repress one another.

Similarly, we can make a quasi-static assumption for the variable $S_A+S_P$ so that

\begin{equation}
S_A+S_P=\frac{\beta+2\rho_0}{d}=C_0
\end{equation}
(basically, we make the system fully symmetrical in $A$, $P$)
This allows using symmetries in the equations to eliminate completely either $A$ or $P$. Keeping for instance $A$,  Meinhardt's reduced model then is:

\begin{eqnarray}
\frac{dA}{dt}&=& \rho_0-d A +f(A,S) \label{Meinhardt_reduced_A}\\
\frac{dS}{dt}&=&\gamma(A-S)+D \Delta S  \label{Meinhardt_reduced_S}
\end{eqnarray}
wifh $f(A,S)=\beta\left (1+\frac{(C_0/A-1)^2}{C_0/S-1}\right)^{-1}=\beta \frac{A^2}{S}\frac{(C_0-S)S}{(C_0-S)A^2+S(C_0-A)^2}$. The simplification of the model is illustrated in Fig. \ref{fig:Mein_Model} .

Notice the similarity with the initial fly model in Eqs.~\ref{fly1}--\ref{fly2}: there still is auto-activation of $A$ and repression by $S$, in particular when $A$ and $S$ are small. But the additional modulation $\frac{(C_0-S)S}{(C_0-S)A^2+S(C_0-A)^2}$ is equal to $\frac{S_PS}{S_PA^2+SP^2}$. This illustrates the symmetry with respect to $P$ and suggests additional non-linear effects when both $A,S$ are close to $C_0$.

A simulation of this model is shown on Fig. \ref{fig:Meinhardt} and indeed recapitulates properties of the full Meinhardt model, see attached Notebook. Interestingly, in the absence of diffusion, the $A/S$ dynamics is a typical relaxation oscillator, as can be clearly seen from Fig. \ref{fig:Meinhardt} B (see below and in the Appendix A for general discussions on relaxation oscillators). $A$  oscillates between two values approximately equal to $0$ and $C_0$, and $S$ slowly relaxes towards $A$, Fig.  \ref{fig:Meinhardt} B. Like standard relaxation oscillators, when $S$ passes a threshold, it induces a "jump" of $A$ towards a new value ($0\rightarrow C_0$ or $C_0\rightarrow 0$) and a symmetric part of the cycle occurs.

One also sees the effect of the $f$ function described above on the nullclines, (i.e.~lines for which respectively $\dot A$ and $\dot S$ are 0 in absence of diffusion) in  Fig. \ref{fig:Meinhardt}) B right. Close to $A\sim 0$, the $A$ nullcline (blue) diverges; this corresponds to a regime where $f(A,S) \propto \frac{A^2}{S}$ so the $f$ term is of order $2$ in $A$, and using  Eq.~\ref{Meinhardt_reduced_A}, one gets in a self-consistent way a small value $A\propto \frac{\rho_0}{d}$, giving a vertical nullcline. Close to $A \sim C_0$ a new  regime occurs: in this regime, assuming $C_0-A$ is small, $f(A,S) \sim \beta$ up to terms of order $2$ in $C_0-A$, so that again using   Eq. \ref{Meinhardt_reduced_A} and definitions of $C_0$, one gets $A=C_0-\rho_0/d$, so that $C_0-A$ is again small in a self-consistent way. This regime essentially is the ``symmetrical'' regime on the posterior variable $P$ of what happens for the anterior variable $A$.

Those two regimes provide the two branches of a relaxation oscillator driving the AP alternation.  When adding diffusion on $S$, a boundary from high to low $A$ is stable and nucleates a moving front stabilizing the pattern. 
 
  \bnormf
\includegraphics[width=\textwidth]{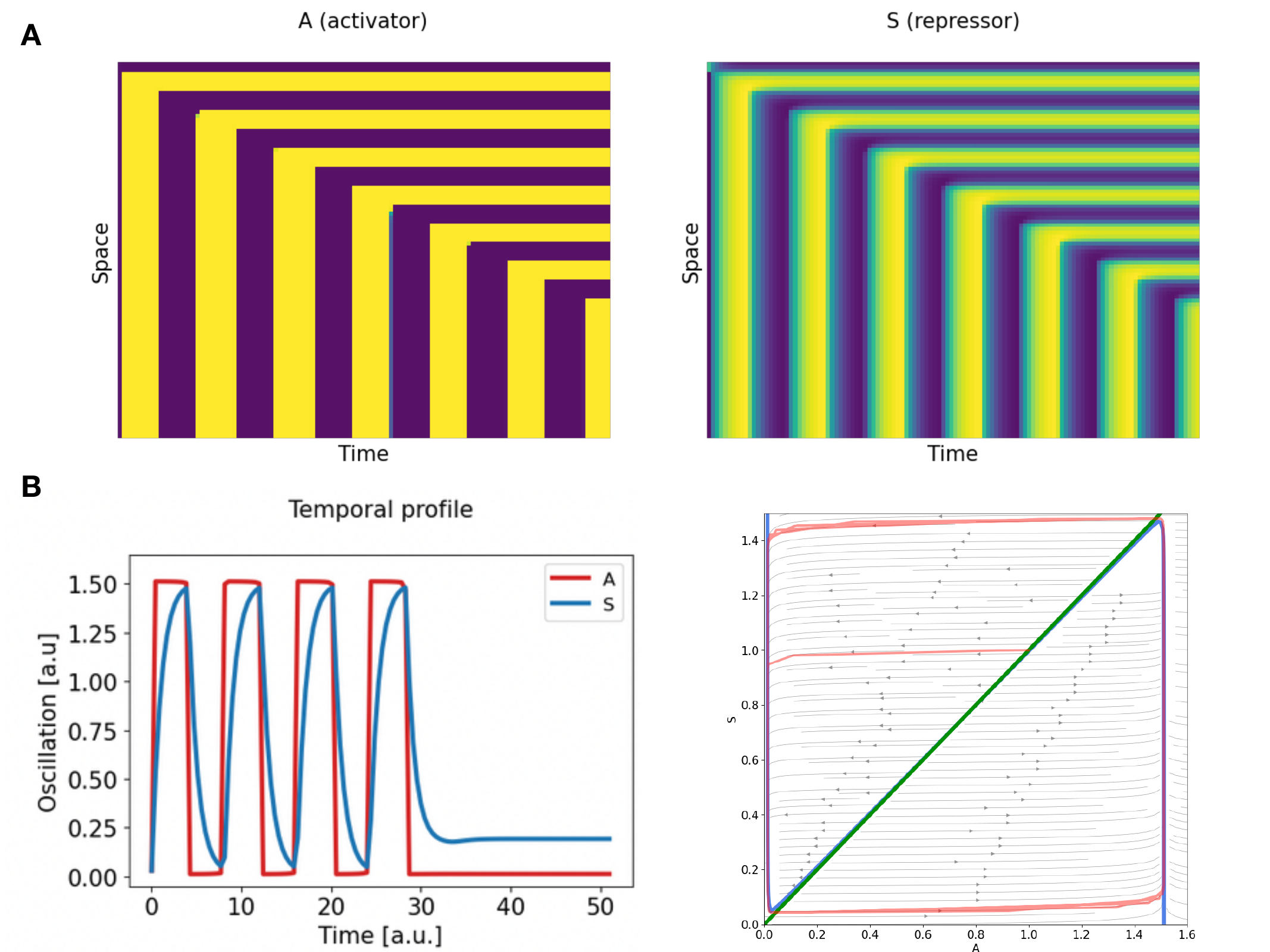}
\caption[Simulation of the reduced Meinhardt model]{Simulation of the Meinhardt model. (A) Kymographs of the variables $A$ and $S$, respectively, obtained with parameter values $\beta=1.5, \rho_0=0.012, d=1, \gamma=0.01, $ and $D=0.01$. The initial condition is an induced boundary ($S=1$ in 3 anterior cells, $S=A=0.1$ everywhere else). (B) Example trajectories of $A$ and $S$ in a cell, and flow diagram of the $A,S$ system in a single oscillating cell, with the limit cycle trajectory in red. Nullclines for $A,S$ are shown, blue for $A$ and green for $S$.}\label{fig:Meinhardt}
\enormf

 \chapsubsec{Generalization : Meinhardt - Van der Pol model}
 
A behavior similar to the  Meinhardt model can be observed with many other (symmetrical) relaxation oscillators, which are better suited for a more precise study of what happens. This was later rediscovered by \cite{Cotterell2015a}  and the associated patterning mechanism was called a "progressive, oscillatory reaction diffusion" (PORD) model (see section \myref{Inverse} below).
 
 Let us for instance consider the following Meinhardt-VanderPol model, based on the addition of a diffusive term to the \textbf{slow} variable of a classical Van Der Pol/Rayleigh oscillator  (see Appendix) :
 
 \begin{eqnarray}
\epsilon\frac{dA}{dt}&=& A-S-A^3/3 \label{Meinhardt_VdP_A}\\
\frac{dS}{dt}&=& \lambda(A-\mu S)+ D \Delta S \label{Meinhardt_VdP_S}
\end{eqnarray}

  \bnormf

\includegraphics[width=\textwidth]{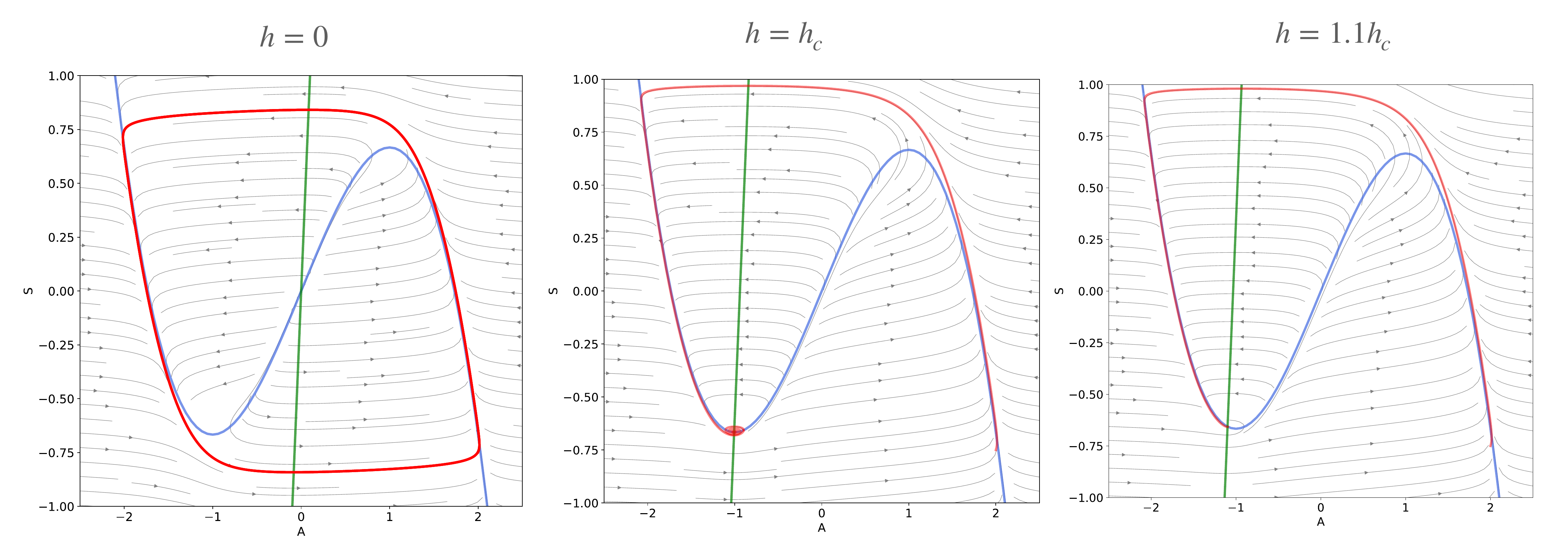}
\caption[Flow-plots of the Meinhardt-Van der Pol model ]{Flow-plots of the Meinhardt-Van der Pol model for different values of the control parameter $h$. nullclines for $A$ is in blue, nullclines for $S$ in green, and trajectories are in red. For $h=h_c\sim\lambda . 0.94$ the system underges a Hopf bifurcation. }\label{fig:MeinhardtVDP-flow}
\enormf

This model has only one non-linearity, the $A^3$ term in Eq. \ref{Meinhardt_VdP_A}.
We can interpret this model "biologically" with A self-activating, and repressed by $S$ (itself activated by A). Instead of the repression by $P$, this model introduces a cubic degradation term for $A$ which makes sure that $A$ non-linearly goes to $0$ once $|A|$ is big enough. Notice however that both $A$ and $S$ can take either positive or negative values and that the initial symmetry of Meinhardt's model is in fact conserved if one flips the signs of $A,S$. Here, only $S$ diffuses (to stay consistent with Meinhardt) and we simulate the system on a line of (discrete) cells and the pattern is stable (consistent with the recent observation that Turing patterns are stable with only one diffusing variable on discrete grids \cite{Wang2022a}).    We also introduce a parameter $\lambda$ allowing to modulate the dynamics of the slow variable (in particular the period of the oscillator).

 Moving now to phase space analysis, when $D = 0$, the system jumps back and forth the sigmoidal branches of the $A$ nullclines (Fig. \ref{fig:MeinhardtVDP-flow} left), like a classical Van Der Pol oscillator (see e.g. \cite{BenderOrzsag} Chapter 10, and Appendix). To understand what happens when there is diffusion, let us treat the $D \Delta S$ term as an external control parameter $h$. Phase plane analysis immediately reveals that when $h$ passes a threshold value (approximately equal to $h_c\sim 0.94 \lambda$), the system $(A,S)$ system undergoes a Hopf bifurcation, due to the fact that the $S$ nullcline moves vertically and intersects one  "bistable" branch of $A$ (the negative $A$ branch in Fig. \ref{fig:MeinhardtVDP-flow} middle). Notice that since the $A,S$ system is fully symmetrical, a similar bifurcation happens when $h<-h_c$, with the system stabilizing on the positive $A$ branch.
 So we expect that wherever the second spatial derivative of $S$ reaches this threshold, the system stops oscillating and, depending on the sign of this second derivative, stabilizes in a branch of either positive or negative A. Also notice that, interestingly, the system stays excitable even for $h>h_c$ (Fig. \ref{fig:MeinhardtVDP-flow} right, see Appendix for the definition of excitability). 
 
 To understand what happens when $D>0$, it is first useful to consider the steady state situation. We see an alternation of stripes of $A,S$, where $A$ jumps from almost constant values and $S$ presents a smoother, oscillatory profile. In particular, for $S$ we get at steady state:
 
 \begin{equation}
     D \Delta S(x) - \lambda \mu S(x)=- \lambda A
 \end{equation}

A crude approximation is to consider that $A$ takes almost constant positive and negative values ($A \simeq \pm A_0)$, then in one stripe (centered with $0$) we expect, solving the equation, that $S \simeq \pm \left( A_0/\mu- S_x \cosh(\sqrt{\frac{\lambda\mu}{D}} x)\right)$ at steady state. At a stripe boundary, $A$ switches sign, so that $S$ has to be equal to $0$ by continuity of its derivatives. This imposes that $A_0/\mu= S_x cosh(\sqrt{\mu/D} (x_0/2))$, and thus defines $S_x$ as a function of $x_0,A_0$ which respectively correspond to the size of the pattern and the scale of $A$ at steady state. $S_x$ and $x_0$ can not be defined by the steady state equation in a self-consistent way, and emerge from the dynamics. Notice that $A$ jumps while $S$ stays continuous, so as a consequence, the control parameter has to be spatially discontinuous at steady state.
 
 It is then useful to plot the dynamics of the control parameter to see how such a discontinuity appears and how the pattern forms. We show kymographs of $A,S$ and rescaled control parameter $|D \Delta S|/\lambda$ in Fig. \ref{fig:MeinhardtVDP-scaling}. We see a "checkerboard" pattern of the control parameter along the front; in particular, at well-defined, discrete times, the control parameter quickly moves above $h_c \sim 0.94$ in blocks, defining stabilized regions.

The precise dynamics explaining stabilization are rather complex, as might be expected for a system defining its control parameter through the second derivative of a bistable variable. To our knowledge, there is no precise mathematical study of this process. We will thus limit ourselves to a qualitative and intuitive description of what happens, Fig. \ref{fig:MeinhardtVDP-pattern}.  Let us focus first on the boundary of a region that has just formed. We see that anterior  to this region (higher $x$ in Fig. \ref{fig:MeinhardtVDP-pattern}) , the control parameter $|h(S)|>h_c$, so that the discontinuity in the pattern is established and stable, Fig. \ref{fig:MeinhardtVDP-pattern} A. This induces a spatial gradient of control parameter $h$: close to this discontinuity, the region is oscillating (like the posterior) but is close to the bifurcation point Fig. \ref{fig:MeinhardtVDP-pattern} B. Such dynamics of the control parameter make sense, since after the jump, there is a discontinuity in $A$ between the stable and the oscillating region, and we thus expect $S$ to follow in a "smoother" way, with an increase in its second derivative.

The absolute value of the control parameter is slowly increasing in this region in a graded way so that oscillations stabilize in more and more cells. Eventually, the posterior oscillation (where the control parameter still is around $0$)  jumps  on the other branch, Fig. \ref{fig:MeinhardtVDP-pattern} C, left. This creates two domains in $A$ (one positive, one negative), between posterior cells which have just jumped and more anterior cells where the oscillation is close/past the bifurcation on the other branch. 

Finally, because of the relaxation-oscillator dynamics, $S$ follows $A$ with delay. This creates a sudden increase of second derivatives of $S$ at the interface between positive and negative $A$ , and eventually a spatial discontinuity in both $A$ and in the control parameter Fig. \ref{fig:MeinhardtVDP-pattern} D-E ensues. This both nucleates the next stable region and stabilizes this region that never jumped, and the process iterates forming a stable alternation between regions of low and high $A$, with $S$ following $A$ in a "smoother" way. Notice in particular that a new block stabilizes in three steps: first, a small stable region is nucleated close to a newly formed boundary  (Fig. \ref{fig:MeinhardtVDP-pattern} A), then the next stable boundary is induced Fig. (Fig. \ref{fig:MeinhardtVDP-pattern} C) and lastly the interior of a newly defined block between two stable boundary stabilizes (Fig. \ref{fig:MeinhardtVDP-pattern} D-E).

 

 \bnormf
 \begin{center}
\includegraphics[width=\sizefig]{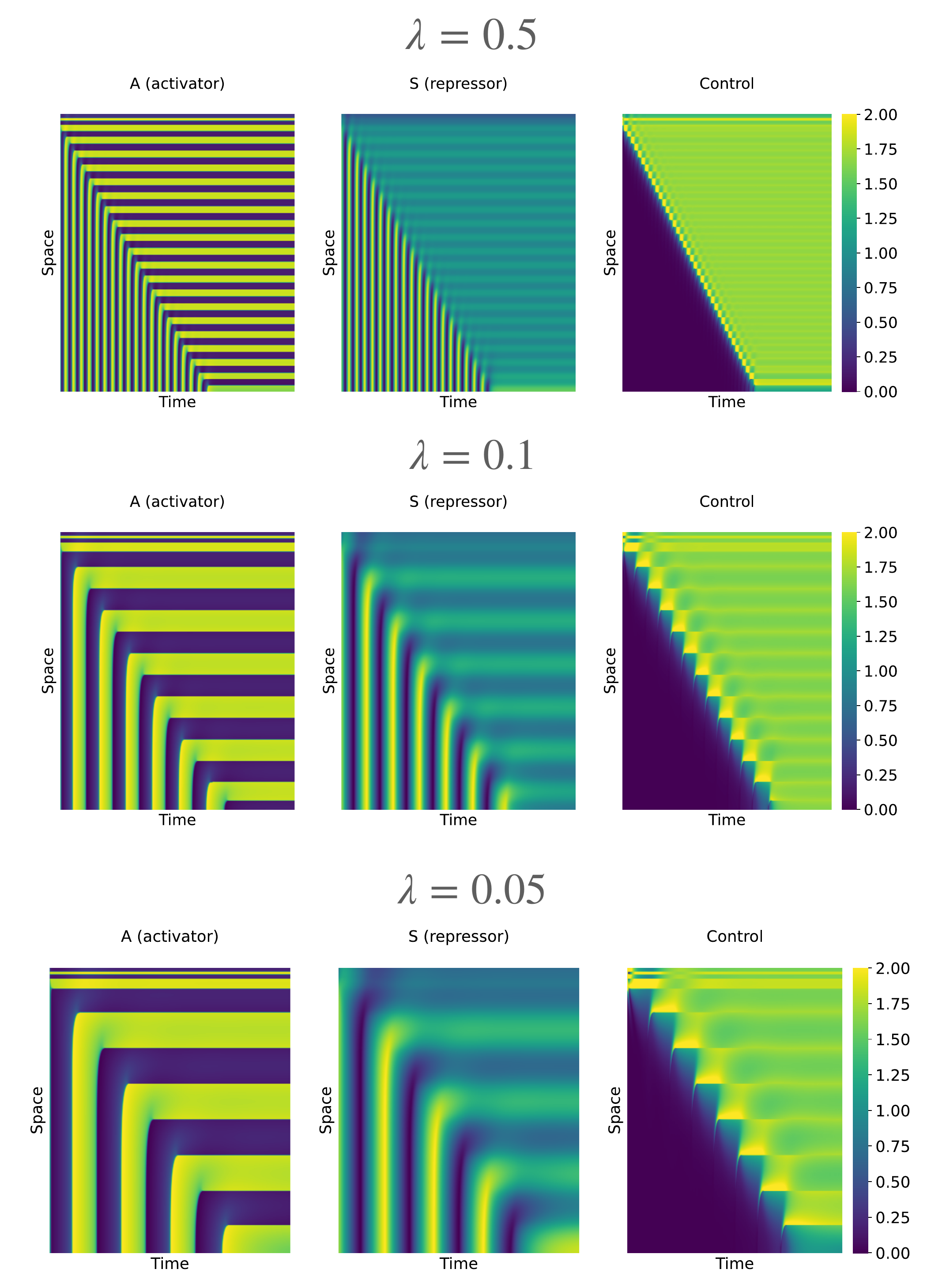}
 \end{center}
\caption[Scaling of the Meinhardt-Van Der Pol model]{Behaviour and scaling of the Meinhardt-VanDerPol model for different values of parameter $\lambda$ from Eqs. \ref{Meinhardt_VdP_A}-\ref{Meinhardt_VdP_S}. Kymographs of the variables $A$ and $S$ are represented. In the third column, we plot $\frac{|D \Delta S|-h_c(\lambda)}{\lambda}$, showing the jump in the control parameter at the front. }\label{fig:MeinhardtVDP-scaling}
\enormf


\bnormf
   \includegraphics[width=\ssizefig]{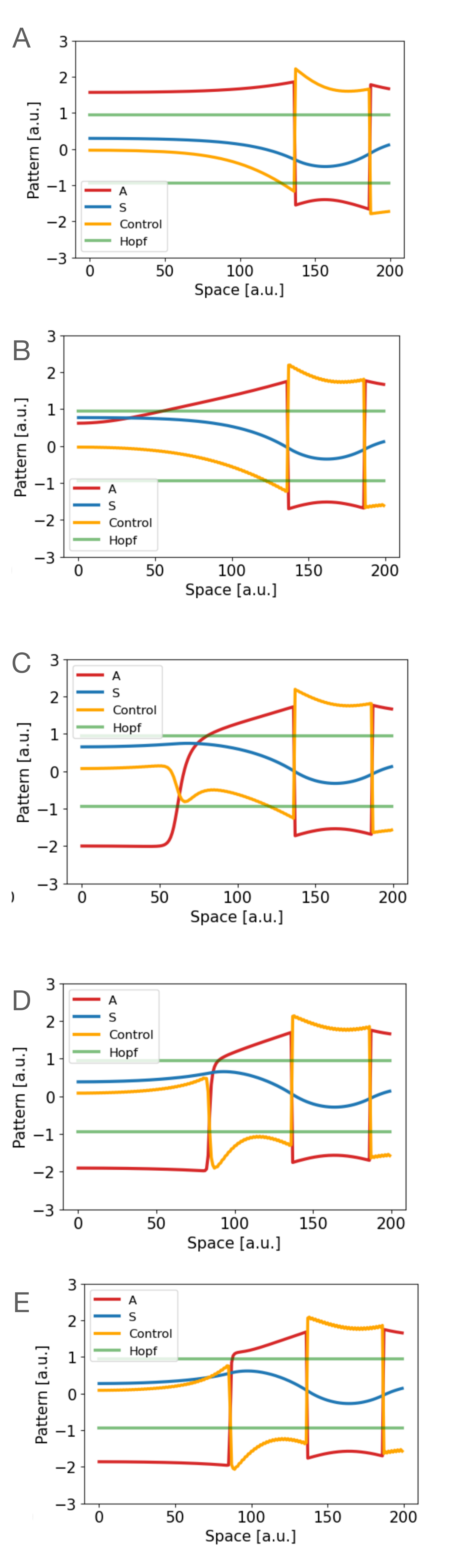}
   \caption[Nucleation of a new border in the Meinhardt model]{Time evolution of the spatial pattern in the Meinhardt model. The posterior is on the left. The range of control parameter for which the system is oscillating is indicated by green lines. In (A-B), a small region left of the boundary stopped oscillating, creating a spatial gradient in S.In (C), the jump of the oscillator in the posterior-most region nucleates a new boundary that moves towards the right. The control parameter crosses the Hopf line, stabilizing the boundary around position 80, and the oscillation stabilizes in an entire block for higher positions (D). The control parameter keeps increasing left of the new boundary (E), leading to a situation symmetrical to (A).}\label{fig:MeinhardtVDP-pattern}
\enormf

 The scaling law of this process is of particular interest. As pointed out by \cite{Cotterell2015a}, the speed of the front is an emerging quantity of the diffusion of the stabilizing zone, induced in our example by the changes in the control parameter $h$. There are a priori at least two possibilities here. First, the speed $v$ could emerge independently from the clock, like in the initial clock and wavefront model, so that the size of the pattern would be $S=vT$.  The other possibility could be that the speed and the clock are coupled by diffusion so that there is a (pattern) wavelength proportional to $\sqrt{DT}$, where $T$ is the period of the clock and $D$ is the diffusion constant. This would then give a wavefront speed proportional to $\sqrt{D/T}$. Going back to simulations, our numerical studies reveal that the wavelength of the pattern is almost exactly proportional to $T$ over more than one order of magnitude of period change (data not shown) so that the wavefront speed does not depend on the period, similar to the former hypothesis. This is visually illustrated in Fig. \ref{fig:MeinhardtVDP-scaling} : the slope of the stable region in the kymographs does not depend much on the control parameter of the period $\lambda$.  This suggests that the front speed is purely diffusion-driven, like many other models in physics and biophysics, see e.g \cite{Radulescu2008}, while the nucleation of the new stable zone is driven by the relaxation oscillation.

 \chapsubsec{Biological Interpretation of the Meinhard's model}
 
  As pointed out by Meinhardt himself :
 
 \begin{quote} "In the model I propose, the oscillation (between A and P), the wavefront (separating the oscillating and the stable pattern), as well as the spatially stable periodic pattern (of A and P), result from one and the same mechanism." \end{quote} 
 

 This simplicity in the equations explaining multiple aspects of the process obviously has a strong appeal for physicists, especially when reduced to two variables. As such it provides important insights into biological mechanisms, both by setting a modeling framework and by suggesting predictions. 
 
 First, the pattern in Meinhardt's model is clearly stabilized by interactions of consecutive domains where $A$ is present/absent. So spatial diffusion is crucial to form and stabilize the boundary. This somehow contradicts the kinematic view of somites formation associated to the robustness to various embryonic manipulations (graft, spatial boundaries), with the caveat that those manipulations are at the tissue scale and might not be the best to falsify local mechanisms at the cellular scale.
 
 Second, as explained above, there is no discrete formation of a block of cells defining somites like in the Cooke and Zeeman model. Rather, somites are assumed to be defined a posteriori as the concatenation of one anterior compartment with a posterior one. Then, the alternation of a APAPAP pattern does not define unambiguously a somite, since boundaries should be defined for the P to A boundary but not for the A to P boundary. Meinhardt, therefore, suggests that there might be a third oscillating variable (called X) so that the real alternation is of the form APXAPX, unambiguously defining the somite boundary. In fact, Meinhardt points out another potential mechanism, where the system might rather detect the \textit{temporal} succession of P to A in opposition to A to P to trigger boundary formation : 
\begin{quote} 
"Imagine a ship in a channel system with locks. A lock can be in two states. Either the lower gate is open and the upper gate is closed or vice versa. In neither state can a ship pass through. But in one state the ship can enter into the lock and after the switch to the other state, the ship can pass. In one state, the transition is prepared but blocked. In the other state, the block is released, the transition can take place, but no preparation of the next transition is possible. For the sequential activation of control genes I assume that, for instance, in the P-state a substance X is produced that activates the subsequent gene, but that its action is blocked. In the A-state, the block is released but X is no longer produced. Only with a P-A transition the activation of the subsequent gene can take place due to the simultaneous release of the block and the presence of the substance X. In contrast, activation of subsequent control genes can not occur if cells remain permanently in the P- or in the A-state."
 \end{quote}

  In modern terms, this describes by essence a phase detector downstream of an incoherent feedforward loop network (see e.g. \cite{Beaupeux2016}), where $P$ activates $X$ but represses its downstream target, while $A$ derepresses the target. $X$ is produced only when $P$ is fading out and $A$ increasing.

Like the Clock and Wavefront model, the differentiation wavefront is emerging from the dynamics. One can first approximate  the wavefront in Meinhardt's model  as the point where the oscillation stops (or the limit cycle disappears), but as seen from simulations, this is not a continuous front, rather, due to the relaxation oscillation, it jumps in a discontinuous way from one boundary to the other, later-on stopping oscillations in-between. This jumping process is not so different qualitatively from the pulses of the clock inducing transitions in the Cooke and Zeeman model. The dynamics of motion are very different though: in the Cooke and Zeeman model, the competency zone for transition to bistability is defined by the external positional information variable $p$, while in the Meinhard't model, it rather is a self-organized 'domino' effect where one stable region nucleates the following one with the help of the ongoing relaxation oscillator and diffusion. This creates a difficulty for scaling/changing the size of the pattern. In particular, in Meinahrdt's model there is no external positional information variable independent from the oscillation. Meinhardt anticipates this potential difficulty by introducing a modulation to his model, adding a spatial dependency in equation $A$ of the form :

\begin{equation}
\frac{dA}{dt} = \rho_0-d_A A +\frac{cA^2}{R(S_A+\Theta(x,t))} \label{Meinhardt_position}
\end{equation}
  
This threshold $\Theta(x,t)$ de facto defines some external positional information in the system,  which can modulate the speed of the clock and as such the size of the pattern. Meinhardt suggests a simple model so that $\Theta$ is essentially monitoring the number of cycles in relation to a morphogen gradient. By adjusting the slope of the morphogen gradient in a size-dependent way, scaling with embryonic size can be obtained. This model can be further adapted to account for further specialization of some segments as a function of time  (e.g. in the case of insects, some segments will give rise to wings while other ones will give rise to halteres, due to the expression of so-called Hox genes).

\chapsec{Cell Cycle model}

In the early 90s, Stern and co-workers proposed that the segmentation clock could be in fact related to the cell cycle \cite{Stern92}. This comes from a series of clever experiments in chick showing very striking features \cite{Primmett1988a,Primmett1989a} :
\begin{itemize}
\item One single heat shock produces several segmental anomalies, restricted to one or two consecutive segments, but separated by 6 to 7 somites - corresponding to roughly 9 hours of development. This suggests the existence of a long temporal cycle implicated in segment formation, with a length corresponding to the time required to form 6-7 somites. Then if this cycle is initially perturbed, the perturbation would be repeated every 6 to 7 somites, corresponding to the period of the oscillator.
\item The 9 hours period was later shown to correspond to the length of the cell cycle, strongly suggesting that it is coupled to somite formation.
\item A single progenitor cell in the tail bud injected with dye gives rise to several clusters of cells in the PSM and in somites, with a 6 to 7 somite periodicity \cite{Selleck91, Stern92}
\end{itemize}

This suggests the following picture: progenitors in the tail bud constantly divide and lay down cells in the PSM in an ordered way so that cells at the same anteroposterior position are roughly  at the same phase of their cell cycle.  A 6-7 somite periodicity thus recapitulates spatially a phae gradient of the cell cycle. Then, the cell cycle is coupled to somite formation, for instance, there might be a special phase $\phi_*$  of the cell cycle for which cells form a boundary when they reach the anterior. Now we need to assume that one cell cycle phase (say $\phi_S$) is specifically sensitive to heat shock (while other phases of the cycle would not be), which could well happen for discrete events in the cell cycle (e.g. a transition between G1 and S/G2/M). So when heat shock occurs, it disrupts all cells in $\phi_S$, not only the older cells in the anterior but also the cells just laid in the posterior a few cell cycles later. When those perturbed cells end up differentiating into somites, theoretically at phase $\phi_*$, their disrupted cell cycle results in segment anomalies. The  cells just posterior to this anomaly were not in phase $\phi_S$ at the time of the heat shock, so are laid down normally and form somites at  $\phi_*$. Then, one full cell cycle later, cells that were again at $\phi_S$ at the time of the heat shock would theoretically reach $\phi_*$ but are disrupted again. This explains why one single heat shock disrupts several segments in a periodic way.



In \cite{McInerney2004}, McInerney \etal  proposed a mathematical implementation of the cell cycle model for somitogenesis. The goal is to understand with a realistic biochemical model how a spatial gradient of cell cycle phases can translate  into blocks of simultaneously differentiating somites. In particular, this model is not concerned with the formation of stripes or AP somite polarity (contrary to Meinhardt's model).  From a modelling standpoint, the challenge is to find how a \textit{continuous} periodic process (such as the cell cycle, with a spatial gradient of phases) can give rise to a \textit{discrete} output (spatially extended somite blocks), and as such, while details differ, this model is in fact very close to the initial Clock and Wavefront vision. This model is also of particular interest from a conceptual standpoint because many subsequent models implement similar ideas with different hypotheses on the nature of the oscillator or of the front.

The model relies on the combination of two continuously moving fronts with a simple, two-component biochemical network, encoded into the following equations :

\begin{eqnarray}
\frac{\partial u}{\partial t }&=&f(u,v) \label{cycle_u}\\
\frac{\partial v}{\partial t }&=&g(u,v)+D\frac{\partial^2 v}{\partial x^2} \label{cycle_v}
\end{eqnarray}

The $f$ and $g$ functions encode generic signaling dynamics where $u$ self-activates, and is activated by $v$, while $v$ is repressed by $u$. After dimensionless reduction, one gets :

\begin{equation}
f(u,v)=\frac{(u+\mu v)^2}{\gamma+\kappa u^2 }\chi_u- \frac{u}{\kappa}
\end{equation}
and 

\begin{equation}
g(u,v)=\frac{1}{\epsilon+u }\chi_v- v
\end{equation}

Two fronts moving with speed $c$ are encoded into a spatial-temporal dependency of the activations on $u,v$  :

\begin{equation}
\chi_u=H(ct-x+x_1) \qquad \chi_v=H(ct-x+x_2) \label{H_uv}
\end{equation}

where $H$ is the Heaviside function. A cell cycle gradient is imposed by the fact that $x_2<x_1$ : so cells become competent to express $u$ before they are competent to express $v$. Practically, the couple $(\chi_u,\chi_v)$ can only take three values $(0,0), (1,0)$ and $(1,1)$. Those three values correspond to three spatially distinct regions of the embryo, respectively corresponding to the posterior of the embryo (region $I$), a somite definition zone (region $II$), and the anterior of the embryo (region $III$).

 \bnormf
\includegraphics[width=\textwidth]{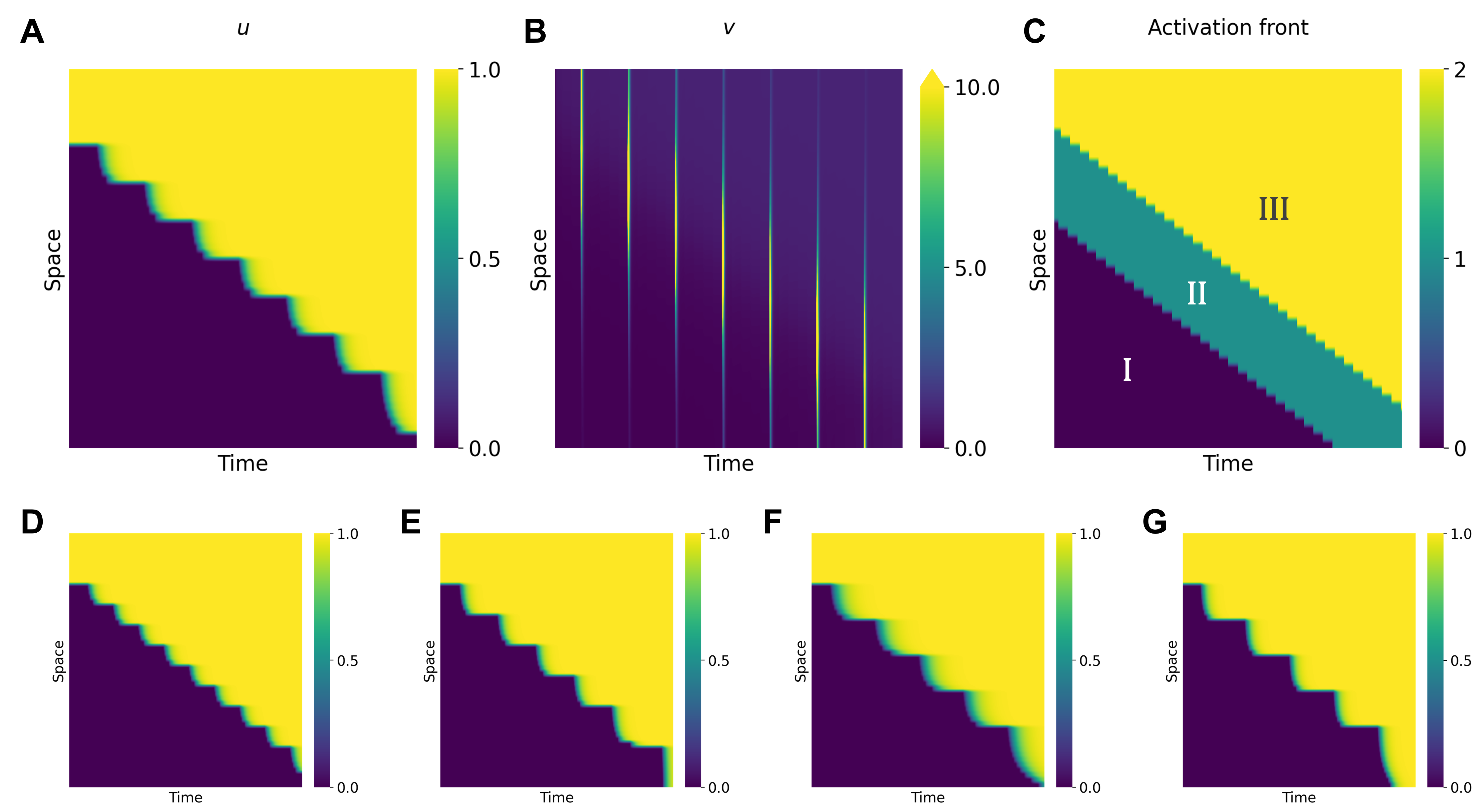}
\caption[Cell Cycle model]{(A-C) Simulation of the cell cycle model for somitogenesis with parameter values $\mu = 0.0001, \gamma = 0.01, \kappa = 10, \epsilon = 0.001, D = 60,$and$ c = 0.00125$. (A) Kymograph of the variable $u$, with blocks of cells moving from low ($u = 0$) to high ($u = 1$) state. (B) Kymograph of the variable $v$ showing spatially extended transient pulses. (C) Propagation of the fronts, shown as the sum of activations $\chi_u + \chi_v$. Three distinct regions are indicated: the posterior ($I$), the somite definition zone ($II$), and the anterior ($III$). (D-G) Keeping the rest of the parameters the same as in (A), we change one parameter value in the simulation. (D)$D=30$. (A). (E) $D=120$. (F) $\kappa = 20$. (G) $\mu = 0.0003$. }
\label{fig:cellcycle}
\enormf

It is useful to first study the behavior of the $u,v$ system for constant values of  $(\chi_u,\chi_v)$ corresponding to different regions.
The posterior of the embryo (region $I$) is the simplest case: $\chi$s functions are $0$ so that the only steady state is $(u,v)=(0,0)$.

In region $II$, the steady state value of $v$ still is $0$, but self-activation of $u$ creates a new stable steady state $u_*=\frac{1}{2}(1+\sqrt{1-4\gamma/\kappa})$. We will also assume for simplicity that $\gamma<<\kappa$ so that $u_*\sim 1$. We thus expect region $II$ to be a region of bistability where $u$ can go to stable values $0$ or $\sim 1$, depending on the initial conditions of both $u$ and $v$. 

In the anterior, region $III$, $v$ can no longer be strictly $0$. Assuming initially $u\sim0$, then we see that $v$ initially rises quickly to a (high) value $\sim 1/\epsilon$. Then since $v$ activates $u$, the value of $u$ starts increasing as well. The final state depends on parameters, but if $\epsilon$ is small enough, $v$ is transiently so high that it strongly activates $u$, which can  become high enough to sustain its own production. In turn, $u$ then squashes down  $v$, to get a steady state not far from the steady state value $u_*\sim 1$, corresponding to a differentiated, somite state. McInerney \etal  proposed that the system is in fact monostable for those parameters, leading to a high, sustained steady state approximately equal to $\sim (u_*,1/(\epsilon+u_*))$. Notice these dynamics also create a transient "pulse" of $v$, before going back to a close to $0$ value of $v$, so akin to an excitable system (see Appendix).

So we see that in region $I$ we expect $u$ to be $0$, in region $II$, $u$ can be essentially $0$ or $1$, while in region $III$, $u$ is essentially $1$.

Now the idea underlying the full model is that the Heaviside function combined to diffusion will induce a sudden transition of $u$ from $0$ to $1$ in a \textit{block} of cells in the region $II$, via a spatially extended pulse of $v$. Let us assume that at $t=0$, $v$ is roughly $0$ everywhere, $u\sim 1$ for $x<0$ and $u=0$ otherwise. Then as time increases, $\chi_v$ is turned on close $x\sim0$, so that there is a sudden pulse of $v$ there. If the diffusion constant is very high, this $v$ pulse is going to diffuse very quickly towards higher $x$, leading the pulse to be spatially extended. What happens then depends on the balance of the parameters, but after some time (say $\tilde t$),  this pulse of $v$ induces a transition from $u=0$ to $u=1$ values in a region close to $x\sim0$. The size of this region depends on parameters. If the diffusion is fast enough, induction occurs in the entire region where $\chi_u=1$,  but if this region is big enough (or diffusion too small), this will happen only in part of the region where  $\chi_u=1$ (see Fig. \ref{fig:cellcycle}).

When $u$ is high enough to self-sustain, it pushes $v$ back towards $0$ in this region. So we end up with an entire region where, after the pulse of $v$, all cells "commit" simultaneously to a high $u$ state, corresponding to discrete somite formation. Once this transition has occurred, $v$ is again $0$ everywhere, $u \sim 1$ for a more extended region, and $u=0$ otherwise, so that this process can start again.

In the initial model, it was assumed that the transition happens in the entire region where $\chi_u=1$ because of fast diffusion. What sets the size of the block then is the time $\tilde t$ for $v$ to activate $u$ everywhere in this region, and the size of the block of the activated bock of cells then is $c \tilde t$. But if diffusion is not fast enough or the region where  $\chi_u=1$  is too big, the $v$ pulse will propagate from $x=0$ and activate cells in a more localized region. The size of the pattern thus is a complex function of all parameters, including diffusion (see different examples in Fig \ref{fig:cellcycle} D-G, and attached Notebook).

In summary, this model allows for the formation of somites by the generation of periodic pulses close to the anterior PSM boundary, synchronously expressed in a field of cells, triggering commitment to somite fate (modeled via a bistable variable $u$). Notice that the dynamics of $u$ thus is very similar to the variable $z$ in the Cooke and Zeeman model, Eq. \ref{cusp}. $v$ plays the same role at the pulsatile clock in Eq. \ref{CW_maths}, interpreted as the cell cycle. It is also worth comparing how primary waves (in the Zeeman sense \cite{Zeeman1974}) are encoded in both models: in the initial Clock and Wavefront model, the $(t,p$) potential associated to the cusp catastrophe was creating an emerging transition from a bistable to a monostable region, while here, a similar primary wave is created by the region II to region III transition, Eq.  \ref{H_uv}, when $v$ is activated and ensures that $u$ is monostable. In other words, the primary wave is defined by $\chi_v$.

The big difference comes from the dynamics of variable $v$. First, similar to Meinhardt's model, diffusion of $v$ is crucial to define the pattern (switching $u$). $u$ also shuts down $v$. This ensures coordination between the state variable and the clock, a possibility we alluded to at the end of the description of the clock and wavefront model.  It is also noteworthy that the oscillator is in fact not explicitly modeled in this $u,v$ model, and rather emerges as a consequence of the sliding window $\chi_v$ which creates a pulsatile window of expression of $v$ in region II.  So there is no explicit need for, say, a posterior oscillation (in the region I) like in Meinhardt's model. It is, in particular, not entirely clear how the differential sliding window would practically connect to the phase of the cell cycle oscillator, and how the initial proposal that heat shocks disrupt specific phases in the cell cycle would be accounted for in this model.

\newpage

\chap{Phase models}
\label{sec:Phase}
On the one hand, the vast number of molecular players implicated in somitogenesis is daunting from a theoretical standpoint, since it is not clear how and what to model in a predictive way. On the other hand, the phenomenology of the segmentation behavior still is relatively simple, with waves of genetic expression sweeping from posterior to anterior, leading to patterning. This suggests first following the spirit of classical models described in Section \myref{sec:pioneer} to focus on rather phenomenological models, not specifically tied to actual genes.  Similar issues arise for oscillators in neuroscience and physiology and motivated the development of a  "phase-based" approach to describe more explicitly the segmentation clock dynamics, which we briefly summarize here (see also Appendix, treatments of various complexities can be found in \cite{Pikovsky2003,Kuramoto,Winfree,Izhikevitch2007}). In line with our previous observation that the clock seems to be tightly connected to Zeeman's primary wave of differentiation, one challenge is to tie those phase descriptions to both clock stopping and patterning.

\chapsec{From chemical equations to phase}

Consider a (biological) oscillator described by equations in the space of its components, e.g. mRNA/protein concentrations :

\begin{equation}
\frac{ d\mathbf{X}}{dt}=F(\mathbf{X}) \label{ODE}
\end{equation}

Given some initial conditions, the system relaxes to the limit cycle, which is a closed curve in the space of concentrations. The position on this curve can thus be indexed by a single parameter. We define the phase $\phi(t)$ of an oscillator by :

\begin{equation}
\frac{ d\phi}{dt}=1 \label{phase}
\end{equation}

and express the phase $\phi$ modulo $T$, where $T$ is the period of the oscillator, to account for the fact that the system is periodic (notice there are other conventions, i.e.  one can rescale time so that the period is either $1$ or $2\pi$). In this formalism, the phase of an oscillator is nothing more than a (rescaled) time variable on the limit cycle. For instance, if we rescale time so that the period of the oscillator is $2\pi$, phase $\pi/2$ means that the oscillator is at the $1/4$ of its cycle, phase $\pi$ means that the oscillator is at half its cycle, and phase $2\pi=0 \quad \mathrm{mod} \quad  2 \pi$ is the initial phase corresponding to the full period. Notice that phases also correspond to positions in the space of protein concentrations, i.e. $\phi(t)=\phi(\mathbf{X}(t))$ where $\mathbf{X}(t)$ is the value of protein concentrations at time $t$ on the limit cycle.

There are now two important observations from the modeling standpoint :
\begin{enumerate}
\item It is possible to extend the definition of phase for points \textit{outside}  of the limit cycle. Imagine for instance that at a given time $t_p$ you first perturb the system, e.g. by making a change $\mathbf{X}(t)\rightarrow \mathbf{X}(t) +  \Delta \mathbf{X}$, then let the oscillator relax. Eventually, the system will go back to the limit cycle, where you have defined a phase using Eq. \ref{phase}. But then, since the phase is nothing more than time,  from this phase on the limit cycle, you can go back in time on the trajectory you have just followed to define a phase corresponding to the initial condition $\mathbf{X}(t) +  \Delta \mathbf{X}$ at time $t_p$. This way, you can define a phase for all vectors $\mathbf{X}$, even outside the limit cycle, defining so-called "isochrons", or lines with identical phases. 
\item for any limit cycle oscillator, the amplitude is stable (so not easily changed by a perturbation) while the phase is neither stable nor unstable  \cite{Pikovsky2003}. Thus, weak perturbations of an oscillator only change its phase.
\end{enumerate}

Those two properties essentially mean that, for many purposes, the behavior of a (perturbed) limit cycle oscillator can be entirely captured by its phase behavior, which remarkably allows us to go from complex dynamics in a high dimensional system to only one phase variable for a given oscillator. For instance, imagine two coupled oscillators, then if their coupling is relatively weak, the perturbations induced by each oscillator onto one another will stay close to the limit cycle in the initial mRNA/protein space, and one can use the isochron theory to translate any coupling into effective phase equations. While it is clear that this substitution is not trivial, and computations of phase responses can be quite tricky (and has to be done numerically for a more complex system, see Appendix A for the Adjoint method and Malkin theorem), some generic simplifications also arise from the periodicity of the coupling and symmetry in the equations \cite{Kuramoto} (see Appendix). One can then use such formalism to study all kinds of effects, from entrainment to changes of the intrinsic period. In summary, if the limit cycle is not too perturbed and the coupling not too strong, all properties of the oscillators  under various hypotheses can then be defined in terms of the phase, which allows for  more powerful treatment, here in the context of segmentation clock/waves.

\chapsec{Clock and Unclock}

It is a good place to briefly mention nuances on the `clock' notion for biology, introduced by Winfree in the context of the circadian clocks \cite{Winfree75}. A `simple clock' is characterized by the fact that
\begin{quote}
"its possible states can be arranged in a recurring sequence, in which each state induces the next in a repeating cycle. The "state space" is a circle and there is no state which is not represented in that space. The clocks of home and industry are of this kind; the only variable quantity in the clock is the angular position of its meshed gears, and to get to a new position it must pass forward or backward through all intermediate positions."
\end{quote}

An example with discrete states is shown in Fig. \ref{fig:clockunclock} A. Phase descriptions introduced above correspond to a continuous limit of such clock definition. Winfree notices that neuronal oscillators represent potential biological examples of such clocks because they can by and large be modeled by one variable that `resets' at the end of its cycle (see Appendix A).

Winfree contrasts this with multiple `unclocklike' behaviors. The first example he gives is limit cycles (possibly in a high dimensional space) Fig. \ref{fig:clockunclock} B. The clearest difference is in the nature of states: in a clock, states are phases, while in a limit cycle, even extended with isochron theory \cite{Winfree}, there are `phase-less' states, such as the center of the cycle (intersection of green and blue nullclines in Fig. \ref{fig:clockunclock} B). One way to explicitly show the `unclock' behavior is to induce a (strong) perturbation of the limit cycle to reach this point. However, finding the `right' perturbations towards the phase-less point can be challenging. An indirect way to prove the existence of such point rather is topological, to get so-called Type 0 phase resetting in response to strong perturbations, which manifests itself by a sudden 2 $\pi$ jump in Phase Resetting Curves (see Appendix for definition) \cite{Winfree75,Winfree,Glass2001}. It sounds a posteriori surprising that limit cycles are not considered as clocks, but as we will see in this review, there are indeed very relevant  differences between phase models and explicit ODE-based models in the segmentation context.

Another possible `unclocklike` behavior in response to the perturbation is a long-term memory effect in response to a perturbation. The harmonic oscillator is not a clock for this reason: it is defined by its amplitude and its phase, and as energy is injected, its amplitude increases and does not return to its initial value. Similarly, any transient stimulus that would impact long-term properties of an oscillator (e.g. the period) excludes the simple clock, see e.g. Fig. \ref{fig:clockunclock} C. Surprisingly, such effects have been observed in multiple biological systems, see e.g. fly circadian clocks \cite{Winfree75}, or for `period' response in human clapping experiments \cite{Thomson2018}.

In the context of somitogenesis, one should also distinguish the individual, cellular oscillators, from the global periodic one. When talking about "the segmentation clock", it is important to specify which level is actually studied, because the `clock' or `unclock' like properties of the oscillatory behavior at the embryonic level might not be the same as the properties at the single cell level (even though they are of course related).

\bnormf
\includegraphics[width=\textwidth]{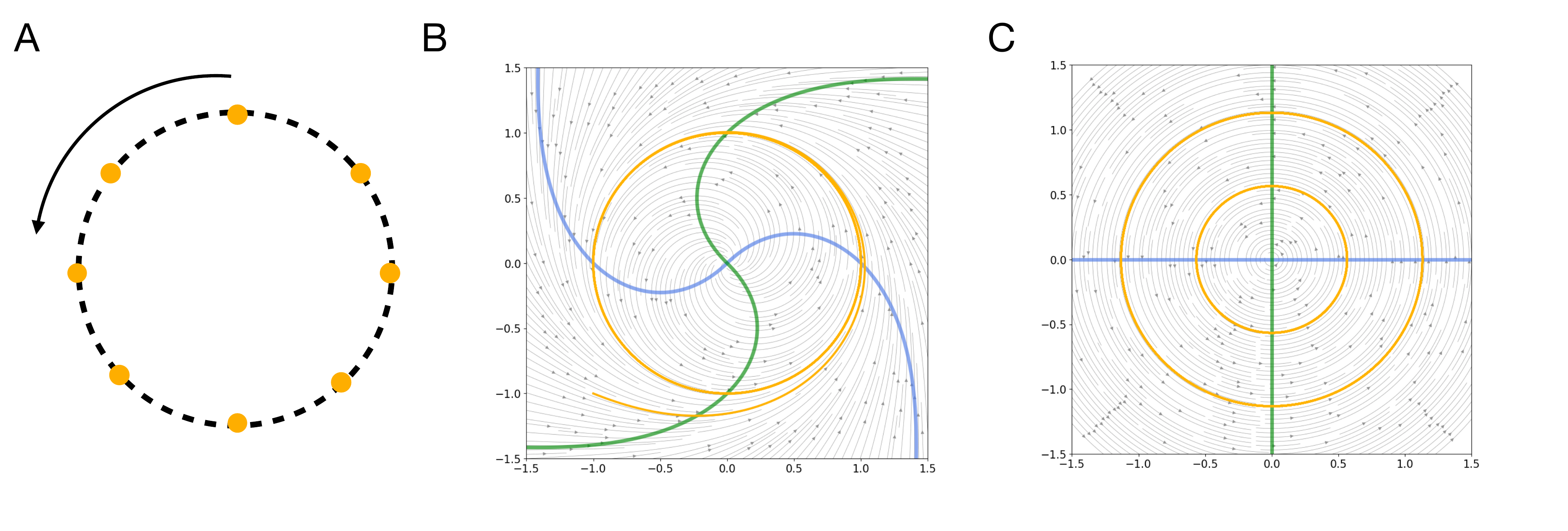}
   \caption[Clock and Unclock]{ Clock and Unclock (A) The standard Clock model: a system where states (discrete or continuous) are visited periodically, in a circular way. States are represented by orange dots. (B) Unclock behavior: a system with a limit cycle (orange trajectory) is not a clock, in particular, because there is a phase-less point at the center (intersection of green and blue nullclines). (C) Unclock behavior: even the simplest harmonic oscillator is not a clock because there are multiple possible cycles (orange lines). In particular, the amplitude of the oscillation can be changed by a transient stimulus. }\label{fig:clockunclock}
\enormf

\chapsec{The Lewis Phase Model (LPM)}
\label{LPM}

We now have the theoretical framework to discuss phase-based models, aiming at phenomenologically describing the dynamics of somitogenesis.
The first historically important model was in fact introduced in the  appendix of  the Palmeirim \etal paper describing the segmentation clock for the first time \cite{Palmeirim1997}. There, Julian Lewis briefly described a simple mathematical model recapitulating the observations of \textit{c-hairy} behavior. This model is important for at least three reasons : 
\begin{itemize}
\item it is the first example of a "phase-model" for segmentation clock, tying clock, pattern formation, and "wavefront"
\item Because of this, many subsequent models of the segmentation clock can be related to and contrasted with this initial model, as will clearly appear in the remainder of this tutorial
\item similarly, many standard observations and calculations (and Ansatz)  for the segmentation clock can be illustrated first on this model
\end{itemize}

For this reason, this model deserves a complete section and analytical study that we perform here. In the remainder of the text, this Julian Lewis Phase Model will be subsequently abbreviated in LPM.

 The LPM assumes that each cell is a phase oscillator, with a local phase $\theta(x,t)$ ($x$ being the position of the cell, increasing $x$ corresponding to the posterior to anterior direction, and $t$ the time). The phase also depends on a variable called maturity $m$, taken to be, in rescaled units, $m=x+t$. In particular, notice that cells of given, constant, maturity $m=m_0$ define a moving front $x=m_0-t$ of speed $-1$. Then the assumption is that the instantaneous angular velocity/frequency of the clock is a simple function of $m$, namely that 

\begin{equation}
\dot \theta = r(m(x,t)) \label{CWPalmeirim}
\end{equation}

where $r(m)$ is a smooth function so that $r(m)\rightarrow 1$ for $m<<0$ (corresponding to the posterior of the embryo) and  $r(m)=0$ for $m >>0$ (corresponding to the anterior of the embryo). A kymograph of this model is shown in Fig. \ref{fig:LPM}A. In this model, each cellular oscillator is slowing down with time, but since $m=x+t$, this slowing down is time and space-dependent, giving rise to a pattern, with a differentiation front moving from top to bottom.

\bnormf
\includegraphics[width=\textwidth]{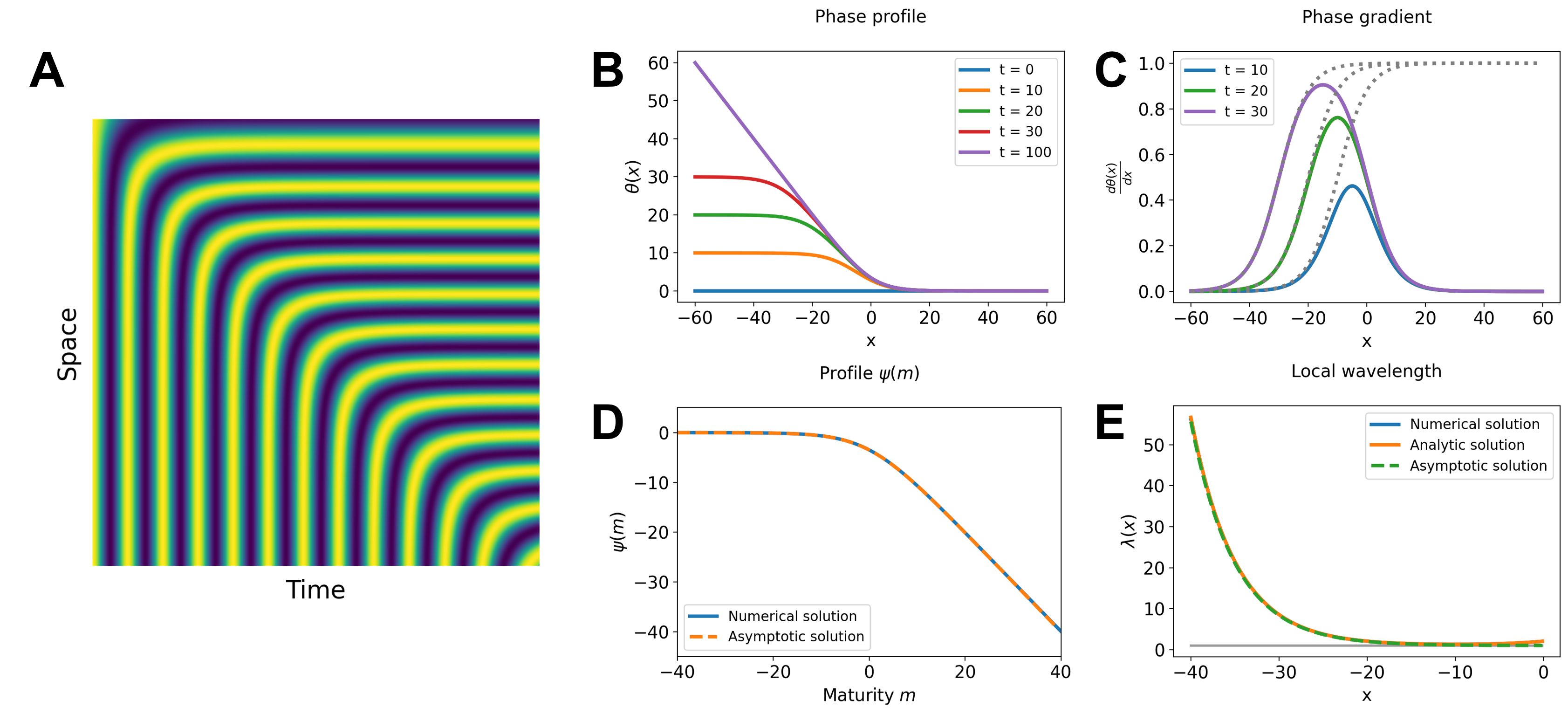}
\caption[Lewis Phase Model]{Simulation and calculations for the Lewis phase model. Antero-posterior axis $x$ increases from bottom to top (A) Kymograph of $\cos(\theta)$ from a numerical simulation with $\sigma = 10$. A spatial region from $-L$ to $0$ (initial front position) is shown. (B) Analytically calculated spatial phase profile $\theta(x)$ at different time points. (C) Analytically calculated phase gradient at different time points and comparison to the approximation for $x<<0$ described in the text. (D) Function $\psi(m) = \phi(m,t) - t$, describing the phase profile for long enough times $t$. For the numerical solution, the phase in the moving frame of reference $\phi(m,t)$ was taken from the diagonal of the kymograph. (E) Divergence of the wavelength $\lambda(x)$. Analytic and numerical solutions calculated from the phase gradient; asymptotic solution calculated with the asymptotic $\psi(m)$.}\label{fig:LPM}
\enormf

To make this intuition more quantitative, it is helpful to see what happens for simple forms of $r$ where all calculations can be done analytically. In the initial Palmeirim paper, no calculation is performed and the model is only simulated, but it turns out it is completely analytic so it is worth solving here explicitly to explain in detail what happens.

Lewis first assumes that 
 
 \begin{equation}
 r(m(x,t))=\frac{1-\tanh(m)}{2}=\frac{1}{1+\exp{2(x+t)}} \label{simpler}
 \end{equation}
  one gets by direct integration (assuming $\theta(x,0)=0$ for all $x$)

\begin{equation}
\theta(x,t)=\frac{1}{2}\left(t+\ln \frac{e^x+e^{-x}}{e^{x+t}+e^{-(x+t)}} \right) \label{CWtanh}
\end{equation}

We then have the following interesting limits :

\begin{itemize}
\item for $x \rightarrow -\infty$, $\theta(x,t)\sim\frac{1}{2} \left(t+\log{e^{-x+(x+t)}}\right)\sim t$, indicating that the clock is oscillating with angular velocity $1$ in the "posterior" region. Conversely, for $x \rightarrow +\infty$ we get $\theta(x,t)\sim\frac{1}{2}\left( t+\log{e^{x-(x+t)}}\right)\sim 0$, so that the phase is stationary in the "anterior" region.  
\item for $t \rightarrow + \infty$ and $x$ negative enough  , $\theta(x,t)\sim\frac{1}{2} \left(t +\log{e^{-x-(x+t)}}\right)\sim |x|$, indicating a stationary phase linear in $|x|$ for long times (Fig. \ref{fig:LPM}B). This indicates  that after a  long enough time, a periodic stationary pattern of size $L= 2\pi$ is reached.
\end{itemize}

A very interesting feature of this model is noticed in \cite{Palmeirim1997}:

 \begin{quote} 
"Note that the model correctly mimics the curious way in which successive waves of \textit{c-hairy1} expression, initially broader than one somite, appear to move rapidly forward in the presomitic mesodern, squeezing up together as they go."
 \end{quote}
 
Such behavior is not predicted by any of the classical models in \myref{sec:pioneer} and thus is accounted for and explained for the first time by the LPM.

The local wavelength at position $x$ is given by the inverse of the phase gradient i.e.
 
 \begin{equation}
 2\pi \lambda(x,t) =\left|\frac{\partial \theta(x,t)}{\partial x}\right|^{-1}=2\left|\tanh(x+t)-\tanh(x)\right|^{-1}
 \end{equation}
 

 Again, from this equation it is clear that for $x$ negative enough and $t\rightarrow + \infty$, we have  $ 2\pi \lambda(x,t) = 2|\tanh(x+t)-\tanh(x)|^{-1}\simeq 2/(1 -(-1))=1$, so that at stationarity the pattern is periodic with size $2\pi$. Conversely, fixing $t$ but in the limit of $x\rightarrow -\infty$ the wavelength clearly diverges with both $\tanh$ going to $-1$,  indicating the (infinitely) broad wave pattern observed corresponding to all oscillators oscillating in phase in the posterior-most region. The transitory region between those two domains corresponds to the behavior described in the quote above. To further  study this transition,  in a region with $x$ very negative (i.e. a region initially in the oscillating phase), we have $\tanh(x)=-1$ which gives :
  \begin{equation}
 2\pi \lambda(x,t) =2\left|\tanh(x+t)+1\right|^{-1}=1+e^{-2(x+t)}
 \end{equation}

 We clearly see from this equation how the wavelength at position $x$  goes from $+\infty$ to $1/2\pi$ as a function of time, and conversely, we see that for a given time, along the $x$ axis, the wavelength diverges exponentially in space with a characteristic length of $1/2$. This length corresponds to the size of the transitory region mentioned above, between the oscillating region and the static region. This characteristic length can be further adjusted by modulating the function $r(m)$, e.g we can generalize by taking  
 \begin{equation}
     r(m)=\frac{1-\tanh( m/\sigma)}{2},
     \label{eq:simpler_wsigma}
 \end{equation}
which gives

   \begin{equation}
 2\pi \lambda(x,t) =2\left|\tanh\frac{x+t}{\sigma}+1\right|^{-1}=1+e^{-2(x+t)/\sigma} \label{wavesigma}
 \end{equation}
 so that as $\sigma$ increases (corresponding to a shallower gradient), the region of divergence of wavelength increases. 
 
 In fact, we get in both cases that 
 
  \begin{equation}
 2\pi \lambda(x,t) =(1-r(m))^{-1} \label{wavelength}
 \end{equation}
 
 so that (the inverse of) the wavelength is a direct measurement of the frequency gradient. This is a very general result that can be derived directly by making a change of variable in equation  \ref{CWPalmeirim} to express the phase in the moving frame of reference $\phi(m,t)$  to get
 
 \begin{equation}
\frac{\partial \phi}{\partial t}+\frac{\partial \phi}{\partial m}=r(m) \label{movingframe}
\end{equation}
 
This change of variable simply adds an advection term $\frac{\partial \phi}{\partial m}$.  Since we expect that $\phi(t,m=-\infty)$ is proportional to $t$ by construction (to ensure  spatially uniform oscillations in this region), after some transitory time, we can hope to find a solution satisfying the following Ansatz :
\begin{equation}
\phi(m,t)=t+\psi(m) \label{Ansatz}
\end{equation}
which nicely separates $t$ and $m$. Such Ansatz is not the full solution of the differential equation with the imposed initial condition: rather, it is an asymptotic behavior that we expect to reach after a transitory time, that will be observed   experimentally  if we wait long enough.



This Ansatz allows us to explicitly solve the full phase profile for any form of $r(m)$. Coming back to Eq. \ref{Ansatz}, we then have $\partial \phi/\partial t=1$ and

 \begin{eqnarray}
\frac{d \psi}{d m} &=&r(m)-1 \label{psiappendix}  
\end{eqnarray}

which can be explicitly integrated to get the value of $\psi(m)$, entirely solving the problem.
$\psi$ represents the stationary phase profile in the moving frame of reference. In particular $\psi$ captures all information on the spatial dependency on the system. So any snapshot property measured experimentally at a given time (e.g. phase gradient in the PSM) are related to $\psi$. For instance, Eq. \ref{wavelength} just expresses the fact that  wavelength is given by 
  \begin{equation}
 2\pi \lambda(x,t) =\left|\frac{\partial \psi}{\partial m}^{-1}\right| 
 \end{equation}

Second,  $r(m)-1$  goes from $0$ to $-1$ as $m$ varies from $-\infty$ to $+\infty$, so that, as we integrate it, $\psi(m)\sim 0$ when $m=-\infty$ to $\psi(m)\sim-m$ when  $m=+\infty$. $\psi(m)=0$ is the region where $\phi=t$, i.e. the oscillating region. $\psi(m)=-m$ gives $\phi(x,t)=t-m=-x$: this is the region where the oscillator freezes to give the pattern. Notice those behaviors are very general and do not assume anything on the shape of $r$ except it asymptotic limits.

 Third, since $\psi$ depends on $m$, this means that the stationary phase profile defined by $\psi$ simply moves with constant speed in the original frame of reference defined by $x$. So $\psi$ captures the transition zone, moving from the region with stationary phase toward the oscillatory region. Fig. \ref{fig:LPM}D represents the shape of $\psi$ for $r(m)$ given by Eq. \ref{eq:simpler_wsigma}.

 Specific forms of $r$ only change the phase profile in the transition zone, as long as the asymptotic values $0$ for $m\rightarrow +\infty$ and $1$ for  $m\rightarrow -\infty$ are fixed. So waves are observed, emerging from the oscillating zone, and traveling from posterior to anterior before stabilizing (notice that  those waves are purely kinematic, they are not corresponding to the propagation of any signal, rather they purely come from the change of frequencies induced by $r$). Waves are locally moving with speed $v=\frac{\partial \phi}{\partial t}/\frac{\partial \phi}{\partial x}=\exp{-2(x+t)/\sigma}$ so that $v=\frac{r}{1-r}$. In particular, the wave has infinite speed in the posterior (corresponding to the infinite wavelength) and stabilizes as they move toward the anterior.

 It is worth computing how many waves are traveling, which is a  simple biological observable (we reproduce with adaptations here a calculation done in \cite{Murray2011}, for which the wave pattern exactly matches the Lewis Phase Model as will be discussed in section \myref{shock}).

One can easily compute the total phase gradient between the oscillating region and the static region, and the number of traveling waves simply is the gradient divided by $2\pi$. We express the total phase difference between the tail bud and position $m_0$ :
   \begin{equation}
 \Delta \psi_0 = -\int_{-\infty}^{m_0} \frac{\partial \psi}{\partial m}\dd m  =\int_{-\infty}^{m_0} \frac{\dd m}{1+\exp{-\frac{2m}{\sigma}}} \label{phase_shift}
  \end{equation}
 
 The problem is that this integral diverges for $m_0\rightarrow \infty$.

But of course, no embryo is infinite, even though it helps simplifying calculations. So the realistic way to perform this calculation is to put a cut-off on positive $m$, since the PSM is of finite size $L$. A reasonable assumption is to assume that $m=0$ (where $r=1/2$) corresponds to mid-PSM, meaning that the "real" PSM starts at $m=-L/2$ and ends at $m=L/2$. One then gets :

    \begin{equation}
 \Delta \psi_0 = \int_{-L/2}^{L/2} \frac{\dd m}{1+\exp{-2m/\sigma}}=\frac{\sigma}{2}\ln\frac{e^{L/\sigma}+1}{e^{-L/\sigma}+1}\simeq L/2 \label{number_waves}
  \end{equation}
  
  assuming $L/\sigma$ large enough (meaning that the PSM length is bigger than the typical length scale for frequency change). The number of waves is $\Delta \psi_0/2\pi$. Notice that $2\pi$ is the length of the pattern in rescaled units, which means that going back to actual physical units, the number of propagating waves is :
  
    \begin{equation}
  N\simeq \frac{L}{2S} \label{n_waves}
    \end{equation}
So from this model the number of waves simply is the PSM length expressed in units of somite size, divided by 2. This is a remarkably simple prediction that works well for many  organisms. For instance, as pointed out by \cite{Murray2011} using data from \cite{Gomez2008}, for zebrafish, the PSM length is $48$ cellular diameters, a somite length is $7$ to $8$ diameters so that the number of traveling waves is around $3$, which visually fits (dynamical) data. For snake \cite{Gomez2008}, $\frac{L}{2S}\simeq 12$, which approximately fits the number of traveling waves observed. 

Intuitively this scaling proportional to $L$ comes from the fact that if $L>>\sigma$, the integrated quantity $\frac{\dd m}{1+\exp{-2m/\sigma}}$ is essentially equal to $1$ for $m$ higher than a few $\sigma$, and so the phase difference with respect to the oscillator basically accumulates almost linearly in time (and space) when the frequency of cells is very small, between positions $m>\sigma$ and $m=\frac{L}{2}$. In other words, the $L/2S$ scaling merely reflects the fact that cells get rather slow over a significant portion of the PSM (here half of it), so that, maybe not surprisingly, the phase difference with respect to the oscillatory region essentially is proportional to the length of the region of significantly slower oscillation. Another way to frame this result is that, if the clock frequency was suddenly dropping as cells enter the PSM, we would expect to have at most $L/S$ waves, because this simply is the number of somites that should come out of a PSM of length $L$. In reality, the slowing down takes some time to be significant, of the order of half the PSM, so the effective phase shift will be of the order of $L/2S$. So we expect this result to be in fact quite generic; as an illustration, a different calculation is also performed in Supplementary Box 2 of \cite{Gomez2008}, accounting for the exponential growth of the PSM but in the end leads to the  exact same result (see Appendix Section \myref{Appendix:growth}).

Those calculations show  that despite its simplicity, the LPM  includes nontrivial  and observable features not present  in the initial clock and wavefront model: both a spatial and temporal dependencies of the clock, which, as cleverly noticed by Palmeirim \etal, explain in a simple way the observed wave pattern. In fact, one could almost say this feature is contradictory with the initial clock and wavefront model, which postulates that the clock and the wavefront are two independent variables. In the LPM model, for finite $\sigma$,  there clearly is nothing such as a discrete wavefront, rather the clock continuously stops, giving rise to traveling waves in the transitory regime. Also, the oscillators eventually stop and form a spatial pattern. Both those aspects are more reminiscent of Meinhardt's model, and indeed suggest that the segmentation process should be considered in its continuity from oscillation to stabilization to be fully understood. In other words, it is important to consider segmentation waves and not only segmentation oscillators. 

 Two differences with Meinhardt's model are nevertheless worth pointing out:
 
 \begin{itemize}
 \item The LPM  is purely cell autonomous, in the sense that the behavior of individual cells is entirely prescribed by the dynamics of variable $m$, which is externally controlled. In Meinhardt's model, the slowing down of the clock depends on the interactions of different oscillators and as such is an integral part of the model. One can not exclude that the maturity $m$ itself is coming from the interactions between cells or within cells, which are not described in the LPM model. We will see an explicit example of this in Section \myref{shock}
 \item The final state of the system $\dot \phi=0$  is rather strange from a dynamical systems standpoint: this literally is a frozen oscillator. As such the model does not converge to a well-defined attractor (contrary to both Meinhardts' model and the catastrophe-centric view from the initial Clock and wavefront framework). One could for instance imagine in this model that even after stopping, the cellular oscillators could be restarted in various ways by taking control of the maturity $m$, or simply that local phases could be shifted after stopping by some other processes.
 \end{itemize}

 \chapsec{Flow-based phase models}

The LPM describes the PSM as a field of passive cells, swept by a frequency gradient. But this picture does not fit the reality of growing embryos, where progenitors divide in the tail bud to generate a flow of PSM cells moving from posterior to anterior relative to the tail bud.

In 2001, Jaeger and Goodwin \cite{Jaeger2001} accounted for this to propose a model based on cellular oscillators recapitulating those developmental features.  In this model, cells exit  the progenitor zone to enter the PSM, and their frequency $\frac{2\pi}{T(\alpha)}$ depend on their `age', $\alpha$, defined as the time from their exit from the progenitor zone. The connection to the LPM model can be made explicit by noticing that, in the moving frame of reference (coordinate $m$), one has the following frequency profiles  for $m>0$

\begin{equation}
r(m)=\frac{2\pi}{T(m/v)}
\end{equation} 
 
 since $\alpha=m/v$ is the age of the cell at relative position $m$, assuming growth occurs at speed $v$, and rescaling time units so that $ r(0)=1$ corresponding to the taibud. A very similar model has been proposed by Kaern \etal  \cite{KAERN2000} within a more general framework of "Flow-Distributed Oscillator" (FDO). This framework explicitly models a flow of oscillating cells injected in a growth zone, so that the phases of cells depend in a simple way on their age from the time of injection. One of the merits of those models is to more explicitly connect the clock, the flow, and growth, thus providing some explanations on observed phenotypes to explain some experimental data.

  For instance, if one assumes that the period of a local oscillator depends on its age, one can simply explain the observation that, if part of the PSM is (artificially) reversed, it will pattern in a kinematic way, with an inverted pattern relative to the normal one. Kaern \etal  also use their flow model to offer an explanation of the periodicity of heat shock phenotypes (that led to the cell-cyle model). In the initial cell cycle model, to observe a periodic pattern of disruption of heat shock, one has to assume that there is a gradient of cell cycle phase within the PSM covering several cycles, so that there are several clusters of cells at the heat-shock sensitive phase $\phi_S$ separated by one cell cycle length (corresponding to 6-7 somites in the chick). Kaern \etal  observe that this explanation is unlikely because one would then need a very long PSM to have up to 4 periodic anomalies due to a single heat shock as observed experimentally. They rather assume that cells in the progenitor zone have a uniform distribution of cell cycle phases and that the heat shock delays mitosis. So at the moment of the heat shock, only a fraction of the progenitors are about to go through mitosis, at phase $\phi_M$ : heat shock delays them and as a consequence fewer cells divide, and thus the PSM growth rate is impacted. Right after the heat shock, progenitor cells slowly recover so that the growth rate transiently increases, before coming back to the normal growth rate when all impacted cells have gone through mitosis and other cells normally divide. In the absence of any coupling/resetting, and assuming the cell cycle progresses uniformly in all cells, the progenitor cells delayed by the first heat shock remain delayed,  so one would see again a decrease in the growth rate at the next cycle, explaining how the defect can be repeated. In reality, one would expect some coupling between cells and noise so that eventually   the initial uniform distribution of cell cycle phases in the progenitor is re-established, but if this takes some time, one expects to observe an oscillatory growth rate of the PSM post heat shock, with a period similar to the period of the cell cycle, thus explaining the periodicity of the anomaly. One issue with this model is that it does not account for the very first  anomaly, which is clearly due to cells already in PSM (and not in progenitor state): Kaern \etal suggest that the initial heat shock also impacts the frequency of those cells, thus implying a direct coupling between segmentation clock and cell cycle.

\chapsec{Delayed  coupled models}


 We now consider elaborations of the LPM  accounting for coupling delays between oscillators (e.g. due to Notch/Delta signaling), presented in  \cite{Morelli2007,Morelli2009,Herrgen2010}.  A line of oscillators (index $i$) is considered and the general time evolution of the phase $\theta$ is given by

\begin{equation}
\frac{d \theta_i}{dt}=\omega_i(t)+\epsilon_i(t)/2 \sum_k \sin[\theta_k(t-\tau_i(t))-\theta_i(t)]+\zeta_i(t)
\end{equation}

$\omega_i(t)$ is a time/space dependent frequency, playing the exact same role as  $r$ in Eq/ \ref{CWPalmeirim}. Morelli \etal  \cite{Morelli2007,Morelli2009} consider the form

\begin{equation}
\omega_i(t)=\omega_\infty(1-e^{-(i-vt)/\sigma})
\end{equation}
where $v$ is speed of the moving frequency gradient, so that $i-vt$ plays the exact same role as the maturity $m$ in Eq. \ref{CWPalmeirim}, except the spatial axis is flipped so $i\rightarrow +\infty$ corresponds to posterior here and very negative $i$ to anterior. For simplicity here, we first count length in discrete units, but Morelli \etal introduce a unit conversion factor $a$. $\sigma$ quantifies the spatial scale of changes of the frequency, similar to Eq. \ref{wavesigma}. Notice that as $i \rightarrow -\infty$, $\omega_i$ would become very negative, so one should rather  assume a cut-off for $\omega_i$, for instance, one can assume that $\omega_i$ is $0$ for $i<vt+f$, with $f\geq 0$ which gives a frequency at the front  

\begin{equation}
\omega_f=\omega_\infty(1-e^{-f/\sigma})
\end{equation}

Similarly, Morelli \etal define a cut-off maximum frequency $\omega_N$ in the tailbud, defined at $i=N+vt$ so that :

\begin{equation}
\omega_N=\omega_\infty(1-e^{-N/\sigma})
\end{equation}

So the PSM is entirely contained in the region $vt<i\leq vt+N$, and its length is $N$ discrete units, or using a scaling factor $a$, a length $L=Na$ in physical units.

The major difference with the LPM relies in the addition of coupling between cells. The sum over $k$ accounts for nearest neighbors interactions, with Kuramoto-like coupling \cite{Kuramoto}. Importantly, there is a delay in the coupling ($\tau_i(t)$), which depends on the local oscillator $i$. A noise $\zeta_i(t)$ is also added for generality, but for now, let us put it to $0$.

The first step is, similar to the LPM, to put oneself into the moving frame of reference, i.e. calling $j=i-vt$, there is a fixed frequency gradient relative to the moving frame of reference  $\omega_j=\omega_\infty(1-e^{-j/\sigma})$. We now define the phase $\phi_j =\theta_i(t)$ in the moving frame of reference, which adds a (discretized) drift term proportional to $v$ in the equation to get (assuming now constant coupling and delays)

\bwt
 \begin{equation}
\frac{d \phi_j}{dt}=\omega_j+ v( \phi_{j+1}-\phi_{j})+\epsilon/2 \left[ \sin[\phi_{j+p+1}(t-\tau)-\phi_j(t)]+\sin[\phi_{j+p-1}(t-\tau)-\phi_j(t)] \right]\label{phidelay}
\end{equation}
\ewt
One subtlety here is that $\phi_j$ represents the phase at position $j$ with respect to the moving frame of reference, so does not represent the phase of an actual, physical oscillator in this model. Rather it is a fictitious oscillator representing the phase of the oscillators $\theta_i$ that we meet at position $i=j+vt$ as we follow the frame of reference moving with speed $v$. One of the major differences with the LPM is that one  has to introduce an integer $p=v\tau$, a length scale combining the delay and the speed of the front. This correction accounts for the fact that during the delay $\tau$, $\phi_j$ has physically moved by a distance $v\tau$  in the fixed frame of reference, and thus is coupled to its physical neighbors at $t-\tau$, which gives this non-local coupling for variable $\phi$, considerably complexifying the analysis. Let us insist again that each physical oscillator $\theta_i$ does not change neighbors, it is only because of the motion of the frequency gradient that the fictitious oscillator $\phi_j$ changes neighbors.

From there, the equation is not solvable in general but one can use classical Ansatz  similar to Eqs. \ref{Ansatz},\ref{psientrainment}, assuming the system globally oscillates and that associated to it one observed a relative phase profile $\psi$ such that

 \begin{equation}
\psi_j=\phi_j-\Omega t \label{AnsatzDiscrete}
\end{equation}

$\Omega$ is the (yet unknown) global (constant) frequency of the process, corresponding to the tail bud frequency. One gets by substitution into \ref{phidelay} 

\bwt
 \begin{equation}
\frac{d \psi_j}{dt}=\Omega-\omega_j- v (\psi_{j+1}-\psi_{j})+\epsilon/2 \left[ \sin[\psi_{j+p+1}-\psi_j-\Omega \tau]+\sin[\psi_{j+p-1}-\psi_j-\Omega \tau]\right] \label{psiansatz}
\end{equation}
\ewt
The big advantage of this equation is that all time dependencies have now been absorbed into the global frequency $\Omega$, so that the right-hand side of Eq. \ref{psiansatz} is a pure function of space (via the index $j$). We expect that at steady state $\frac{d \psi_j}{dt}=0$ which gives :

\bwt
 \begin{equation}
 \Omega=\omega_j+ v (\psi_{j+1}-\psi_{j})+\epsilon/2 \left[ \sin[\psi_{j+p+1}-\psi_j-\Omega \tau]+\sin[\psi_{j+p-1}-\psi_j-\Omega \tau]\right] \label{AnsatzDelay}
 \end{equation}
 \ewt
This equation couples the global frequency of the process to the phase profile $\psi_j$. Now the last step here, very similar to the LPM , is to consider what happens in the tail bud, corresponding to the limit $j \rightarrow + \infty$ (in the notation of \cite{Morelli2009}). There, we assume that the tail bud is reached and that there will be a homogeneous phase so that all $\psi_j$ are identical with frequency $\omega_N$. This gives the self-consistency relation 
 \begin{equation}
 \Omega=\omega_N-\epsilon \sin (\Omega \tau) \label{DelayOmega}
 \end{equation}

\bmarf
\includegraphics[width=\textwidth]{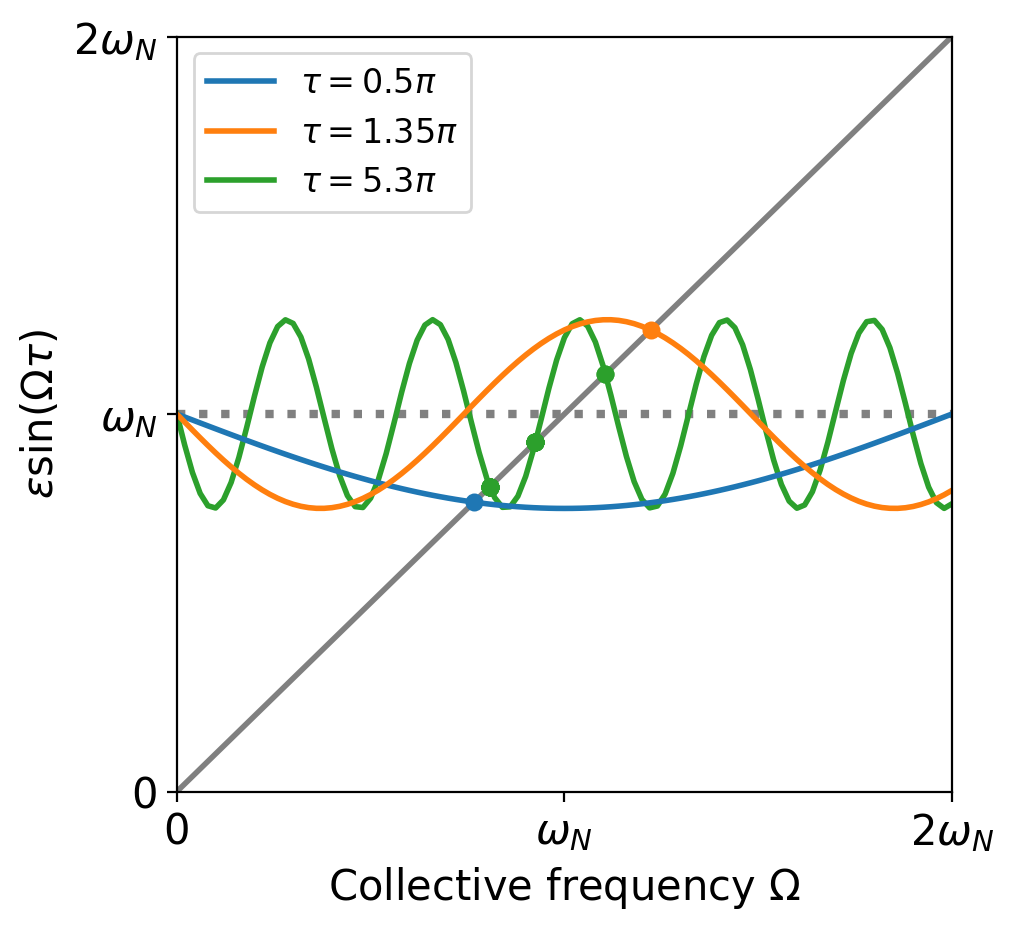}
\caption[Solving the transcendental equation \ref{DelayOmega}]{Solving the transcendental equation \ref{DelayOmega}. Graphically, the root(s) of the equation are found at the intersection of functions $y = \omega_N-\epsilon \sin (\Omega \tau)$ and $y = \Omega$. One way to numerically solve the equation is an iterative procedure: starting from an initial condition $\Omega^0$, calculate $\Omega^{k+1} = \omega_N-\epsilon \sin (\Omega^{k} \tau)$ until a desired precision is reached. If there are multiple solutions, the procedure has to be repeated with different initial conditions (i.e. uniformly spaced on the relevant interval). The roots found are indicated with dots; $\epsilon = 0.25$.}
\label{fig:transcendental}
\emarf
 Eq. \ref{DelayOmega} defines self-consistently a collective frequency of somitogenesis, which depends in a non-trivial way on the intrinsic tail bud frequency $\omega_N$, on the coupling strength $\epsilon$ and on the delay $\tau$. It is clear that if the delay $\tau$ or the coupling strength $\epsilon$ are very small, $\Omega \sim \omega_N$ corresponds exactly to the LPM. So significant deviations are expected for longer delays and stronger couplings. Equation \ref{DelayOmega} is a transcendental equation with typically several solutions (as illustrated in Fig. \ref{fig:transcendental}). Assuming $\epsilon$ still is relatively small, we see that for short delays $\tau$, the global frequency is lowered compared to $\omega_N$ (meaning that delays slow down the intrinsic oscillators). Increasing the delay flips the sign of the sine (for delays roughly bigger than $\frac{\pi}{\omega_N}$), there, the global frequency is increased, which means that the delays now speed up the intrinsic tail bud oscillators. From this theory, one expects a relative maximum change of period (compared to the intrinsic period $\omega_N$) of about $20\%$ in both directions (depending on the delay).

A continuous limit can also be obtained for this case, assuming a typical distance $a$ between oscillators, calling $x=ia/L$ and defining $\Psi(x) =\psi_i$ while taking the limit $a\rightarrow 0, N\rightarrow \infty$, keeping $L=Na$ constant.
While the discretized drift term $v (\psi_{j+1}-\psi_{j})$ simply becomes $v \Psi'(x)$, one must be careful with the delay terms. The reason is that there is an intrinsic length scale in the system $v\tau$ combining the speed and the delay, as can be seen with the $p$ terms in Eq. \ref{phidelay}. This distance should certainly not go to $0$ as we take the limit $a\rightarrow 0$. More explicitly  we have : 

\bwt
\begin{eqnarray}
\phi_{j+p\pm 1}-\phi_j &=&\Psi(x+v\tau\pm a)-\Psi(x) \\
&=& \Psi(x+v\tau\pm a)-\Psi(x+v\tau)+\Psi(x+v\tau)-\Psi(x)\\
&\simeq & \Psi(x+v\tau)-\Psi(x) \nonumber \\
&& \pm a \Psi'(x+v\tau)+\frac{a^2}{2}\Psi''(x+v\tau)
\end{eqnarray}
\ewt
From this, we can Taylor expand both $\sin$ terms in Eq. \ref{AnsatzDelay} close to $\Psi(x+v\tau)-\Psi(x)-\Omega \tau$ to get the continuous self-consistent equation for $\Omega$:

 \begin{eqnarray}
 \Omega &=&\omega(x)+ v  \Psi'(x)+\epsilon\sin[  \Psi(x+v\tau)-\Psi(x) -\Omega \tau] \nonumber \\
 & - &\frac{\epsilon a^2}{2} \sin[ \Psi(x+v\tau)-\Psi(x) -\Omega \tau] [\Psi'(x+v\tau)]^2 \nonumber \\
 &+& \frac{\epsilon a^2}{2}\cos[ \Psi(x+v\tau)-\Psi(x) -\Omega \tau] \Psi''(x+v\tau) \label{Omegacontinuous}
 \end{eqnarray}
 Notice the linear terms have canceled out so that the discretization has a very small influence on $\Omega$. This formula has been derived in \cite{Ares2012} for a coupling function $h$ more general than the $\sin$ coupling assumed here (the reasoning is completely equivalent).
 Again Eq.\ref{Omegacontinuous} is highly nonlocal, coupling phases at position $x$ and $x+v\tau$, so specific assumptions must be done to solve this equation. However, like the discrete case, one can simply assume that for $x\geq1$ the phase profile is flat to account for the tailbud dynamics so that all derivatives are $0$ and we get the exact same relation as before
 
  \begin{equation}
 \Omega=\omega(x=1)-\epsilon \sin (\Omega \tau)  \label{OmegaProof}
 \end{equation}

Also, just like one considered the limit $m\rightarrow \infty$ in the Lewis Phase Model, one can consider the equivalent limit $x\rightarrow \infty$ in this model where $\omega=0$. In the limit $a\rightarrow 0$ we get 
\begin{equation}
 \Omega = v  \Psi'(x)+\epsilon\sin[  \Psi(x+v\tau)-\Psi(x) -\Omega \tau]
\end{equation}
for which one obvious solution is $\Psi(x)=\frac{x \Omega}{v}$, indicating the formation of a pattern of size $S=vT$, as expected from dimensional analysis \cite{Ares2012}. Numerical studies of the delayed models (continuous and discrete) can be found in \cite{Morelli2009, Ares2012}.

Taking a step back, the phenomenology of this model, with traveling kinematic waves,  is qualitatively similar to the Lewis Phase Model, as can be seen from the analogous mathematical treatments of both models. The Ansatz used are very similar in spirit, allowing to derive a differential equation connecting the stationary spatial phase profile to the moving frequency profile: Eq. \ref{psiappendix} for the Palmeirim model, and the more complicated Eq. \ref{Omegacontinuous} for the delayed model. To be more quantitative in this comparison, it is useful to combine Eqs. \ref{DelayOmega}, \ref{Omegacontinuous}, \ref{OmegaProof} to write :
\begin{equation}
 v  \Psi'(x) = \omega(x=1)-\omega(x)+\epsilon \mathcal{H}\{\Psi\}
 \end{equation}
where $ \mathcal{H}$ is a complicated functional of the delay, $a$, $\Omega$ and of $\Psi$. Comparing with Eq. \ref{psiappendix} , we see that   $\omega(x=1)-\omega(x)$ is completely equivalent to $1-r(m)$ in rescaled units (a minus sign comes from the different directions of propagation of the fronts) and in particular, the dominating term does \textit{not} depend on the modified frequency $\Omega$. So all influences of delays, couplings, etc... captured in the $\mathcal{H}$ term  gives only a (small) perturbation of order $\epsilon$ for the phase profile compared to the much simpler Lewis Phase Model. Practically, it might be rather difficult to see a difference between the theory with and without coupling/delays by focusing only on the shape of the phase profile.

 Conversely, the clear new prediction of the delayed model is the influence of delay and coupling strength on the global frequency of the system, which is amenable to experimental verification. Indeed Herrgen \etal \cite{Herrgen2010} estimated parameters of the delayed model from zebrafish data. Mutants of Notch pathway were considered, as well as DAPT (a Notch inhibitor) treated embryos, and it appears that for those mutants, clock period and segment length are increased compared to WT, with a maximum change of around $10-20 \%$. This is consistent with the delayed theory in the sense that modifications of Notch pathway are expected to change $\epsilon$, and the magnitude of the change is consistent with the analytical theory. From there, saturating DAPT concentration was assumed to put $\epsilon$ to $0$, allowing the authors to estimate that the uncoupled period (corresponding to $\omega(x=1)$ in Eq. \ref{DelayOmega}) was $18\%$ higher than the observed segmentation period ($28$ min at $28 $ C for zebrafish). This gives a coupling strength $\epsilon=0.07 \pm 0.04 \mathrm{min}^{-1}$ and a minimal coupling delay $\tau=21 \pm 2 \mathrm{min}$. Notice that the delay is very close to the actual period of the oscillator.  It is remarkable that the effects predicted by delayed coupling theory are consistent in magnitude with experimental data. 


\chapsec{Doppler period shift }

\label{sec:Doppler}

The phase-based models presented so far assume the existence of a moving external frequency gradient, moving at speed $v$, which gives rise to a steady state moving phase profile (defined by Eqs.\ref{psiappendix} for the LPM, and self-consistent relation \ref{Omegacontinuous} for the delayed coupling theory).

It is, however, well-known that the size of the PSM  slowly varies as a function of time. This should have a direct observable consequence of speeding up the segmentation period. To see this intuitively, one can first notice that irrespective of the model considered, if the system at the embryo level is truly periodic, the period of formation of new segments is exactly the same as the period of the oscillation in the tailbud. But let us now assume that the PSM starts shrinking, in the sense that anterior cells stop oscillating earlier than expected for a fully periodic system. Then, since clock stopping means segment formation, the next segment is expected to form earlier than in a normal situation, thus the period of segmentation should decrease. Alternatively, from the front standpoint,  in the  periodic case,  the speed of the moving frequency gradient  (e.g. like in the LPM) is constant, so that an observer at the front will meet a peak of the wave with the same frequency as the tail bud oscillation. But if the front speeds up, an observer at the front meets a peak of the wave more frequently. This means that the frequency of segment formation is increasing, or that the segments will form with a shorter period than the tail bud oscillation. This is reminiscent of the Doppler effect in physics and has been studied experimentally \cite{Soroldoni14}. As we will see, biology is more complicated because it turns out that this Doppler effect is partially compensated. To understand what happens we start with a dedicated mathematical study  detailed and expanded from \cite{Jorg2014,Jorg2015} (see also Appendix B2) before we discuss experimental data and their implication.

\chapsubsec{Contributions to Doppler shift}

We start with the expression phase profile in the moving frame of reference with constant speed $v$ (equivalent to the LPM model,  Eq.  \ref{movingframe})

\begin{equation}
\frac{\partial \phi}{\partial t}+ v\frac{\partial \phi}{\partial x}=\omega(x,L=\bar{x(t)}) \label{doppler}
\end{equation}

Here, we have defined the PSM length $L=\bar{x}$, which can change as a function of time. We added a $L$ dependency in $\omega(x,L)$, which now defines the frequency of the oscillator at position $x$ when the PSM has size $L$.  This will account for possible changes in the frequency profile with PSM size.  For instance, in \cite{Soroldoni14,Jorg2015} it is assumed that   $\omega(x,L)$ is a function of the \textit{ratio} $x/L$, i.e. the frequency gradient $\omega$ scales with the PSM size. However, other functional forms might be possible so we will stay generic for now.
 The tail bud is at $x=0$.   For simplicity, we also assume that for $x<0$ all cells are effectively synchronized, in particular, $\left. \frac{\partial \phi}{\partial x}\right|_{x=0}=0$, and that $\omega(0,L)=\omega_0$ does not depend on $L$, defining a constant reference angular velocity/frequency. Notice that here, $x$ is the coordinate in the moving frame of reference (corresponding to variable $m$ in Eq. \ref{movingframe}). 

The observed angular velocity at the front is by definition :

\begin{equation}
\Omega_A=\frac{d}{dt} \phi(\bar{x}(t),t)=\left.\left(\frac{\partial \phi}{\partial t}+ \frac{d \bar{x}(t)}{dt} \frac{\partial \phi}{\partial x} \right)\right|_{\bar{x}(t)}\label{frequencyfront}
\end{equation}

and the period of segmentation is $2\pi/\Omega_A$. If $\bar{x}$ does not depend on time, we know from our previous Ansatz that $\left.\frac{\partial \phi}{\partial t}\right|_{\bar{x}}=\omega_0$ corresponding to the frequency in the tail bud (See e.g. Eqs. \ref{Ansatz} and \ref{AnsatzDiscrete}). But if $\bar{x}$ is changing, we can already see that there is one  added contribution coming from  $\frac{d \bar{x}(t)}{dt}$, which is mathematically similar to the Doppler effect in physics (see the end of this section for a discussion of the common points and  differences with physics). We will see that other contributions arise, in particular, there will be a correction to $\left.\frac{\partial \phi}{\partial t}\right|_{\bar{x}(t)}$.

Since the PSM size is changing, we can no longer use an Ansatz assuming there is a moving phase profile with constant shape. Fortunately, it is possible to explicitly integrate the equations, by considering a fixed position $z=x-vt$  in the static frame of reference. A cell at this position enters the PSM ($x=0$) at $t_0=-z/v$.
So coming back to the absolute position $z$ we get from Eq.\ref{doppler}:
\begin{equation}
\frac{\partial \phi (z,t)}{\partial t}=\omega(z+vt,\bar{x(t)})
\end{equation}

We directly integrate this equation keeping $z$ constant to get


\begin{eqnarray}
\phi(z,t) &= &\tilde\varphi(z)+\int_{t_0}^t \omega(z+vt',\bar{x}(t'))dt' \nonumber \\
&=& \varphi(t-\frac{x}{v})+\frac{1}{v}\int_0^{x}\omega\left( x',\bar{x}\left(t-\frac{x-x'}{v}\right)\right)dx' \label{phi_doppler}
\end{eqnarray} 
where $\tilde\varphi(z)$ is the initial phase of the cell as it enters the PSM and $\varphi$ the same function expressed as a function of $(t-\frac{x}{v})$. It is convenient to simplify this term by  immediately using  the boundary condition $\left. \frac{\partial \phi}{\partial x}\right|_{x=0}=0$, meaning a flat phase gradient in $0$. We have from \ref{phi_doppler}
\begin{equation}
0=\left. \frac{\partial \phi}{\partial x}\right|_{x=0}= -\frac{1}{v} \dot \varphi+ \frac{1}{v}\omega(0,\bar{x(t)})
\end{equation}

so that 
\begin{equation}
\dot \varphi=\omega(0,\bar{x(t)})=\omega_0
\end{equation}

i.e. as expected the tail bud simply oscillates with frequency $\omega_0$, and we thus have (taking as an arbitrary initial condition $\varphi(0)=0$)

\begin{eqnarray}
 \phi(x,t) &=& \omega_0(t-\frac{x}{v})+\frac{1}{v}\int_0^{x}\omega\left( x',\bar{x}\left(t-\frac{x-x'}{v}\right)\right)dx' \\
 & =&\omega_0 t +\frac{1}{v}\int_0^{x}(\omega\left( x',\bar{x}\left(t-\frac{x-x'}{v}\right)\right)-\omega_0)dx' \label{psi_complete}\\
  & =&\omega_0 t + \psi(x,t) \label{psi_def}
 \end{eqnarray} 

defining

\begin{eqnarray}
\psi(x,t)=\frac{1}{v}\int_0^{x}(\omega\left( x',\bar{x}\left(t-\frac{x-x'}{v}\right)\right)-\omega_0)dx' \label{psi_doppler}
\end{eqnarray}

Eq.\ref{psi_complete}-\ref{psi_def} are equivalent to the now familiar Ansatz defined e.g. in Eqs. \ref{Ansatz}, \ref{AnsatzDiscrete}, since we see from Eq. \ref{psi_def} that $\psi(x,t)$ is the phase profile relative to the tail bud phase.
In previous derivations, this phase profile did not depend on time,  e.g. integrating  Eq. \ref{psiappendix} of the LPM with proper units we had  in a similar way : 
\begin{eqnarray}
\psi(x)=\frac{1}{v}\int_0^x (\omega(x')-\omega_0) dx
\end{eqnarray}
 The big difference is that $\omega$ now depends  on time in a very complex way: we see from Eq. \ref{psi_doppler} that the relative phase profile $\psi$ depends  on all past PSM lengths $\bar{x}\left(t-\frac{x-x'}{v}\right)$. So this problem is highly non-local; below we will  see some simplifications that can arise for special forms of $\omega$.
It is useful to define the time derivative of this phase profile 
\begin{equation}
\Lambda(x,t)=-\frac{\partial \psi(x,t)}{\partial t}=-\frac{\partial}{\partial t} \frac{1}{v}\int_0^{x}\omega\left( x',\bar{x}\left(t-\frac{x-x'}{v}\right)\right)dx'  \label{Lambda_Doppler}
\end{equation}
Notice that as expected, $\Lambda$ is $0$  for constant $\bar{x}$

This allows us to simply compute from Eq. \ref{psi_def}
\begin{equation}
\frac{\partial \phi}{\partial t}=\omega_0-\Lambda(x,t) \label{dphit}
\end{equation}
and
\begin{equation}
v\frac{\partial \phi}{\partial x}=\omega(x,\bar{x}(t))-\omega_0 +\Lambda(x,t)\label{dphix}
\end{equation}
[Here we used the fact that the integrated $\bar{x}$ term  in Eq. \ref{psi_doppler} depends on $\left(t-\frac{x-x'}{v}\right)$ so that for this term $\partial_t \bar{x}=-1/v \partial_x \bar{x}$, which allows  to express $\frac{\partial\Psi}{\partial x}$ as a function of $\Lambda$  ]

 Now injecting  this  into  Eq.  \ref{frequencyfront} we get :

\begin{eqnarray}
\Omega_A & =& \omega_0-\Lambda(\bar{x},t)+\frac{\dot{\bar{x}}}{v}\left[\omega(\bar{x},\bar{x})-\omega_0+\Lambda(\bar{x},t)\right] \\
&=& \Omega_P+\Omega_W+\Omega_D \label{eq:Omega}
\end{eqnarray}
defining
\begin{equation}
\Omega_P=\omega_0
\end{equation}
\begin{equation}
\Omega_W=-\Lambda(\bar{x},t)=\left.\frac{\partial \psi(x,t)}{\partial t}\right|_{x=\bar{x}}
\end{equation}
\begin{equation}
\Omega_D=\frac{\dot{\bar{x}}}{v}\left[\omega(\bar{x},\bar{x})-\omega_0+\Lambda(\bar{x},t)\right]=\dot{\bar{x}}\left.\frac{\partial \psi(x,t)}{\partial x}\right|_{x=\bar{x}} \label{omegadoppler}
\end{equation}

There are three more or less intuitive contributions in this sum. The first contribution is $\Omega_P=\omega_0$ : this is the usual angular frequency in the tail bud (posterior), which is the "standard" term when the PSM size is constant.

$\Omega_D$  is the so-called "Doppler" effect. It is proportional to the shrinkage speed of the PSM, and composed of two terms. First, $\omega(\bar{x},\bar{x}) - \omega_0$ quantifies the difference in angular frequency between front and tail bud. The contribution to the Doppler effect is rather intuitive: since cells at $\bar{x}$ oscillate with angular frequency $\omega(\bar{x},\bar{x})$, within a small time interval $dt$, there is a (negative)  phase difference accumulation relative to the tail bud equal to  $(\omega(\bar{x},\bar{x})-\omega_0)dt$. Now when the PSM shrinks by a small quantity $d\bar{x}$, this defines a small time  $d\tau=d\bar{x}/v$ for which those cells do not accumulate this phase difference, and thus, per unit of time a relative increase of phase $\frac{\dot{\bar{x}}}{v}(\omega(\bar{x},\bar{x})-\omega_0)$, which corresponds to an increase of angular velocity as expected. So this term exactly is the intuitive Doppler contribution described in the preamble of this section.

The less intuitive contributions come from  $\Lambda(x,t)=-\frac{\partial \psi(x,t)}{\partial t} $ both independently in $\Omega_W$ and as a term in the Doppler term. By definition, $\Lambda(x,t)$ is the change of the phase profile due to the shrinking of the PSM. Since the phase profile is directly related to the wavelength of the pattern, it is called  the "dynamical" wavelength in   \cite{Jorg2014,Jorg2015}. $\Lambda$ is particularly difficult to compute (analytically) since it contains information about the entire history of PSM shrinkage, so extra assumptions are needed to compute it. In the Appendix B2, we compute $\Lambda$ for two cases: when the frequency gradient is instantaneously scaling with PSM size (frequency scaling), which is the hypothesis made in \cite{Soroldoni14, Jorg2014, Jorg2015}, and when the frequency gradient is left unchanged and a front of $0$-frequency  moves towards the anterior without changing it (frequency cropping).

Experimentally, Soroldoni \etal \cite{Soroldoni14} visualized the segmentation clock dynamics in a zebrafish embryo. Using moving Regions of Interest (ROIs), they monitored oscillations both in the posterior and  anterior end of the PSM for almost 20 cycles and noticed that for 9 oscillations in the posterior, one observes 10 oscillations in the anterior, suggesting a non-stationary process. The period of segment formation is however the same as the period of the oscillator in the anterior, suggesting they are indeed the same process.

The difference between anterior and posterior periods is accompanied by a shrinkage of the PSM by about 60 \% while 13 segments are formed, allowing to quantify the Doppler shift. A phase map is then experimentally derived to visualize the profile $\psi(x,t)$. It is very clear that these profiles change with time: for instance, over 500 mins (corresponding to 12 oscillations in the posterior), the phase difference $|\psi(\bar{x},t)-\psi(\bar{x},0)|$ decreases from more than $5\pi$ to roughly $3\pi$ (Fig S5 in \cite{Soroldoni14}). It is quite noticeable though that there are different regimes for $\psi$, for instance by visualizing $\partial \psi/\partial t$, one sees that initially, $\partial \psi/\partial t$  (corresponding to the term $\Lambda$) is non-zero in a broad anterior region of the embryo, before reaching $0$ later on after $300$ mins. Quantitatively, the posterior frequency  $\Omega_P$ can be directly measured from the moving ROI in the posterior, and is around $0.15 min^{-1}$ for most of the time, slowly decreasing to roughly $0.12 min^{-1}$. To estimate other contributions, one has to experimentally measure $\partial \psi/\partial t$  and  $\partial \psi/\partial x$, which can be done from kymographs of the oscillations. The Doppler contribution $\Omega_D=\dot{\bar{x}}\left.\frac{\partial \psi(x,t)}{\partial x}\right|_{x=\bar{x}} $ varies with time between $0.025 min^{-1}$ and  $0.05 min^{-1}$, so quite significantly compared to $\Omega_P$. The Dynamic wavelength $\Omega_W=\left.\frac{\partial \psi(x,t)}{\partial t}\right|_{x=\bar{x}}$ is of the order of $-0.03 min^{-1}$ and gets to $0$ after 300 mins. Those values are summarized on Fig. \ref{fig:Doppler}.

 So the Doppler contribution and the Dynamic wavelength are rather high in magnitude (roughly $1/3$ of the posterior frequency) but almost compensate  so that the segmentation period is only roughly 10 \% shorter than the posterior period. To understand how this compensation happens, going back to Eq. \ref{eq:Omega}, one can make the following assumptions from the data \cite{Jorg2015}:

\begin{itemize}
    \item a constant shrinking PSM rate $\beta=\frac{\dot {\bar x}}{v}=\frac{\bar v}{v}$
    \item a `scaling' frequency gradient with PSM size, i.e. $\omega(x)=\omega_0 U(x/L)$ where $L$ is the current size of the PSM
\end{itemize} 

 One then gets a more compact expression for $\Omega_A$ :
 
 \begin{equation}
     \Omega_A=(1+\beta)(1-\Delta) \omega_0 \label{eg:doppler_delta}
 \end{equation}
introducing $\Delta=\beta\int_0^1 U(x)(1+\beta \xi)^{-2}d\xi$ (see Appendix B)

We see better what happens from this expression. There are two effects. The  $1+\beta$ corresponds to a `traditional` Doppler effect. In the Appendix we show that this contribution would exactly be the Doppler effect corresponding to a shrinking PSM with a $0$ frequency in the anterior, thus shortening the period. However, there is an added factor $(1-\Delta)$, which clearly goes in the other direction, i.e. lengthening the period. This is due to the dynamical wavelength effect \cite{Soroldoni14}, originating in this model from the frequency scaling \cite{Jorg2015}. The origin of this  effect is a bit easier to understand from the frequency standpoint: if the frequency profile scales with the PSM size, in means that as the front moves, the frequency observed right at the front is actively lowered compared to a situation where the frequency gradient is held constant (see e.g. frequency cropping calculation in Appendix). As noticed in \cite{Jorg2015}, a similar effect can be obtained for a fixed observer in classical waves equation with a refractory index increasing with time, which gives local frequency going to $0$ in the limit $t\rightarrow \infty$.

Experimentally, $\beta$ approximately is $1.15$, meaning that the PSM shrinks as fast as the tail bud grows. That means that without frequency gradient scaling, we would (initially) expect $\Omega_A$ of the order of $2\omega_0$, much faster than what we observe. But, fitting the frequency gradient with a functional form $U(x)=\sigma + (1-\sigma)\frac{1-e^{k(x-1)}}{1-e^{-k}}$ for $0\leq x \leq1$, one finds experimentally $\sigma=0.34$ and $k=2.07$, which gives $\Delta \sim 0.45$, so that the $(1-\Delta)$ factor almost exactly compensates  $(1+\beta)$  \footnote{It is also interesting to see that $\Delta$ depends only very weakly on the shape of the gradient, varying $\sigma$ gives values of $\Delta$ between $0.4$ and $0.5$}. 
 
 In physics, compensation of big numbers to give relatively smaller ones are often called "unnatural" because they suggest some arbitrary parameter fine-tuning. Here, the observed compensation might in fact suggest an active scaling mechanism so that indeed $\omega(x,L)$ is a pure function of $x/L$, the relative position of a PSM cell, as hypothesized in \cite{Jorg2015} to fit the data.  Such a hypothesis further accounts for some nontrivial features of segmentation, for instance, the segment length reaches a maximum for some intermediate segment numbers \cite{Jorg2015}, before shrinking down at a later time.

\chapsubsec{Doppler effect or Doppler period shift? }

 In physics, the Doppler effect is associated with a wave propagating towards a moving observer, or, similarly, a  wave source moving relative to a static observer. This is the familiar experience of an ER vehicle moving rapidly in the street: when the vehicle moves towards an observer, the pitch of the sound is higher than when it moves away from the observer. Another example in physics is the famous "red shift" perceived in light emitted from galaxies further away, indicating that they are moving away from us (and demonstrating universe expansion). More quantitatively: assuming a standard wave equation in one-dimension with constant  speed $c$ given by:
\begin{equation}
\frac{\partial^2 u}{\partial t^2}=c^2\frac{\partial^2 u}{\partial x^2} \label{dopplereffect}
\end{equation}

A propagating solution  from left to right with angular velocity $\omega$ takes the form 

\begin{equation}
u(x,t)= u_0 \sin(\omega (t -x/c)+\phi_0)
\end{equation}

defining a phase 

\begin{equation}
\phi(x,t)=\omega (t -x/c)+\phi_0
\end{equation}

Now assuming an observer moves as a function of time, position $\bar{x}(t)$, the observer sees the phase $\bar{\phi}(t)=\phi(\bar{x}(t),t)$ and the corresponding observed frequency $\Omega$ is :

\begin{eqnarray}
\Omega=\frac{d \bar \phi}{dt} & =&\left.\left(\frac{\partial \phi}{\partial t}+ \frac{d \bar{x}(t)}{dt} \frac{\partial \phi}{\partial x} \right)\right|_{\bar{x}(t)}\\
& =&\omega(1-\frac{1}{c} \frac{d \bar{x}(t)}{dt})
\end{eqnarray}
We indeed see that if $\frac{d \bar{x}(t)}{dt}<0$, meaning that the observer is moving towards the wave, $\Omega>\omega$ (corresponding for instance to the higher ER vehicle pitch), and conversely is lower if the observer is moving further from the wave. Dynamical wavelength effects can also be observed in physics if $\omega$ in Eq. \ref{dopplereffect} is time-dependent (e.g. a time-dependent refraction index) as discussed in \cite{Jorg2015}.

So the mathematical formalism is clearly similar, however, there is an important difference in terms of the actual meaning of the equations, especially in the biological context. In the classical examples of Doppler effects mentioned above (ER vehicle moving, redshift), there is a local observer moving relative to the source of the signal and 'sensing' the frequency $\Omega$. In a biological context, there is a priori no such \textit{local} observer moving with respect to the source of the oscillation (e.g. the tailbud): rather the front is a macroscopic variable, defined by a moving location $\bar{x}$, that might only be defined on average, see e.g. the complicated motion of the bifurcation in the Meinhardt model.  Remarkably though, there is an associated \textit{macroscopic} observable at the embryo level: the frequency of segment formation, which is measured by the experimentalist (again with multiple caveats, since e.g. boundary formation is a discrete process). So maybe one could rather refer to this phenomenon as a Doppler segmentation period shift,  indicating this is an emergent effect at the embryo level to be contrasted with a Doppler effect usually tied to an observer moving relative to the wave.





\bnormf
\includegraphics[width=\textwidth]{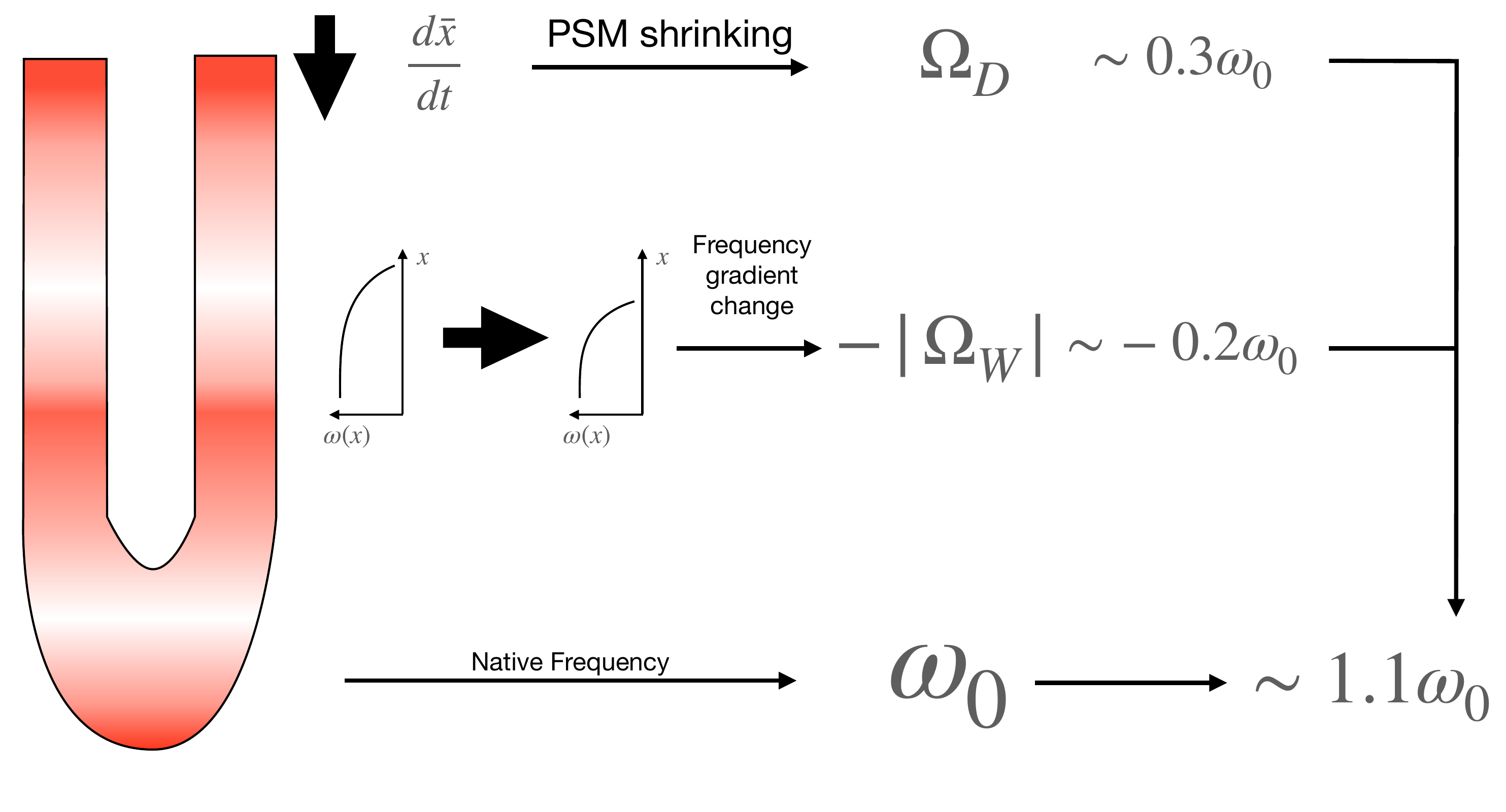}
\caption[Contributions to the Doppler Shift of Segmentation period]{Contributions to the Doppler Shift of Segmentation period in zebrafish}\label{fig:Doppler}
\enormf

\chapsec{Wavefront as a phase shock emerging from coupling}
\label{shock}
In all phase models described so far, the oscillators are controlled by an imposed frequency gradient and associated front, similar to the original clock and wavefront model. Biologically, this could correspond to a morphogen gradient (such as FGF)  controlling clock frequency and imposing a threshold for clock stopping. However, recent models have suggested different views where both the observed frequency gradient and the clock stopping are much more connected to the clock itself.

A remarkably simple model has been proposed by Murray \etal \cite{Murray2011,Murray2013} with spatially coupled phases  :
\begin{equation}
\frac{\partial \theta}{\partial t}= \omega + A \frac{\partial ^2 \theta}{\partial x^2}-B\left( \frac{\partial  \theta}{\partial x} \right)^2 \label{Murray}
\end{equation}

This model can be derived as a Taylor expansion of oscillators spatially coupled with a Kuramoto-Sakaguchi coupling of the form $H(\Delta \theta)=a \sin(\Delta \theta)+b(\cos\Delta \theta -1)$ : the coupling term at position $x$ would then be (Taylor expanding at order $2$ in $dx$)
\bwt
\begin{eqnarray}
   && H(\theta(x+dx)-\theta(x))+H(\theta(x-dx)-\theta(x)) \\&=& a \left( \sin(\theta(x+dx)-\theta(x))+\sin(\theta(x-dx)-\theta(x)) \right) \nonumber \\
   &+&b \left( \cos(\theta(x+dx)-\theta(x))-1+\cos(\theta(x-dx)-\theta(x))-1 \right)\\
   & \sim& a (\theta(x+dx)+\theta(x-dx)-2\theta(x)) \nonumber \\ && -\frac{b}{2}\left((\theta(x+dx)-\theta(x))^2+(\theta(x-dx)-\theta(x))^2\right)\\
     & \sim& a dx^2 \frac{\partial ^2 \theta}{\partial x^2}-b dx^2\left(\frac{\partial  \theta}{\partial x} \right)^2
\end{eqnarray}
\ewt
so that we see that $A=a dx^2 $ and $B=b dx^2$.

$\omega$ represents the frequency in the posterior part of the PSM. The $A$ term is a classical phase diffusion term. The novelty arises from the quadratic $B$ term. If $B>0$, we see that any local phase gradient tends to slow down the local oscillator. So imagine that we have a  region with a sharp phase gradient next to a region with a flat phase profile: we would then expect that the oscillators at the boundary between the two regions would get very slow, thus expanding the region of high phase gradient (see Fig. \ref{fig:Murray}). In other words, regions with high phase gradients would tend to propagate, suggesting a plausible mechanism for wavefront formation.
 
\bnormf
\includegraphics[width=\textwidth]{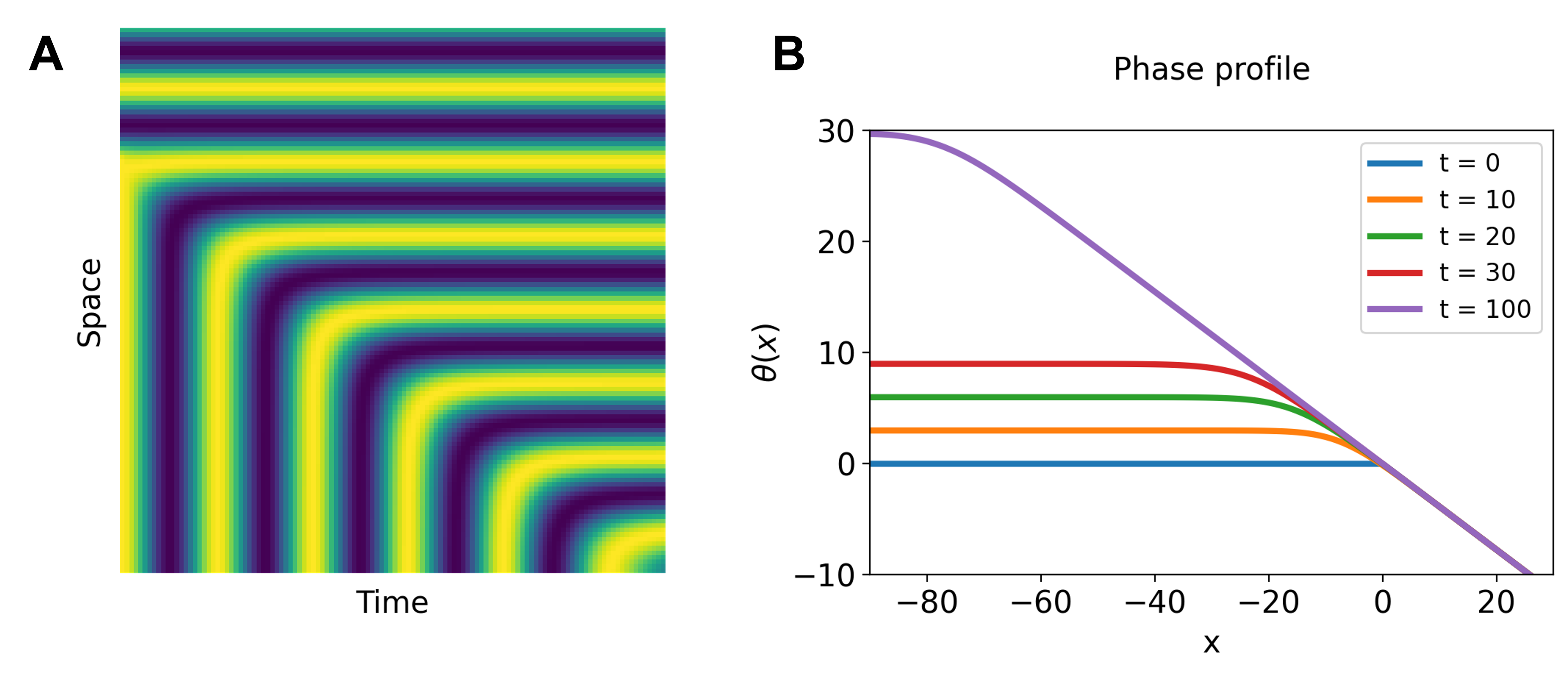}
\caption[Murray model]{Simulations of the Murray model : kymographs on the left and phase profiles for different times on the right. }\label{fig:Murray}
\enormf
 
To be more quantitative, Eq. \ref{Murray} clearly has a steady state solution, that we interpret as the pattern in the anterior part of the embryo. Assuming that the phase is stationary and linear spatially as $x\rightarrow -\infty$ (so corresponding to anterior, like in the delayed phase model from \cite{Morelli2009}), we get 
\begin{equation}
\left. \frac{\partial \theta}{\partial x}\right|_{x\rightarrow -\infty} = \sqrt{\frac{\omega}{B}}
\end{equation}

In the posterior part of the embryo, we assume synchronous oscillations, so that

\begin{equation}
\left. \frac{\partial \theta}{\partial x}\right|_{x\rightarrow +\infty} = 0
\end{equation}

 Defining as a new variable  the gradient of the phase $\Psi=\frac{\partial \theta}{\partial x}$ we see that $\Psi$ has to interpolate between $0$ and a finite value. In fact, as intuitively explained above, one will observe a propagation of the region with non-zero $\Psi$, towards the region of $0$ $\Psi$, corresponding to the stabilization of the phase gradient giving rise to a periodic pattern! More precisely we get for $\Psi$ :
\begin{equation}
\frac{\partial \Psi}{\partial t}+2B\Psi \frac{\partial \Psi}{\partial x}= A \frac{\partial^2 \Psi}{\partial x^2}
\end{equation}

This equation is well-known in physics and is called Burgers' equation \cite{Burgers1948}. The left-hand side of this equation looks similar to standard advection equations such as Eq.  \ref{movingframe} or Eq. \ref{doppler} but with a speed proportional to the value  of the propagating field $\Psi$. The Burgers' equation has been applied to various contexts, e.g. traffic flow modeling, or boundary propagation and there are standard methods so solve it analytically \cite{Wang1996}. A stochastic version of Burgers' equation is the famous Kardar-Parisi-Zhang  model of stochastic growth of surface \cite{KPZ}, which is also related to stochastic models of particle motions in 1D such as the asymmetric exclusion process \cite{Derrida1993}. 
Burgers' equation is clearly non-linear, but can still be solved using the same stationary Ansatz as before :

\begin{equation}
\Psi(x,t)=w(x-vt)=w(m)
\end{equation}
assuming a propagation of the pattern from left to right with (yet unknown) speed $v$.
We then get
\begin{equation}
-v w'+2B w w'=A w''
\end{equation}

which directly integrates into an order one differential equation

\begin{equation}
-v w+B w^2=A w' +C  \label{w1storder}
\end{equation}

Now we use the boundary condition that $w=w'=0$ at $x=+\infty$ and $w'=0,w=\sqrt{\frac{\omega}{B}}$ at $x=-\infty$. This directly gives $C=0$ and the value of the speed $v=\sqrt{\omega B}$.  With the value of $v$ it is straightforward to integrate Equation \ref{w1storder} and one gets immediately (properly shifting the initial value for $m$)

\begin{equation}
w(m)=\frac{\sqrt{\omega/B}}{1+e^{\sqrt{\omega B}m/A}}
\end{equation}

Now $\Psi=\partial \theta/\partial x=w(m)$  with $m=x-vt$ so that we recover the simple equation

\begin{equation}
\frac{\partial \theta}{\partial m}=\frac{1}{v}\frac{\omega}{1+e^{ v m/A}}
\end{equation}

where we made the dependency on $v=\sqrt{\omega B}$ explicit. Rescaling time and space to that $\omega=v=1$, we then get 
\begin{equation}
\frac{\partial \theta}{\partial m}=\frac{1}{1+e^{ m/A}}
\end{equation}

which is completely equivalent to Eq. \ref{psiappendix} with
\begin{equation}
r(m)=\frac{1}{1+e^{-m/A}}=\frac{1}{1+e^{(-x+t)/A}}
\end{equation}

This exactly is the frequency profile assumed in the  LPM  Eq. \ref{simpler} ! (modulo a flipping of the direction of the $x$ axis). Thus this model is indistinguishable from the LPM by only considering stationary propagation of the front.

To really contrast this model with the LPM, one should test the mechanism underlying front propagation related to the $B$ term, in particular its influence on the front speed.  The first interesting prediction relates to the scaling of the pattern with the clock period \cite{Murray2011, Murray2013}. Since $v=\sqrt{\omega B}$, and the somite size $S=2\pi v/\omega$, we get :

\begin{equation}
S=2\pi v \sqrt{B/\omega}=\sqrt{2 \pi B} \sqrt{T} \label{somitesize}
\end{equation}

where $T$ is the period of the tail bud oscillator. So if we were to modify the period, one would get a scaling law of power $1/2$ for the somite size with respect to the period, while if speed and period vary  independently we expect $S$ and $T$ to be proportional, so rather a power $1$.

Another way to test the model is to directly change $B$. In particular, if we locally decrease $B$, we expect a locally smaller speed, and thus a pattern with  a smaller wavelength. Conversely, if $B$ is locally increased, the front speed is higher and the pattern should have a bigger wavelength.  As said above, we expect coupling to be related to Notch signaling, so that those predictions on the $B$ term could be tested by performing transplants of cells with different Notch levels. Such transplants have been done in \cite{Horikawa2006}, where cells expressing high levels of Delta ligands induce locally smaller somites, so consistent with a smaller $B$ term. However, we notice that those experiments are difficult to interpret in the present context without more direct connections to the $A$ and $B$ terms. In particular, since there are two coupling terms $A$ and $B$, with $A$ acting to keep oscillators in synchrony while $B$ tending to increase phase gradient, it is unclear what would be the net effect of the modification of Notch signaling pathway. But those experiments combined with this modeling have immense merit to point out the potential role of Notch lateral inhibition in destabilizing the phase gradient to generate a wavefront and stabilize the pattern.

Experimentally, one can estimate relatively simply the different parameters $\omega,A,B$ by connecting them to the physical observables we already discussed, respectively the tail bud frequency, the size of the transition zone from oscillation to dynamic pattern, and the speed of the front/size of the pattern. Based on data from \cite{Giudicelli2007}, $\omega$ obviously is the angular velocity in the tail bud, corresponding to $0.21 {min}^{-1}$ in zebrafish \cite{Murray2011, Murray2013}. Since $B$ is related to the speed $v$ of the front and $\omega$, it can be estimated from somite size $S=vT$ to get $B=S^2/{2\pi T}$: with a somite length of roughly 6 cell diameters ($cd$), one finds that $B = 0.19 cd^2. min^{-1}$. The most difficult parameter to estimate is $A$, which relates to the typical length scale of the phase gradient $L_{exp}$. Using a linear approximation, one finds $L_{exp}=4A/\sqrt{\omega B}$, and experimentally one finds that $L_{exp} \sim 48 cd$, which gives $A\sim 2.4   cd^2 min^{-1}$. Murray \etal \cite{Murray2011, Murray2013} also inferred parameter values for different species using published data \cite{Gomez2008}, in particular, rescaling the period to generate the dimensionless equation

\begin{equation}
\frac{\partial \theta}{\partial t}=\gamma \frac{\partial^2 \theta}{\partial t^2}-16 \gamma^2 \left(\frac{\partial \theta}{\partial t}\right)^2+1
\end{equation}

with $\gamma=S/4L_{exp}=\pi B/8A$. The rescaled speed is $4\gamma$, and the number of waves in the PSM can be computed from Eq.\ref{number_waves} and is $1/8\gamma$. One then finds that mouse and chick have relatively close $\gamma$ of around $0.06$, giving two waves, for zebrafish $\gamma=0.04$, giving slightly more than 3 waves and for snake $\gamma=0.02$ giving around 7 waves.


\chapsec{Phase-amplitude coupling and excitability for oscillation arrest}
\label{sec:phaseamplitude}

In \cite{Shih2015}, we observed that, within one future presumptive somite,  the apparent wave of clock arrest does not move smoothly from anterior to posterior  part, as expected  in classical clock and wavefront or phase models (such as the LPM or the Murray model from the previous section). Rather, the Notch wave sweeps from posterior to anterior, so that it eventually stops in the anterior part of each somite. This means that, as measured in terms of Notch signaling maximum activity, the oscillation cycle dies out first in the \textbf{posterior part} of a given somite and not in its anterior part where the front should first pass. We named this process  "clockwave stopping", to stress out the motion of that last wave. So there is some interesting paradox between scales here, suggesting that phase dynamics alone is not enough to account for the difference between the local (somite, posterior to anterior) scale and global (tissue, anterior to posterior) dynamics of clock stopping. 

A possible explanation is that the clock stopping is coupled to some other observable moving in the same direction as the oscillation wave, from posterior to anterior. For instance, it is well known  that the amplitude of the cellular oscillations increases from posterior to anterior, as described in \cite{Shih2015}, Fig. 4. To explain the local vs global paradox in the direction of the clock stopping, it is then natural to assume that the overall oscillatory signal (combining phase and amplitude) controls some bifurcation leading to clock arrest.

This can be captured by the following simple phenomenological model, that was used in \cite{Shih2015}, Fig. 4. Adding a variable $A$ for amplitude and defining the overall clock signal $s(A,\theta)=A(1+\cos \theta)$, we write in the PSM frame of reference:

\begin{eqnarray}
    \dot \theta  & =& \omega(m) \label{eq:phase-ampt} \\
    \dot A &=& \Theta(m) \lambda_0 A  \label{eq:phase-ampa}
\end{eqnarray}
As usual, $\omega(m)$ accounts for the slowing down of the clock with variable $m=x+t$ in rescaled units.  The precise functional form does not matter as long as $\omega$ goes to $0$ for big $m$.  $\Theta(m)$ is the Heaviside function equal to $1$ if $m>0$ and $0$ otherwise.  So for tail bud cells ($m<0$), $\lambda(m)=\Theta(m) \lambda_0 =0$ and $A$ stays constant (and non zero). Then when cells are injected in the PSM (i.e. $m>0$), $\lambda=\lambda_0$, the amplitude increases in PSM (as the clock is slowing down). So those two equations simply capture  both the clock slowing down and the amplitude increasing. 
To account for the clockwave stopping, we impose the following stopping conditions :

\begin{itemize}
    \item as soon as $s(A,\theta)> A^*$, the bifurcation is crossed 
    \item once the bifurcation is crossed, the oscillation completes its cycle up to phase $\theta=\pi$

\end{itemize}
So the primary wave here is defined by the condition $s(A,\theta)> A^*$. The second condition phenomenologically reproduces  the situation after a bifurcation from an oscillating to an excitable system, such as the system on  the right-most panel of Fig. \ref{fig:MeinhardtVDP-flow} (see Appendix for discussion on excitable systems). The idea is that a new fixed point appears on (or very close) to the cycle, because of relaxation-like dynamics the overall oscillations need to be fully completed before the system stabilizes.

 Fig. \ref{fig:WaveStopping} shows a simulation of this simple model, which strikingly recapitulates many features of the zebrafish clock arrest, in particular a "sawtooth" pattern at the front, and the maximum intensity in the anteriormost part of each somite, corresponding to the location where the clock stops last at each wave. We represent dynamics in three cells in Fig. \ref{fig:WaveStopping} D : green is a posterior somite cell of a given somite, while blue and orange respectively are posterior and anterior cells of the next somite.  Notice that the blue and orange cells go through an additional cycle compared to the green cell and that the orange cell, despite being very close to the green one, cycles once more with higher intensity. This possibly solves the Meinhardt paradox between AP and PA boundaries: from the orange cell standpoint (anterior A), there is an additional cycle in the blue cell vs the green one, both being committed to posterior fates. So because of this asymmetric wave dynamics, one would simply need a mechanism to coordinate the epithelialization of cells locally going through the exact same number of oscillations to define a proper somite. More generally, this suggests that excitable dynamics close to the front might indeed play a role, not only to explain the wave pattern but also to control the bifurcation itself. In fact, as will be described in section \myref{SNICsection}, such bifurcation could also be more directly coupled to the slowing down dynamics itself. We will discuss further evidence for such excitability of the segmentation oscillators in section \myref{section:hack}.


\bnormf
\includegraphics[width=\textwidth]{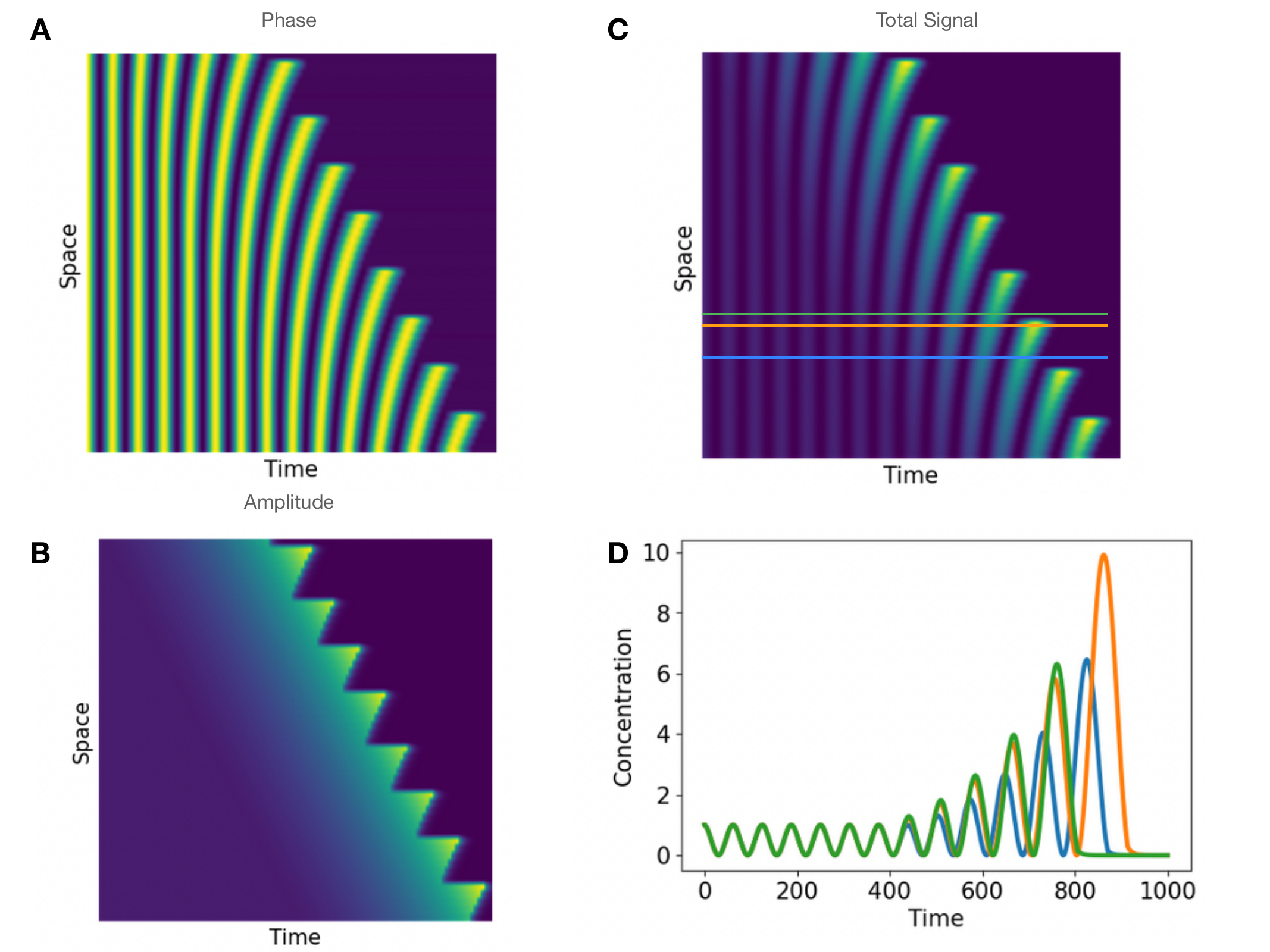}
   \caption[Phase Amplitude model]{Simulation of Eqs. \ref{eq:phase-ampa}-\ref{eq:phase-ampt}, showing a sawtooth pattern at the front and recapitulating the posterior to anterior stopping dynamics in a given somite. (A) Simulated Phase (B) Simulated Amplitude (C) Total Signal. Notice the strong asymmetry and the "sawtooth" pattern. Green, orange, and blue lines correspond to positions displayed on panel B, the green cell is assumed to be one somite earlier than the blue and orange ones (D) Simulated oscillations in three cells indicated in (C).  }\label{fig:WaveStopping}
\enormf

\newpage
\chap{From systems biology to geometric models}

In our discussion so far, we have considered relatively abstract models, e.g. focused on the dynamics of the phase in the previous section. But many "low-level" components (genes, coupling effects, possibly mechanical effects)  are driving the segmentation process.  

With the advances in systems biology (both theoretical and experimental), it is possible to  detail more explicitly processes implicated in segmentation. It would be a tremendous achievement to  follow a bottom-up approach,  identifying precisely all components of the system, modeling them, and from there inferring the observed dynamics. We are still not there, and the historical approach has been much more top-down, with at least two modeling steps : 
\begin{itemize}
    \item relate phase dynamics to possible biochemical mechanisms
    \item relate possible biochemical mechanisms to actual genes
\end{itemize}

We make  here an explicit distinction between biochemical mechanisms and genes because the same biochemical process (defined in a loose sense, say for instance a negative feedback oscillator, or a bistable system) could possibly be implemented with different genes in different species, explaining interspecies variability mentioned earlier \cite{Krol2011}. This can lead to confusion (common to many systems biology modeling work) on what is actually modeled and predicted. For instance, we have already seen in the previous section that very similar phase dynamics can be modeled with different biochemical processes (e.g. via spatial coupling terms), so a detailed gene-based model reproducing  only the correct wave pattern might not necessarily be insightful without specific experimental tests.  This raises a common problem for all models based on explicit gene dynamics: they can be falsified on at least three levels (the genes themselves, the biochemical process they implement, and the emerging phase dynamics). For this reason, in this section, we focus on models exploring specific biochemical mechanisms inspired by actual genetic interactions, eventually circling back onto "geometric models" not tied to specific genes.

\chapsec{Simple delayed oscillators }

Many models of segmentation clocks are based on so-called negative feedback oscillators. There is a large (experimental and theoretical) literature on such oscillators in a broader systems biology context, notably  for circadian clocks modeling \cite{Smolen2002,Goldbeter2002,Ruoff2001} and cell cycle \cite{Tyson2003}, see also \cite{Negrete2020} for a recent review on developmental timing.

Many molecular networks implicated in somitogenesis could plausibly implement such negative feedback. For instance, \textit{her1} and \textit{her7} in zebrafish appear to both oscillate and negatively regulate their own expression \cite{Oates2002, Holley2002}  as well as the ortholog of \textit{her1} in mouse, \textit{Hes7} \cite{Bessho2001,Bessho2003}. \textit{Hes1} in mouse is also clearly implicated in a negative feedback loop \cite{Hirata2002}.

When modeling negative feedback loop models to get oscillations, it is tempting to write a system of RNA ($m$))/protein ($p$)regulating its own production in the following way (using notations from \cite{Lewis2003, Monk2003}):
\begin{eqnarray}
\frac{dm}{dt} & = & f(p)- cm =u(m,p) \\ \label{moscillator}
\frac{dp}{dt} & = & am- bp = v(m,p) \label{poscillator}
\end{eqnarray}

with $f'(p)<0$ to model negative self-interaction. However, it can be proved that such a system can \textit{not} oscillate: assume there is a closed orbit $S$ in 2D, then the flux $\oint_S (u(m,p) \mathbf{m}+ v(m,p) \mathbf{p})\cdot \mathbf{n} \mathrm{d}s$ is $0$ since $\mathbf{n}$ is always normal to the tangent of the orbit. But from the divergence theorem, this integral is also equal to $\iint\limits_R  (f'(p)-b)  \mathrm{d}m  \mathrm{d}p$ which is strictly negative inside the domain $R$ limited by $S$, so there is a contradiction, and thus there can not be any closed orbit. This simple reasoning is called Poincar\'e-Bendixson criterion (See discussion in \cite{Murray2002}). 

There are several ways to fix this.  For instance, negative feedback loops in gene regulatory networks are not necessarily direct and could be implemented via extra steps (e.g. a phosphorylation cascade, a spatial segregation step, or an extra transcriptional step where a gene activates its own repressor instead of repressing itself directly). Mathematically, this increases the dimensionality of the system which is no longer limited by the Poincar\'e-Bendixson criterion but also becomes more complicated to study. Examples of such models include the Goodwin model \cite{Ruoff1996},  or the MFL model \cite{Francois2005}, both of which are still simple enough to be studied analytically.  Alternatively, the addition of positive feedbacks allows us to circumvent the Bendixson criterion to get oscillations with a 2D model (see e.g. \cite{Vilar2002}). 
\label{delayed}
Coming back to the segmentation oscillator, direct binding of \textit{Hes7} protein to its own promoter has been shown \cite{Bessho2003}, excluding the presence of a clear intermediate step and  there is no evidence either for positive feedback in this part of the network. A parsimonious modeling choice rather is to assume the presence of explicit delays in the negative feedback loop \cite{Smolen2002,Monk2003,Lewis2003, Jensen2003, Tiana2007}, i.e. modifying Eq. \ref{moscillator} such that :
\begin{equation}
\frac{dm}{dt}  =  f(p(t-\tau))- cm \label{delayed_mRNA}
\end{equation}
Such delays could be due to the combination of various processes, including transcription/translation kinetic delays. An even simpler model consists in making a quasi-static approximation on either $m$ or $p$ to get an effectively reduced system of the form 
\begin{equation}
\epsilon \frac{db}{dt} = f(b(t-1))- b \label{rescaled}
\end{equation}
This equation is identical to the famous Glass-Mackey oscillator \cite{Mackey1977, Glass1988}, proposed in a broader physiological context to model phenomena ranging from dynamical respiratory to hematopoietic diseases. In this equation, the time is expressed in units of the delay  $\tau$. $\epsilon$ quantifies the ratio between the delay and the time scale of accumulation of $b$. For simplicity, we will study first this simplest version of the delayed model, but we refer to \cite{Momiji2008} and below for more explicit versions of this model in the context of somitogenesis.

\chapsubsec{Conditions for oscillations}

\label{oscillationshopfdelay}
We will be especially interested in the limit of small $\epsilon$, where the delay is much longer than typical degradation kinetics. Simple intuitive results can be immediately derived. In particular, to understand why there is an oscillation, it is insightful to take the limit $\epsilon\rightarrow 0$ which gives \cite{Lewis2003}

\begin{equation}
b=f(b(t-1))
\end{equation}

If $f$ is a  (well-chosen) decreasing function of $b$ , one can define two values  $b_1$ and $b_2$ such that $b_1=f(b_2)$ et $b_2=f(b_1)$. In such a situation, starting say in $b_1$, the system stays in $b_1$ for a time of order $1$, then, since $b=f(b(t-1))$ switches to $b_2$, stays there for a time of order $1$, before switching back to $b_1$. This implements a simple "toggle" oscillator of period $2$, or in the original units, twice the delay of the oscillator \cite{Lewis2003}. This yields an immediate prediction: if one can modify the delay, then one should be able to change the clock period.  In fact, one can easily derive the conditions on the parameters to get oscillations \cite{Hayes1950, Glass1988} to confirm the relevance of this limit. In a nutshell, the repression function $f$ needs to be "strong" enough (e.g. in terms of cooperativity of the negative feedback), correspondingly $\epsilon$ needs to be small enough, and the delay $\tau$ big enough to have oscillations.  In the Appendix B3, we derive the full conditions to have oscillations.

\chapsubsec{Cycle description}

\bnormf
\includegraphics[width=\textwidth]{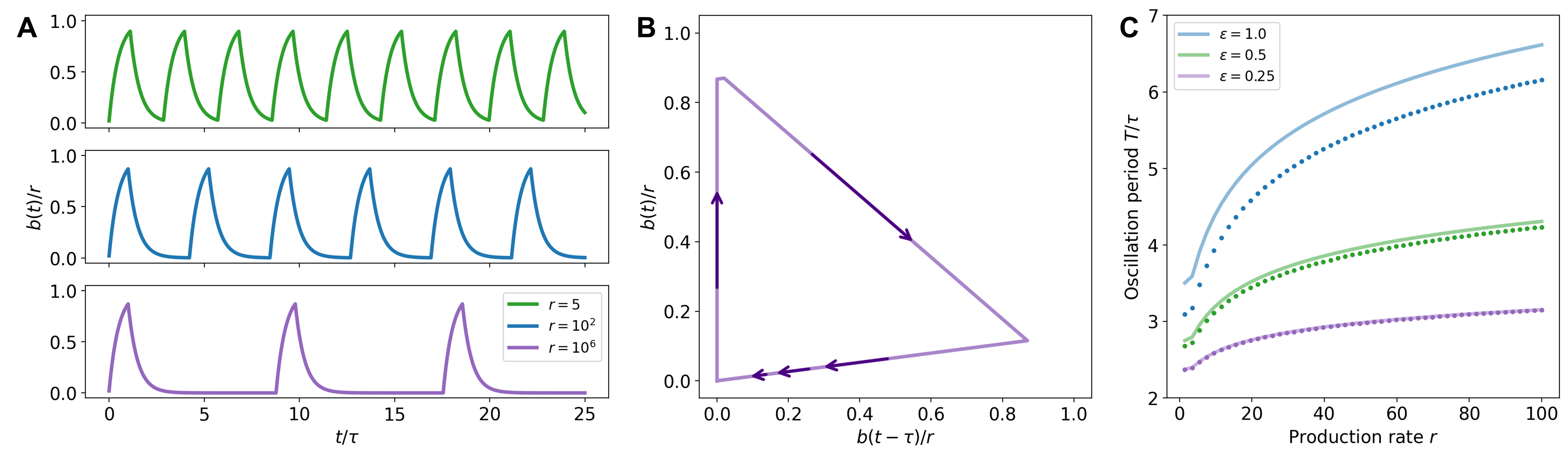}
\caption[Delayed Oscillator]{A) Numerical trajectories of the delayed oscillator \ref{rescaled} with step-like repression $n\rightarrow \infty$ and $\epsilon = 0.5$. (B) Reconstruction of the limit cycle in 2D: $b(t)$ as a function of $b(t-\tau)$ for the purple trajectory in (A), production rate $r = 10^6$. The arrows correspond to a fixed time interval $\Delta t$, indicating a slowing down of the trajectory as it approaches the origin. (C) Period of the simulated oscillations $T(r)$ compared to the approximate solution \ref{period_delayed}. For numerical calculation of the period, the time points where the trajectory derivative changes sign (the oscillator switches between "on" and "off" and vice versa) were detected.}\label{fig:delayed}
\enormf

In the very strong repression limit, taking $f$ as a step function ($f(b)=r$ if $b<1$, $0$ otherwise), the limit cycle can be simply described with piece-wise linear equations with two different phases: a pure degradation phase when transcription is off, and a pure production phase when transcription is on. One can then analytically describe the limit cycle \cite{FrancoisThesis}.
We get
\begin{equation}
   \epsilon \dot b^{-}=-b^{-}
\end{equation}
in the degradation phase (as long as $b(t-1)>1$) and
\begin{equation}
   \epsilon \dot b^{+}=r-b^{+}
\end{equation}
in the production phase (as long as $b(t-1)<1$) where we use $\pm$ superscript to indicate which phase (degradation vs transcription) the system is.

Let us start at a point in the cycle where $b^{-}$ is crossing $1$ (which means that $b>1$ for $t<0$). We then get for $t \in [0,1]$ 
\begin{equation}
b^{-}(t)=e^{-t/\epsilon}
\end{equation}

At $t=1$, $b^{-}(1-1)<1$ so that $f$ flips to $r$. We then get 
\begin{equation}
b^{+}(t)=r+(b_1-r)e^{-(t-1)/\epsilon} \label{bhautdelai}
\end{equation}
with $b_1=e^{-1/\epsilon}$. This remains valid until $b^{+}(t_1-1)=1$. One can obtain $t_1$  by Taylor expanding Eq. \ref{bhautdelai}  :
\begin{equation}
t_1=2+\epsilon \ln\left(\frac{r-e^{-1/\epsilon}}{r-1}\right)\sim 2+\epsilon \ln \left(\frac{r}{r-1}\right)
\end{equation}

at lowest order in $\epsilon$.
Then $b$ decreases until $1$, reached at $t=t_2$, closing the cycle. The equation for $b$ between $t_1$ et $t_2$ is :

\begin{equation}
b^{-}(t)=b_{max} e^{-(t-t_1)/\epsilon}
\end{equation}

with $b_{max}=b^{+}(t_1)$. By a Taylor expansion, one finds the value 
\begin{equation}
t_2=t_1+\epsilon \ln b_{max}  \simeq  2+\epsilon \ln\left(\frac{r^2}{r-1}\right)
\end{equation}
at lowest order in $\epsilon$ (notice that $b_{max}\sim r$ up to an exponential correction in $e^{-1/\epsilon}$).

It is also useful to extract the duration of the "on" and  "off" phases, in particular 

\begin{eqnarray}
    t_{on} &=&\tau(t_1-1)=1+\epsilon \ln\left(\frac{r}{r-1}\right)\\
    t_{off}&=&\tau(t_2-t_{on})=1+\epsilon  \ln r
\end{eqnarray}

There is a simple intuition here: $t_{off}-1$ essentially is the time to exponentially decrease due to degradation from $b_{max}\simeq r$ to $1$, while $t_{on}-1$ is the time to  increase from $0$ to $1$ in the production phase. Notice in particular that $t_{off}$ increases with $r$ (it takes more time to decrease from a higher value) while $t_{on}$ decreases with $r$ (it takes less time to reach $1$).

Getting back to the initial time units, one gets for the overall period:

\begin{equation}
T=2\tau+\epsilon\tau \ln\left(\frac{r^2}{r-1}\right) \label{period_delayed}
\end{equation}

This equation nicely combines the two-time scales of the system: the (dominating) delay $\tau$, fixing the period, and $\epsilon\tau$ corresponding to the transient phases of the system with magnitudes dominated by production and time scale by degradation.

In \cite{Negrete2021} a generalized version of this calculation is derived to compute the perturbations due to multiplicative noise, i.e. an equation of the form 

\begin{equation}
    \epsilon \frac{db}{dt} = f(b(t-1))- b + b\eta
\end{equation}

where $\eta$ is a standard Gaussian noise  $<\eta(t)\eta(t')>=\sigma_0^2 \delta(t-t')$, and $f$ the same  step function as above, i.e. $f(b)=r$ if $b>1$, $0$ otherwise. The main motivation for such calculation comes from experiments, where it has become recently possible to monitor in real-time concentrations of fluorescent proteins, which allows for direct comparison between theory and data. It is argued in  \cite{Negrete2021}  that a multiplicative noise is best suited to account for data. We summarize the steps and results of this calculation in Appendix B4.
It is possible then to go back to experiments and to fit those distributions, with very good agreement.  In particular, in presence of noise $b_{max}$ is the addition of a noisy production term with $b_{min}$, so we expect it to have not only a higher average value but also higher variance, which is clearly seen experimentally. One can also use similar self-consistent approximation to get distributions of $t_{on},t_{off}$ and numerically estimate the expectation value and variance of the times for each phase. Compared to the deterministic case described above, there is a correction in the average due to the noise autocorrelation.

\chapsubsec{Infinite period bifurcation for increased production rate}
\label{delay_increase}
In the limit of big production rate $r$, we have seen above that $t_{on}\simeq 1$ while $t_{off}\simeq 1 + \epsilon \ln (r)$. In particular if the production rate $r$ is becoming very big, we might get a strong asymmetry between the two phases of the cycles, with a much longer off phase.
It then makes sense to assume $r$ is very big,  to define a corresponding small parameter $\epsilon_r=1/r$, and to introduce the new dimensionless variable $c=\epsilon_r b$. Again there is a degradation and a production phase such that :

\begin{equation}
   \epsilon \dot c^{-}=-c^{-}
\end{equation}
in the degradation phase (as long as $c(t-1)>\epsilon_r$) and
\begin{equation}
   \epsilon \dot c^{+}=1-c^{+}
\end{equation}
in the activation phase (as long as $c(t-1)<\epsilon_r$). Notice that $c$ is necessarily bounded between in the interval $[0,1]$.

Assuming the limit cycle starts at $c_{max}\simeq 1$, the off phase clearly takes a time of order $-\ln \epsilon_r$ to reach the threshold value $\epsilon_r$. The on phase has a duration roughly equivalent to the delay, and if $\epsilon$ is small enough, $c$ has ample time to get back to a value of order $1$. So we get to a situation where there is a (relatively) short transcription phase where the system resets from $0$ to a high value $c_{max}$, followed by a slow decay from  $c_{max}$ to $0$. This behavior and the logarithmic period divergence are very reminiscent of the classical "fire and integrate" neuronal model in the homoclinic bifurcation regime (see \cite{Izhikevitch2007} and Appendix A). Interestingly, in the limit $\epsilon_r\rightarrow 0$, we thus get an infinite period bifurcation with excitability, where it takes an infinite time to reach $0$ from a positive value, but where an infinitesimal fluctuation pushes the system to an excursion. See Fig. \ref{fig:delayed} for a simulation of this system and a comparison between the simulated period and the analytical one from Eq. \ref{period_delayed}


\chapsubsec{"Realistic" delayed models and experimental validations}

In \cite{Monk2003, Momiji2008}, Monk and co-workers study delayed models in a broader systems biology context (see also \cite{Tiana2007}). It is proposed that delayed negative feedback loops are at the core of several systems, not only \textit{Hes1} oscillator in somitogenesis, but also the oscillations of the oncogene \textit{p53} \cite{BarOr2000} and the immune NK$\kappa$-B system \cite{Hoffmann2002}. A common feature of those systems are "ultradian" oscillations (i.e. shorter then the 24-hour circadian period), where one gene represses its own expression via multiple slow steps. As pointed out in \cite{Monk2003}, the idea that a gene can repress itself to generate an oscillation is first due to Goodwin \cite{Goodwin1965}, but the so-called Goodwin oscillator requires very strong, almost step-wise non-linearity to oscillate, which appears unrealistic. In comparison, a delayed model appears more plausible experimentally, especially since there are natural biological delays, e.g. transcription and translation.

It is however important to check if the experimentally measured parameters are in fact consistent with the regime of oscillations observed and the period measured. In  \cite{Monk2003, Momiji2008} a two-step model is studied by combining a delayed equation for mRNA production (i.e. Eq. \ref{delayed_mRNA}) with a rate equation for production  (Eq. \ref{poscillator}). In mouse, half-lives of both \textit{hes1} mRNA and proteins are slightly higher than $20$ minutes, and a reasonable estimate for transcript elongations (accounting for delays) is another 20 mins. Numerically, this gives an oscillatory period of roughly 2 hours, with a 20 mins shift between RNA and protein peak, in excellent agreement with experiments! This is a similar order of magnitude as the calculation in Eq. \ref{period_delayed} with an $\epsilon$ of order $1$ (which assumes fast dynamics on either the protein or the mRNA); interestingly, it is also found that the oscillation period does not depend much on the repression threshold but depends much more on the half-life of either proteins or RNA (period ranging from 100 to 140 mins with either half-life ranging from 10 to 30 mins; see also \cite{Jensen2003}).

This leads to a direct experimental prediction: changing the mRNA half-life should impact the segmentation period. To do this, in \cite{Hirata2004}, mice with $30$ mins half-life for \textit{Hes7} are generated. It is then observed that segmentation is in fact impaired after 3 to 4 cycles. Importantly, \textit{Hes7} loss of functions presents no segment at all, which suggests that mutants with longer half-period are still functional in the beginning of the segmentation. The authors interpret this phenotype as a "dampening" of the oscillation:  from the theoretical analysis (e.g. Eq \ref{epsilon_crit}), a too-long half-life would indeed lead to oscillation death via a Hopf bifurcation. Another route to test those models is to modify delays. In \cite{Takashima2011}, it is observed that introns in the \textit{Hes7} sequence could lead to overall delays (introns are part of the RNA that have to be removed prior to translation). A reporter gene is constructed, with and without introns, and it is found that while the gene with intron has similar kinetics to the endogenous \textit{Hes7}, expression of the gene without introns is advanced by roughly 20 mins. In fact, it turns out that decreasing the $\sim$ 40 mins delay from \cite{Hirata2004} to 20 mins indeed abolish the oscillations. 

While those changes are consistent with a delayed model, they are not true validation since they both destroy oscillations. A better route to validate the model is to modify the period without breaking segmentation. This is done in \cite{Harima2013}, where the number of introns is simply \textit{reduced}. There are three \textit{Hes7}  introns in mice. If only the third intron is conserved,  one or two extra somites form in the cervical region compared to WT (for more caudal regions, segmentation is impaired and all vertebrae are fused). With the help of a reporter similar to \cite{Takashima2011}, it is found that the one intron construct turns on \textit{Hes7} 5 mins earlier than WT, suggesting shorter delay and subsequent shorter period (leading to more somites). The oscillation itself can be monitored with luciferase, and the one-intron clock indeed oscillates with a period of 115 mins vs 126 mins for the wild type. Consistent with this shorter period, the somite size are reduced by roughly 10 \%.
\chapsubsec{Experiments: Delay or phase shift ?}

\label{sec:delayorshift}
The modifications detailed in the previous section are consistent with delayed models, but it is not clear that other models might not explain experiments equally well. A recurring issue in the overall discussion is an ambiguity on the meaning of "delay". Very often,  the delays described experimentally  correspond to delays between expression peaks. In other words, they correspond to phase shifts between different oscillating components. But the existence of such phase shifts does not imply that the core mechanism is driven by a delayed differential equation similar to  Eq. \ref{delayed_mRNA}, nor that a delay in expression corresponds to an explicit delay in a process corresponding to parameter $\tau$.   

Such questions are especially crucial when relating theory to experiments to understand the origin of the segmentation period \cite{Matsuda2020a,Lazaro2022}. For instance, the delayed transcription model was tested in zebrafish by considering mutants of \textit{her1} and \textit{her7} \cite{Hanisch2013}. Their respective gene sequences have very different lengths, so if delays come from the time to transcribe genes, one should observe very different segmentation clock periods in mutants. But their period is the same, indeed suggesting the "delays" (or rather phase shifts) come from other processes.




 Another issue is that delayed models can "easily" oscillate, so are possibly less realistic and might miss important properties. For instance, in the circadian clock context, in \cite{Smolen2002} it has been shown that the delayed model bypasses the need for a positive feedback loop, which might in fact be biologically relevant \cite{Tyson1999}. Also notice that when a positive feedback loop is present, the divergence of the flow is not necessarily negative so 2D models can oscillate. We will see below that segmentation oscillators present features reminiscent of excitability and relaxation dynamics, which indeed typically require such positive feedback loops, and can thus give different bifurcation patterns and oscillatory behaviours.
 From a practical standpoint, delayed models are also often more difficult to study mathematically (see e.g. \cite{Francois2005, Kotani2012}), because they explicitly introduce tight couplings between variables expressed at very different  (delayed) times.

\chapsec{Molecular (clock and) Wavefront models}

The models in the previous section mostly focused on the modeling of the oscillation itself. But as clearly discussed in the section \myref{sec:Phase}, segmentation is not only about the clock: it also is about the transition from an oscillatory system to a stable pattern of gene expression accompanied by somite formation. So a full model of segmentation should not only focus on the oscillatory mechanism but also account for such transition, like the original clock and wavefront or Meinhardt models. In particular, the experimental realization that FGF drives the dynamics of the wavefront led to the proposal of different models more explicit from the molecular standpoint on FGF influence on the oscillation.

\chapsubsec{Cycle model with FGF gradient}

Following papers showing the clear influence of FGF on the front determination \cite{Dubrulle2001,Dubrulle2004}, Baker \etal updated the cell cycle model from \cite{McInerney2004} to include a control for the determination front, assumed to be done by an FGF morphogen gradient (while also being more agnostic on the clock nature).  In brief, a new variable $w$ is added to account for FGF, with the following dynamics :

\begin{equation}
\frac{\partial w}{\partial t}=\chi_w-\eta w + D_w \frac{\partial^2 w}{\partial x^2}
\end{equation}

This is a classical production/degradation/diffusion system, with 
\begin{equation}
\chi_w=H(x-x_n-c_n t)
\end{equation}
where $x_n,c_n$ are constant and $H$ the Heaviside function similar to the cell cycle model  \cite{McInerney2004}  (Eq.\ref{H_uv}). We thus see that the production region of FGF moves as a function of time with speed $c_n$. In a stationary regime, we thus expect to observe a gradient of $w$ moving in the direction of increasing $x$, which accounts for the experimental dynamics observed in \cite{Dubrulle2004}.

Then the exact same model as  \cite{McInerney2004}  for oscillations coupled to "pulse" generations is used (Eqs.\ref{cycle_u}-\ref{cycle_v}, with identical forms for $f,g$), but the model is modified to redefine the activation of genes $u,v$ in a region of lower FGF (i.e. past the front), namely Eq.\ref{H_uv} is replaced by :

\begin{equation}
\chi_u=H(w^*-w) \qquad \chi_v=H(t-t_w(w^*,x)-t_s) \label{H_uvw}
\end{equation}

The $\chi_u$ term is relatively straightforward to interpret: $u$ can get activated only when $w$ is lower than a constant $w^*$, defining the determination front. The $\chi_v$ looks very different from the term in Eq. \ref{H_uv}: $t_w(w^*,x)$ is defined at the time when $w$ reach $w_*$ at position $x$, and $t_s$ accounts for the period of the segmentation clock. So it is assumed that the system gets activated at the front, but the $v$ pulse is delayed for a (segmentation) clock period to activate.

It turns out both those terms are in fact mathematically equivalent to the terms in Eq. \ref{H_uv}, so the model is not fundamentally different but simply ties parameters of the initial model  \cite{McInerney2004}   to the dynamics of $w$. To see this, let us define $x^*(w^*,t)$ the position at which $w$ reaches $w^*$ at time $t$. Since the gradient of $w$ is moving with speed $c_n$ we thus have

\begin{equation}
    x^*(w^*,t)=x^*(w^*,0)+c_n t \label{x_w}
\end{equation}

With this, we can redefine the $\chi_u,\chi_v$ production terms in terms of $x$ to get
\begin{equation}
    H(w^*-w)=H(x^*(w^*,t)-x)=H(c_n t+x^*(w^*,0)-x)
\end{equation}

so that $\chi_u$ is completely equivalent to the initial cell cycle model by taking $x^*(w^*,0)=x_1$ in Eq. \ref{H_uv}.

Noticing that from Eq. \ref{x_w} we have

\begin{equation}
c_n t_w(w^*,x)=x-x^*(w^*,0)
\end{equation}
we get 
\begin{eqnarray}
\chi_v&=&H(t-t_w(w^*,x)-t_s) \nonumber \\&=& H(c_n(t-t_w(w^*,x)-t_s)) \nonumber \\
&=&H(c_n t- x+x^*(w^*,0)-c_n t_s)
\end{eqnarray}

we see that we get again a similar equation to Eq. \ref{H_uv} by choosing $x_2=x_1-c_n t_s$. This also makes clear that the $x_2$ position where the pulse of $v$ is activated simply is the position reached by the front after a period of the clock $t_s$.

One interest of the model though is that it offers clear tests of what FGF concentration changes should entail on the front dynamics since they directly affect variable $w$. For instance, to model perturbations by a bead soaked with SU5402 similar to \cite{Dubrulle2001}, one can assume that $w$ is transiently going down (due to a sink term for instance). This directly affects the terms $\chi_u$ and $\chi_v$ which are activated later than they should be, so that one transient big somite is formed in this model. However, notice that the initial experiments from \cite{Dubrulle2001} are not quantitative and that there likely is a rich dynamics of FGF inhibition (e.g. diffusion), which could be important to explain other experimental aspects, such as the formation of several smaller somites anterior to the bead.

\chapsubsec{Somite determination: RA-FGF bistability downstream of coupled Notch/Wnt oscillator}

The model discussed in the previous section is really much more focused on the front dynamics than on the clock. For instance, the clock is modeled with pulses of $v$, themselves activated only \textit{past} the front so that there is no real account for the multiple oscillatory pathways in the PSM way before the front.

In a series of papers, Goldbeter and co-workers rather suggested detailed models of the entire process from oscillation to somite formation \cite{Goldbeter2007, Goldbeter2008}. Following descriptions of oscillatory pathways, a first model was built to couple three oscillators on three different pathways \cite{Goldbeter2008}. Three" modules" are explicitly considered and modeled: a Notch module relying on a negative feedback loop  where \textit{Lfng} prevents its own activation, a negative feedback loop in the Wnt pathway implicating \textit{Axin2} via Protein-protein interactions, and a negative feedback loop in the FGF pathway. Interestingly, each of those modules is proposed to oscillate  by themselves, with close but different periods, but it is hypothesized that they are coupled in the embryo to form a global oscillator.

Downstream of this global oscillator, Goldbeter and Pourqui\'e \cite{Goldbeter2007} proposed that the bistability of the traditional clock and wavefront model (Eq. \ref{cusp} and Fig. \ref{CW_maths}) is implemented by mutual repression between FGF and Retinoic Acid (RA). Both proteins are known to form an anteroposterior gradient with inverted polarity (Fig. \ref{fig:MolPlay}). Mutual repression between genes indeed is a classical motif displaying bistability \cite{Gardner2000, Cherry2000}. Just like the classical Clock and wavefront model, the clock needs to deliver a synchronous signal in the future presumptive somite to trigger segmentation, and it is proposed that \textit{Mesp2} (which is not explicitly modeled) could also play a role in this synchronous activation. Remarkably, this model provides a simple explanation of the time dependency defining the bistability region in Eq. \ref{cusp} : it simply arises here from the dynamics of the FGF mRNA and proteins, which degrade as cells get more anterior \cite{Dubrulle2004}. Consistent with the bistability hypothesis, RA response elements  (RARE) sharply turn on at the determination front \cite{Vermot2005}. However, as said earlier the precise role of RA is unclear since mutants of RA still form somites \cite{Vermot2005,Vilhais-Neto2010}, but other proteins could play a similar role.

\chapsec{Somite AP patterning: Inverse problem approach}
\label{Inverse}
As said in the introduction of this section, there are multiple levels to consider to describe the process, and it could be useful to first focus on possible biochemical processes/networks rather than on specific genes. A popular route in systems biology is to frame this as an "inverse" problem question, i.e figuring out (in a relatively unbiased way) what kind of networks :
\begin{itemize}
\item gives rise to oscillations
\item stabilizes into patterns of genetic expressions, e.g. similar to \textit{c-hairy} in chick or pair-rule genes in insects.
\end{itemize}
The Meinhardt model offers an example of such a model but  presents issues, in particular, because it heavily relies on diffusion/cell couplings to stabilize a pattern. So one can wonder if other solutions exist. A popular way to address such problems is to computationally explore the space of models performing specific functions \cite{Salazar-Ciudad2001,Fujimoto2008,Cotterell2010,tenTusscher2011}

\chapsubsec{Clock and switch model}
\label{ClockSwitch}
We used simulated evolution of models of gene networks\cite{Francois2004, Henry2018},  an approach that led to great success to predict general features in many contexts from biochemical adaptation \cite{Francois2008} to immune recognition \cite{Lalanne2013}. Evolution is simulated as a standard mutation-selection process on a population of networks, modeled mathematically using ordinary differential equations. Random mutations include changes in parameters and changes in network connectivity (addition/removal of genes in the network and interactions between them), see \cite{Francois2014} for a review of the method and  \cite{Henry2018} for a recent Python-based implementation.
We used this method to evolve networks giving rise to spatial patterns under the control of either static or dynamic morphogens \cite{Francois2007}, in the absence of cellular communications or diffusion.  Importantly, the only evolutionary pressure (or fitness) was to create patterns of gene expression of one specific gene $E$  along a one-dimensional axis, so we neither imposed constraints on the type of genetic architecture we were looking for, nor on the type of patterns (e.g. on the size or nature of stripes).

Evolution is done under the control of a moving input parameter of the form $G=H(x-vt)$, in the initial work $G$ was effectively modeling a sliding gradient, but in the following, we will show results for $H$ being a step (Heavyside) function for simplicity. A simple network architecture leading to many domains of expression of $E$ spontaneously evolves Fig. \ref{fig:evolved}.  The simulated evolutionary pathway was highly reproducible from one simulation to the other, Fig. \ref{fig:evolved} A, in the sense that successful simulations systematically follow similar evolutionary trajectories, that we describe below.

The first step in evolution is for a gene $E$ to become bistable via a positive feedback loop. In terms of equations, this simple network can be described with :

\begin{equation}
\frac{dE}{dt}=\rho  \mathrm{max}\left(G, \frac{E^n}{E^n+E_0^n}\right)-\delta_E E
\end{equation}

The $\mathrm{max}$ term encodes a transcriptional dynamics akin to a logical OR, meaning that gene $E$ can be activated either by $G$ or by itself (via a Hill function). Such logic is necessary to ensure that gene $E$ can be first activated, then is able to self-sustain when the $G$ morphogen has disappeared, Fig. \ref{fig:evolved} A and B Step 1, leading to bistability. In systems biology, such mechanisms are usually called "switch" and generally rely on self-activation like here, or on mutual repressions \cite{Gardner2000,Cherry2000}, (see also \cite{JutrasDube2018} for a detailed discussion on how to model bistable systems).

Then, the second step is the evolution of a stripe module, where a gene $R$ limits the extension of $E$  Fig. \ref{fig:evolved} A and B Step 2,  giving equations of the form : 

\bwt
\begin{eqnarray}
\frac{dR}{dt} &=&\rho_R G-\delta_R R \\
\frac{dE}{dt} &=&\rho  \mathrm{max} \left(G, \frac{E^n}{E^n+E_0^n}\right)\frac{1}{1+(R/R_0)^m}-\delta_E E
\end{eqnarray}
\ewt
This stripe module relies on what is called an incoherent feedforward loop \cite{Shen-Orr2002,Milo2002}, because the influence of  G is positive on both R and E, but then R "incoherently" represses E. Concretely, the E switch is turned on by G, then off by R,  ensuring a localized expression of E. Such a module is heavily studied in multiple (evolutionary) contexts, see e.g. \cite{Schaerli2018}.

From there, new regulations evolve upstream of $R$, e.g. new repressions. Remarkably, eventually, evolution finds that a very simple way to generate a complex pattern of expression of $E$ is to close a negative feedback loop with delay on $R$, Fig. \ref{fig:evolved} A and B Step 3, to get a system of Equations of the form :

\bwt
\begin{eqnarray}
\frac{dR}{dt} &=&\rho_R \frac{G}{1+(R/R_*(t-\tau))^p}-\delta_R R \label{Rosc} \\
\frac{dE}{dt} &=&\rho  \mathrm{max}\left(G, \frac{E^n}{E^n+E_0^n}\right)\frac{1}{1+(R/R_0)^m}-\delta_E E \label{Eosc}
\end{eqnarray}
\ewt
Looking in detail, Eq. \ref{Rosc} is very similar to Eq. \ref{rescaled}, and with proper parameters (selected by evolution) gives rise to oscillations of $R$, as long as $G$ is high enough. Then $R$ influences $E$ from Eq. \ref{Eosc}: when $R$ is low, the $E$ switch is activated by $G$ but when $R$ is high, $E$ is abolished and the switch is turned off. This means that the system alternates between regions where $E$ is active or inactive depending on the value of $R$. Lastly, when $G$ eventually disappears, the final state of the system depends on the remaining activity of $E$ :  if $E$ is high enough, it sustains its own expression and gives rise to a local gene expression, Fig. \ref{fig:evolved} C. Because the time of disappearance of $G$ depends in a linear way on the position in the embryo, at steady state, one obtains an alternation between regions of sustained $E$ and regions where $E$ could not maintain itself due to $R$. Thus the network effectively turns the $R$ temporal oscillations into a striped pattern of $E$.  Interestingly, when those simulations are repeated, one sees the same kind of evolutionary pathway emerging again and again leading to an oscillator combined with a bistable system, with only changes in the precise implementation of the different parts of the network. For instance, in some cases, we saw the emergence of a Repressilator \cite{Elowitz2002} as the controlling oscillator.

So  the evolved network implements phenomenological dynamics closer to Meinhardt's model than the traditional clock and wavefront model: in particular, just like the Meinhardt model, the clock dies out, and a bistable component discretizes the phase of the clock into a pattern, that we interpret as providing AP polarity to somites. But there are key differences. In Meinhardt's model, the exact interactions between A and P underlie both oscillations and patterning, and the front emerges from non-homogeneous initial conditions and cell-to-cell interactions. Conversely, here  this evolved "clock and switch" model does not require any cell coupling to produce a pattern. It is a purely \textbf{local} model. Also, the striped pattern emerges from the interplay of two, independent, genetic modules (first the switch due to the bistability of E, second the clock due to the negative feedback of R), under the control of a moving morphogen $G$.

The nature of the wavefront (or primary waves) in such a clock and switch model is worth discussing.  In phase models, there is an unambiguous position where the frequency of the oscillators goes to $0$, but we already saw that there is some ambiguity on the front definition for more explicit models (like Meinhardt's), which we face again here. In the clock and switch model, there is a spatially extended transition \textit{zone},  from the oscillatory behavior to the bistable behavior, controlled by $G$. In fact, there are at least \textit{two} primary waves. It is because the generic way to go from an oscillatory to a bistable expression is through \textit{two} bifurcations, as illustrated in Fig \ref{fig:evolved} A, and those will typically not happen for the same values of $G$ without symmetries in the equations. 

As $G$ decreases, the system first goes through a Hopf bifurcation, killing the oscillation, then through a saddle-node bifurcation, creating a bistable system. In between those two bifurcations, there is a region of $G$ where there is only one single fixed point for $E$. If the system spends too much time in this region, any level of molecular noise would jiggle $E$ around so that $E$ would not be able to remember the phase of the clock (position on the cycle).  So, for the bistable system to keep information about the phase of the oscillation, there are only two possibilities: either noise should be strongly suppressed in this region (e.g. with the help of some extra variable reinforcing the phase encoding), or the transition should happen rapidly (e.g. $G$ presents a step-like behavior, effectively collapsing the two primary waves into one, as simulated here). The first scenario appears more consistent a priori with data since there is a clear gradient in the PSM of various observables (from morphogens to frequencies), suggestive of a rather gradual process. Notice the two bifurcations correspond to the two modules that are evolved: the clock needs to die for the bistable system to turn on, and the morphogen should adjust those transitions.

\bnormf
\includegraphics[width=\textwidth]{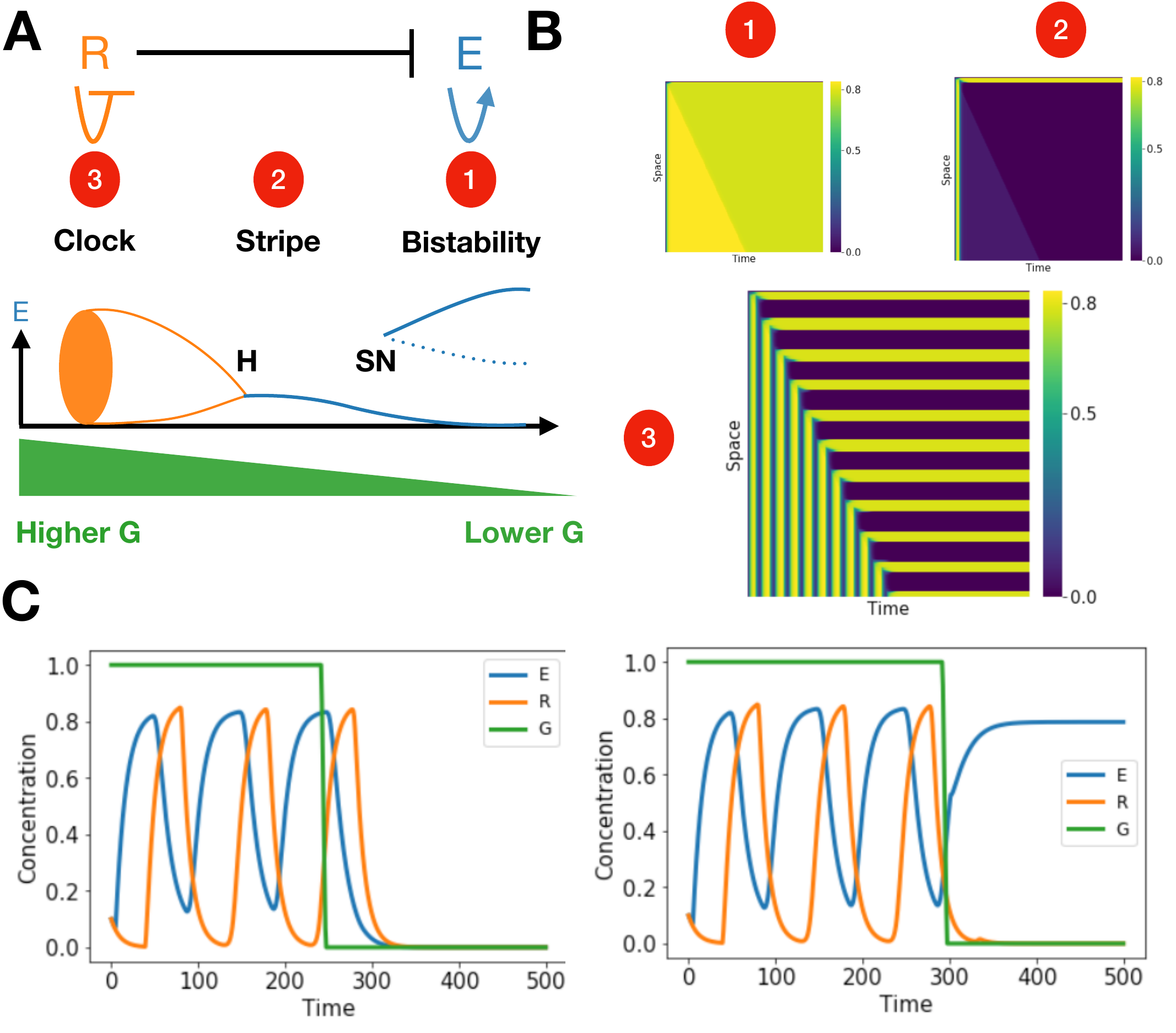}
\caption[Evolved clock and switch model]{(A) Evolved network, corresponding to Eqs. \ref{Rosc}-\ref{Eosc}. $R$ represses $E$ and self-represses with delay. $E$ self-activates. Circle numbers indicate the order of appearance of each interaction during evolution (see corresponding kymographs in B).  We show below the network a scheme of the associated bifurcation diagram: when activator $G$ decreases, the system goes through a Hopf bifurcation (H) and then a saddle-node bifurcation, i.e. transitioning from a clock to a bistable system. Notice the modular structure of the network recapitulating both the  bifurcation diagram and the evolutionary steps. (B) Kymograph for $E$ under control of a moving, step-wise $G$ dynamics, for the 3 evolutionary steps corresponding to the addition of three interactions indicated in A  (C) Two single-cell trajectories with different dynamics of $G$ for the final evolved network. Depending on the timing of $G$ disappearance, the system ends up either in a high or low $E$ state }\label{fig:evolved}
\enormf

\chapsubsec{Inferred PORD and hybrid models}

For small enough networks, another way to figure out network topologies performing a specific function simply is a direct enumeration, a method pioneered in \cite{Cotterell2010}. The advantage of this method is that it provides a more extensive sense of simple possible mechanisms, with the drawback that solutions might not be "evolvable". Also notice that by design such an approach can not explore networks with more than a couple of nodes because of combinatorial explosion.

When this approach is applied to stripe formation in a somitogenesis-like context, Cotterell \etal \cite{Cotterell2015a} first recover the clock and switch mechanism described in the previous section (that they call "Clock and Gradient Model") but they also exhibit another mechanism that they call "progressive, oscillatory reaction diffusion" (PORD) model. Equations for this mechanism are 

\begin{eqnarray*}
    \frac{dA}{dt} &=& ReLu_s(k_1 A -k_2R+F+\beta)-\mu A \\
    \frac{dR}{dt} &=& \frac{k_3 A}{1+k_3 A}-\mu R+D \Delta R
\end{eqnarray*}

where $ReLu_s (x)=ReLu(\frac{x}{1+x})$ and $ReLu(x)=max(x,0)$ is the rectified linear function (commonly used in machine learning). $F(x,t)$ depends on both time and space and is assumed to be a sliding morphogen gradient, encoding FGF. $A$ and $R$ are respective activator and repressor. $R$ is clearly activated by $A$ only. The $RELU_s$  function for $A$ thus encodes a saturating activating function, positively influenced by $A$ (and $F$) and negatively influenced by $R$. So the network dynamics with $A$ self-activating and $R$ diffusing and repressing $A$ is very similar to the reduced version of Meinhardt model Eqs. \ref{Meinhardt_reduced_A}-\ref{Meinhardt_reduced_S}, see phase space and kymographs for a system with constant F in Fig. \ref{fig:PORD}, to be compared with Figs. \ref{fig:MeinhardtVDP-flow}, \ref{fig:MeinhardtVDP-pattern}. It should however be mentioned that this PORD model is more realistic than the Meinhardt model in the sense that it relies on more explicit transcriptional biochemistry, in particular with fewer symmetries in the equations. 

Otherwise, the phenomenology of this model is largely identical to the Meinhardt model: in absence of diffusion, the system displays relaxation oscillations, with A switching rapidly between $0$ and its maximum value, and $R$ more smoothly oscillating. When diffusion is included, a pattern emerges from short-range interactions, where bistable stripes of $A$ form, and activate $R$ which diffuses and stabilizes the pattern. This  requires no external morphogen gradient, even though a modulation by morphogen gradient $F$ is included (an external control could also be added to the Meinhardt model to regulate other aspects such as stripe size, see Eq. \ref{Meinhardt_position}). Again, as we discussed in the section on the Meinhardt model,  similar behaviors can be obtained with various relaxation oscillators.

\bnormf
\includegraphics[width=\textwidth]{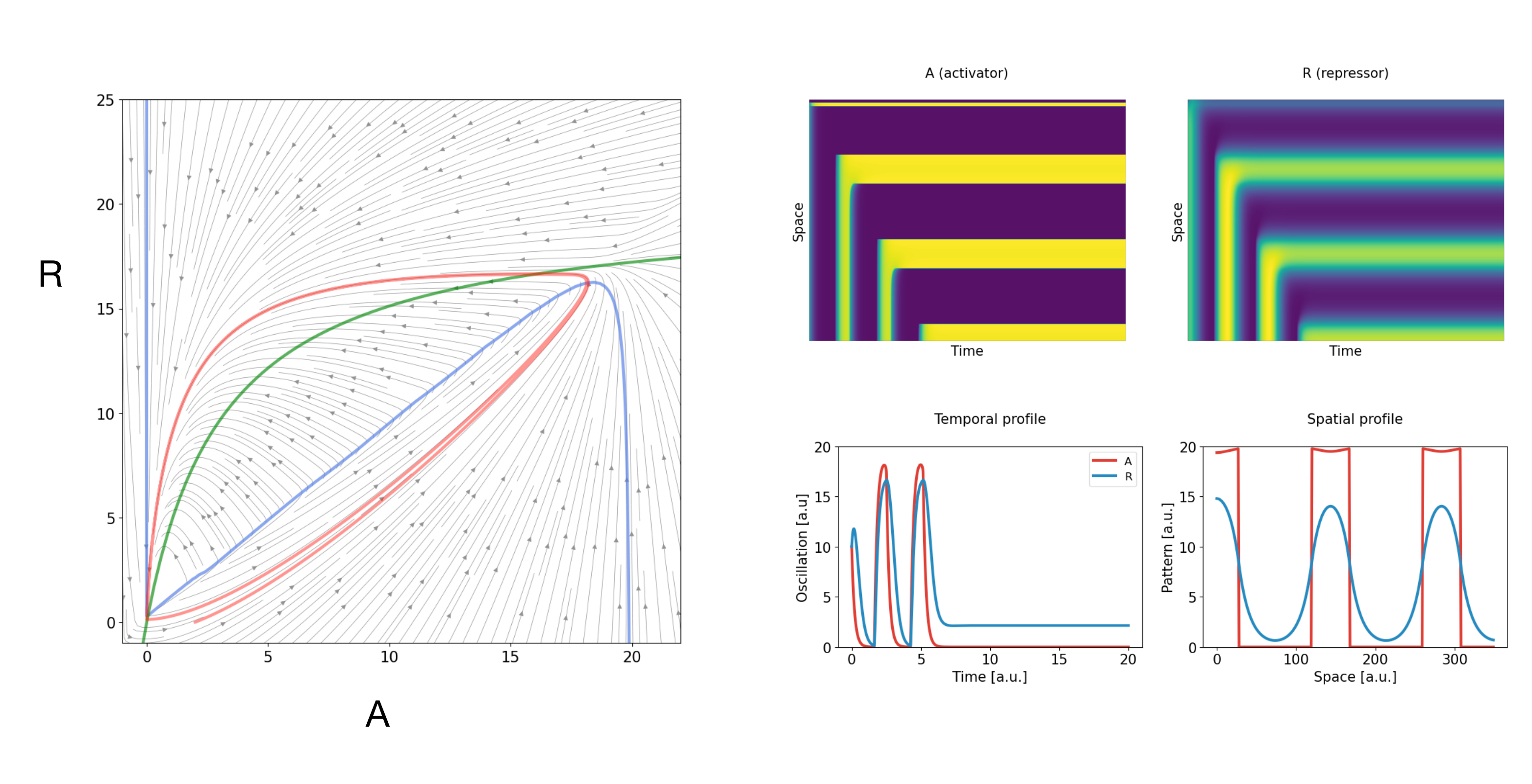}
\caption[PORD model]{ PORD model with constant $F$. On the left, the phase portrait is represented, with $A$ nullclines in blue and $R$ nullclines in green, and the limit cycle is in red. On the right, we show kymographs, temporal behavior in one posterior cell, and final pattern. The behavior of this system is qualitatively similar to the Meinhardt-VanderPol model see Figs. \ref{fig:MeinhardtVDP-flow}, \ref{fig:MeinhardtVDP-pattern} }\label{fig:PORD}
\enormf


There are several possible experimental evidence for the FGF influence. For instance, in this model $F$ can lead to local self-sustained activation of $A$. This appears consistent with the effect of local cutting in the embryo, which induces a FGF response and a stripe of \textit{Lfng}.  Because of the way the pattern is formed (similar to the "domino" effect mentioned for the Meinhardt model), local changes of FGF might "propagate" and influence pattern formation very far from the initial perturbation. This prediction is validated by the fact that a cut in the embryo gives a diagonal, striped pattern far from the cut.

It is, however, not entirely clear how the imposed motion of the initial gradient of $F$ in \cite{Cotterell2015a} couples to the wavefront, since both those speeds are governed by different, uncoupled parameters ($v$ is imposed externally, and the wavefront progression is a purely local variable, as detailed in \cite{Cotterell2015a}). Pantoja-Hernandez \etal \cite{Pantoja2021} include a more explicit mechanism for the coupling between a moving morphogen and the relaxation-diffusion dynamics. They consider a model of the form 

\begin{eqnarray}
    \frac{dA}{dt} &=& P_A(A,R,\beta)-\mu_A A+D_A \Delta A \\
    \frac{dR}{dt} &=& P_R(A)-\mu_R R+D_R \Delta R
\end{eqnarray}
where

\begin{eqnarray}
P_A(A,R,\beta) &=& \frac{\beta +(A/K_1)^{n_1}}{1+(A/K_1)^{n_1}+(R/K_2)^{n_2}} \\
P_R(A) &=& \frac{(A/K_3)^{n_3}}{1+(A/K_3)^{n_3}}
\end{eqnarray}

which again gives rise to relaxation oscillations. They show that for large values of $\beta$ compatible with oscillations, alternating spatial patterns do not appear, while they can form spontaneously for smaller values of $\beta$. They thus suggest that a moving wavefront of $\beta$ might control somite formation: for higher $\beta$, no pattern forms, but when $\beta$ is  graded spatially and reaches smaller values, the Meinhardt-PORD mechanism  kicks in and a pattern stabilizes. So one gets here an external control of the transition from a pure oscillatory system to pattern formation, reconciling gradient-induced transitions with pattern formation, into a "hybrid model". It is suggested that this provides a more robust way to control pattern formation.

\chapsubsec{Experimental evidence for a switch}

Such systems-level approach yields important predictions that can be tied to specific gene dynamics. AP identities within somites are plausible candidates for the two steady states found in the Clock and Switch model, and should be associated with specific genes. Notice that anterior and posterior fates within somites have mutually exclusive markers, so that the evolved positive feedback loop on $E$ (Eq. \ref{Eosc}) could be a mutually repressing system \cite{Cherry2000, Gardner2000}. Experimental evidence is consistent with the idea that anterior-posterior somite fates are bistable identities discretizing the clock phase. Genes of the Notch signaling pathway are indeed implicated both in the clock and in anteroposterior definition, e.g. in zebrafish, \textit{deltaC} oscillates and becomes restricted to half a somite, consistent with the behavior of gene $E$ \cite{Giudicelli2004}.  \textit{Dll1} in mouse appears to be necessary and sufficient for the definition of posterior somite fate \cite{Saga2001}, and is maintained after somite boundary formation. It should nevertheless be pointed out that in the clock and switch model,  $E$ does not necessarily have to be expressed in the entire PSM: it is only required that it oscillates right before the clock stops to encode the pattern. \textit{Mesp2} precisely displays such behavior:  it is controlled by the clock, expressed right at the front, and localizes in the anterior part of the somite \cite{Saga2012}  (see also \cite{Oginuma2010} for a study combining theory and experience showing how a spatial \textit{Mesp2}  gradient can be established within one future somite in response to the clock). Interestingly, anteroposterior markers are expressed stably before boundary formation, and there is multiple evidence that they control genes implicated in segmental border formation \cite{Nakajima2006}. If somite cells are indeed bistable, one expects that those fates could be stably induced  independently of the clock, which has been realized in an in vitro system for \textit{Mesp2}  and \textit{Uncx4.1} \cite{Miao2022}.

Lastly,  evidence from some mutants is also very consistent with the clock and switch model. Convergence-extension mutants in zebrafish have two-cell wide somites, and, quite spectacularly, display single-cell wide rows,  expressing  anterior or posterior markers in alternation  \cite{Henry2000}. Furthermore, those extremely narrow somites nevertheless segregate.  This excludes any predefined length scale in the process (e.g. due to diffusion) that would define somite length. This is rather consistent with a cell-autonomous, local  mechanism, where the final pattern is more dependent on the time course of a morphogen like $G$ which, here, would lay down alternating fates. It is also quite spectacular that only two fates are expressed in such mutants, showing that no stably expressed "third" state is necessary for somite formation, and that the clock information is clearly discretized, again completely consistent with the clock and switch  model.

\chapsec{Landscape geometry of segment formation}
\label{SNICsection}

\bnormf
\includegraphics[width=\textwidth]{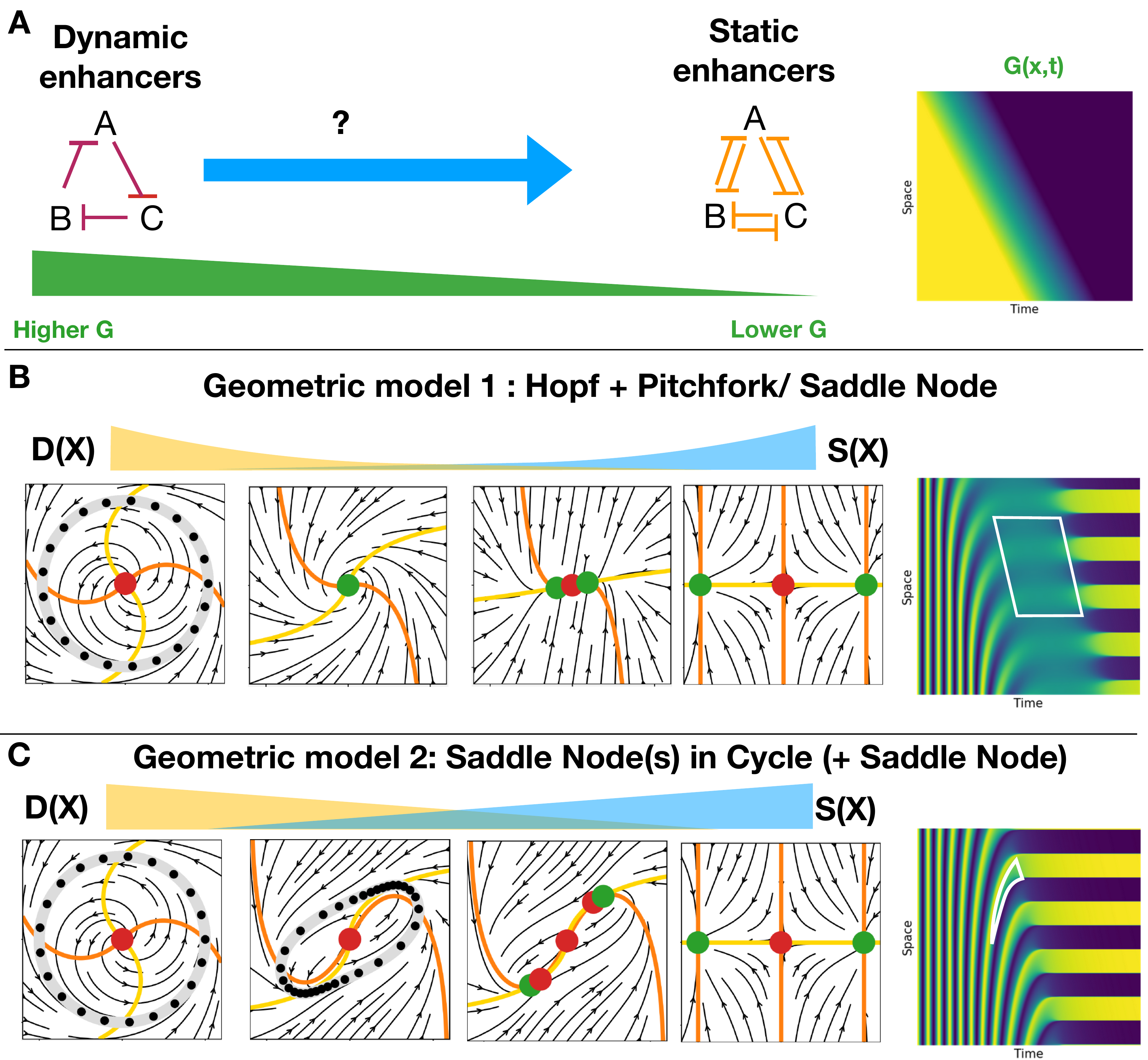}
\caption[Building a Geometric model]{(A) Building a geometric model: a system continuously translates from a dynamical, oscillatory state, towards a multistable one, under the control of a sliding morphogen $G$. (B) If the transition is non-linear, the system goes through a Hopf then a saddle-node/pitchfork bifurcation. Corresponding flows in the phase plane are represented for lower values of $G$ from right to left. The limit cycle is shown in light gray in the 2D plane, stable fixed points in green, and unstable fixed points in red.  A kymograph for $x$ is shown on the right, with the "blurred" transition zone highlighted. (C) If the transition is more linear, the oscillation disappears through a SNIC bifurcation, where new fixed points appear on the cycle (giving rise to excitability). Notice the sharper kymograph, we also highlight the asymmetry in the wave. }\label{fig:geom_model}
\enormf

More and more data are accumulated on embryonic development (e.g. via single cell RNAseq \cite{Heumos2023}), and low dimensional representations of data using unsupervised machine learning techniques (such as UMAP \cite{McInnes2018}) are becoming increasingly popular. Mathematical constraints on  low-dimensional systems aiming at modeling development are under intense scrutiny. Models are generally of the so-called  'Morse-Smale' type and  can be fully characterized by the way attractors (corresponding to cellular states) appear or disappear \cite{Rand2022}. One can also characterize the way morphogens can modulate cell fates, and derive experimental predictions from small, "gene-free" models. Recent examples of this approach include predictions on \textit{C. Elegans} vulva formation, and  neural progenitor differentiation \cite{Corson2017,Saez2021}.

This is very reminiscent of the initial clock and wavefront model, based on a landscape description (e.g. Fig. \ref{fig:CW_Waddington}). Assuming that in each cell there is indeed a well-defined oscillator in the posterior and that oscillations eventually stabilize into different fates, we expect from dynamical systems theory that there is only a limited number of possible effective models describing the cells' behaviors, and associated bifurcations (in the dynamical system sense).

Concretely, how can such bifurcations occur? Developmental systems are regulated by different enhancers at a different time of development. Many dynamical features of the development are correlated to changes in enhancer controls, coordinating the expression of genetic 'modules' \cite{Mau2022,Lalanne2022}. This has been clearly shown experimentally for gap genes control in fly (where different enhancers correspond to different initial and final positions of gap gene \textit{Kni} \cite{El-Sherif2016}), and led to the suggestions that in other insects, there could be both "dynamic" and "static" enhancers, the former responsible for the initiation of the patterns and the later for its stabilization. A model of \textit{Tribolium} segmentation based on those ideas recapitulates well both the normal developmental transition and mutants \cite{Zhu2017}.

Elaborating on those ideas, we  studied the transition of a system from a set of enhancers controlling an oscillation toward another set of enhancers controlling a bistable system \cite{Jutras-Dube2020}, Fig. \ref{fig:geom_model}. 
This can be done with explicit, gene-network models, or with purely geometric descriptions of a 2D flow.  Interestingly, the geometric description captures well all the interesting features of  more explicit gene network models, as we now describe.

In detail, let us consider two 2D flows, one corresponding to an oscillator $D(X)$ and the other corresponding to a multistable system $S(X)$, where $X$ is a 2D vector. We assume that the dynamics of the system are given by the following equation :

\begin{equation}
\dot X= \Theta_d(G)D(X)+\Theta_s(G)S(X)
\end{equation}

where the weights $\Theta_d$ and $\Theta_s$ are functions of an external parameter $G$. For instance, if we take $\Theta_d(1)=\Theta_s(0)=1$ and $\Theta_d(0)=\Theta_s(1)=0$, then system $X$ moves from the oscillator $D$ to the multistable system $S$ as $G$ decreases from $1$ to $0$. Corresponding attractors are illustrated in Fig. \ref{fig:geom_model}. Practically, $G$ corresponds to a morphogen traveling over the embryo, so takes the form $H(x,t)=H(x-t)$ where $H$ is a decreasing function from $1$ to $0$. Other enhancers can be of course added in a similar way.

We used various polynomials for $\Theta$s and standard models for both the oscillator and the multistable system. In particular, for the oscillator $D(X)$ we can use a Poincar\'e model best written in polar coordinates  (Fig. \ref{fig:geom_model} B and D, left-most flow plot):
\begin{eqnarray}
\dot r & =&r(1-r) \\
\dot \theta & = &1 
\end{eqnarray}
 while possible equations for the bistable system is  (Fig. \ref{fig:geom_model} B and D, right-most flow plot)
 
\begin{eqnarray}
\dot x & =& x(1-x^2) \\
\dot y & = & -y 
\end{eqnarray}
ensuring that $(-1,0)$ and $(1,0)$ are stable fixed points.

Let us immediately point out that flows here are very symmetrical, but the symmetry can be broken without losing the general properties observed \cite{Jutras-Dube2020} (see also discussion in section \myref{GeomtoPhase}). We also checked that the bifurcation diagrams observed and the general properties of the system are largely preserved when using different attractors (e.g. different types of non-linear oscillators).

Such a system recapitulates the sequence of two primary waves (Hopf+saddle-nodes) first obtained through computational evolution, by taking non-linear functions  $\Theta$ and further stabilizing the origin (Fig. \ref{fig:geom_model} B). However, we also obtained a much more generic behavior, when the transition from the oscillatory attractor to the bistable one is linear (e.g. with $\Theta_d(G)=1-\Theta_s(G)=\theta_G$)

\begin{eqnarray}
\dot x & =& \theta_G(1-r) x -\theta_G y+ (1-\theta_G)x(1-x^2) \\
\dot y & = &  \theta_G(1-r) y+\theta_G x-(1-\theta_G)y 
\end{eqnarray}

with $r=\sqrt{x^2+y^2}$.

It is useful to start with an intuitive explanation of what happens then. If the $\Theta$s vary smoothly with $G$, e.g. $\Theta_G=G$, the dynamics of the system around a critical value $G_c~1/2$ is a "perfect" mix of an oscillator and a multistable system. Right below $G_c$, the system does not oscillate, but by continuity, half of the dynamical flow (corresponding to the limit cycle) still is present in the dynamics, meaning that for some initial conditions, the system might still follow part of the cycle. This is reminiscent of what is observed in excitable systems which have a single fixed point/steady state, but when perturbed suitably, come back to it through long excursions very similar to cycles  (see Appendix). Such systems are very common and have been well-studied in theoretical neuroscience  \cite{Izhikevitch2007}.

With this geometric modeling, we in fact obtained  excitable so-called type I oscillators \cite{Izhikevitch2007}, which arise through a bifurcation called a SNIC (Saddle-Node on Invariant Cycle, see Appendix). SNIC bifurcations exactly correspond to the intuitive picture of a system being at the same time oscillating and stable: they arise when a fixed point appears on a limit cycle or, conversely, when a pair of stable/unstable fixed points cancel out on a closed trajectory to give rise to oscillations (explaining the name of the bifurcation). Right after a SNIC, the system is normally monostable, but  in our case, if both $D$ and $S$ are symmetrical enough, a saddle-node bifurcation generates a second fixed point, giving rise to bistability encoding fates, Fig. \ref{fig:geom_model}, Geometric Model 2. So similar to the Clock and Switch model there still are two primary waves, but with a SNIC bifurcation instead of a Hopf one to destroy the oscillator.

By varying the functions $\Theta(G)$, one can easily compare different bifurcation scenarios with the same "boundary conditions", i.e. dynamic and static flows $S,D$. Kymographs are shown in the right column of  Fig. \ref{fig:geom_model}B-C for two scenarios. One can then compare a Hopf+saddle-node scenario to a SNIC+saddle-node one. In short, the SNIC bifurcation scenario is in general more robust to all kinds of perturbations (noise, shape of traveling morphogen $G$) than the Hopf scenario (as quantified in \cite{Jutras-Dube2020}), suggesting that it might be evolutionarily adaptive. The main reason is that at the Hopf bifurcation, the dynamics tend to "collapse" at the fixed point so that the phase information from the cycle can be easily lost through perturbations (as is visible in the highlighted blurry phase in  the kymograph for the Hopf model in Fig. \ref{fig:geom_model} B). Conversely, in the SNIC scenario, since fixed points appear on or close to the cycle, information about the phase is more directly encoded and less easy to perturb (as can be grasped from the overall sharper kymograph  in Fig. \ref{fig:geom_model}  C for the SNIC scenario). 

Strikingly, the behavior of the system close to the SNIC bifurcation is qualitatively reminiscent of what is observed in actual cells. As we get closer and closer to the SNIC bifurcation in this model, the period of the cycle diverges to infinity (explaining why it is sometimes called Saddle Node with Infinite Period bifurcation (SNIPER) \cite{Strogatz2018}). So SNIC/SNIPER bifurcations induced by the transition to bistability offer a very natural explanation for the considerable period slowing down and the associated kinematic waves observed in embryos (in particular zebrafish and snake). Another interesting aspect is that close to the bifurcation, the SNIC system behaves like a relaxation oscillator, which gives rise to rather asymmetrical wave profiles, looking more like a sawtooth profile (see e.g. the highlighted "`teeth" in the yellow phase of the model 2 kymograph  in Fig. \ref{fig:geom_model} C). Such asymmetrical waves are observed experimentally \cite{Shih2015} and motivated the phenomenological coupled phase-amplitude model (which assumed a form of  excitability) described  in Section \myref{sec:phaseamplitude}. This further provides a possible answer to the somite polarity conundrum, like explained at the end of section \myref{sec:Phase} : because of those asymmetrical waves, an anterior/posterior compartment boundary in a future somite is not comparable to a posterior/anterior compartment boundary between somites.  Lastly, the SNIC scenario could be explicitly revealed in mutants of the Wnt signaling pathway. In mouse, a gain of function of $\beta$-\textit{catenin} is associated with more waves in the presumptive PSM, making it more "zebrafish"-like, which would mean here imposing a more gradual bifurcation. Interestingly, Wnt mutants are also associated with richer wave patterns in insect segmentation, suggesting a possibly conserved mechanism.

\chapsec{Clock and switch modulated }

In the geometric picture described above, the period change and corresponding polarity of wave patterns naturally emerge from a limited number of assumptions on the transition between attractors. Clark \cite{Clark2021} revisited the problem of polarity using a more coarse-grained description.  He combines a phase model with a boolean switch under the control of a linearly decreasing "timer", effectively measuring the time a cell takes to mature. This is a local model, where the timer is mathematically equivalent to a sliding spatial morphogen since the temporal decay of the timer results in a spatial gradient along the PSM. Such a model therefore is a discrete version of the clock and switch model presented in section \myref{ClockSwitch}. Because of its discrete nature, there is only one primary wave in this model, transitioning directly from oscillation to bistability, which, as detailed above, is not a generic case. Clark proceeds to add degrees of freedom and couplings to the system, to account for other aspects. For instance, if the clock changes the elongation rate at some phase, one can get at the wavefront an asymmetry in the wave profile, that we can thus associate with somite polarity as pointed out above. Various modulations can also be obtained if the clock influences the timer itself: for instance if the phase of the clock modulates the timer decay, the spatial and temporal dynamics of the phase can be modulated. Notice this effectively means that the cycle in the tail bud (when there is oscillation but no timer) is of a different nature than the cycle in the PSM (where the coupling of the clock to the timer generates a different wave pattern). This is also reminiscent of the model in the previous section, where a change of dynamics is induced by the transition from dynamic to static enhancers: the timer there would thus be very similar to a control parameter. By adding more couplings and more timers, one can then obtain increasingly complex modulations and wave patterns.

\chapsec{From geometric back to phase models ?}
\label{GeomtoPhase}

Geometric descriptions such as the one described in section \myref{SNICsection} have a natural connection to phase models, since the polar angle in the phase plane defines a periodic variable, akin to a phase.  It is then tantalizing to start from the normal form of bifurcation and see if it can be modulated to reproduce the phenomenology of the geometric models. To illustrate this, let us start from the normal form of a SNIC (see Appendix) that introduces a non-linear term in the equation for the polar angle :

\begin{equation}
\dot \theta = 1- a \cos(\theta)  \label{SNICphase}
\end{equation}
The implicit assumption here is that close to the bifurcations, the system quickly relaxes to its limit cycle so that only the angular dynamics are modulated, see explicit solution of Eq. \ref{SNICphase} in Appendix. It is important to point out that although $\theta$ in Eq. \ref{SNICphase} is an angle, it does not correspond to the phase of the oscillator, which is always linear in time (Eq. \ref{phase}). $\theta(t)$ rather defines a new limit cycle, of constant radius $1$ for simplicity, but with local accelerations and slowing-downs of the polar angle in the phase plane.

For $a<1$, the system displays oscillations, but as $a$ gets closer to $1$, the system spends more and more time close to $0$, until the period diverges at $a=1$ and the system stabilizes at $\theta=0$.  The model described by Eq. \ref{SNICphase} accounts for a (generic) change of limit cycle close to the SNIC bifurcation (see "quadratic fire and integrate model" in Appendix).

To further match the geometric clock and switch model, one should look for the simplest way to add bistability. It is easily done with an additional factor 2 in Eq. \ref{SNICphase} :

\begin{equation}
\dot \theta = 1-a \cos( 2\theta)\label{SNICphase_bistable}
\end{equation}
 which ensures that both $\theta=0$ and $\theta=\pi$ are fixed points for $a=1$ (we could even get to an arbitrary number of stable states with the same trick). One can then simulate the behavior of this system with a moving gradient of $a=1-g$  similar to what is done in the previous section, Fig. \ref{fig:Phase_model_SNIC}. We see as expected the transition from oscillations to bistability, discretizing the phase of the oscillator via two simultaneous SNIC bifurcations, see kymograph on Fig. \ref{fig:Phase_model_SNIC} C, top left. As expected for a SNIC, we see a clear period divergence and associated waves, similar to the geometric model described in the previous section.

Because of its compactness, the  polar equation \ref{SNICphase_bistable} presents many implicit symmetries, and as such is not generic. For instance, the fixed points are symmetrical with respect to the origin. One possibility is to break the symmetry to change the positions of the fixed point e.g. by substituting  $2\theta$ in Eq. \ref{SNICphase_bistable} by a term of the form :

\begin{eqnarray}
f(\theta,\alpha) & = & 2 \alpha \theta \quad  \mathrm{   if }   \quad  0<\theta <\pi \nonumber \\
	& = & 2 \alpha \pi + 2(2-\alpha)(\theta-\pi)  \quad  \mathrm{   if }  \quad \pi<\theta <2\pi
\end{eqnarray}
That maps the circle on itself for $0<\alpha<2$. The fixed points are defined by $\cos f(\theta,\alpha)=0$, e.g. for $\alpha>1$ $\theta=0$ or $\theta=\pi/\alpha$.
Importantly, this symmetry breaking introduces an asymmetry in the cycle: before the bifurcation, the oscillation is smoother close to one fixed point and more spike-like close to the other Fig. \ref{fig:Phase_model_SNIC} D. This results in an asymmetry in the wave profile, where there is an equivalent asymmetry between the boundary between the future zone Fig. \ref{fig:Phase_model_SNIC} C, top right. Again, such asymmetry could be used biologically to distinguish between the AP and the PA boundaries. We also notice that increasing $\alpha>1$ increases the ratio of cells going to the fixed point $\theta=0$.

Lastly, in this simple model, the two SNICs happen for the same value of $a$, another symmetry that is not generic. It is more realistic biologically to assume that one saddle-node bifurcation happens first on the cycle (the subsequent one being a saddle-node bifurcation). This can be done by modulating the $a$ term as a function of the phase, e.g. 
\begin{equation}
\dot \theta = 1-(a+\epsilon \sin^2(\theta)) \cos( f(\theta,\alpha))\label{SNICphase_asymmetric}
\end{equation}

While the last phase equation \ref{SNICphase_asymmetric} looks less elegant than Eq. \ref{SNICphase_bistable}, it is more realistic with the addition of two symmetry breakings: the asymmetry in the position of the fixed point ($\alpha$) and the asymmetry in the timing of bifurcations ($\epsilon$) term. The $\epsilon$ term here ensures in particular that the saddle-node bifurcation happens last at $\theta=0$, and thus reinforces the $\theta \neq 0$ fixed point. It is clearly visible in Fig. \ref{fig:Phase_model_SNIC} C, bottom right, where we see that increasing both $\alpha$ and $\epsilon$ gives rise to rather complex, asymmetric and polarized wave patterns (compare e.g. with top left panel). Notice in particular that the wave asymmetry/polarity mostly arises from the symmetry breaking of the fixed points ($\alpha>1$). This simple model thus suggests that the wave patterns visible in the embryo might in fact be a generic consequence of symmetry breakings in the (two) bifurcations (or corresponding primary waves) leading to the stopping of the clock and definition of somite anterior-posterior fates, without the explicit need of ad-hoc controls/feedbacks.

\bnormf
\includegraphics[width=\textwidth]{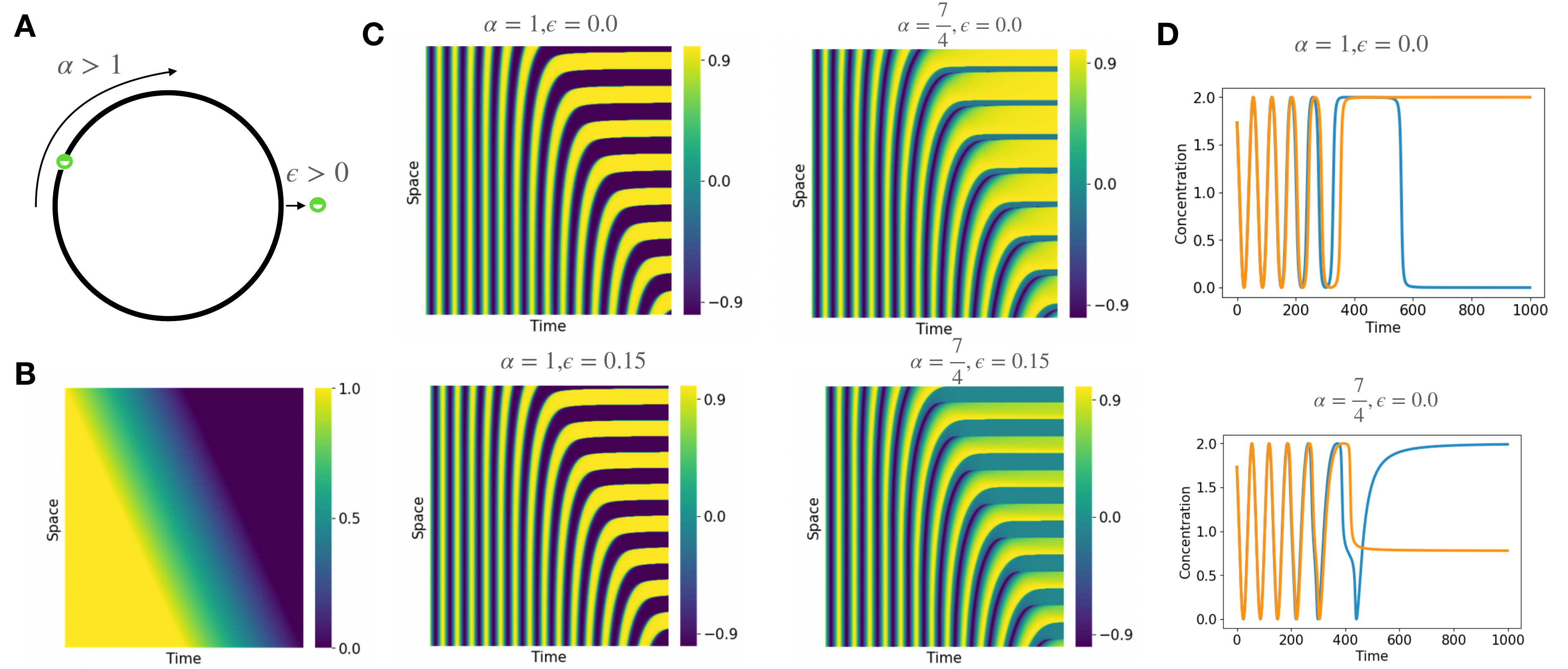}
   \caption[Phase model with SNIC]{Simulation of the Phase model with SNIC defined by Eq. \ref{SNICphase_asymmetric}. (A) Symmetry breakings on the position and appearance of the Saddle Node in Cycle due to parameters $\alpha$ and $\epsilon$. $\alpha>\pi$ moves the second saddle towards the bottom of the cycle, while $\epsilon>0$ delays the appearance of the saddle at $\theta=0$. (B) Control parameter dynamics $a(x,t)$ used for the simulation. We used a sliding dynamics similar to Jutras-Dub\'e, eLife, 2020 (C) Kymographs of the system (we show $\cos(\theta)$)  under control of panel (B) for different values of $\alpha,\epsilon$ (D) Oscillation as a function of time in two cells for two different parameter sets. Notice that for $\alpha=1$ the oscillations are more 'squared' shape close to the bifurcation, while for $\alpha>1$, the cycle is more asymmetric}\label{fig:Phase_model_SNIC}
\enormf

\newpage

\chap{Hacking the segmentation clock}
\label{section:hack}
All models described so far have been mostly based on the description of what happens in embryos, with few additional indications from mutants. In recent years,  progresses in multiple experimental techniques have allowed getting a much finer resolution and control of the system.  Various reporters offer versatile and precious tools to figure out the precise segmentation dynamics, combined with other experimental techniques such as microfluidic controlled cultures, optogenetics, and more recently stem-cell derived systems.  This led to considerable experimental and theoretical advances in our understanding of the system that we now summarize.

\chapsec{Monolayer PSM cell cultures : the $\alpha$ two-oscillator model}

\label{mPSM_theory}
In \cite{Lauschke2013}, a new culture technique was proposed, where a mouse tail bud was extracted and plated, and oscillations were monitored in real-time using a reporter for Lfng (LuVeLu). Following plating, cells move radially, leading to a monolayer of PSM cells (mPSM), with the presumptive tail bud at the center, Fig. \ref{fig:alpha_culture}. Remarkably, those mPSM cultures display robust oscillations. Qualitatively, cells are initially synchronized, then a radial period/frequency gradient self-organizes  until an almost radial wavefront regresses. This leads to the expression of \textit{Mesp2} and even the formation of radial somitic boundaries. The period of the oscillations and the phase difference between the presumptive tail bud and the wavefront are very similar to what is observed in embryos, suggesting that mPSM cells faithfully recapitulate the embryonic process, but with a different geometry and boundary conditions.

\bnormf
\includegraphics[width=\textwidth]{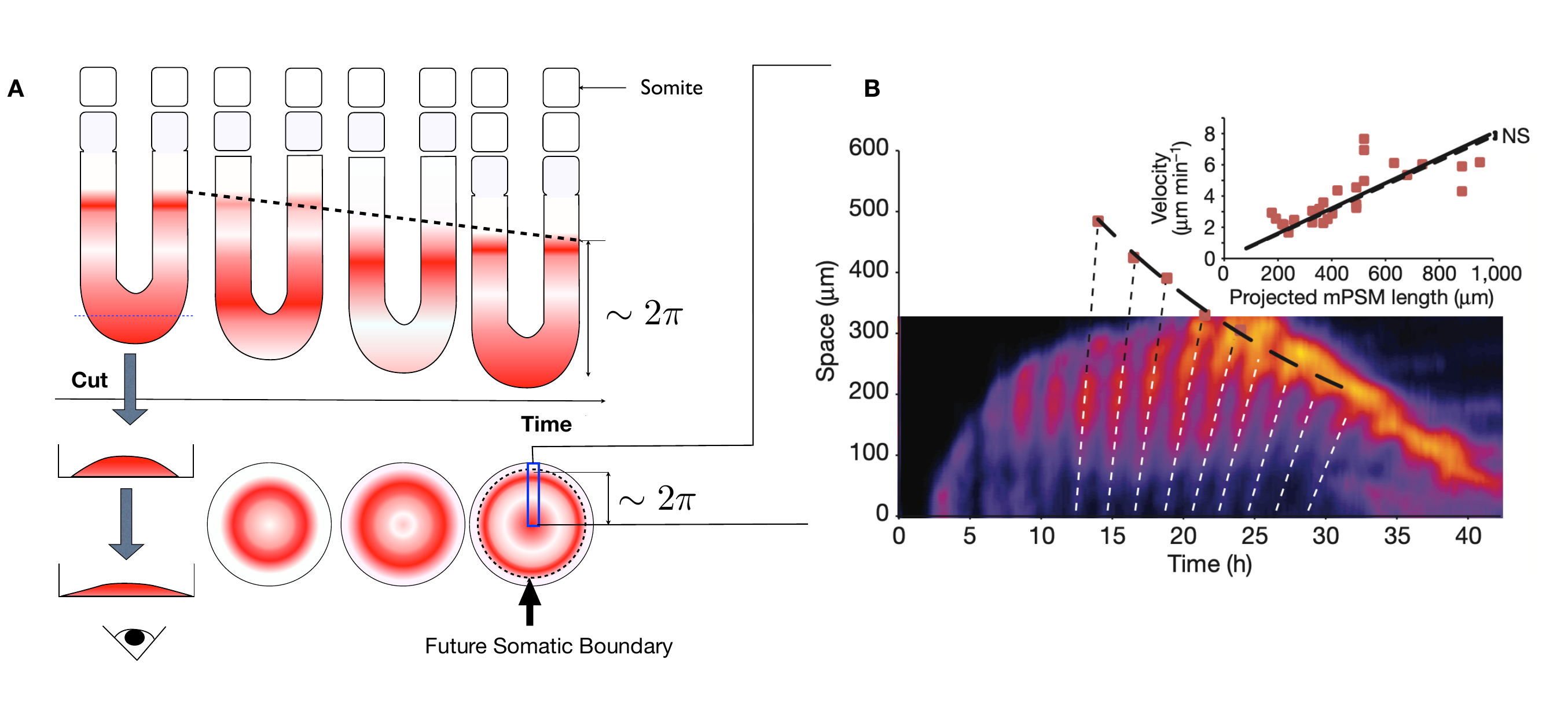}
\caption[mPSM culture]{(A) Principle of the mPSM culture: a tail bud is cut, plated and oscillations are visualized in the dish (B) Experimental data and kymograph for a line of cells in the culture. One sees an exponential shrinking of the oscillatory field, with a phase gradient scaling as a function of time.}\label{fig:alpha_culture}
\enormf

In particular, since cells are simply plated, there is no equivalent of embryonic growth in mPSM. So those cultures allow decoupling the wavefront dynamics from the embryonic growth.  If the wavefront and period gradient are normally controlled by a morphogen, where higher concentration corresponds to faster oscillation, one would expect the emergence of kinematic waves, with a phase gradient increasing as a function of time, but importantly, no propagation of the wavefront (since the morphogen defines one single radius where the frequency is going to $0$). This fits the scaling law where $S=vT$: if $v=0$, then $S=0$, meaning here that we would not see any pattern formation. So the very fact that one sees regression and somite boundaries suggests that this picture is incorrect. One could then expect that for some reason, such morphogen would decrease as a function of time, which would explain regression. However, being more quantitative, mPSM cells display a fascination scaling property suggesting a more 'self-organized' picture :
\begin{itemize}
\item at each cycle, the new "somite" boundary roughly form at  80\% of the radius of the oscillating zone. Since the front regresses towards the center, this means that the somite size is decreasing exponentially as a function of time
\item the phase difference between the center of the oscillation and the boundary of the oscillating zone (i.e. presumptive somite) is essentially constant, and roughly equal to $2\pi$, meaning that  only one wave is propagating in the mPSM at any single time
\item in fact, the entire phase gradient appears to roughy scale linearly with the size of the oscillating zone. This means that the system is essentially scale-free: a smaller mPSM behaves exactly like a bigger one, Fig.  \ref{fig:alpha_culture} B.
\end{itemize}
This suggests at first that there could be an active scaling process, where mPSM would be able to somehow "measure" in real-time its own length and actively scale a frequency/period gradient accordingly. Such a process would be remarkable by itself, suggesting a flurry of new tissue-level interactions, see e.g. \cite{Ishimatsu2018}. It is also consistent with observations made over the years in multiple systems: for instance, as mentioned previously the frequency gradient in snake and zebrafish appear essentially identical in relative PSM units \cite{Gomez2008}, and a similar active scaling of frequency gradient is necessary to fully account for the Doppler Period Shift described above \cite{Soroldoni14}.

 However, careful examination of the mPSM dynamics suggests another possibility. While it takes a few oscillations before the appearance of the somite boundary, the apparent scaling process starts way \textbf{before} the formation of the first somite boundary.  By this, we mean that the dynamics of the phase gradient are consistent with the existence of a  shrinking "ghost" PSM,  much bigger than the culture, Fig.  \ref{fig:alpha_culture} B. This excludes that the scaling process observed mainly comes from some biochemical process performing a complex PSM measurement/scaling. Instead, this suggests the observed slowing down/phase gradient establishment finds its origin in more \textbf{local} dynamics, from which the observed scaling emerges.
 
 It is useful to turn to mathematical modeling to understand what happens. The effective scaling of the phase gradient suggests an exponential phase dependency where calling $x$ the distance to the center of the culture, one has 
 
 \begin{equation}
 \phi(x,t)=t-\lambda x e^{\alpha t}  \label{AnsatzmPSM}
 \end{equation}
 This equation looks very similar to what we obtain from the Ansatz in the moving frame of reference \ref{Ansatz}, with a first time proportional to $t$, accounting for tail bud oscillation, and the second term accounting for a linear phase gradient, with an exponential dependency accounting for 'scaling'.
To account for the front propagation, which experimentally occurs for a phase difference of approximately $2\pi$ with the tail bud, we define the position of the front $x^*$ such as
 \begin{equation}
 \phi(x^*,t)=t-2\pi  
 \end{equation}
 
 We immediately get 
 \begin{equation}
 x^*(t)=\frac{2 \pi}{\lambda}e^{-\alpha t} \label{shrink_alpha}
 \end{equation}
  indicating that the front regresses with an exponential dynamics, which is indeed observed experimentally.  It is also useful to rewrite Eq.  \ref{AnsatzmPSM} into
\begin{equation}
 \phi(x,t)=t- \frac{1}{2\pi}\frac{x}{x^*(t)}
 \end{equation}
 which expresses the fact that the phase of the clock depends only on the relative position of a cell within the mPSM $\frac{x}{x^*}$, again consistent with the scaling observation.
 
In the spirit of previous models, it is then useful to take the time derivative of Eq. \ref{AnsatzmPSM} to get :

 \begin{equation}
 \frac{\partial \phi}{\partial t} = 1-\alpha \lambda x e^{\alpha t}  = 1-\alpha \frac{1}{2\pi} \frac{x}{x^*(t)} \label{alpha_x}
 \end{equation}

which is similar  to Eq. \ref{doppler} but with $v=0$ (consistent with the absence of growth in mPSM). An important difference is that Eq. \ref{alpha_x} describes what happens not in an abstract, moving frame of reference, but rather in a given cell at constant position $x$.  Furthermore, the dynamics is observed even if $x^*(t)$ is higher than the culture size, so that it can not correspond to a "physical" measurement of the mPSM radius, but rather emerges from an unknown dynamical process.

To make further progress, and suggest a possible, mostly local, mechanism, we notice that $\lambda x e^{\alpha t}=t-\phi(x,t)$ which means that we also have :
 \begin{equation}
 \frac{\partial \phi(x,t)}{\partial t} = 1-\alpha(\phi_0(t)-\phi(x,t)) \label{alpha_model}
 \end{equation}

or
 \begin{equation}
 \frac{\partial (\phi - \phi_0)}{\partial t} = \alpha(\phi-\phi_0(t)) \label{coupled_model}
 \end{equation}

where we defined $\phi_0(t)=t$ the phase of the tail bud.
This very simple equation captures all aspects of the phase dynamics, with only one scale-free parameter, $\alpha$, related to the effective shrinking rate of the PSM as  visible from Eq. \ref{shrink_alpha}. Furthermore, this simple equation bypasses the need for any active scaling mechanism: a cell at position $x$ can adjust its frequency just based on the knowledge of its local phase on the one hand, and of the phase of the tail bud on the other hand. This model can be made purely local if we assume that, in a single cell, there are not one but \textit{two} oscillators: one "reference" oscillator, corresponding to $\phi_0$ (assumed to be synchronized over the entire PSM), and the local Notch oscillator, corresponding to $\phi$. In such a situation, a cell just needs to effectively compute the phase difference between $\phi_0$ and Notch and to slow down accordingly in a cell-autonomous way. Notice that $\phi-\phi_0=0$ is an (unstable) fixed point of the dynamics, so that the model can be extended to include the transition from tail-bud to PSM via a small initial phase-shift (see Supplement of \cite{Lauschke2013}).

Starting with a small, non-zero initial condition, Eq. \ref{coupled_model} suggests that the phase difference diverges exponentially, which neither makes mathematical or biological sense. For this reason, we introduce a stopping rule, consistent with the experimental observations in \cite{Lauschke2013}: a well-defined phase difference $\phi_*$ defines the front, then the clock stops. We can now formalize the full $\alpha$ model :
\begin{itemize}
    \item We assume the existence of a reference oscillator $\phi_0(t)$, with the same period as the segmentation period
    \item Oscillations in the PSM are cell autonomous, and the phase within a single cell follows the following dynamics with respect to the reference oscillator
    
     \begin{equation}
 \frac{\partial (\phi - \phi_0)}{\partial t} = \alpha(\phi-\phi_0(t))
 \end{equation}
    
    \item The oscillation stops when $|\phi - \phi_0|=\phi_*$
    
\end{itemize}

This model is summarized in Fig.  \ref{fig:alpha_model}, with corresponding kymograph.

\bnormf
\includegraphics[width=\textwidth]{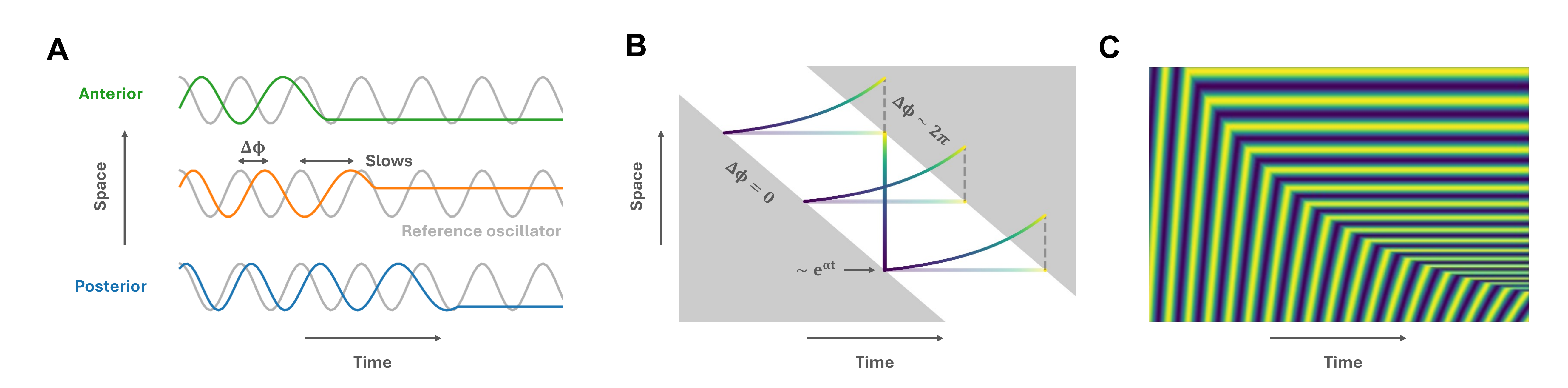}
\caption[$\alpha$ model]{Illustration of the $\alpha$ model, Eq. \ref{alpha_model} (A) Each cell has two oscillators, the reference one, with constant frequency, and Notch. As the phase shift between the two oscillators increases, the Notch oscillator further slows down (B) This gives rise to non-linear (exponential) dynamics of the phase shift between the oscillators. Notch stops when some critical phase shift is reached. (C) If the initial phase gradient is approximately linear, the phase gradient `scales' on a kymograph.}\label{fig:alpha_model}
\enormf

Interestingly, the existence of such reference oscillator $\phi_0$, which here derives from purely theoretical considerations, is completely consistent with the phenomenology of other oscillatory pathways than Notch. Niwa \etal \cite{Niwa2011} have observed FGF oscillatory dynamics (via its target, phosphorylated \textit{Erk} - \textit{pERK}) are much more synchronized in the PSM than Notch oscillation, so that FGF could play the role of $\phi_0$.  By comparison of the dynamical expression (waves) of \textit{Hes7} with the pERK oscillations, they suggest that a future somite could be defined in the following way: first a broad anterior stripe of Notch signaling (NICD) is formed due to the wave dynamics. Then when \textit{pERK} turns on, and the overlap between \textit{pERK}  and NICD defines the future somite, so that when \textit{pERK}  turns off, NICD activates \textit{Mesp2}. This is completely consistent with the idea that a given phase difference between two pathways defines the segmentation front. Another candidate for the $\phi_0$ reference oscillator in mouse is the Wnt signaling pathway, since real-time monitoring of oscillations reveals that   \textit{Axin2}  oscillation is  synchronized as well in the entire PSM \cite{Sonnen2018}. Because of this, Notch and Wnt oscillations appear in phase opposition in the posterior close to the tail bud, and exactly in phase in the anterior close to the front, again suggesting that phase coordination between two oscillators might define the differentiation front.

One could argue that the addition of another oscillator in the system adds more problems than it solves. However, as said above, it is very clear that there are oscillators of different natures driving the system (possibly including the cell cycle, see discussion), and has been already hypothesized in earlier models e.g. by Goldbeter and Pourqui\'e \cite{Goldbeter2008}. The introduction of a second oscillator allows for a parsimonious explanation of the ubiquitous  scaling dynamic, without the need to account for additional mechanisms actively adjusting frequency gradient within the PSM. Noteworthy, the idea that phase shifts could define positional information in developing systems had been proposed and discussed in the late 60s by Goodwin and Cohen \cite{Goodwin1969}, although their mechanism relies on the propagation of waves at different speeds from a common pacemaker, with very fast frequencies compared to morphogenetic time scales.

\chapsec{Entraining the segmentation clock}

It is well known since classical observations of coupled pendulum clocks by Huyghens \cite{Yoder1988} (recently studied in \cite{Goldsztein2021}) that non-linear oscillators can synchronize, with one another or in response to external perturbations. In a nutshell, the theory of phase response (see Appendix A) predicts that periodic stimulation of an oscillator entrains it, meaning that a fixed phase relationship between an external signal and the oscillator of interest is established \cite{Pikovsky2003,Cross2005,Izhikevitch2007}. So the question is: can we entrain an embryonic oscillator as complex as the segmentation clock \cite{Juul2018}? Does classical phase response theory apply to embryos?

\bnormf
\includegraphics[width=\textwidth]{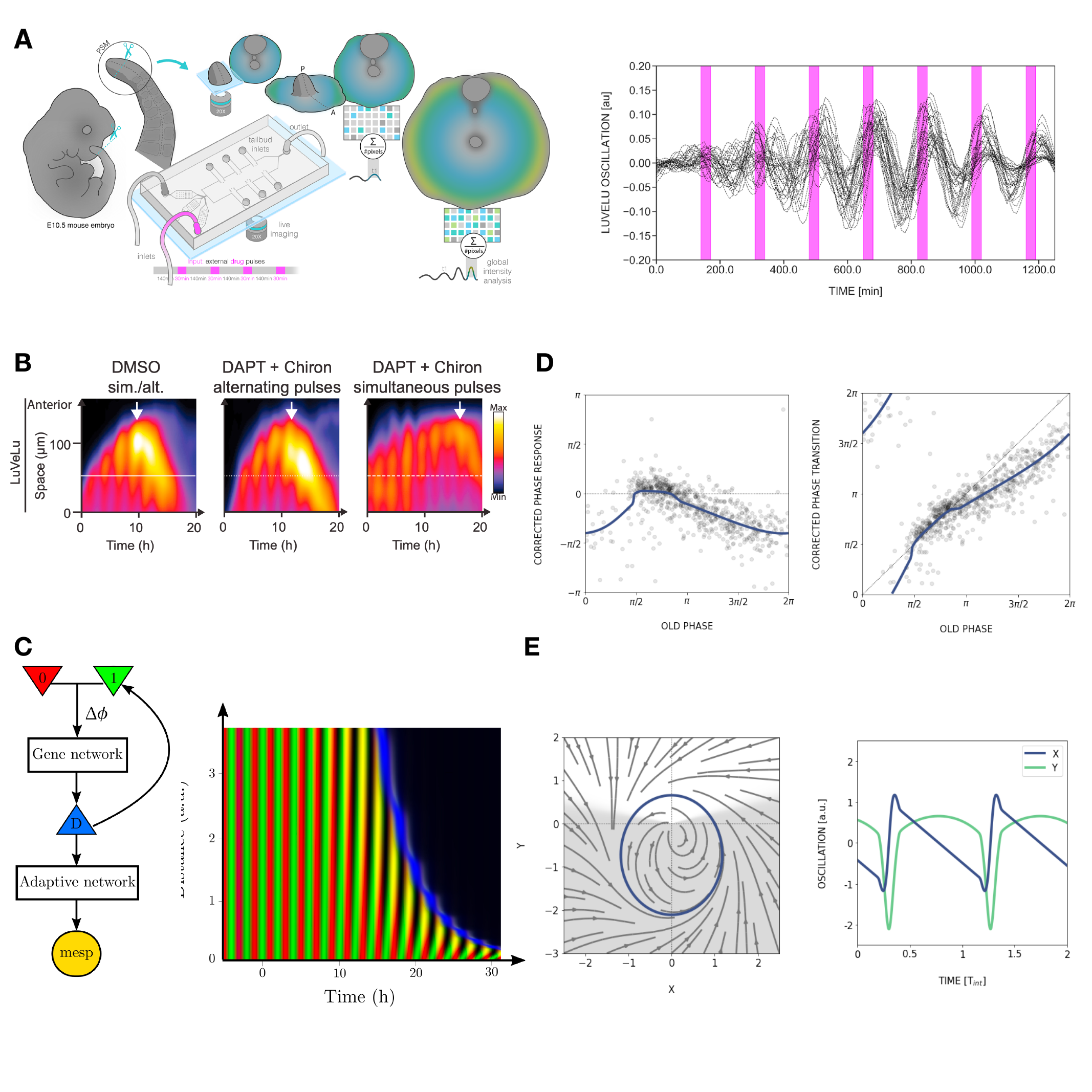}
\caption[Entraining the segmentation clock. ]{Entraining the segmentation clock. (A) A microfluidic set-up allows entraining oscillations in tail bud explants with the help of Chiron (a Wnt activator) and/or DAPT (a Notch inhibitor). Reproduced from Sanchez, eLife, 2022 (B) Periodic Chiron and DAPT pulses can entrain the segmentation oscillator. When they are simultaneous, front regression is impaired and the entire PSM keeps oscillating. (C) $\alpha$ model combined to an adaptive module reproduces block-like expression at the front. (D) Phase response and transition curves derived from experiments (after period correction), and corresponding fit with the ERICA model (E) ERICA model. On the left, the elliptic cycle is shown with the sped-up sector in gray. On the right, $x,y$ oscillations for this model as a function of time, show a fast activation phase followed by a slow decay (visible on $x$), and a pulsatile behavior for $y$ in the negative values. }\label{fig:Entrainment}
\enormf

\chapsubsec{Testing the two-oscillator model}

The first experimental indication of segmentation clock entrainment was done in zebrafish \cite{SozaRied2014}: using periodic heat shock induction, it is possible to (re)synchronize cells in the PSM to rescue segmentation of \textit{delta} mutant. However, in the absence of real-time monitoring of intrinsic oscillations,  this does not establish entrainment. Sonnen \etal  \cite{Sonnen2018} further  developed a microfluidic strategy to first visualize, then entrain the segmentation clock, visualizing both Wnt and Notch in cultured mouse tails. Their microfluidic device allows to submit the embryo to periodic pulses of different chemicals and to study the phase response, Fig \ref{fig:Entrainment} A.

Periodic pulses of chemicals (Wnt activator Chiron or Notch inhibitor DAPT) are imposed with the same period as the segmentation, but variable phase relations. Overall fluorescence level is measured, and both \textit{Axin2} and \textit{Lfng} reporters are clearly entrained in response to perturbations Fig \ref{fig:Entrainment} A. This is all the more remarkable that segmentation emerges from (coupled) cellular oscillators at different instantaneous frequencies in the entire PSM, and there could be many fundamental reasons why the emergent oscillation might not behave similarly to a single phase oscillator (for more discussion on this see section \myref{Arnold}).  Even more remarkably, the general phenomenology of segmentation is preserved in any setting: waves of Notch fluorescence sweep across the embryo, while \textit{Axin2} displays a more on/off oscillation pattern, and segmental boundaries form. Furthermore, entrainment occurs irrespective of the signal used (Chiron or DAPT), revealing that the two oscillating systems, despite their different behaviors and wave patterns, are coupled.

This entrainment strategy is then used to take control of both Notch and Wnt oscillators independently, breaking the natural connections between them Fig \ref{fig:Entrainment} B. Remarkably, when DAPT and Chiron pulses are used to entrain both oscillators (in the anterior PSM) out of phase, segmentation essentially proceeds normally. But when DAPT and Chiron pulses are used to entrain both oscillators in phase (similarly to what happens in the posterior part of the PSM close to the tailbud), the segmentation process is impaired and the entire PSM seems to be maintained closer to a "tail bud state". Notch phase gradient in the PSM is considerably flattened, the regression of the front is severely delayed, and no segmental boundary is formed. This indeed confirms directly that the phase relationship between the Notch and Wnt oscillators plays a role in modulating the front, as predicted by the two-oscillator model from \cite{Lauschke2013}. Interestingly, the front and polarity markers are also impaired. In those embryos, \textit{Mesp2} still is expressed and regresses "normally", i.e. with similar timing and speed as in control embryos. This means that \textit{Mesp2}  activation is not directly related to oscillation nor is it a consequence of its stopping. Strikingly though, the expression of \textit{Mesp2} is not functional, in the sense that there is no obvious further localization of \textit{Mesp2} stripe, and consistent with this, posterior markers of somites (such as \textit{Uncx4.1}) are not expressed at all in those embryos. There is also no spatial pattern of \textit{Axin2} in the presumptive posterior somite region.

Those experimental results raise an intriguing theoretical question: if positional information is defined by the phase difference of two oscillators, what kind of network structure would allow for the decoding of such information? In \cite{Beaupeux2016}, \textit{in silico} evolution was used to derive possible gene networks explaining this phenomenon. Evolution then selects for incoherent feedforward loops combined with positive feedbacks to control the concentration of a gene $A$ that depends on which oscillator peaks first (see e.g. \cite{Mangan2003, Lalanne2013} for how incoherent feedforward loops can detect patterns in temporal signals). Interestingly, those networks translate a phase difference $\Delta \phi$ into a (linear) variable $A (\Delta \phi)$, so that when the $2\pi$ circle is completed, such $A$ necessarily has to "reset" quickly from the value  $A(2\pi) =a_0+ \alpha 2 \pi$ to its value $A(0)=a_0$. If such a gene exists, it would slowly ramp within the PSM, before rapidly coming back to its initial value at the front. Such a sudden change provides a possible signal for the front, and as such for the $2\pi$ rule experimentally observed. It is also not difficult to build simple models where the drop of $A$ is detected to trigger the production of a gene at the front (such as \textit{Mesp2}) that would stop both oscillations and recapitulate many observations from \cite{Lauschke2013}, Fig \ref{fig:Entrainment} C. Interestingly, a recent model for zebrafish single-cell oscillation explains the oscillatory behavior by using such a variable slowly ramping up before collapsing at the front \cite{Rohde2021}.

\chapsubsec{Quantifying entrainment regions of the segmentation clock }
\label{Arnold}

More detailed entrainment experiments allow us to first quantify the entrainment properties of the system, then probe some internal dynamical properties of the segmentation clock \cite{Sanchez2021}. Using a similar setup as in \cite{Sonnen2018}, the segmentation clock can be systematically and reliably entrained using DAPT pulses. Again, the segmentation process itself is largely unperturbed, in the sense that one still observes the formation of a Notch phase and periods gradient within the entire PSM but remarkably, the entire process is entrained by the period of the external pulses, with periods ranging from 120 to more than 180 mins, a considerable period increase compared to the intrinsic 140 min period in mouse. Those experiments allow us to quantify the phase response of the segmentation clock and infer the internal properties of the system.

It is well known that the proximity of an oscillator to some bifurcations constrains its phase response curve (PRC), the main reason being that close a bifurcation, one can remove many degrees of freedom and describe the oscillations in a compact way (e.g. with a normal form). In Appendix A, we illustrate this by providing derivations of phase response curves for different oscillators. There are also several general mathematical results on the way oscillations can disappear \cite{Keener81} (briefly mentioned in the Appendix) and typologies established for systems biology oscillators \cite{Tsai2008}. While there is no one-to-one correspondence between a phase response shape and a bifurcation \cite{Ermentrout2012}, two generic scenarios can be observed  \cite{Keener81}. Systems close to Hopf bifurcations (where oscillations disappear by a collapse of the amplitude and finite period)  are very "symmetrical", presenting sinusoidal phase response curves (which also typically leads to Kuramoto coupling between oscillators).  Practically, this means that the same external perturbations produced at different phases of the cycle can either advance or delay the clock.

At the other end of the spectrum, systems close to an infinite period bifurcation are very asymmetrical: phase response curves are almost of constant sign, meaning that a given external perturbations will always advance (or delay) the oscillator, irrespective of the phase of the cycle at which it is performed. Most non-linear oscillators are expected to be in between those two cases, and there also are exceptions depending on the parameter regime and the symmetry of the oscillators. Nevertheless, in light of our discussion on the various models underlying somitogenesis, it is worth asking what kind of phase response is observed for the global segmentation clock.

Strikingly, the response of the segmentation clock to DAPT perturbations inferred from entrainment experiments is mostly negative, meaning that the clock is always delayed by DAPT pulses. For this reason, it is much easier to slow down the segmentation clock rather than to speed it up, explaining the very asymmetric entrainment range. Interestingly, an asymmetry in the response of the clock has also been observed in a completely different set-up where different tail-buds are connected \cite{roth2023unidirectional}.
Coming back to entrainment, the phase response curve (PRC) derived at different periods have similar shapes but are translated vertically with respect to one another, which is not accounted for by the classical PRC theory. The origin of such an effect is unknown. It is important to point out in particular that those phase responses are computed on the entire embryo, and the signal extracted to compute them is likely coming mostly from cells close to the front, which are not the same from one cycle to the other. In other words, the culture might "adapt" to the perturbations, so that the entrainment might reveal global feedback within the system.

Concretely, a vertical shift of PRC has been observed in the context of cardiac cell oscillations \cite{Kunysz1995}, and has been associated there to an internal change of the clock period in response to perturbations. So to explain the data, we assumed that for some unknown reason, the segmentation system is able to adjust its own period to the entrainment one, which allows collapsing the PRCs at different entrainment periods, Fig. \ref{fig:Entrainment} D, taking as a reference PRC the one for the natural period around 135 min. Combining multiple entrainment experiments at different periods allows now for a  reliable estimation of the global PRC.

\chapsubsec{Modelling the segmentation cycle from the PRC: the ERICA model}

From there, we proceeded to build the simplest possible model of the PRC data. Inspired by minimal modeling developed by others, e.g. \cite{Webb2016}, we looked for a minimally modified version of the Radial Isochron Cycle (RIC) model. The RIC is a generic model of a symmetrical oscillator close to a Hopf bifurcation, which can be derived from standard Stuart-Landau oscillators. We start from the phase response curve of such a model, known to take the form $Z(\theta)=\sin(\theta)$, which is odd as a function of $\theta$. A simple way to break this symmetry is to go to the second order and consider an asymmetric response curve of the form :

\begin{equation}
Z(\theta)=\sin\theta + \lambda \sin^2 \theta=\sin\theta +\frac{1}{2}\lambda(1-\cos2 \theta) \label{PRC_ERIC}
\end{equation}
where $\lambda$ quantifies the symmetry breaking. In the limit where $\lambda\rightarrow 1$, $H$ is very asymmetrical with $H$ very small for $\theta\in[-\pi,0]$ and sinusoidal for $\theta\in[0,\pi]$

Our proposal then is to use  Eq. \ref{PRC_ERIC} as a constraint to engineer a limit cycle in this plane, i.e. a function $r(\theta)$.  To do this, we know (see e.g. \cite{Kuramoto,Izhikevitch2007}) that the infinitesimal PRC is $grad \quad \theta(\vec{r})$ where  $\theta(\vec{r})$ is the phase of the limit cycle at position $\vec{r}$. Assuming we work in polar coordinates, and that the perturbation is in the $x$ direction, the PRC thus equals to $-\sin(\theta(\vec{r}))/r$.  Dropping the minus sign and imposing that this PRC is equal to $Z$ in \ref{PRC_ERIC} gives the simple relation between $r,\theta$
\begin{equation}
r_c(\theta) = \frac{1 }{1 +\lambda \sin \theta}
\end{equation}

which defines an ellipse with the principal axis in the $y$ direction. Thus this ellipse with radial isochrons defines a PRC similar to \ref{PRC_ERIC} (with an extra minus sign in front of the expression). Notice that for $\lambda=0$, we recover the unit circle. 

 To generalize this and impose dynamics in the entire plane, we thus consider the following system :

\begin{eqnarray}
\dot r &=& \dot r_{c}+ r(r_{c}-r)\\
\dot \theta &= &1
\end{eqnarray}
    
where     $r_{c}(\theta) = \frac{1 }{1 +\lambda \sin \theta}, \dot r_{c}(\theta) = -\lambda \frac{\cos \theta}{(1+\lambda\cos \theta )^2}$. This replaces the circular orbit of the RIC oscillator with an elliptic orbit, defined by $r_{cycle}$. It is not difficult to see from this expression that $r=r_{c}(\theta)$ is a stable orbit. 
We name this model Elliptic Radial Isochron Cycle or ERIC. An important feature of ERIC is that the angle in the plane is the phase of the oscillator, in particular since the phase is defined by the planar angle, the PRC following a horizontal perturbation of size $\epsilon$ towards the right can be computed in a straightforward way and is:

\begin{equation}
   PRC_{\epsilon,\lambda}(\theta)= \cot ^{-1}(\epsilon (\csc (\theta )+\lambda )+\cot (\theta ))-\theta
\end{equation}

It is not difficult to check that

\begin{equation}
\left.\frac{\partial PRC}{\partial \epsilon}\right|_{\epsilon=0}=-\sin^2(\theta)(\csc (\theta )+\lambda )=-(\lambda \sin^2 \theta +\sin(\theta))
\end{equation}
which has the desired asymmetric form for the infinitesimal PRC: for $\lambda>0$, it is in particular much flatter for $\theta \in [-\pi,0]$ than for $\theta \in [0,\pi]$ 

One issue with this model is that the "flat" part of the PRC always occurs for an interval of size $\pi$. To allow for more flexibility here, we then introduce a second modification, introducing a "speeding factor" $s$ so that 
\begin{equation}
    \dot \theta=s(\theta)
\end{equation}

This keeps isochrons radial but changes their spacings. We name this class of model Elliptic Radial Isochron Cycle with Acceleration or ERICA. For simplicity, and to keep the system analytical, we restricted ourselves first to $s$ functions linear by piece. We reproduce the full system of equations in the Appendix B5. 

We then used Monte Carlo simulations to fit the PRC inferred from the data. We find in particular that $\epsilon = 0.43$, $\lambda = 0.53$, $s_* = 5.64$. This corresponds to the elliptic limit cycle shown in Fig. \ref{fig:Entrainment} E. All those inferred values are "big" in the sense that they indicate that the fitted perturbation is strong, the eccentricity of the ellipse is high, and the speeding factor (over more than half of the cycle) is big, meaning that, as expected, the system appears rather far from the RIC model generally associated to a Hopf bifurcation. 

It is informative to plot the behavior of this optimized limit cycle: it bears resemblance to negative feedback oscillators with (long) delay, e.g. looking at variable $x$ we see a short activation phase, followed by a slow decay \ref{fig:Entrainment} E, right panel.  Such a system where one variable slowly varies and then quickly resets is reminiscent of integrate and fire networks, corresponding to normal forms of SNIC bifurcations (see Appendix). Consistent with this, the behavior of variable $y$ also looks similar to the polar angle close to an infinite period bifurcation, see e.g bottom of Fig. \ref{fig:Phase_model_SNIC} D.  Indeed, in phase space, Fig. \ref{fig:Entrainment} E, left panel, the system spends much of the time on the blue line in the white region, so essentially decreasing $x$, then quickly resets in the gray region. So again a parsimonious model of phase response suggests that the segmentation clock might be close to an infinite period bifurcation.

\chapsec{Inferring coupling rules : walkie talkie model}

\label{sec:walktalk}
The knowledge of response function can also be used to infer effective coupling rules between oscillators (see section \myref{sectionPRC} in Appendix).  In \cite{roth2023unidirectional}, it was observed that upon activation of Notch signalling pathway (via  chemical inhibition of degradation NICD), presomitic cell cultures are typically slowed down. The effect of the slowing down seems to depend on the phase when the chemical is applied : for some phases, there is no effect while for other phases there is a maximum effect. This effect is thus very reminiscent of the phase response curve observed in entrainment experiments described above, where the oscillator is either unchanged or slowed down depending on the phase \cite{Sanchez2021}, and again rather is suggestive of more 'pulsatile' behaviour, characteristic of oscillators close to infinite period bifurcation. This can be used to model what the authors call a 'walkie-talkie' model of cellular coupling. There are two assumptions :
\begin{itemize}
\item cells are only responsive during some fraction of the cycle, where they slow down (receiver phase)
\item cells positively signal during another fraction of the cycle, then sending a signal to their neighbouring cells (sender phase)
\end{itemize}
Based on mathematical arguments, it is argued that within one cycle, the receiver phase must immediately follow the sender phase to get eventual synchronization between cells. Also, because cells only 'slow' down, this gives a 'brake' model where a cell delayed with respect to another one will tend to slow down its neighbours.
It is also observed that synchronization of multiple, randomized oscillators with such coupling rules would look very differently compared to a standard Kuramoto model. Taking a system of random oscillators with uniformly distributed phase, upon synchronization, cells coupled with Kuramoto model synchronize to a global phase that is random, and uniformly distributed over a cycle. By contrast, simulations show that cells coupled via walkie-talkie would synchronize to a well-defined phase. This is validated experimentally by looking at the phase synchronization for different group of PSM cells, which is indeed peaked to a well defined value.

\chapsec{Exploring cell communications/coupling with optogenetics}

More direct quantification of the impact of cell communications and cellular coupling on the clock has become possible in recent years. In \cite{Isomura2017} an optogenetic system is introduced, allowing to both induce and visualize the oscillating protein \textit{Hes1} in cultured human cells (\textit{Hes1} spontaneously oscillates in many cellular types \cite{Hirata2002, Shimojo2008}, the cell culture of interest oscillates here with a period of around 2.5 hours). While individual cells oscillate in an asynchronous way in constant dark, periodic pulses of light  (as short as a couple of minutes) entrain those culture's cells. Interestingly, modulation of the entrainment period shows a diverse pattern of entrainment, including a 2:1 pattern when the period of the pulse is twice as long as the endogenous period. While oscillations are clearly rather stochastic,  the optogenetic perturbation is very strong and tends to move the oscillator very close to one specific phase of the oscillator (around $\phi_*\sim 3\pi/2$ in their units) (so called 'Type 0' resetting \cite{Winfree})

This result is especially interesting given the shortness of the light pulse compared to the period of the clock: this indicates that the optogenetic system has a very strong, non-linear effect on endogenous oscillation. With this knowledge in hand, one can refine the system to build "sender" cells for which the Delta ligand \textit{Dll1} is sensitive to the optogenetic input, and "receiver" cells without optogenetic sensitivity  where \textit{Hes1} oscillations can be monitored.  Periodic stimulation of the sender cells can indeed lead to the entrainment of the receiver cells, although only for periods of entrainment close to the endogenous period. Again,  stimulation of sender cells massively appears to `reset' the receiver cell close to a given phase. From there, one can quantify how perturbations are transmitted in a signaling cascade from one cell to the other. The entrainment phase of the \textit{Hes1} oscillation is roughly 50 mins earlier when entrained by endogenous NICD compared to entrainment by the neighboring \textit{Dll1}, which the authors interpret as the delay for \textit{Dll1} to activate NICD (we refer to section \myref{sec:delayorshift} for a discussion on the interpretation of such delays).

In \cite{Yoshioka-Kobayashi2020}, a new  "Achilles" reporter is further developed to monitor \textit{Hes7} (a protein in the Notch signalling pathway) oscillations in mouse, with an unprecedented single-cell resolution. This is used to observe what happens in the context of \textit{Lfng} knockout, where  oscillations have smaller amplitude, shorter period (150 mins vs 170 mins for WT), and are noisier. The authors test what happens when different types of cells are first dissociated and then coupled to one another. When \textit{Lfng} KO cells are put in a WT cells environment, they do not seem to couple with them, while when WT cells are put in a \textit{Lfng}-KO environment, they tend to follow the (noisy) \textit{Lfng}-KO cells, with a $\pi/4 $ phase shift. The sender-receiver system described above is also used to induce \textit{Dll1} in one cell and see what happens in the receiver cell in both contexts : it was found that \textit{Lfng}  in the sender cell increases the phase shift between sender and receiver, and when in the received system, it increases the amplitude.
To explain those effects, the authors used a modified version of the delayed model similar to Eq. \ref{rescaled} accounting for cellular coupling with delay :

\begin{equation}
    \frac{d X_i}{dt}= \nu \frac{K_1^m}{K_1^m+X_i(t-\tau_1)^m} \frac{K_2^n}{K_1^m+Z_i(t-\tau_2)^2} - rX_i
\end{equation}

where $Z_i=1/N\sum_{j} X_i(t)$ is a mean-field variable averaging values of $X$ in neighbouring cells (index $j$). The system is simulated on a hexagonal lattice. It is then found that maximum coordination of the oscillations between cells (including maximum amplitude) requires that both delays are essentially equal $\tau_1=\tau_2$. This makes intuitive sense: this would make all cells more or less identical so that the system could effectively be reduced to a single delayed equation similar to Eq.\ref{rescaled} but with a slightly more complex non-linearity, and with a maximal production rate of $\nu$. Any change of delay $\tau_2$ would induce a maximal transcription rate lower than $\nu$, thus decreasing the amplitude, while also creating a phase shift between cells, as is observed. It is thus proposed that \textit{Lfng}  might play a role in adjusting the delay $\tau_2$ coordinating the communication between cells, to an optimal value $\tau_1$ consistent with the intrinsic time scale of the negative feedback oscillations within one cell. Another interesting effect worth mentioning is that, as $\tau_2/\tau_1$ is becoming different from 1, oscillators in the model are synchronized but phase dispersion is increasing.

\chapsec{In vitro mouse segmentation clock}

An interesting variation of the mPSM culture is introduced in \cite{Hubaud2017}. There, the cultures are exposed to a cocktail of Wnt activator, FGF4, and BMP inhibitor, which appears to globally maintain the entire culture into an oscillatory state, phenomenologically similar to what is observed in the mouse tail bud. Those cultures oscillate for over 20 cycles with periods similar to the mouse segmentation clock, do not display any FGF or Wnt gradient (indicative of differentiation into somites), and, consistent with this, do not form segments. A fast wave is propagating in the entire culture (indicative of a small phase gradient); propagation of this wave is unperturbed by the removal of the center and the introduction of a physical separation in the explant, which argues against the existence of a local pacemaker and suggests that all cells have the potential to oscillate. Cells are then dissociated and seeded on a dish at different concentrations. While no oscillations are observed for a few cells per dish, at confluency (i.e. when all the dish is covered by cells) the system starts oscillating, with a period similar to the full explant, which suggests a quorum sensing effect to control oscillations. However, single cells can become oscillatory by a simple change of substrate glass coating (from fibronectin to Bovine Serum Albumin), changing at the same time the shape and behavior of cells from motile and extended to round and static. Indeed, the authors further show that YAP signaling (usually implicated in cellular mechanosensing) needs to be turned off for oscillations to occur.

The authors aim to explain theoretically how one can go from a quiescent state to an oscillating system with large amplitude, by the continuous change of one single parameter (proposed to be related to mechano-transduction). The system is proposed to be excitable, and a model based on the FitzHugh-Nagumo system is proposed 

\begin{eqnarray}
\tau_u \dot u & =& u(u-a)(1-u)-v +I +\epsilon \label{hubaud_u} \\
\tau_v \dot v &=& u - gv\label{hubaud_v}
\end{eqnarray}
The control parameter $I$ is assumed to go from $1$ to $0$ when conditions change. $\epsilon$ is a noise term.

\bnormf
\includegraphics[width=\textwidth]{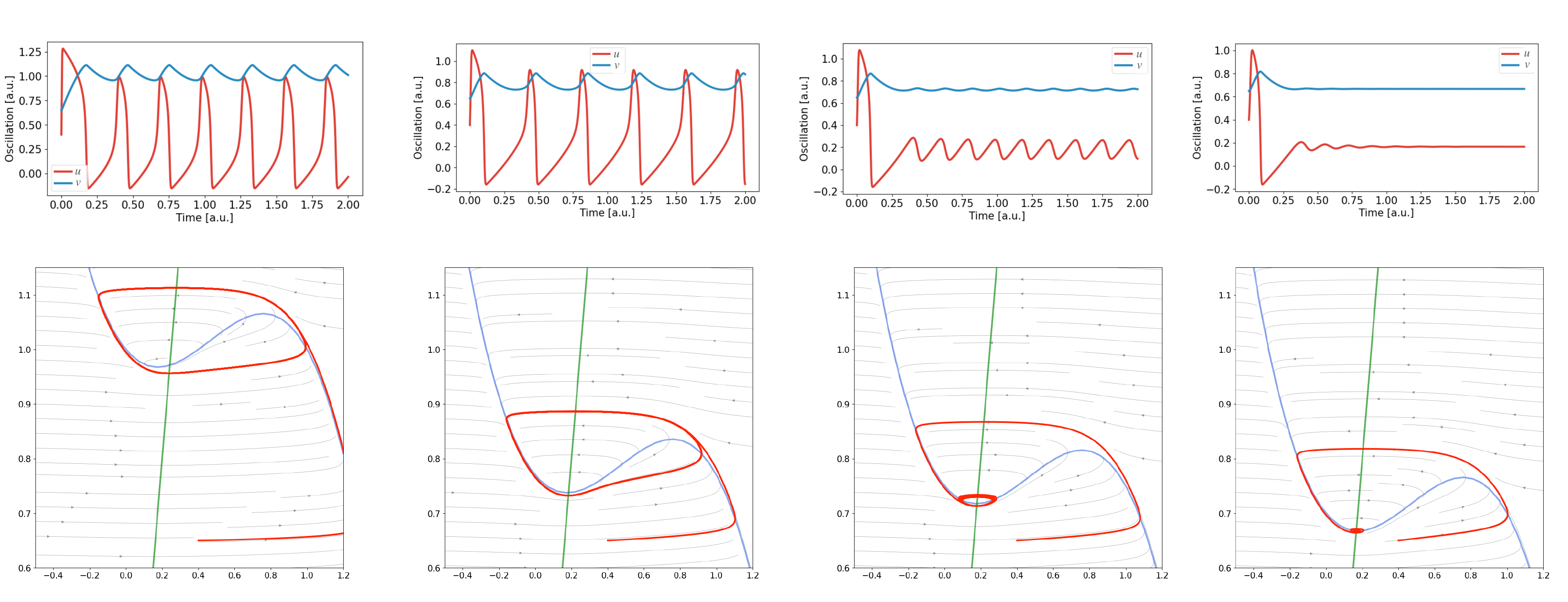}
\caption[FitzHugh-Nagumo oscillator] { Behaviour of the FitzHugh-Nagumo oscillator used in Hubaud, Cell, 2017. $v$ null-cline is in green, $u$ null-cline is in blue.As the control parameter $I$ is decreased (here $I$ takes successive values $1,0.77, 0.75,0.7$ from left to right), the model goes through Hopf bifurcation and thus is of Type  II (see definition in Appendix). Notice how the period first increases as $I$ decreases, then how the oscillation amplitude suddenly decreases right at $I=0.75$ just before the Hopf bifurcation. If the system is initialized just on the right of the fixed point, it goes through a "pulse" before stabilizing. Parameters are $\tau_u=0.25, \tau_v=30, a =0.4, g=0.25, \epsilon=0$.   } \label{fig:TypeII_hubaud}
\enormf

Below the bifurcation, the system stabilizes at a fixed point, but there is a "ghost" cycle so that a noisy stimulus can "kick" it and lead to one single oscillation as can be seen in Fig. \ref{fig:TypeII_hubaud} right. As the control parameter increases, the system starts oscillating, with a fixed amplitude. The experimental system shows several features compatible with this model. First, when non-oscillatory cells are transferred to oscillatory conditions, they start oscillating immediately and synchronously, consistent with the interpretation that they start from the \textit{same} point on the cycle. Second, the dependency of frequency on the control parameter can be tested by varying doses of YAP inhibitors. It then appears that close to the threshold of oscillation, the number of LuVelu cycles over a given window of time is indeed reduced, consistent with a stochastic excitable system. The model presented in the paper presents Type II excitability (see Appendix for definitions), meaning that the system goes through a Hopf bifurcation. However, experimentally the amplitude of the oscillation appears fixed, and for some experiments at least, oscillations appear to have periods much longer than control, which is possibly more consistent with Type I excitability and could be easily accounted with small changes in the model (see Appendix for discussions of this effect).

\chapsec{Zebrafish cultures}

 Webb \etal \cite{Webb2016} were able to culture isolated cells from zebrafish tailbud, and to track the activity of \textit{Her1} using fluorescent proteins. When cells are cultured in serum only, the reporter shows a few oscillations before dying out. But when \textit{Fgf8b} is added to the culture, multiple oscillations in individual cells are recorded, very similar to what happens in the cultures described in the previous section. Strikingly, single cells from the tailbud indeed oscillate, albeit with a slightly longer period (1.5 fold compared to the segmentation clock period), and with much noisier behavior than cells within the embryo. Again, contrary to what happens in the embryo, cells never slow down: they likely keep some tailbud identity (possibly due to the presence of \textit{Fgf8b}) and do not go through any further segmented fates. However, cells appear to alternate between two dynamical regimes: either they oscillate with a relatively constant period, or they stay in a quiescent state without oscillation. The authors interpret it as a system close to a Hopf bifurcation (but not excitable), modeled with a system of the form  (in polar coordinates)  :
\begin{eqnarray}
    \dot \theta & = & \omega \label{Oates_hopf1} \\
    \dot r & = &\mu r (1-\frac{b}{\mu}r^2) \label{Oates_hopf2}
\end{eqnarray}

The behavior of the system depends on the value of $\mu$. If $\mu<0$, the only steady is $r=0$ and there is no oscillation. If $\mu>0$, steady state is $r=\sqrt{\frac{\mu}{b}}$ and the system thus oscillates with angular frequency $\omega$. There is a Hopf bifurcation at $\mu=0$ when the fixed point at the origin changes stability.

To interpret their data, the authors suggest that $\mu$ is randomly changing in time (e.g. via a standard 1D random walk) so that alternation between oscillating/non-oscillating behaviors is entirely controlled by the sign of $\mu$. Since it is a Hopf bifurcation, the period and the amplitude at the transition have very different behaviors in this model compared to the excitable model described in the previous section. Here, at the bifurcation, the period is finite and constant over some range of the control parameter $\mu$, and amplitude changes with the control parameter. Indeed, precise quantification between successive cycles suggests that there is no significant period change at the bifurcation, while the amplitude appears to widely vary, consistent with the Hopf model.

The observation protocol was refined by Rohde \etal  \cite{Rohde2021}, allowing for a much better overall resolution, both in whole embryos and in cell cultures. A \textit{Mesp2} reporter was further added to follow the dynamics of segmentation. In zebrafish embryos, cells follow a pattern consistent with previous observations (e.g. \cite{Delaune2012, Shih2015}),  but are monitored here over a much longer time scale. As cells move towards the anterior, they go through multiple oscillations (up to ten are routinely observed !). Both the amplitude and the period of oscillations increase in a non-linear way, with at least an increase of factor 3 for the period as cells move close to the front. Lastly, the wave of the oscillation encodes a pre-pattern within one somite, with \textit{Mesp2}  eventually expressed in the future rostral part. This wave behavior is fully consistent with previous observation in zebrafish \cite{Shih2015}, with the \textit{Mesp2} expression dynamics observed in mouse \cite{Oginuma2008, Oginuma2010}, where \textit{Mesp2} could play the role of a bistable rostral marker \cite{Meinhardt1986,Francois2007}. The last peak of \textit{Her1}  and \textit{Mesp2}  expression are visually separated by exactly one somite spatially, i.e. one cycle temporally. Interestingly, the qualitative dynamics of dissociated cells in low-density culture follow the exact same pattern, consistent with what happens in the embryo: dissociated cells oscillate for up to 8 cycles, with increasing period and amplitude, before eventually expressing \textit{Mesp2}. The ratio of durations of each cycle w.r. to the previous one is higher than $1$ and similar to what is observed in the embryo. The clearest difference between cultured vs embryonic cells is the overall time scale : oscillations in cultured cells are overall roughly twice as slow as the embryonic ones, similar to what was already observed in \cite{Webb2016}.

Those observations suggest again a  cell-intrinsic, kinematic program, where oscillations slow down (and amplitude increase) in a cell-autonomous way. Rohde \etal \cite{Rohde2021} suggest a simple yet elegant model, based on a delayed oscillator, but with an increased production rate. As stated in section \myref{delay_increase}, an increase in production rate leads to an increase in the overall period, eventually leading to an infinite period bifurcation. The oscillation shape looks increasingly like a pulsatile oscillator with a short "on" phase followed by a longer "off" phase. This matches the experimental observations made both in culture and embryos, where it appears that the "degradation" part of the cycle increases its duration from one cycle to the other while the "activation" part stays more or less the same.

Rohde \etal \cite{Rohde2021} suggest that this increased production rate acts akin to a cell-autonomous timer of unknown origin, increasing linearly with  time and as a consequence lengthening the period. To explain the oscillation death, they suggest that this timer collapses once it reaches some predefined threshold value. We notice that it is unclear why such a variable would suddenly collapse. An idea could be that, similar in spirit to the phenomenological model presented in Eq. \ref{eq:phase-ampa}, once the amplitude reaches some threshold, some feedback turns on and the system goes through a bifurcation. Noteworthy, the idea of  a non-linear, intrinsic slowing down of the oscillation  is also at the basis of the $\alpha$ model developed to explain scaling dynamics in mouse mPSM cells \cite{Lauschke2013}, e.g. Eq. \ref{coupled_model}. In fact, as detailed in \cite{Beaupeux2016}, gene networks computing phase difference naturally display increased amplitude, followed by a sharp collapse at the front, because this simply recapitulates the "resetting" of a phase difference to $0$ once a 2$\pi$ cycle is completed. Such collapse strikingly recapitulates the timer behavior postulated in \cite{Rohde2021} possibly suggesting that the  increased production rate might correlate to some phase difference. The scaling dynamics inferred from \cite{Lauschke2013} suggests additional feedback where such phase difference further activates the slowing down  of the clock (possibly here through an increase of production rate), which remains to be  checked experimentally.

\chapsec{Stem-cell systems}
\label{Section:synthetic}
Following considerable progress in stem cell culture techniques and controlled differentiation ("embryoids", "gastruloids") multiple groups have recently developed new cultures to isolate and reconstruct models of mammalian PSM 
\cite{Veenvliet2020,vandenBrink2020,matsuda2020,Sanaki2022,Budjan2022,Miao2022, Yaman2022}. This allows for the exploration and visualization of the mechanisms underlying somite formation in dishes with great flexibility and versatility \cite{matsuda2020,Budjan2022, Miao2022,yamanaka2023}. 

Technically, those artificial systems are derived from embryonic stem cells (which allows for maximal flexibility in terms of genetic manipulation), through a clever combination of biochemical modulation (induction of Wnt, downregulation of BMP) and embedding of cells in a Matrigel matrix with well defined biomechanical properties which are crucial to induce somite-like structures \cite{Veenvliet2020,Sanaki2022}. A precise (and scaled) symmetry breaking first occurs \cite{Merle2023} so that culture separates into an undifferentiated "proto" tail bud connected to a growing, differentiated tissue. Importantly, this approach can also be started from artificially induced pluripotent stem cells, which allows for the studying of (reprogrammed) human tissues.

Single-cell RNA seq analysis of those systems shows that differentiated cells appear to qualitatively recapitulate similar programs to cells in the embryo moving from posterior to anterior within the PSM \cite{Veenvliet2020,vandenBrink2020, Budjan2022, Miao2022}, and one indeed observes somite-like structures, either arranged like a "bunch of grapes" \cite{Veenvliet2020} or in a more sequential way following growth \cite{vandenBrink2020,Sanaki2022, Yaman2022}.  During growth, there clearly is a transition from an oscillatory state (for Notch)  to a differentiated one following \textit{Mesp2} expression \cite{Sanaki2022,Yaman2022}, however, there is only a very shallow spatial phase gradient in the oscillatory region, which is different from what is observed in embryos.   Budding somite-like structures nevertheless present relatively well-defined antero-posterior identities within somites (e.g.  localized \textit{Uncx4.1} expression) contrasting with what is observed for the grape-like structures induced by \textit{Noggin} signaling in \cite{Dias2014}.

Multiple studies are ongoing to use such systems to uncover minimal mechanisms underlying somite formation. Starting from human pluripotent stem cells, Miao \etal \cite{Miao2022} generate spreading organoids (a bit akin to what is done in \cite{Lauschke2013,Hubaud2017}) and observe \textit{Hes7} concentric waves of oscillations for 4 cycles of 4-5 hour period each, followed by simultaneous, global \textit{Mesp2} expression in the culture. Later on, tissue segregates into "rosettes" structurally very similar to somites and called "somitoids". Interestingly, it can be shown that "somitoids" neither require the clock nor \textit{Mesp2} expression to form, but strongly depend on the expression of molecules implicated in biomechanics (such as myosin contractility), which suggests that rosette formation could be a late, self-organized biomechanical process.

Normal somitoids nevertheless express classical somite anteroposterior marker, but in an unexpected way: each individual somitoid expresses either anterior (\textit{Mesp2}) or posterior (\textit{Uncx4.1}) marker, but never both genes in the same somitoid, contrary to what happens in the embryo.  Reaggregation experiments were performed based on the expression of \textit{Mesp2}  when it first peaks after clock stopping: cells with high \textit{Mesp2} form "anterior-like" rosettes, while cells with low \textit{Mesp2} form "posterior-like" rosettes, which suggests that the antero/posterior identity is defined once and for all right after clock stopping (similar to the clock and switch model). The fact that somitoids have well-defined anteroposterior identities (and in particular that no somitoids present the normal balance of  fates) is also consistent with the idea that such polarity could play a role in boundary formations since it appears that boundaries are clearly induced between anterior and posterior domains (of various size and shapes). Interestingly, it is shown that initial \textit{Mesp2}-positive cells eventually segregate together via cell motions, which would contribute to and reinforce such boundary formation.

The same type of culture can be made in Matrigel. There is a spontaneous symmetry breaking with, on one end, an undifferentiated part with a genetic expression similar to posterior PSM, from which somite-like structures sequentially bud off (called "segmentoids") on the other end. Strikingly, several segmentoids can form in parallel, each of them displaying normal somite-like features, including proper anteroposterior polarity. Dynamics within a given somitoid recapitulate known features of  segmentation, where future segmentoids are defined by a broad stripe of \textit{Mesp2} positive cells narrowing into half a segmentoid. It is argued that \textit{Mesp2}  is first expressed in a salt and pepper way in the entire broad stripe, without clear anteroposterior bias, and that later on \textit{Mesp2}  positive cells migrate towards the anterior part, without induction of new cell fates. So in this view, the clock would define the entire future somite via a salt and pepper expression of \textit{Mesp2}  but then part of those cells would turn off (or never express \textit{Mesp2}), giving a mixture of anterior and posterior cells within a somite. Cells would then segregate to get proper AP polarity within a somite. We notice, however, that in the absence of any positional information, there is no reason why all \textit{Mesp2}  positive cells would segregate towards the anterior (vs any other random directions), which thus does not fully exclude a role of the clock (or of other processes) to set up a segmentoid wide polarity.
From a theoretical standpoint, it should be pointed out that cell sorting presents  clear advantages in terms of robustness: if somite AP polarity does not need to be perfectly set up by the clock, a mechanism such as the noisy Hopf bifurcation to bistability scenario from Fig. \ref{fig:geom_model} could be realized, since cell sorting would take care of somite polarity formation.

\chapsec{Randomization }

\chapsubsec{Self-organization}
Gastruloid and somitoid systems clearly self-organize, but their spatial structure still emerges from a coherent growth process. It raises the question on what happens if one completely wipes out both growth and spatial organization. Tsiairis and Aulehla \cite{Tsiairis2016} studied mouse cultures where PSM cells were dissociated, mixed together, then plated. Remarkably, the system is able to self-organize to give waves and dynamical phase gradients.

After global cell synchronization (roughly 5 hours after plating), multiple foci form in the culture. Those foci oscillate in synchrony, with a period matching the embryonic segmentation period, and are faster than the rest of the culture. There is no obvious structure of length scale of those foci, although the more cells there are initially, the more foci form, and there is no less than 100 $\mu m$ between foci. Remarkably this can be done with cells from different embryos, but when taking cells from a single embryo, 4 to 5 foci form. Then, around each focus,  phase and period gradients appear and sharpen, very similarly to what happens in mPSM cultures \cite{Lauschke2013}, although the overall phase gradients develop over a much smaller length scale ($<70 \mu m $ for those reaggregates which is less than the minimum mPSM length after several oscillations). Importantly, this dynamics happens even if only posterior tail bud cells are used, indicating that some of the cells drastically slow down.

Varying modalities of those reaggregates reveal clear feedback at the collective level. For instance, one can label posterior and anterior cells from different PSMs, randomize them and reaggregate them. After 22 hours, originally posterior cells are all together inside foci, while originally anterior cells are excluded from the foci. This happens despite the fact that all those cells initially resynchronize with a similar period, which means that some internal variable keeps longer-term memory of their initial state. Along the same line, the global synchronization period essentially is a weighted average of the initial period of cells, i.e. when the proportion of more anterior cells is increased, the synchronization period is longer. We notice that this is consistent with the proposal from entrainment experiments that some internal feedback allows the clock to adjust its own internal period in response to external pacemakers (which would be here the average initial period, that would then self-sustain).

\chapsubsec{Coupling rules}

Reaggregates can also be used to study more carefully the coupling of oscillators. In \cite{Ho}, a new Randomization Assay For Low input (RAFL) protocol is designed to study how tail bud oscillators synchronize. In a nutshell, two tail buds (labeled $A,B$) are dissociated, then reaggregated together, while a fraction of cells of each embryo is also reaggregated independently as references. The phases of the mixture reaggreagate $\phi_{AB}$ and of the references reaggregates $\phi^r_A, \phi^r_B$ are then monitored  . After synchronization, one can then systematically study how the synchronized phase $\phi_{AB}$ depends on the reference phases $\phi^r_A, \phi^r_B$, Fig \ref{fig:RAFL} A.
Again a Kuramoto coupling rule predicts  $\phi_{AB}$ to be the average of $\phi^r_A, \phi^r_B$. This is not what is observed; rather a 'Winner-takes-it-all' effect is seen, where the phase of the aggregate aligns to the phase of either reference oscillators,  Fig \ref{fig:RAFL} B.

From a theoretical standpoint, one can then write a minimal model of cell synchronization in this assay. Calling $\phi_A,\phi_B$ the coarse-grained phases of cells in the reaggregates respectively coming from embryos $A,B$, and assuming that interactions between cells average out, we get \cite{ermentrout1992stable,Kuramoto}(see Appendix)

\begin{align}
\dot{\phi}_A &= \omega_A + c \: \Gamma(\phi_B-\phi_A) \\
\dot{\phi}_B &= \omega_B + c \: \Gamma(\phi_A-\phi_B)
\end{align}
with
\begin{equation}
H(\Delta\phi)=\frac{1}{2\pi}\int_0^{2\pi} Z(u) Q(\Delta\phi+u)du
\end{equation}

$Z$ and $Q$ correspond respectively to the phase response of one oscillator and to the signal emitted from the other one. This model generalizes the 'walkie-talkie' model discussed in section \myref{sec:walktalk} to arbitrary forms of coupling. The effect is captured in a single coupling function $\Gamma$, which shape depends on $Z$ and $Q$. Notice, in particular, that different functions $Z,Q$ can result into the same global coupling function $\Gamma$.

A 'Winner-takes-it-all' dynamics implies that either $\Gamma(\phi_B-\phi_A)$ or $\Gamma(\phi_A-\phi_B)$ is identically $0$ as $\phi_A,\phi_B$ converge to the common phase observed $\phi_{AB}$. The simplest functions $\Gamma$ with such properties giving stable in-phase synchronization with a coupling constant $c>0$ are :
\begin{align}
\Gamma^+(\Delta\phi) &= \max(0,\sin(\Delta\phi)) \\
\qquad\mathrm{or} \qquad\nonumber\\
\Gamma^-(\Delta\phi) &= \min(0,\sin(\Delta\phi)).
\end{align}
Those functions are called "Rectified Kuramoto" (ReKu) coupling, by analogy with "Rectified Linear Units" in machine learning already mentioned.
There is a simple prediction associated to those coupling : if cells are coupled by $\Gamma_+$, the oscillator "ahead" (within a half cycle $\pi$) is accelerating the oscillator "behind" to win, while if cells are coupled by $\Gamma_-$, the oscillator "behind" is slowing down the oscillator "ahead" to win. Notice again that this generalizes the "accelerator/brake" models suggested in \cite{roth2023unidirectional}. In Fig \ref{fig:RAFL} C, we represent $\Gamma^+$, and a polar plot of how the "ahead" oscillator pulls the "behind" one.

Strikingly, experiments show that the oscillator ahead is always winning, thus suggesting that the $\Gamma^+$ coupling is the correct one, Fig \ref{fig:RAFL} D. Conversely, experiments from \cite{roth2023unidirectional} suggest a braking model, so that one would expect a $\Gamma^-$ coupling. The discrepancy could come from the fact that more posterior (tail bud) cells are used in \cite{Ho}, and more anterior (PSM) cells are used in \cite{roth2023unidirectional}, in line with the ideas that some qualitative transitions happen when cells exit the tail bud into the PSM. Interestingly though, in both experiments, one gets a constant sign response, which might suggest that the internal geometry of the oscillator, possibly close to an infinite period bifurcation, does not change.

\bnormf
\includegraphics[width=\textwidth]{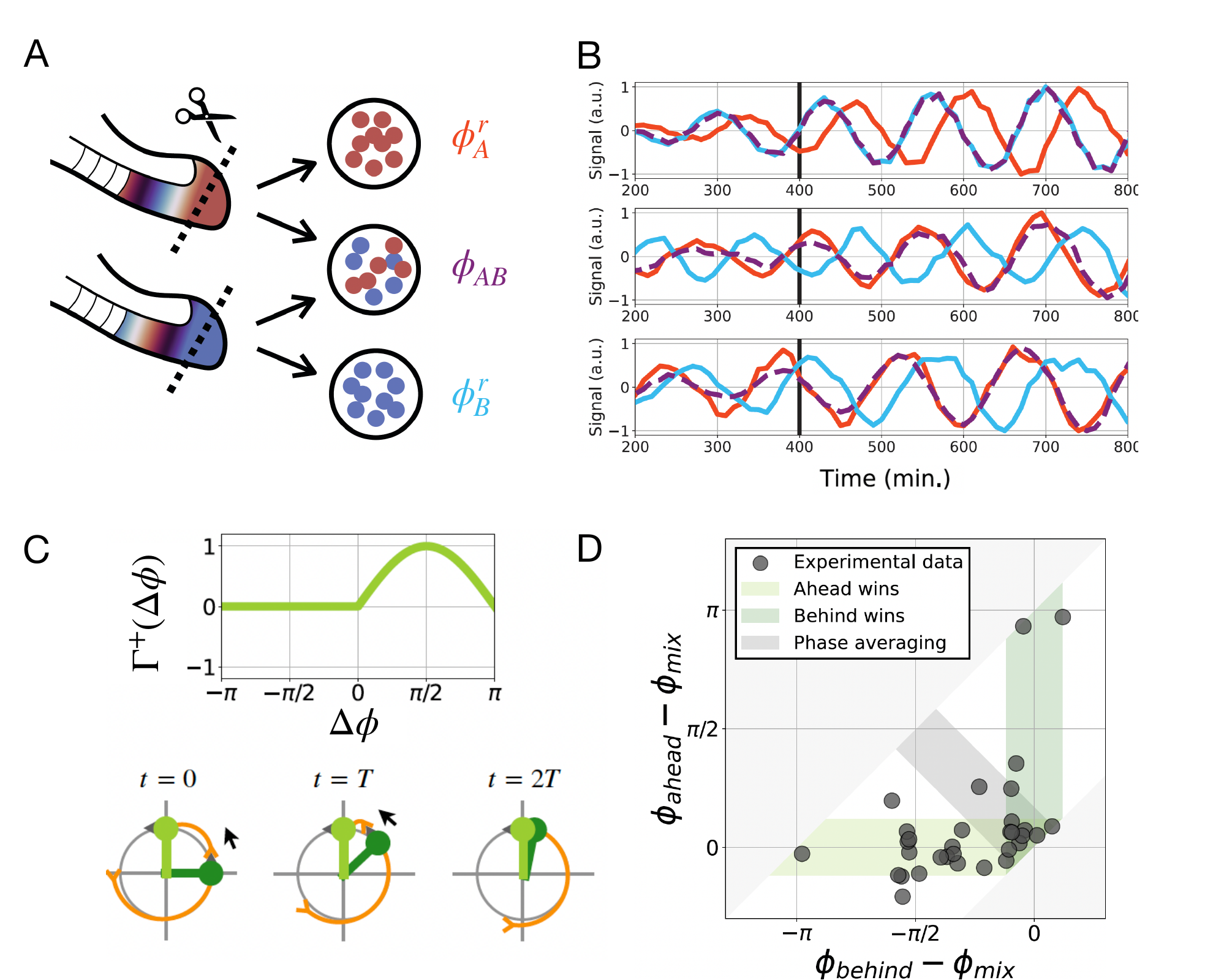}
\caption[Randomization Assay to infer coupling rules]{ Randomization assays to infer coupling rules. (A) Schematic of the assay: tailbud cells from two embryos are aggregated independently (defining references phases $\phi_A^r,\phi_B^r$ or together, defining phase $\phi_{AB}$). (B) Three different experiments. $\phi_A^r$ is displayed in red,$\phi_B^r$  in blue and $\phi_{AB}$ in purple dashed line. In the three cases, the phase of the mixed reaggregate is identical to one of the reference oscillators. (C) Illustration of the proposed Rectified Kuramoto coupling $\Gamma^+$, with polar plots illustrating how oscillators synchronize over 2 periods. The 'ahead' oscillator (light green) is effectively pulling the 'behind' oscillator (dark green). (D) The phase of "ahead" or "behind" oscillators is measured in experiments. So $\phi_{ahead},\phi_{behind}$ correspond to either $\phi_A^r$ or $\phi_B^r$ depending on which is ahead/behind, and $\phi_{mix}$ is the phase of the aggregate $\phi_{AB}$. In a "Winner-takes-it-all" situation, either $\phi_{ahead}-\phi_{mix}$ or $\phi_{behind}-\phi_{mix}$ should be $0$. By plotting one as a function of the other, one can see which oscillator (ahead or behind) is winning. Experimental data are represented and concentrated on the horizontal axis, suggesting that the ahead oscillator is always winning. Notice that because of periodicity, the points on the vertical axis in the top right correspond to a $\pi$ phase shift and are ambiguous. Those points are very close to the point in the bottom left due to periodicity.  }\label{fig:RAFL}
\enormf

\newpage
\chap{Theoretical challenges and future insights}

In this tutorial, we reviewed models of different processes connected to somitogenesis (e.g. the clock itself, the coupling of individual oscillators, or the sequence of bifurcations). Following Anderson's observation that "More is Different" \cite{Anderson1972}, many aspects of segmentation might arise in a non-trivial way from the coupling of subprocesses. A well-known example (mentioned in this tutorial) is the coupling of individual oscillators, that "emerge" into a global oscillator with properties defined only at the higher level (e.g. the modified period of Eq. \ref{DelayOmega}).  Similarly, while it is clear experimentally that somitic boundaries can form without the clock, the fact that the clock always stops prior to boundary formation suggests the existence of strong couplings between processes giving rise to embryonic patterning. In that case, the "arrow of explanation" should go from the higher level (i.e. embryonic/tissue) to the lower one  (cell/genes) \cite{Woese2004}, and for this reason, we find it useful to go back to the  notion of primary/secondary waves to categorize and discuss models.

\chapsec{Categorizing models: primary waves and bifurcations}

In simple terms, primary waves correspond to moving fronts where bifurcations --in the dynamical system sense -- occur. Thus, a primary wave defines the (moving) boundary between  regions of the embryo with different attractors. While in his initial discussion Zeeman was mostly concerned about steady states \cite{Zeeman1976}, there is no reason not to include other types of attractors such as oscillators. Secondary waves are due to processes downstream of primary waves in the absence of any new bifurcation.  By carefully listing the possible primary waves and their order, one can distinguish at least four categories of models, Fig. \ref{fig:WaveScenario}, with some properties summarized in the Table below.

\begin{center}
\footnotesize
\begin{tabular}{lllll}
\toprule
Model type & Primary wave I  & Primary wave  II & Origin of boundary formation & Timing of boundary definition\\
\midrule
\midrule
Clock and wavefront & Wavefront   & N/A & Clock pulses & Before primary wave \\
&&&&\\
Clock and Switch & Hopf or Infinite Period  & AP Bistability & wave in AP polarity  ? & Primary wave II \\
&&&&\\
Switch/Clock/Switch & Two attractors  & Homoclinic & wave in AP polarity  ?& Primary wave III (AP bistability) \\
&&&&\\
Meinhardt-PORD & Clock $\rightarrow$ A or P &  N/A & AP polarity + X factor ? & Primary wave\\
\bottomrule
 \end{tabular} 
 \label{summary_properties}
\end{center}

\chapsubsec{Initial Clock and Wavefront model : anticipating the somite primary wave}

In the initial clock and wavefront model, a primary wave is defined by the region where the system goes from bistable to monostable, corresponding to a transition from undetermined to somite (Fig. \ref{fig:WaveScenario} A). A similar primary wave is observed in the cell cycle model, and in the RA/FGF Goldbeter-Pourqui\'e model \cite{Goldbeter2008}.  A recent model with similar features in terms of the primary wave is the "Clock-dependent oscillatory gradient" (COG) model, where a spatial gradient of Erk/FGF is modulated by the clock, and defines boundary formation \cite{Simsek2018,Simsek2022}. An important aspect of those models is that they do not require the clock to stop to have patterns, rather the clock is only initiating the undifferentiated to somite transition close to the primary wave, periodically defining blocks of cells. Just like the clock does not have to stop,  in those models one can find conditions where somite boundaries can form in the absence of the clock. For instance, one can in principle change the $z$ variable from its high to low value in the classical clock and wavefront model (corresponding to the undifferentiated vs somite state, Eq. \ref{CW_maths}), or in the COG model one can trigger boundary formation in the absence of a clock with a pulsatile external inhibition \cite{Simsek2022}. Those models are not concerned with somite AP polarity, although as shown in this review (Fig. \ref{fig:CW_maths_2} C), in the regime of the slow clock, one can get interesting interactions between the primary wave and the clock giving rise to a jagged pattern, that could be leveraged to define such polarity. Therefore one needs to assume that, later on, a second primary wave should occur to define somite AP polarity (Fig. \ref{fig:WaveScenario} A).

\chapsubsec{Clock primary wave}

In all other models discussed in this review, the clock is stopping, so there should be a primary wave corresponding to a bifurcation (Fig. \ref{fig:Mesp2Geometric} B-D). This is the first main difference with the Clock and Wavefront model,  where oscillations can in principle continue beyond somite formation. The simplest case would be that there is a Hopf bifurcation\cite{Francois2012,Jorg2015}.  A primary wave via an infinite period bifurcation offers an alternate scenario, recapitulating the slowing down prior to the bifurcation \cite{Jutras-Dube2020}. It is also noteworthy that delayed models with increased production rates bring the system close to infinite period bifurcations. Phase models such as Lewis' are not easily interpreted in terms of primary waves, because they circumvent the limit cycle formalism by design (see the 'Clock and Unclock' section) so there are no bifurcations, and all oscillators continuously slow down. Although one could mimic a primary wave by having a sudden drop of the frequency to $0$ in Eq. \ref{simpler}, it is ill-defined from a dynamical standpoint since  it would freeze the system on a line of continuous states. Another possibility is to add another variable, such as in the phase-amplitude model Eqs. \ref{eq:phase-ampa}-\ref{eq:phase-ampt}, which assumes a primary wave from an oscillatory relaxation-like system to an excitable one (Type I or Type II). Hybrid geometric phase models (Eq. \ref{SNICphase_bistable})  can more easily account for saddle-node bifurcations on the cycle.  Murray \etal model \cite{Murray2013} offers an interesting solution for phase models where a primary (shock) wave emerges  from the coupling of oscillators.

\chapsubsec{Adding AP polarity primary waves}

Multiple models aim at explaining AP patterning within somites. In such models, clock stopping usually defines one fate, so at least one more fate is required (with the two fates corresponding to A and P), and thus a second bifurcation is required to get bistability. This defines another primary wave so that the overall segmentation process combining clock stopping and somite AP patterning requires \textit{two} primary waves \cite{Francois2012a}. The simplest model accounting for those dual primary waves is the clock and switch model, which can be realized via a Hopf bifurcation \cite{Francois2012} or a SNIC bifurcation \cite{Jutras-Dube2020}, subsequently followed by a saddle-node bifurcation (Fig. \ref{fig:WaveScenario} B).  Notice that a difference between the Hopf and the SNIC scenario lies in the nature of the first fate when the clock dies : in the Hopf scenario, the first fate appears where the cycle collapses (explaining the 'blurred' region in kymographs, Fig. \ref{fig:geom_model} B) and then later evolves to define an anterior or posterior fate. Conversely the SNIC scenario, the first fate appears on the cycle, leading to more robustness in patterning, Fig. \ref{fig:geom_model} C.

It is often assumed that the primary wave for clock stopping happens before the primary wave for AP definition, but in principle, one could imagine an alternative scenario with a first primary wave where one fixed point corresponding to a future fate (say A) appears while the cycle still exists, then later on the primary wave for clock stopping occurs via a homoclinic bifurcation towards this fate, then a third primary wave appears to define the second fate (P) (Fig. \ref{fig:WaveScenario} C). It is relatively straightforward to follow a strategy similar to the one in \cite{Jutras-Dube2020} to engineer the geometry of such 'Switch/Clock/Switch' model, see  Fig. \ref{fig:Homo}.  While this model might appear at first more convoluted, with three primary waves, it combines three independent attractors (A,P and Clock) and thus is more modular.  This example also  illustrates how listing possibilities in terms of primary waves help uncover new models.

\bnormf
\includegraphics[width=\textwidth]{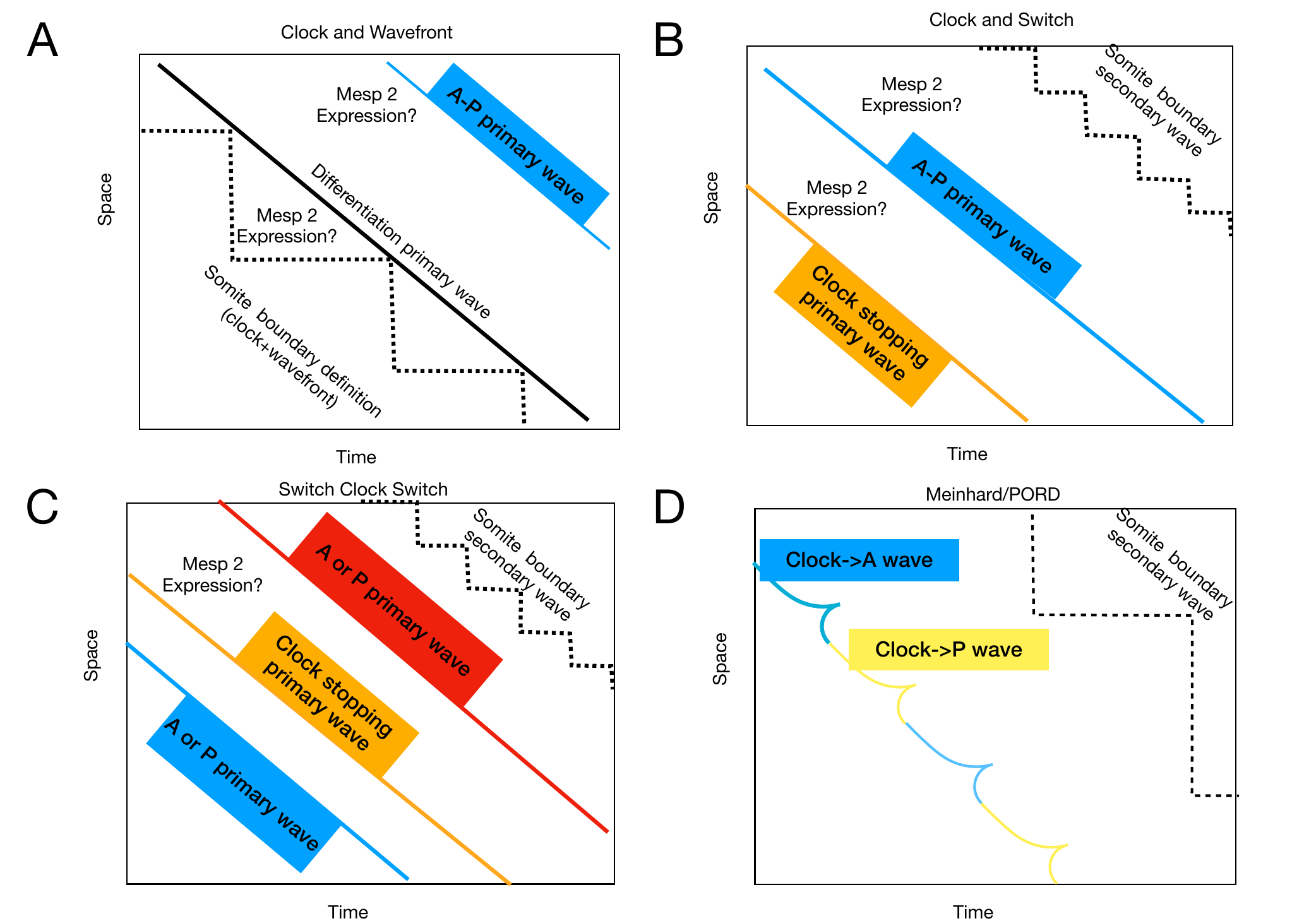}
   \caption[Four possible scenarios for primary/secondary waves in somitogenesis] { Four possible scenarios for primary/secondary waves and somite boundary formation. Each panel represent propagation of different waves using same space/time conventions as kymographs. See  the discussion of each scenario in the main text. }\label{fig:WaveScenario}
\enormf


\chapsubsec{Combining primary waves}
If there are indeed several primary waves (e.g. clock stopping and AP polarity), an intermediate region should exist between those waves . Consistent with this view, there are multiple, specific genes tied to a transition region between the oscillatory region and the stable one, such as \textit{Mesp2}, or \textit{Ripply}. Conversely, some models essentially couple together the primary waves from clock stopping and for AP definitions, e.g. \cite{Clark2021}. Interestingly, the two primary waves can occur simultaneously in SNIC-based models  if the underlying equations are symmetrical, as observed in \cite{Jutras-Dube2020}, see also Eq. \ref{SNICphase_bistable}. Such symmetry is not generic, but one could imagine some evolutionary pressure leading to such symmetry (see e.g. \cite{Johnston2022} for arguments that parsimony in evolution might in fact generally favour symmetries). 
The Meinhardt and PORD models represent special cases combining two primary waves, but instead of having a sequence of clock-stopping wave/AP bistability, they display in fact two alternating primary waves, where two different, but symmetrical Hopf bifurcations lead to A and P states in alternation (Fig. \ref{fig:WaveScenario} D). The waves in such models present much more complex dynamics, with nucleation of stable zones at boundaries between A and P fates, and subsequent propagation in between. Notice that cell coupling is crucial to explain both primary waves for those models.

\chapsubsec{The origin and timing of discrete boundaries}

To wrap up this section, let us summarize the precise origin of the discrete somite pattern in those different models. For the clock and wavefront type models, there is only one final attractor for each cell, and the somite boundary is defined by the position where the periodic influence (e.g. pulses) of the clock is not strong enough to overcome the barrier between the high and low $z$ state, which defines periodic, discrete regions.  Thus the positioning of a somite boundary depends a lot on the details of the potential close to the wavefront, and on how the clock modulates the dynamics close to it. Boundaries are defined early, \textbf{before any primary wave}, because of the premature "catastrophic jump" induced by the clock.
 For all other models, the discrete nature of somites is a direct consequence of the eventual bi- or multi- stability of the cell state. Because there are two attractors corresponding to anterior A and posterior P, there are natural alternations between those two regions, defining presumptive boundaries at steady state. By definition, the PA boundary is the somite boundary,  and its positioning depends on the way those two final states are reached. The role of the clock is to set the initial conditions leading toward those two AP states and thus boundaries are defined  \textbf{after the last primary wave}. In fact, the problem is not to define discrete boundaries in such models, but rather, as pointed out first by Meinhardt \cite{Meinhardt1982}, to distinguish between within AP somite boundary and proper PA somite boundaries. Meinhardt suggested the existence of an intermediate state, alternatively one could imagine a transient inhibitor of boundary formation at the AP boundary, e.g. controlled by the clock, or by some asymmetry in the wave pattern. As discussed, this could arise from specific temporal dynamics, e.g. in relation to the excitability of oscillators or to SNIC bifurcations. 

\bnormf
\includegraphics[width=\textwidth]{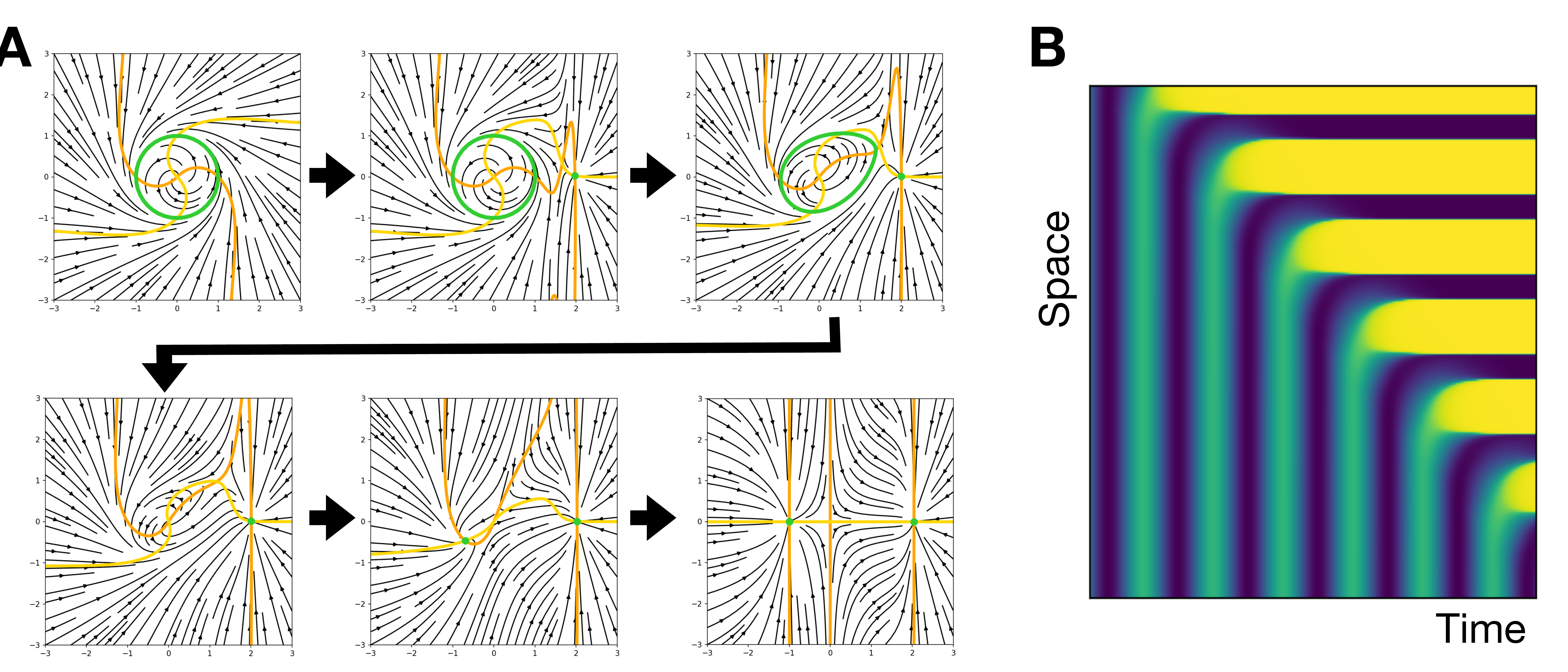}
   \caption[A possible  'Switch/Clock/Switch' scenario]{ A system where the AP primary wave (where another fixed point appears)  occurs before the clock stopping primary wave (A) Evolution of the flow as a function of time. As the control parameter is varied, one gets from a situation with one cycle, to one cycle + one steady state  (second panel), to a monostable system via a homoclinic bifurcation (4th panel), then to a bistable system with a saddle-node bifurcation (5th panel) (B) Corresponding kymograph assuming a graded control parameter similar to Fig. \ref{fig:geom_model} }\label{fig:Homo}
\enormf

\chapsec{Reconciling models with experimental observations}

To go further, we first need to look for experimental evidence favoring or disproving different models, in particular on the nature of bifurcations happening in the embryo.

\chapsubsec{Clock stopping and bistability primary waves}

It is easiest to first discuss the waves associated to AP patterning. It can be parsimoniously explained by bistability (e.g. between \textit{Mesp2} and \textit{Uncx4.1}). Stem cell systems are very informative regarding this \cite{Miao2022}: the fact that somitoids can form expressing either \textit{Mesp2} or \textit{Uncx4.1} is clearly suggesting the existence of two mutually excluding fates, thus bistability. Also, the fact that there is no need for a strong phase gradient to form a somite-like structure suggests that a simple alternation of fates following clock stopping is sufficient, again consistent with the clock and switch model.

The nature of the primary wave for clock stopping is less clear. It seems well-accepted that cellular oscillators slow down to define waves of genetic expression, that the oscillation eventually stops in the anterior, and that the slowing down of individual oscillators imprints some anteroposterior patterning  in the embryo. Clock stopping suggests that there is indeed a primary wave associated with a bifurcation.  Waves come from a period increase of oscillators moving towards the anterior. They are often explained by graded "delays" along the anteroposterior axis \cite{Ay2014}, but as discussed in this review those delays might be phenomenological. Period increases could correspond to a system transitioning towards an  infinite period bifurcation (e.g. SNIC). SNICs occur through localized slowing down in a small part of the limit cycle, and would precisely manifest itself through increases of phenomenological "delays" between pulses of the cycle. Infinite period bifurcations also parsimoniously explain why the oscillator period is increasing before the primary wave of clock-stopping.  Conversely, Hopf bifurcation occurs with a constant frequency close to the bifurcation, so it would seem curious  (but not impossible) that the period of the oscillators changes drastically right where such bifurcation occurs. 

More elements from various experimental and theoretical works are indeed more consistent with SNIC than Hopf. For instance, there are multiple indications that segmentation oscillators can be put into an excitable regime, which is associated with considerable period changes close to the bifurcation because the system spends much time in a small region of the phase space (see Appendix \footnote{it should be mentioned here that although Type II  systems get through a Hopf bifurcation, they behave very similarly to Type I SNIC oscillators with a period increase as the system gets closer to the bifurcation. In both cases, the period increase comes from the fact that the system spends more and more time close to a small region in the phase space so that Type I and II are essentially indistinguishable there and phenomenologically behave like SNICs. Only very close to the bifurcation do Type II oscillations get more sinusoidal, but with very low amplitude,  see Appendix A  }). Stem cell systems appear to be excitable \cite{Hubaud2017}, but it is not clear if this excitability is associated in any way with the bifurcation at the front.  Single-cell transcriptomics in zebrafish reveals that transcription of the core \textit{hes} oscillator occurs in bursts  \cite{eck2024}, consistent with models tuned closed to infinite period bifurcations.  Entrainment experiments suggest constant-sign phase responses, again consistent with such bifurcations \cite{Izhikevitch2007}.   Lastly, SNICs appear as the most parsimonious way to turn an oscillator into a multistable pattern \cite{Jutras-Dube2020}, meaning that infinite period bifurcations naturally occur in presence of two primary waves (clock stopping and AP). A transition towards a SNIC further encodes asymmetries in wave patterns very similar to what is observed experimentally, possibly allowing for a distinction between AP and PA boundaries.Based on those observations, it seems to us the combination of the clock stopping and of a primary wave associated with bistability would thus be most consistent with the Clock+Switch scenario of (Fig. \ref{fig:WaveScenario} B or C)

\chapsubsec{Somite boundary definition and Mesp}

It seems at first that the role of the broad expression of genes like \textit{Mesp2} in mammals within a future presumptive somite would rather be more consistent with the original clock and wavefront model than with a clock and switch model (e.g. \textit{Mesp2} would mediate the change of state $z$ in Eq. \ref{CW_maths}). However, \textit{Mesp2} presents several domains of expressions in two consecutive stripes. It is graded within one future somite, with higher expression in the anterior \cite{Takahashi2000,Morimoto2005}, and almost complete absence in the posterior. In a clock and wavefront model, as said above, the somite boundary is defined before the wavefront, right where the clock signal exactly compensates the potential barrier between the high and low $z$ states. It would thus presumably correspond to the \textit{Mesp2} posterior boundary.  This domain appears quite fuzzy experimentally, which seems contradictory to the idea that it precisely defines somite boundaries.

Conversely,  \textit{Mesp2} anterior expression seems more crucial since the future somite boundary appears anterior to cells where it is most strongly expressed \cite{Saga2007} and this domain has been shown to play a role in setting up AP polarity \cite{Oginuma2010}. This suggests that \textit{Mesp2} expression, and subsequent somite boundary formation, could rather be a secondary wave associated with AP polarity. It is relatively easy to model a behavior similar to \textit{Mesp2} as a secondary wave of a clock and switch dynamics : for instance, one can simply assume that \textit{Mesp2} is downstream of the clock but expressed for roughly half a cycle and only close to  the first primary wave  (technically when the control parameter reaches a given threshold), see Fig. \ref{fig:Mesp2Geometric} A. In such a simple model, we first see a broad expression domain within a future presumptive somite, that localizes to the anterior past the bifurcation. This model also parsimoniously explains why \textit{Mesp2} concentration would decrease in some cells while being maintained in others Fig. \ref{fig:Mesp2Geometric} B. Of note, a posterior-type cell close to the AP boundary expresses Mesp2 for a long time (orange cell in  \ref{fig:Mesp2Geometric} B) while a  posterior-type cell close to the PA boundary expresses Mesp2 for a very short time (blue cell in  \ref{fig:Mesp2Geometric} B), so that such difference of timing could be used to define the AP vs PA boundary. We also notice that in such a situation, the dependency on the control parameter gives a behavior for \textit{Mesp2} similar to the last wave observed in the zebrafish clock  (Fig. \ref{fig:WaveStopping}).

\chapsubsec{Coordination of primary waves}
\label{sec:coordination}
 Several models implicitly or explicitly coordinate or even collapse the two primary waves \cite{Clark2021, Jutras-Dube2020}. This is not generic in terms of dynamical systems theory, but such tight coordinations of waves could certainly provide biological robustness , and thus might have been selected by evolution.  It could be that  \textit{Mesp2} (or another gene) also plays a role in primary waves' coordination. For instance, one could imagine that once expressed, \textit{Mesp2} "speeds up" bifurcation crossings, by stopping the clock and inducing the AP primary wave almost simultaneously. This would be reminiscent of the way the clock triggers early the "catastrophic jump"  in the clock and wavefront model. \textit{Mesp2} is known to negatively regulate its own expression  \cite{Takahashi2000}, likely through \textit{Tbx6}, which is likely implicated in clock regulation \cite{Hubaud2016, Zhao2015, Yabe2023}, suggesting indeed that \textit{Mesp2} downregulation induces the primary wave associated to clock stopping. In zebrafish, \textit{Ripply} could play a similar role to \textit{Mesp2} in mouse. Interestingly, it has been suggested that a wave of \textit{Ripply} turns off the clock by degradation of  a bistable \textit{Tbx6}  \cite{Yabe2023}. High concentration of  \textit{Tbx6} would correspond to an oscillatory state, while low concentration would correspond to a monostable non-oscillatory state \cite{Yabe2023}, thus reminiscent of the behaviour of the $z$ variable in the original clock and wavefront model. One possibility here could be that the primary wave for AP polarity is triggered first, which manifests itself through \textit{Ripply} expression, itself graded because of the clock waves. Later on \textit{Ripply} stops the clock, so that in this picture the primary wave for somite boundary formation coincides with the AP polarity, and the primary wave for clock stopping happens downstream to it, thus possibly realizing a "Switch/Clock" scenario of Fig. \ref{fig:WaveScenario} C.

A last issue related to the AP polarity within somite and boundary formation is the role of space/time in this process. The fact that one can obtain two-cell-wide somites in zebrafish mutants \cite{Henry2000} suggests that two AP fates are necessary and sufficient to define proper somite polarity and boundary. This means that there is no possibility for a hypothetical spatial gradient to make the distinction between the AP and the PA boundary (unless such a gradient would be subcellular). Rather, this suggests that the polarity effect comes from the \textbf{temporal} dynamics, e.g. via some coordinated timing of AP segregation as detailed for the phase-amplitude model in Section \myref{sec:phaseamplitude}. This can be realized through single-cell oscillators with an asymmetrical temporal profile such as the ones shown in Fig. \ref{fig:geom_model}, or again an effect similar to what is observed in Fig. \ref{fig:WaveStopping} or Fig. \ref{fig:Mesp2Geometric} where the anterior-most cells get activated last within a given somite.

\bnormf
\includegraphics[width=\textwidth]{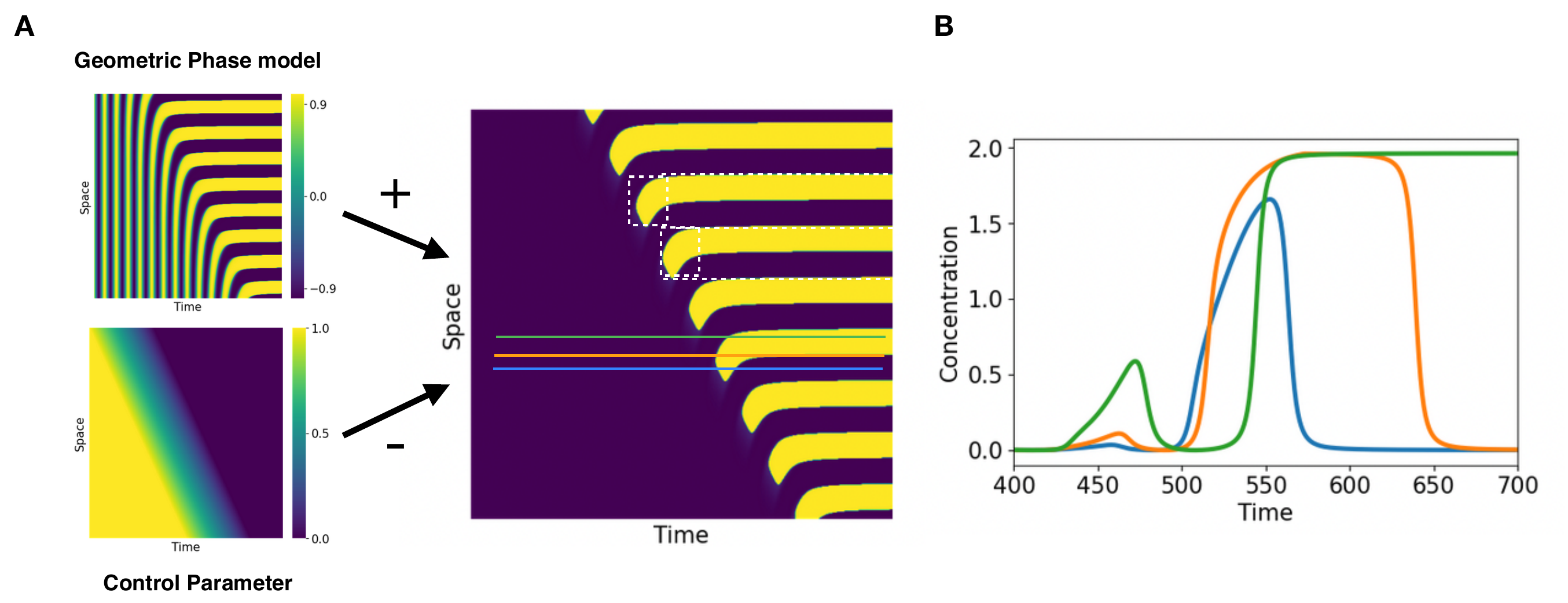}
   \caption[Geometry of a Mesp-2 like secondary wave]{ (A) Generation of a Mesp2-like secondary wave by simply combining a purely geometric model $\theta(x,t)$ with its associated control parameter $a(x,t)$. Kymograph for such secondary wave $Mesp2(\theta,a)$ on the right of the panel, showing broad expressions followed by restriction in the anterior. Simulation is available in attached Notebook. We highlight blocks of cells presumably corresponding to the same presumptive somite. Green, orange and blue lines correspond to positions within the same future presumptive somites (B) temporal profiles of the Mesp2 like secondary wave at positions indicated in (A) }\label{fig:Mesp2Geometric}
\enormf

\chapsubsec{Which control parameters ?}

Assuming primary waves corresponding to bifurcations are identified, the question of their control parameters still is largely open. Are the waves triggered by external signals? local cues? cell-to-cell interactions?

First, it is now clear that single cells outside of embryos can recapitulate the behavior within the embryo, going through oscillations, stopping and leading to \textit{Mesp2} expression \cite{Rohde2021}. This is also consistent with what is seen in somite-related organoids, where segmentation can be entirely recapitulated without many features observed in the embryo (such as well-defined PSM waves). This confirms the classical "kinematic" view of somitogenesis, where local cues are necessary and sufficient for cells to move from tailbud to somite fates. All of this seems to exclude a crucial role of cellular coupling by itself to trigger primary waves; conversely, it suggests the existence of local (i.e. cellular) control parameters. This does not mean that cell coupling plays no role in the embryo: it could certainly modulate oscillations, play a role in some noise-correcting mechanisms, or coupling of different primary waves.

Many models assume that the first primary wave (clock stopping or somite boundary definition)  is controlled by one or several morphogen gradients (such as \textit{FGF8}). It is, however, clear now that at the very least such control is not simply imposed externally on the cell (like classical positional information in the French Flag Model), but rather is regulated either at the cellular or the tissue level. Experiments in zebrafish suggest that the slowing down of the oscillations is cell-autonomous, excluding full control (but not some modulation) by an exogenous morphogen. A timer independent from the clock, turning on once the cell exits the tail bud is the simplest scenario \cite{Francois2010,Rohde2021}. One issue is that such a timer can not easily explain scaling experiments in mouse, where cells effectively  stop much later than they should \cite{Lauschke2013}. The dynamics of the front in explants rather suggest that slowing down might come from some internal feedback \cite{Lauschke2013,Sanchez2021,Clark2021}, e.g. of the \textit{Notch} oscillator on itself, and more precisely of some phase variable.  There could also be more complex controls, for instance, the oscillatory signal itself could induce the primary wave, like in the phase-amplitude model in Section \myref{sec:phaseamplitude}. Other works have suggested more local computations where cells "compare" their (oscillatory) state to the neighbors \cite{Boareto2021}, or possibly measure specific features of local morphogen gradients such as their fold-change \cite{Simsek2018, Simsek2022} (see also the recently proposed 'neighbourhood watch model' in another developmental context \cite{Lee2022}). Such mechanisms do not seem consistent a priori with the fact that single cells can recapitulate bifurcations seen in the embryo. However, since epithelialization defining somites obviously requires several cells, the study of single cells dynamics might not allow us to determine where a boundary forms at a tissue level. The primary wave and their control parameters might be defined at the single cell level, but could be coordinated at the tissue level  by some local computations.

In many models, at least two (temporal) coordinates are needed to define the clock stopping primary wave: for instance, in \cite{Rohde2021} one needs a clock and a timer, while in \cite{Lauschke2013, Sonnen2018} one needs two oscillators. To some extent, both ideas are elaborations of the original vision of the wavefront in \cite{Cooke1976}, where the positional information variable was initially assumed to be laid off by some independent temporal process. But phenomena such as temperature compensation or scaling in development imply that those two temporal components have to be coupled one way or another, the question being if they are both downstream of the same global control (e.g. metabolism) or if they are actively feeding back on one another. The cell cycle could also play some role as the second temporal coordinate. The initial cell cycle model was motivated by puzzling, periodic, somite boundary phenotypes associated with cell cycle period in heat shock experiments. Those phenotypes still are unexplained, and the cell cycle model somehow lost traction once it was discovered that there is a Notch oscillation controlling waves and boundaries. However, if segmentation requires a  second (reference?) oscillator, the cell cycle is an obvious candidate, which would likely explain the heat shock phenotypes. The existence of a graded cell cycle phase of a reference (cell cycle) oscillator within the PSM could also explain some differences in the wave patterns between mouse and zebrafish (where there is no known Wnt oscillator).

The control parameters for the primary wave of AP patterning are even less clear, possibly because so far experiments have not systematically dissected the nature of bifurcations. In most models, the same control parameter as the clock-stopping primary wave is used to explain the AP primary wave \cite{Francois2012}, but this is not an absolute necessity. That said, the relatively linear developmental trajectory  of cellular fates from oscillation to somite formation suggests at the very least that control parameters are tightly coupled.

\chapsec{Towards a universal model of the cellular dynamics during segmentation ?}

A general problem of quantitative/systems biology is to relate the forest of interacting genes to high-level function \cite{Gunawardena2013}. In different contexts, several theorists have moved the "systems" focus up in scale, e.g. building network-based models \cite{Clark2019,Verd2018}, or even gene-free, geometric models \cite{Corson2017,Corson2017a,Jutras-Dube2020,Saez2021}. A justification of this approach comes from complex systems theory: it is now well established that generic, low-dimensional coarse-grained dynamics typically emerge from complex biochemical networks \cite{Gutenkunst2007,Proulx-Giraldeau2017}, irrespective of the molecular details. Such an effect can possibly be leveraged and amplified by evolution, locking the dynamics on a low-dimensional manifold \cite{Furusawa2018}. In light of the primary/secondary wave discussion, the question is to figure out if one can describe somitogenesis in a simple, compact, general way, e.g. as a system not far from some bifurcation with few control parameters and feedbacks.

Importantly, such reduction is naturally low-dimensional, allowing for simple categorizations and geometrical insights. As long as we have the right low-dimensional representations, one should be able to recapitulate all behavior and phenotypes by modulating parameters, as recently shown in the context of \textit{Drosophila} gap gene network \cite{Seyboldt2022}. Fig. \ref{fig:Mesp2Geometric} illustrates how to tie a specific geometry to actual gene expressions like \textit{Mesp2}. More generally, if one can derive all genetic expression from two coupled variables such as $\theta(t)$ (a polar coordinate) and $a(t)$ (control parameter) in Fig. \ref{fig:Mesp2Geometric}, then by definition we have captured the entire process. Importantly,  such low-dimensional representations are more fundamental because they are simpler and have broader explanatory power,  providing quantitative explanations for apparently unrelated aspects.  Here, examples include models such as the one in  Fig. \ref{fig:Phase_model_SNIC}, where a SNIC bifurcation is explained by the presence of two primary waves. Then it manifests itself through both slowing down of oscillators and spatial wave asymmetry, properties that we get here 'for free' without any extra hypothesis. Along similar lines, it was already suggested multiple times that epistatic interactions are natural consequences of low dimensional parameter dependency \cite{Corson2012,Husain2020}.

In addition, we also expect the existence of feedback, connecting all observables of interest (oscillation, differentiation, and possibly elongation \cite{Clark2021}). Thus, importantly, control parameters might be (slowly) changed by the dynamics, leading to a self-organizing/referencing process.

 More practically,  what would be the minimal, self-organized model consistent with the mostly cell-intrinsic dynamics observed experimentally \cite{Clark2021,Rohde2021} ?  To get there, we need to identify bifurcations, their order, their control parameters, then possibly couple them in some way.   As an illustration, let us make a couple of assumptions. Focusing on the clock stopping bifurcation, let us assume it is an infinite period one. As said above, we have little idea of the actual nature of the control parameter, but for an illustration, let us assume that it is set by some intrinsic temporal processes, eg the phase difference between two oscillators, one being \textit{Notch} \cite{Sonnen2018} and the other being more synchronized within the embryo. Initially, we assume this control parameter is high, and when it reaches $0$ the bifurcation occurs. If so, a slight slowing down of the \textit{Notch}  clock would get the control parameter closer to $0$. But as the control parameter goes closer to $0$, the period would further increase due to the proximity of the infinite period bifurcation. As a consequence, \textit{Notch}  would slow down even more. This gives a snowball effect, with a divergence of the cellular period in a finite time. 
 
 More quantitatively,  if there is indeed an infinite period bifurcation, we know that the period depends in a generic way on the control parameter of the bifurcation (see Appendix A).  We define the phase difference to the reference $\Delta \phi = \phi-t $, where $t$ is the phase of the reference oscillator. Let us now assume that the control parameter is $\Delta\phi-\phi_*$, where  $\phi_*$ a critical phase such that at the bifurcation the control parameter is $0$. We thus expect in general for the \textit{Notch} phase of a single cell that 
 \begin{eqnarray}
    \frac{d \phi}{dt} &\sim & f(\Delta\phi-\phi_*) \label{SNIC_Notch} \\
    \implies \frac{d \Delta \phi}{dt} &\sim& f(\Delta\phi-\phi_*)-1
\end{eqnarray}
 
 where $f$ captures the control parameter dependency (typically very close to a SNIC bifurcation $f(x)\propto \sqrt{|x|}$), and $f(0)=0$ indicates an infinite period bifurcation (see also measurement in  \cite{Giudicelli2007}, Fig. 2B).

We know  from experiments that the tail bud cells can  maintain their state. This suggests that in the tail bud the control parameter is stable, i.e.  $\frac{d \Delta \phi}{dt} \sim 0$ for $\Delta \phi=0$. Thus, $f(-\phi_*)\sim 1$. This indicates that the reference oscillator goes at the same frequency as \textit{Notch} in the tail bud. Furthermore the tail bud defined by $\Delta \phi=0$  appears as a fixed point for the slowing-down dynamics, which ties the property of a dynamical system to a well-known and observed cell fate.
 
 Now Taylor expanding $f$ close to the tail bud where $\Delta\phi =0$, we get :
 \begin{equation}
    \frac{d \Delta \phi}{dt}=f(-\phi_*)+\Delta\phi f'(-\phi_*) -1 \sim \Delta\phi f'(-\phi_*) \label{SNIC_alpha}
\end{equation} 

using the fact that  $f(-\phi_*)\sim 1$. Eq. \ref{SNIC_alpha} captures the snowball effect described above in a compact way, and clearly corresponds to Eq. \ref{coupled_model} with $\alpha=f'(-\phi_*)$.

Taking now a step back, those simple calculations show that the  phenomenological $\alpha$ model, that was inferred from `scaling' experiments, in fact naturally derives from two generic hypothesis 
\begin{enumerate}
\item the sensitivity of the single cell oscillator period to the control parameter of the bifurcation (most easily realized through an infinite period bifurcation) 
\item the feedback of the state of the oscillator on its own bifurcation control parameter (here, the fact that the control parameter is related to the phase shift $\Delta \phi$).
\end{enumerate}
For this reason, the dynamics described by Eq. \ref{SNIC_alpha}  are very generic as soon as both hypotheses hold. For instance, the control parameter could be another variable, e.g. the amplitude of the clock. In that case, if there is an infinite period bifurcation, one prediction would be that the amplitude of the oscillator strongly feeds back on its period. In \cite{Rohde2021}, a timer variable is used to account for the period slowing down and collapses at the front to account for the clock stopping. This looks very similar to the effect of a control parameter on a bifurcation, and as already mentioned very similar to what would happen if this control parameter was a phase difference suddenly jumping from $2\pi$ to $0$, see e.g. \cite{Beaupeux2016}. If this hypothetical timer variable is in turn regulated by the oscillators, then a behavior similar to  Eq. \ref{SNIC_alpha} should hold.  Lastly, it should be pointed out that an explicit example of such coupling between oscillations and control parameter to regulate timing can be found in a related context, namely Hes oscillations during neurogenesis.  Oscillating Hes1 represses its post-transcriptional repressor, microRNA mir9 \cite{goodfellow2014}. This nevertheless leads to a slow accumulation of mir9, which, past a threshold, abolishes oscillations and triggers differentiation. This provides a timing mechanism, proposed to give robustness at the tissue level \cite{Phillips2016}, with the difference that oscillations are more stochastic and that oscillations disappear via a Hopf bifurcation (i.e. with a constant period and decreased amplitude) .


\chapsec{Unknown and Known limitations, blind spots }

\chapsubsec{Finding the right rules}
Even if we believe a simple, geometric model with feedback exists, we are still far from knowing which true geometry is underlying the dynamics, and how to test for it.
The first question is how to connect the control parameter and state of the system to actual observables. Eq. \ref{SNIC_Notch} holds for a single-cellular \textit{Notch} oscillator, and as seen from recent experiments, the dynamics in a single isolated cell recapitulates the dynamics in the embryo. But similar equations can be written for \textit{Wnt}  oscillator in mouse: in particular, we know from entrainment experiments that the \textit{Wnt}  oscillator at the embryonic level can be entrained by perturbations on \textit{Notch} signaling, showing that coupling works in both directions, but possibly at the global, embryo level (with possible influences on intrinsic period \cite{Sanchez2021}).
If we believe that the control parameter is related to phase differences, one issue is that Equations such as Eq. \ref{SNIC_Notch} are not $2\pi$-periodic, which suggests other processes might be implicated to account for maturation/number of cycles aside from the phase difference (global reference oscillator, \textit{Wnt} slowing down, variable amplitude, etc...). There could also be here roles for external morphogens, e.g. to modulate control paramerters. It is in principle possible to write general forms for such models, however, the simplest models do not easily recapitulate all fundamental aspects of somitogenesis or of entrainment. For instance, simple entrainment models recapitulate phase waves but with a uniform period \cite{Juul2018}.

\chapsubsec{Role of cell coupling}

As said above, the dynamics of cells in the embryo can be qualitatively captured in different types of cell cultures (single cells, stem cell medium, somitoids, mPSM), but there are quantitative differences. This suggests that some amount of coupling (between cells in particular) is necessary to account for the full observed dynamics within the embryo, e.g. to coordinate waves. There is still much work to do to precisely quantify what happens. We already mentioned optogenetics as an exploration tool, but it is also possible to directly culture cells/embryos and to study the propagation of the waves \cite{Hubaud2017}; reaggregates also provide a possible direction \cite{Tsiairis2016,roth2023unidirectional,Ho}. Importantly, just like the frequency of the cells clearly varies in space and time, it could well be that coupling properties vary as well. Also entrainment experiments \cite{Sanchez2021} reveal global coupling at the embryo level (e.g. between Notch and Wnt),  still unexplored. Finally, the boundary formation has been tied to the gradient of FGF between cells modulated by the clock\cite{Simsek2022},  but it is not clear how this connects to the bifurcations at the single cell level (such as clock stopping and AP polarity definition).

\chapsubsec{Role of mechanics and cellular motions}

Another direction is the exploration of the strictly mechanical properties of the system.  At the very least, physical forces appear to play a role in the robustness of the pattern: for instance, it was recently shown that some left-right asymmetry in the embryo at the clock level could be corrected by cellular rearrangements \cite{Naganathan2020}. Such rearrangements are coordinated, and differ depending on the PSM region \cite{yin2019}. We also already mentioned the "somites without a clock" mechanism \cite{Dias2014}, and somitoids/segmentoids \cite{Miao2022}, suggesting that boundary formation can be triggered independently/downstream of the clock. A rare example of models explicitly modeling boundary formation can be found in Glazier \etal \cite{Glazier2008}. The model is based on a generalized Glazier-Graner-Hogweg (or Cellular Potts) model \cite{Glazier1992} and works downstream of most phenomena (clock, primary waves) discussed in this review. The clock acts as an external pacemaker, sequentially committing cells to be anterior or posterior type within one somite. There further is an implicit coordinate system within one somite, both proximal-distal and anteroposterior. In this model, rearrangements, somite formation, and rounding naturally occur through the definition of 10 different cellular fates corresponding to different levels of adhesion molecules (such as ephrins). The model recapitulates the segregation of single somites (with proper AP polarity), and mutants can also be modeled, such as the half-sized somites from \cite{Horikawa1999} due to the segregation of AP half somites. More precise measurements are now possible \cite{Piatkowska2022}, showing that epithelialization is a much earlier process than previously thought and, interestingly, AP markers within somites are shown to be graded and in some cases absent from core somites, thus confirming the existence of another set of local coordinates and associated local regulations. That said, it remains to be seen how such a fine level of patterning reconciles with the two-cell wide somites observed in \cite{Henry2000}.

Other physical aspects remain relatively unexplored. Physical rigidity of the tissue might play a role, as exemplified by the Matrigel concentration dependency in stem-cell systems, and as suggested by the implication of the YAP signaling pathway in the initiation of the oscillations \cite{Hubaud2017}. In organoid systems, the density of cultures seems to relate to phase gradient formation and FGF gradient (see \cite{Yaman2022} Fig. 5B and S5B), suggesting another direct physical input on the clock. 
It is also well known that cells in the PSM move in a complex way \cite{Benazeraf2017,Romanos2021}. Their motility/diffusivity  is regulated by FGF, which influences elongation \cite{Benazeraf2010} (see also \cite{mkrtchyan2015embryonic,regev2022} for models of elongation).  The establishment of AP polarity within a somite might be further reinforced by cell sorting in response to salt-and-pepper activation of \textit{Mesp2} within a future presumptive somite \cite{Miao2022}, again suggesting at the very least a physics-based refinement system. Any kind of local or global oscillatory coupling is expected to be impacted by such cell motility \cite{Uriu2013}. In a recent study, Uriu \etal \cite{Uriu2021} developed an explicit model of PSM with motile cells and advection. Each cell is modeled as a phase oscillator with Kuramoto coupling,  under the control of an imposed frequency gradient, and with some additional phase noise. Just like previous phase models \cite{Morelli2009}, segmentation is modeled as a "freezing" of the phase at a given position. Desynchronization/resynchronization experiments under influence of DAPT \cite{Riedel-Kruse2007, Liao2016} are then studied within this framework. Multiple features are reproduced, such as intermingled segments (i.e. spatial mixtures of frozen oscillators at different phases). Those come from the nucleation of phase vortices in the posterior (following resynchronization), that get advected towards the anterior and fixed by the front.

\chapsubsec{Higher level processes}
Lastly, it is also important to connect segment formations to two higher-level processes. The first one is embryonic growth: in all existing models, growth is an external driver of the system, largely independent from the segmentation clock itself. This is unlikely to be true: as said above, at the very least, growth speed and segmentation clock period should be coupled in some way. FGF is known to influence both growth and segmentation, but higher-level processes or regulations must ensure phenomena such temperature or cell number compensation. Axis elongation is likely driven by a jamming transition \cite{Mongera2018} , thus also coupling growth to cell motility. Another related aspect is the coordination of phase gradient with growth and PSM length. The $\alpha$ model explains the apparent scaling of the phase gradient, but it is known that the phase shift between the posterior and the front slowly changes as a function of time, which requires additional regulations. There could still be a slower, active process, explaining dynamical wavelength/scaling of frequency gradient \cite{Ishimatsu2018}.

Another important and related process is the Hox gene system. Hox genes constitute a highly conserved  group of genes, responsible for determining the anteroposterior identities of cells in pretty much all metazoans. Hox gene expression directs the identity of the future vertebrae. Interestingly, the developmental stage where somites form and Hox genes are expressed corresponds to the neck of the so-called 'hourglass model' \cite{Duboule1994}, where all vertebrates embryos are very similar, probably due to strong constraints on the underlying developmental mechanisms \cite{Duboule2022}. Somite boundary formation and embryonic elongation are coupled to Hox gene expressions and to the associated Hox gene 'timer' \cite{Iimura2007, Iimura2009} but many aspects of this coupling remain unclear. Of note, the problem of segment formation in relation to Hox genes had been already studied by Meinhardt \cite{Meinhardt1986}, and more recent computational explorations have indeed suggested some connections between segmentation and local differentiation \cite{tenTusscher2011}.


\chapsec{Theory, learning and Evolution}

 It becomes increasingly clear that many developing systems rely on genetic oscillators for patterning \cite{DeSimone2021, Negrete2020,DiTalia2022} and  a better description and understanding of those other systems might suggest new directions for somitogenesis. Along the same line, new theoretical ideas might also come from an evolutionary developmental ("evo-devo") perspective. As mentioned in the introduction, many other animals  form segments  sequentially,  recapitulating many features described in this review. For instance, arthropods have a segmentation clock, that likely evolved independently from the vertebrate one \cite{Clark2019}. A two-enhancer model proposed for \textit{Tribolium} segmentation \cite{Zhu2017} served as an inspiration for the model proposed in \cite{Jutras-Dube2020}, suggesting that infinite bifurcations might be the core mechanism leading to segmentation in multiple animals. Interestingly, SNIC bifurcations have been suggested to regulate another global developmental oscillator in the nematode \textit{C. elegans} \cite{Meeuse2020}. A generic feature of excitable systems close to SNICs is that they allow for freeze/restarts on the cycle, which might be ubiquitous in many biological oscillators \cite{Hubaud2017, Jia2022}. Other interesting examples include annelids which form segments sequentially but for which a segmentation clock has yet to be discovered \cite{Dray2010,Balavoine2014}, and \textit{Amphioxus}, a segmented animal belonging to a family close to vertebrates \cite{Schubert2001}. Interestingly, Amphioxus somites bud off directly from the tail bud, without the equivalent of the undifferentiated presomitic mesoderm \cite{Schubert2001}, which is in fact reminiscent of the segmentation dynamics in somitoids \cite{Sanaki2022,Yaman2022}. Somitoids morphology is itself very reminiscent of \textit{Tribolium} segments, which might suggest that stem-cells systems leverage deep, evolutionarily conserved mechanisms. There are even examples of similar processes outside of the animal kingdom. The pattern of root branching in plants is regulated by oscillatory genetic expression coupled to growth at root tip \cite{Moreno-Risueno2010}. Similarly,  in  the \textit{Neurospora Crassa} fungus, classical race-tube experiments allow for simple visualization of the coupling of filamentous growth with the periodic sporulation pattern regulated by circadian clocks \cite{Casselton2002}, possibly implementing a clock and switch model.

Further inspiration might come from the current boom in machine learning, in particular reinforcement learning oriented towards biomimetism. We already mentioned how evolutionary simulations selecting for pattern formation spontaneously converge towards a clock and switch model, and, remarkably, follow an evolutionary pathway transparently recapitulating both the (simulated) dynamics and the bifurcation diagram (bistability evolving first, then oscillation). More elaborated simulations are possible, e.g. combining segment formation with hox gene patterning \cite{tenTusscher2011}. Other evolutionary simulations have suggested that a timer could control sequential activation of \textit{Hox} genes, with \textit{Caudal}  as a potential timer candidate \cite{Francois2010}. It is then striking that \textit{Caudal} has been shown to regulate wave propagation in \textit{Tribolium} segmentation clock \cite{Zhu2017}, and similar wave propagation in \textit{Drosophila} has been indeed associated with segment polarity both in theoretical \cite{Rothschild2016} and experimental \cite{Clark2017} work. Is \textit{Caudal}  an ancestral control parameter of the bifurcation, in connection with Hox genes patterning?

With more powerful numerical tools, more complex (evolutionary) scenarios will be explored, and compared to biology. For instance, explorations of models of gene networks with epigenetic modifications already reveal a generic pattern where cellular reprogramming is driven by transient oscillations, subsequently converging to an unstable manifold before transitioning to fixed points \cite{Matsushita2022}. Thus the clock and switch mechanism is universal, and further self-organizes in this example. This is reminiscent of the idea discussed above that the system  dynamics feedback on its control parameter to induce differentiation: here, the epigenetic control appears to play the role of the slow-varying control parameters modulated by the oscillator. Lastly, there are more and more attempts to generate and explore simulated models of artificial life. An example of interest is LENIA, which generalizes Conway's game of life, and is able to generate numerical "creatures", some of them clearly segmented \cite{Chan2018}. It is even possible to use modern learning techniques biased towards innovation to systematically generate and explore numerical models of morphogenesis \cite{Etcheverry2020}. The rules used in such models are very far from actual biology, but one can imagine that such optimization methods with more realistic cellular automata mimicking cellular/developmental interactions  might eventually allow us  to simulate, explore, and mathematically study increasingly realistic models of development, converging towards universal models with direct application to biology.

\chap{Supplementary Materials}

Python Notebooks to generate most simulation figures are available at the following url \url{https://github.com/prfrancois/somitetutorial}. 
Supplementary movie 1 illustrating the landscape changes for the Clock and Wavefront model can be found at \url{https://github.com/prfrancois/somitetutorial/blob/main/Movie_landscape_Clock_and_Wavefront.mp4}, with corresponding code in Notebook  \url{ https://github.com/prfrancois/somitetutorial/blob/main/Zeeman_Cooke.ipynb}

\textbf{Supplementary Movie 1} : A movie illustrating the landscape changes for the Clock and Wavefront model. Anterior is on the top, posterior in the bottom. The state of the system is depicted with a disk. The left colum is the situation without a clock: the system continuously transits from the right well to the left well. The right column is the situation with the clock: the system is periodically pushed towards the left, committing 'blocks' of cells together.

\chap{Acknowledgements}

We would like to thank Alexander Aulehla, Olivier Pourqui\'e, Sharon Amacher, Eric Siggia, David Rand for useful discussions over multiple years. We thank Massimo Vergassola, Ina Sonnen, Usha Kadiyala and one anonymous reviewer for useful comments. This work was funded by a Simons Investigator in Mathematical Modelling of Living Systems award, a NSERC Discovery grant award, and a Fonds Courtois award to PF,  a NSERC CREATE in Complex Dynamics award and a FRQNT PhD fellowship award to VM.

\newpage

\startappendix


\appsec{Some dynamical systems theory for biological oscillators}
\appsubsec{Bifurcations and excitable systems}

In this section, we introduce some general dynamical systems notions associated with oscillators. We refer to \cite{Strogatz2018, Murray2002} for general introductions.
We consider a general dynamical system, modeled by ordinary differential equations similar to Eq. \ref{ODE}. 

\begin{equation}
\frac{ d\mathbf{X}}{dt}=F_\mu(\mathbf{X}) \label{ODE_param}
\end{equation}

Here we have added a $\mu$ dependency, indicating that the ordinary differential equations depend on parameters, e.g. in a biological context, biochemical rates. Now let us assume Eq. \ref{ODE_param} describes an oscillating system, one can wonder what happens to the oscillation when such parameters are varied. This is the essence of the more general bifurcation theory. It is especially relevant in the segmentation context since it is clear that cellular oscillations vary as a function of time and space, so one might hope to concisely describe the behavior of the system using bifurcation theory. Notice that such ideas have been recently rejuvenated in the context of so-called Morse-Smale systems to describe dynamics of differentiation (see e.g \cite{Rand2021, Saez2021}).

\textit{Hopf and Saddle node of cycles}
In particular one can wonder how oscillations disappear when parameters are varied. For many well-known systems, oscillations  disappear via a so-called Hopf bifurcation. Such bifurcations typically occur when the real part of two complex conjugate eigenvalues of the linearized system at a fixed point cancels out (see \cite{Marsden76} for precise mathematical statements and generalization on the Hopf bifurcation theorem). An example of such real part cancellation associated with Hopf bifurcation is given  below for the delayed negative feedback system.

Remarkably, close to the Hopf bifurcation, it is usually possible to find a new (polar) system $(r,\theta)$ of coordinates so that the dynamics of the system can be described by the following equations (essentially identical to Eqs. \ref{Oates_hopf1}-\ref{Oates_hopf2}):

\begin{eqnarray}
 \dot r & = & r (\mu-r^2) \label{Hopf1}\\
    \dot \theta & = & \omega \label{Hopf2}
\end{eqnarray}

Eqs. \ref{Hopf1}-\ref{Hopf2} are the so-called normal form of bifurcation. Those describe universal behaviors observed following cancellation of the real part of the conjugate eigenvalues at a fixed point (one of the mathematical challenges is to show that such forms generalize to higher dimensional systems, which is Hopf's major contribution \cite{Marsden76}).

Bifurcation happens at $\mu=0$. For $\mu>0$, the system is oscillating with frequency $\omega$ and amplitude $r=\sqrt{\mu}$. For $\mu<0$, $r\rightarrow 0$ so that there is no oscillation and the system spirals to the origin which is a stable fixed point. Notice that as $\mu\rightarrow 0$ the oscillation amplitude decreases but the frequency stays the same. This bifurcation is called Hopf supercritical.

Another Hopf bifurcation takes the form :

\begin{eqnarray}
 \dot r & = & -r (\mu-r^2) \label{Hopf1_u}\\
    \dot \theta & = & \omega \label{Hopf2_u}
\end{eqnarray}

This case reverses the stability of the limit cycle and the origin. For $\mu>0$, there is an unstable limit cycle and the origin is stable, for $\mu<0$ the origin is an unstable fixed point. This is called a Hopf subcritical bifurcation. A priori such bifurcations might seem less intuitive and frequent because the cycle is unstable, but in biology, very often some limiting steps will prevent the flow from diverging to infinity and there will be another stable limit cycle around the unstable one.

To see what can happen then, let us add the next $r$ even non-linearity  in Eq. \ref{Hopf1_u} and rescale/shift parameters to get another normal form :

\begin{eqnarray}
 \dot r & = & r (\mu-(r^2-r_c^2)^2) \label{Hopf1_s}\\
    \dot \theta & = & \omega \label{Hopf2_s}
\end{eqnarray}

If $\mu$ is very big, the origin is unstable and there is a stable limit cycle (this can be seen since for big $r$, $\dot r \sim -r^5$ meaning that the flow at infinity is attracted towards a non zero solution). Decreasing $\mu$, as soon as $r_c^4>\mu>0$, the origin becomes stable and there are now two limit cycles, of amplitude $r_\pm^2=r_{c}^2\pm \sqrt{\mu}$. The outer cycle of radius $r_+$ is still stable, and the inner cycle of radius $r_-$ is unstable.
  When $\mu<0$, the two limit cycles cancel out, and we are left only with the stable fixed point at the origin. Such bifurcation is called a saddle bifurcation OF limit cycles, and is another way cycles can disappear. This is our first example of global bifurcation: it is global because it somehow creates entire limit cycles "from scratch", e.g. not through the expansion of a local smaller cycle with a control parameter $\mu$, which is what happens for the Hopf bifurcations. Also, when the bifurcation happens, the frequency of both cycles still is $\omega$.

Notice though that such bifurcation is more "complex" from a physics standpoint, requiring more non-linearities to create both cycles. In biological models, one would see the following sequence of bifurcation as one parameter is varied: starting from a stable fixed point, a supercritical Hopf bifurcation first creates a stable limit cycle, then from the same fixed point a subcritical Hopf bifurcation would create an unstable limit cycle and stabilize again the fixed point, finally the two limit cycles would cancel out to go back to a single stable fixed point. Such a sequence of bifurcations has been proposed to explain how a stable point could co-exist with a limit cycle, e.g. in the context of light-induced oscillation death for circadian clocks \cite{Leloup2001}.

\textit{Infinite period bifurcations}

Another example of global bifurcations can be built, again elaborating on Eqs. \ref{Hopf1}-\ref{Hopf2} :

\begin{eqnarray}
 \dot r & = & r (1-r^2) \label{SNIC1}\\
    \dot \theta & = & \mu+1+\cos\theta \label{SNIC2}
\end{eqnarray}

Here, the control parameter is included in the polar angle equation.  When $\mu>0$, $\dot \theta>0$ and we have a limit cycle of radius $1$. But when $\mu=0$, the point of the cycle where $\theta=\pi$ (corresponding to $(x,y)=(1,-1)$ in cartesian coordinates) is becoming a fixed point, and for $\mu<0$, there is a pair of stable/unstable fixed points at $r=1$ and $\theta=cos^{-1}\left(-(\mu+1)\right)$. So, intuitively, there is a pair of unstable/stable fixed points appearing IN the cycle. This bifurcation is called a Saddle Node on Invariant Cycle, or SNIC.  Again, this is a global bifurcation because when $\mu$ goes from negative to positive values, a limit cycle with amplitude $1$ appears "from scratch".

An important property of SNICs is that the period diverges when $\mu\rightarrow 0^+$ (because $\dot \theta \rightarrow 0$). For this reason, this bifurcation is sometimes called Saddle Node with Infinite Period, or SNIPER \cite{Strogatz2018}. Also, because of this divergence, when $\mu>0$ but stays small, the cycle spends much time close to the region where $\theta=\pi$ where the fixed point appears at $\mu=0$. For this reason, one often talks about a "ghost" fixed point, where the cycle spends most of the time, before doing a big, rapid excursion on the limit cycle $r=1$ to come back. Such ghost fixed points associated with saddles might in fact be observed in developmental contexts more general than somitogenesis: slowing down the flow might provide a way to build biological timers \cite{Tufcea2015,Negrete2021}.
Being more quantitative, one can directly integrate Eq.\ref{SNIC2} to get :
\begin{equation}
    \theta(t)=2 \arctan \left( \sqrt{\frac{\mu+2}{\mu}}\tan \frac{\sqrt{\mu(\mu+2)}}{2} t\right) \label{eq:solution_SNIC}
\end{equation}

From this expression, it is clear that $\theta$ is periodic with period $\frac{2\pi}{\sqrt{\mu(\mu+2)}}$, which indeed diverges for $\mu \rightarrow 0$. Also, for any  value of $t$ far from $\frac{ n \pi}{\sqrt{\mu(\mu+2)}}$, the term inside the $\arctan$ is of the order of $\pm 1/\sqrt{\mu}$, meaning that $\theta \sim \pi$, so that the cycle spends most of the time there. When $t$ is a multiple of $\frac{\pi}{\sqrt{\mu(\mu+2)}}$, the $\tan$ terms diverge and $\theta$ cycles around the circle in a time of order $1$.

For this reason, it is useful to focus on the slow region,  calling $x=\theta-\pi$, and Taylor expanding the system close to $\pi$ (for small $x$), we get after some standard manipulation and rescaling  (to get rid of a  $2$ factor) :
\begin{equation}
\dot x =  \mu+x^2 \label{QFI}
\end{equation}

This equation is of particular interest in neuroscience, where it is called the "quadratic integrate and fire" model \cite{Izhikevitch2007}. There, it is assumed that when $x$ passes a threshold  $x_{p}$, the system quickly resets to a lower value $x_{r}$, which is phenomenologically identical to the rapid excursion on the limit cycle $r=1$ to escape then go back close to the "ghost'' fixed point. This gives a very simple example of a model oscillating with "spikes", since Eq. \ref{QFI} diverges in finite time. In fact one can even take $x_p = +\infty$ and $x_r=-\infty$ so that the solution for Eq.\ref{QFI} simply is 

\begin{equation}
x(t)=\sqrt{\mu} \tan \sqrt{\mu} t \label{xSNIC}
\end{equation}

 which oscillates between $\pm \infty$ with period $\pi/\sqrt{\mu}$, so very similar to the full solution \ref{eq:solution_SNIC}. Again for $\mu \rightarrow 0$ one clearly recovers an infinite period as expected from a SNIC.

There is a last case relevant to our discussion. Consider a system with two attractors, a limit cycle and a fixed point, with a saddle between them. When parameters are varied, the saddle and the cycle might coalesce, such as the unstable direction of the saddle cycles back into its stable direction. An example is represented in Fig. \ref{fig:Homo}. Such bifurcation is called a homoclinic bifurcation and is another type of infinite period bifurcation. Similar to the SNIC, a quadratic integrate and fire model of a homoclinic bifurcation can be built \cite{Izhikevitch2007} : 
\begin{equation}
\dot x =  -1+x^2 \label{QFIH}
\end{equation}

The stable fixed point is at $x=-1$, and the unstable fixed point is at $x=1$. We will assume that when $x$ goes to $\infty$, it resets at $x_r=(1+ \mu)$. $\mu$ plays the role of the control parameter, since when $\mu=0$ the oscillator resets on the saddle (unstable fixed point), giving a homoclinic orbit.
The solution of Eq. \ref{QFIH} for $x>1$ is 
\begin{equation}
x(t)=-\coth(t-t_0) \label{xHomoc}
\end{equation}
where $\coth u=\frac{\cosh(u)}{\sinh(u)}$, the ratio of hyperbolic cosine  and sine. $t_0$ is defined by $x(0)=x_r=(1+ \mu)$, which gives 
\begin{equation}
t_0=\frac{1}{2}\ln \frac{\mu+2}{\mu}
\end{equation}
The period is found by noticing that $x(t)$ diverges when $\sinh(t-t_0)=0$, i.e. $t=t_0$, so that $t_0$ is in fact the period. As $\mu\rightarrow 0$, it diverges logarithmically.

In Table \ref{TableBif} we summarize the standard behavior of those different bifurcations as a function of the control parameter.

\begin{table}
\begin{center}
\begin{tabular}{|c| c| c|}
\hline
Bifurcation & Period & Amplitude \\
\hline
 Hopf supercritical & roughly constant & $\sqrt{\mu}$ \\
 Saddle node of cycle & roughly constant &   roughly constant\\
 SNIC & $1/\sqrt{\mu}$ & roughly constant ($x_p$)\\
 Homoclinic & $-\ln \mu$ & roughly constant  ($x_p$)\\
 \hline
\end{tabular}
\end{center}
\caption{Standard phenomenology for limit cycle disappearance as a function of the control parameter $\mu$}\label{TableBif}
\end{table}

Anticipating the section on phase response, it is useful to consider now what happens when 'integrate and fire' oscillators are perturbed. Since those models are essentially one-dimensional, perturbations can only increase or decrease $x$. In both SNIC and homoclinic cases (Eqs. \ref{QFI}-\ref{QFIH}), $x$ is a monotonically increasing function of $t$ (before the reset), which means that a perturbation of a given sign (say an increase of $x$) invariably leads to a perturbation of the phase of the oscillator of a fixed sign (e.g. always a phase advance for an increase of $x$), or, in technical terms, a constant sign PRC (see section \myref{sectionPRC}). This is an unusual situation, since in most oscillators (close to a Hopf bifurcation), the phase response is typically a sinusoidal function of the phase, so a given perturbation can lead to phase advances or delays depending on when the perturbation is imposed.
In both SNIC and homoclinic cases, one can easily compute the infinitesimal phase response corresponding to a small change of $u$. Phase is defined by $d\phi/dt=\omega$, so one can simply write in both cases:
\begin{equation}
\frac{d\phi}{dx}=\frac{d\phi}{dt}\times \frac{dt}{dx}=\frac{\omega}{\dot x}
\end{equation}
For SNIC, the period is $\pi/\sqrt{\mu}$, i.e. $\omega=\frac{2\pi}{T}=2\sqrt{\mu}$, and adding a $\sqrt{\mu}$ factor so that the infinitesimal perturbation $dx$ scaled with $x$ in the limit $\mu\rightarrow 0$, we get 

\begin{equation}
\sqrt{\mu} \frac{d\phi}{dx}=\sqrt{\mu}\frac{\omega}{\dot x}=\frac{2\mu}{\mu+\mu \tan^2 (\sqrt{\mu} t)}=2\sin^2 (\phi/2)
\end{equation}
defining the phase $\phi=0$ for $t=0$, and the fact that $\phi=\omega t=2\sqrt{\mu}t$.

For homoclinic, we get

\begin{equation}
\frac{d\phi}{dx}=\frac{\omega}{\dot x}=\frac{\omega}{-1+\coth^2(t-t_0)}=\omega \sinh^2{\frac{t_0}{2\pi}(\phi-2\pi)}
\end{equation}
defining the resetting phase $\phi=0$ for $t=0$, and the fact that $\phi=\omega t=\frac{2 \pi t}{t_0}$

Notice that both phase responses are of constant sign (positive) as expected, but have very different behaviors, in particular, the phase response of the homoclinic bifurcation is discontinuous at the resetting point (which makes intuitive sense since the flow moves from the stable to the unstable branch of the saddle there).

\textit{Connecting bifurcations}
There are general mathematical results for how oscillatory systems change as one parameter is varied \cite{Yorke78, Keener81}. Quoting Keener \cite{Keener81}, the general theorem follows :
\begin{quote}
    If a vector field $X_\mu$ has a closed orbit $\Gamma_\mu$, then as $\mu$ changes either
    \begin{itemize}
        \item $\Gamma_\mu$ remains a closed orbit
        \item the period of $\Gamma_\mu$ becomes infinite; or
        \item $\Gamma_\mu$ shrinks to a fixed point
    \end{itemize}
\end{quote}

The two latter situations respectively correspond to infinite period and Hopf bifurcations. 
Concretely, this means, that one would expect for instance oscillators born via a Hopf bifurcation to die either via a Hopf bifurcation or an infinite period bifurcation. Notice however the above theorem does not apply to the saddle-node of cycles because there are two closed orbits.

 \textit{Excitability, Type I and Type II oscillators }
 
 The Van Der Pol oscillator is ideal to understand the notions of excitability, with the associated sub-categories of Type I and Type II oscillators. To illustrate this, let us consider two slightly modified types of Van Der Pol oscillators. 
 
 The first one is similar to the Meinhardt-VanDerPol one introduced in the main text  and is of Type II (see below)
\begin{eqnarray}
\epsilon \dot x & = & x-x^3/3 -y  \label{eq:VdP_typeIIx}\\
\dot y & =& \lambda(x-\mu y +h)  \label{eq:VdP_typeIIy} 
\end{eqnarray}
 The behavior of this model is similar to the FitzHugh-Nagumo model used in \cite{Hubaud2017}, mentioned in the main text.
 
We contrast this model with a saturating nullcline/activation function for $y$  (which is of Type I, see below):
 
\begin{eqnarray}
\epsilon \dot x & = & x-x^3/3 -y  \label{eq:VdP_typeIx}\\
\dot y & =& h+\lambda(\tanh(\mu x)-y) \label{eq:VdP_typeIy} 
\end{eqnarray}
 We argue that the latter system is more "biological" because of the sigmoidal activation of $y$ by $x$

We have introduced in both cases a control parameter $h$. The behaviors of both models are illustrated in Figs. \ref{fig:TypeI}-\ref{fig:TypeII}. When the control parameter is small enough, both models present very similar, relaxation type oscillations, as expected.

To contrast Types I and II, it is more convenient to start describing the second model, with saturating activation, corresponding to Eqs. \ref{eq:VdP_typeIx}-\ref{eq:VdP_typeIy}, Fig.\ref{fig:TypeI}. As $h$ increases, the $y$ nullcline (green on Fig. \ref{fig:TypeI}) moves up and gets closer and closer  to the local minimum of the $x$ nullcline. As a consequence, the frequency of the oscillation decreases, and the oscillation becomes more and more pulsatile, with a constant amplitude. This leads eventually to a SNIC bifurcation when the null-clines cross. At the bifurcation, we know the frequency is $0$ (and the period infinite). Below the bifurcation, if the system is initialized just on the right of the fixed point, the system goes through one "pulse" of the oscillation before relaxing to the fixed point, which is the hallmark of excitability. If we were to decrease $h$ from there (i.e. now from right to left in Fig. \ref{fig:TypeI}), the system oscillates with $0$ frequency (infinite period), then the period will decrease. An excitable system becoming oscillatory with a $0$ frequency is called Type I in neuroscience \cite{Izhikevitch2007}, and is clearly characterized here by a SNIC.

\bnormf
\includegraphics[width=\textwidth]{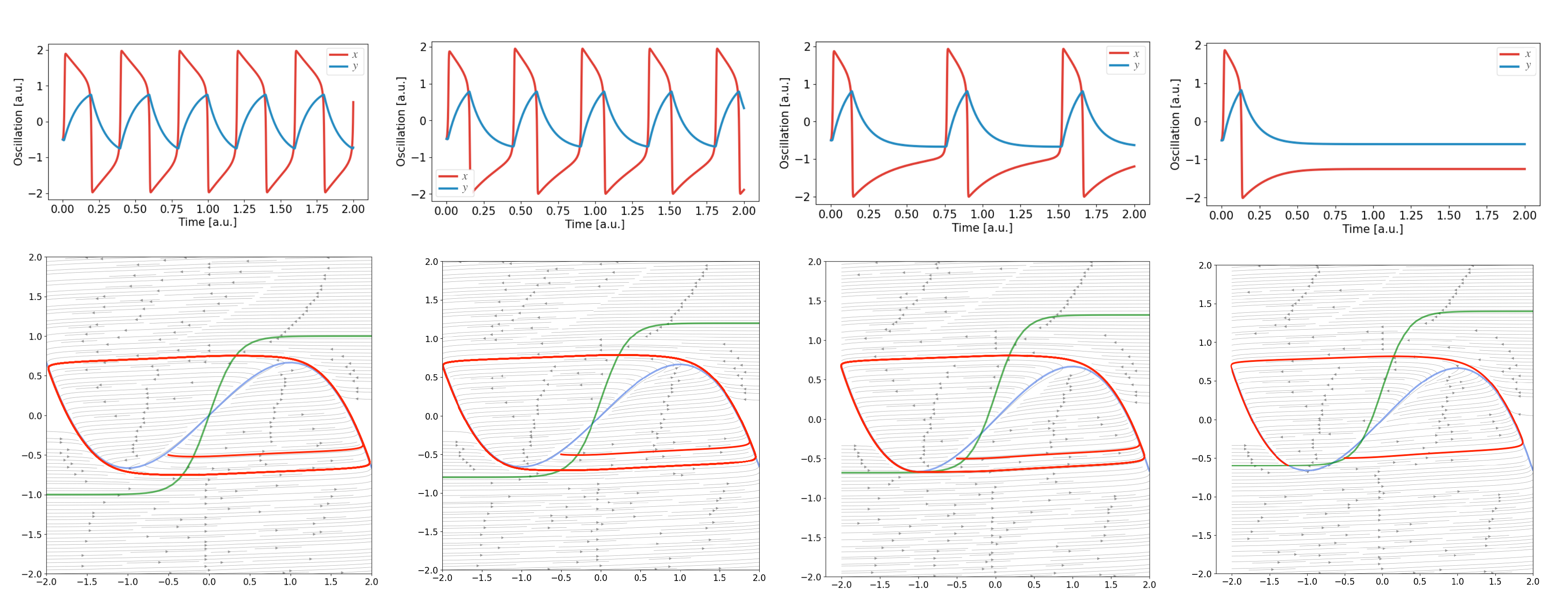}
\caption[Type I excitability] { Oscillator going through a SNIC with Type I excitability, corresponding to Eqs. \ref{eq:VdP_typeIx}-\ref{eq:VdP_typeIy}. Parameters used are $\lambda=0.1,\mu=3, \epsilon=1/3$. The top row represents the behavior of the system with an initial condition on the right of the local minima of the $x$ nullcline. The bottom row shows the 2D flow. From left to right, bifurcation parameter $h$ takes value $h=0,0.02,0.032,0.04$, leading to a SNIC bifurcation. $y$ null-cline is in green, $x$ null-cline is in blue. } \label{fig:TypeI}
\enormf

For the other model without saturation, Fig. \ref{fig:TypeII}, as $h$ increases, the $y$ null-cline is now moving towards the left. Initially, the behavior is qualitatively similar to the Type I oscillator, with an increase in the period, and a conserved amplitude. But then, as the system gets very close to the bifurcation the limit cycle suddenly collapses around the fixed point  (Fig. \ref{fig:TypeII}, 3rd column), becoming more sinusoidal and leading to a Hopf bifurcation. Below the bifurcation, just like Type I,  if the system is initialized just on the right of the fixed point, the system goes through one "pulse" of the oscillation before relaxing to the fixed point, thus again showing excitability. Again, if  we were to decrease $h$ from there (i.e. now from right to left in Fig. \ref{fig:TypeII}), the system starts oscillating again, but with a non $0$ frequency (finite period), because it is a Hopf bifurcation. An excitable system becoming oscillatory with a non $0$ frequency is called Type II in neuroscience \cite{Izhikevitch2007}, and is thus characterized by a Hopf bifurcation. Notice here that if the system now goes to even smaller $h$, there is a sudden increase in both period and amplitude of the oscillation. This is called a "Canard explosion" \cite{Murray2002,Wechselberger2007}. We notice in particular that in those models,  Type I and Type II oscillators become qualitatively different only  very close to the bifurcation. In both cases, when the control parameter $h$ is reduced, we first see an increase of period with fixed amplitude, thus closer to the generic SNIC/Type I scenario. While it is possible to get a smoother transition for Type II oscillators without Canard explosions close to the bifurcations in other types of models (see e.g. \cite{Izhikevitch2007} Figure 4.33), the oscillations would then stay rather sinusoidal for a broader range (as expected from a Hopf bifurcation) and it is not clear if below the bifurcation a perturbation would give a pulsatile response.

\bnormf
\includegraphics[width=\textwidth]{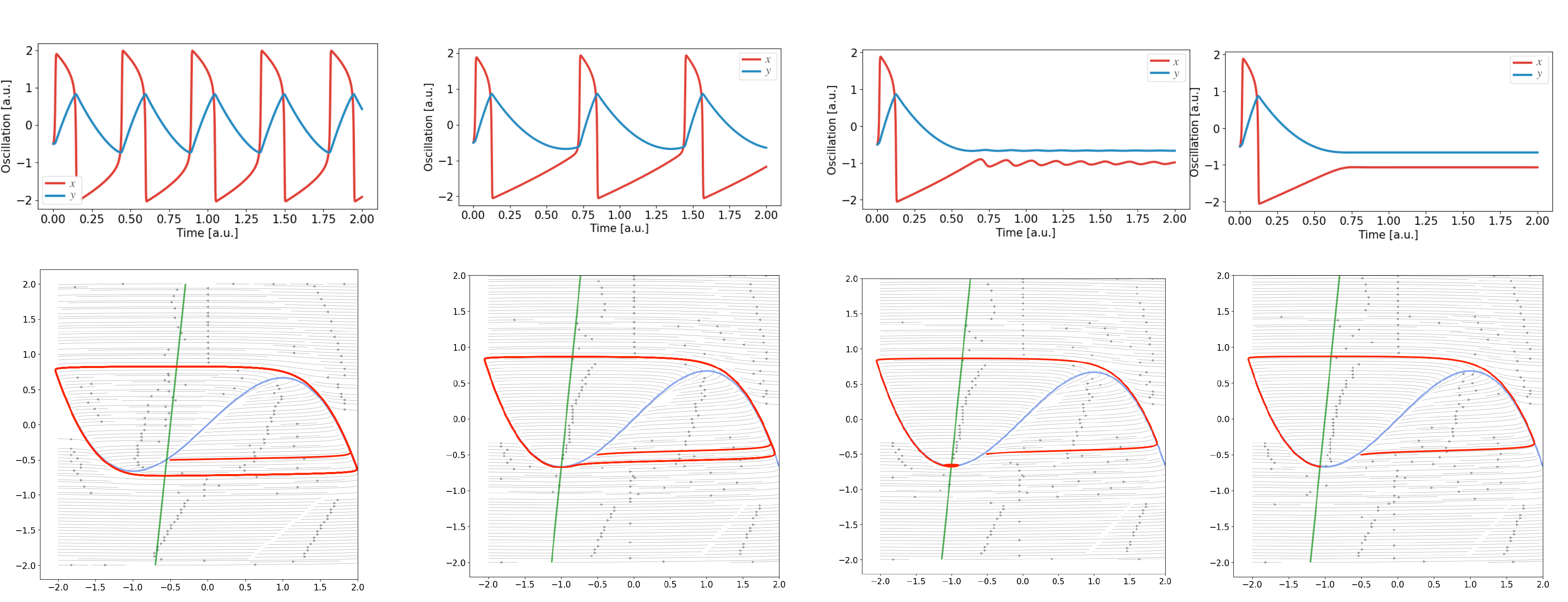}
\caption[Type II excitability]{ Oscillator going through a Hopf with Type II excitability, corresponding to Eqs. \ref{eq:VdP_typeIIx}-\ref{eq:VdP_typeIIy}.The top row represents the behavior of the system with an initial condition on the right of the local minima of the $x$ nullcline. The bottom row shows the 2D flow Parameters used are $\lambda=0.05,\mu=0.1, \epsilon=1/3$. From left to right, bifurcation parameter $h$ takes value $h=0.5,0.93,0.935,1$ suddenly leading to a Hopf bifurcation.$y$ null-cline is in green, $x$ null-cline is in blue.  } \label{fig:TypeII}
\enormf

\appsubsec{Perturbing the phase}

\textit{Phase Response Curves}
\label{sectionPRC}
In this section, we introduce the general framework to study oscillator synchronization using a phase-based formalism, which allows in particular to derive many results related to entrainment.

Let us consider some arbitrary initial phase $\phi_{init}$ (corresponding to a position $\mathbf{X}_{init}$ on the limit cycle) and let us exactly perform a change $\mathbf{X}_{init}\rightarrow \mathbf{X}_{init} +  \Delta \mathbf{X}$. By definition, we then reach a new phase  $\phi_{new}$.
We define
\begin{equation}
\phi_{new}= PTC(\phi_{init},\Delta \mathbf{X})=\phi_{init} + PRC(\phi_{init},\Delta \mathbf{X}) \quad \mathrm{mod}  \quad 2 \pi \label{PC}
\end{equation}

Eq \ref{PC} introduces two new functions \cite{Winfree}: the Phase Transition Curve (PTC) and the Phase Response Curve (PRC). Both quantities depend on the initial phase $\phi_{init}$ and on the perturbation $\Delta \mathbf{X}$ (notice that while we discuss phase changes, the perturbation still is defined in the initial space of protein concentrations).  Intuitively, the Phase Transition Curve simply is the phase change induced by the perturbation $\Delta \mathbf{X}$, while the Phase Response Curve is the phase difference induced by the same perturbation. Importantly though, knowledge of either the PTC or the PRC ensures that we can fully describe the behaviour of the system to any perturbation. This is a considerable reduction in complexity since we no longer have to consider the full system of differential equations defining the dynamics of  $\mathbf{X}$ but can focus on the study of PRC and of the PTC.

Further simplifications occur if the perturbation $\Delta \mathbf{X}$ is small. In such a situation, we expect the PRC to be (at most) linear in $\Delta \mathbf{X}$. So we can define the linear response or sensitivity function 

\begin{equation}
PRC(\phi,\Delta \mathbf{X}) \simeq \mathbf{Z}(\phi) \Delta \mathbf{X}  \label{PRC_Kuramoto}
\end{equation}

Now imagine that $\Delta \mathbf{X}$ is no longer a single perturbation, but rather a train of (small) pulses at different times $t_n$, so that $\Delta \mathbf{X}=\epsilon h \mathbf{p}(t_n)$, where $h=t_{n+1}-t_n$ is small, and $p$ some arbitrary time-varying function. Assuming again that the system stays close to the limit cycle, we thus get that
\bwt
\begin{equation}
\phi(t_{n+1})=h+ PTC(\phi(t_n),\Delta \mathbf{X})=(h+\phi(t_n)+\epsilon \mathbf{Z}(\phi(t_n))  h \mathbf{p}(t_n) )\quad \mathrm{mod}  \quad 2 \pi
\end{equation}
\ewt
Taking the difference and the limit $h\rightarrow 0$ we then get

\begin{equation}
\frac{d\phi}{dt} = 1+\epsilon \mathbf{Z}(\phi) \mathbf{p}(t) 
\end{equation}

This equation will be the basis for all phase models and expresses the response of the oscillator to an arbitrary signal $p(t)$. Such an equation is valid if  the perturbation is small enough to stay in the linear regime, close to the limit cycle, so that the response of the oscillator is well approximated by $ \mathbf{Z}(\phi) $.

An important particular case is found when the perturbation is itself periodic, with frequency $\omega =1 - \epsilon \Delta  $ with small $\epsilon$, so that one can write $\mathbf{p}(t)=Q(\phi,\omega t)$, with $Q$ $2\pi$-periodic in both variables (we keep a $\phi$ dependency to indicate that the perturbation might also depend on the location on the limit cycle). 

 In that case, the standard approach is to define 
 
 \begin{equation}
 \psi=\phi -\omega t \label{psientrainment}
 \end{equation}

Notice that $\psi$ is the difference between the phases of the current oscillator and of the perturbation, and it is completely equivalent mathematically to study the dynamics of $\psi$ instead of $\phi$. This approximation also is very similar in spirit to the Ansatz of Equation \ref{Ansatz}: in both cases, we derive equations for the difference between the phase and a "reference" oscillator (the tail bud oscillator $t$ in Eq. \ref{Ansatz} and the perturbing oscillator $\omega t$ here). Like before, focusing on $\psi$ allows us to derive a few simplifying assumptions leading to a closed form and very intuitive results.

We  get by substitution 
\begin{equation}
\frac{d \psi }{dt} = \epsilon( \mathbf{Z}(\psi+ \omega t) Q(\psi+\omega t,\omega t) +\Delta)
\end{equation}

While it does not seem like a big simplification at first, this equation shows that $\psi$ varies slowly (with a time scale of order $1/\epsilon$). Furthermore, assuming $\psi$ is almost constant, all (faster) time dependencies on the right-hand side of this equation are periodic, with period $T=\frac{2 \pi}{\omega}$. We can thus safely average over one period to get the averaged change of $\Psi$

\begin{equation}
\frac{d \psi }{dt} = \epsilon \Delta + \Gamma (\psi)  \label{Psi}
\end{equation}

with 

\begin{equation}
\Gamma (\psi) =\epsilon \frac{1}{T} \int_0^T \mathbf{Z}(\psi+ \omega t) Q(\psi+\omega t,\omega t) \mathrm{d}t \label{Gamma}
\end{equation}
which is well a defined quantity since all functions in the integral are periodic with period $T$.

Eq. \ref{Psi} gives a closed form, allowing us to fully predict the time evolution of the phase as a function of the frequency mismatch ($\epsilon \Delta$) and of $\Gamma$. There are several cases depending on the value of $\Delta$ and $\Gamma$, but as an important particular case, steady state (if it exists) satisfies
\begin{equation}
\epsilon \Delta + \Gamma (\psi^*)=0  
\end{equation}

This corresponds to a constant phase shift $\psi^*$ between the oscillator and the external signal of period $T$. In that case, the oscillator is entrained by the external signal, and phase-shifted by $\psi^*$. Notice in particular that the phase shift depends on $\epsilon \Delta$, the frequency mismatch between the oscillator and the external periodic signal.

Finally, it can also be useful to come back to the initial phase $\phi$ to get

\begin{equation}
\frac{d\phi}{dt} = 1+\Gamma(\phi-\omega t)
\end{equation}

showing explicitly how the time evolution of the phase of the oscillator simply depends on the phase \textit{difference} between the current phase and the one of the external periodic perturbation. Notice, however, that this equation has been obtained through averaging equation \ref{Gamma} so, strictly speaking, is only valid after a few periods of the periodic perturbation.

This approach can be generalized to the coupling of two or several oscillators. For instance, considering the coupling of two identical oscillators  with slightly different frequencies, it is not difficult to see that one would get equations of the form

\begin{eqnarray}
\frac{d\phi_1}{dt} &=& \omega_1+\Gamma(\phi_1-\phi_2) \label{phi1}\\ 
\frac{d\phi_2}{dt} &=& \omega_2+\Gamma(\phi_2-\phi_1)\label{phi2}
\end{eqnarray}

which allows to solve for the phase difference between both oscillators 

\begin{eqnarray}
\frac{d(\phi_1-\phi_2)}{dt} &=& \omega_1-\omega_2+\Gamma(\phi_1-\phi_2)-\Gamma(\phi_2-\phi_1) \label{coupling}
\end{eqnarray}

Solving such equations requires the knowledge of the function $\Gamma$. In the context of two coupled oscillators, considering oscillator $1$ to fix ideas, this function combines in a rather elaborate way both the linear response of the oscillator $ \mathbf{Z}(\phi_1) $ and the influence of the perturbation $Q(\phi_1,\phi_2)$. Clearly both functions will strongly depend on the details of the model. Let us mention several classical examples:
\begin{itemize}
\item Pulse coupling: imagine that  $Q(\phi_1,\phi_2)=A\delta (\phi_2) \mathbf{x}$, where $\delta$ is the Dirac function, meaning that, without loss of generality, oscillator $2$ "kicks" oscillator 1 in a constant direction (unit vector $\mathbf{x}$) every time it reaches phase $0$. Then using equation \ref{Gamma} with $\Psi=\phi_1-\phi_2$ and $\phi_2=\omega t$ we immediately get

 \begin{equation}
\Gamma(\phi_1-\phi_2)= A \frac{\epsilon}{T} Z_{x}(\phi_1-\phi_2)
 \end{equation}

The main interest of this equation is to show that for pulse coupling, the function $\Gamma$ coincides with the linear response in the direction of the perturbation.
\item  since $\Gamma$ is periodic, a very standard approximation is to consider up to the first Fourier mode of $\Gamma$, i.e. 

\begin{equation}
\Gamma(\phi_1-\phi_2)=\epsilon_{12}+\lambda \sin(\phi_2-\phi_1+\Psi_{12})
\end{equation}

An even simpler form for oscillators close to Hopf bifurcation is the familiar Kuramoto coupling:

\begin{equation}
\Gamma(\phi_1-\phi_2)=\lambda \sin(\phi_2-\phi_1)
\end{equation}

with $\lambda>0$. In that case, notice that if the mismatch $\omega_1-\omega_2=0$, both oscillators end up in phase.
\item more non-linear models can be observed, in particular in neuroscience. For instance, for  two oscillators close to a SNIC bifurcation with pulse coupling one gets :

\begin{equation}
\Gamma(\phi_1-\phi_2)=\lambda \sin^2(\phi_2-\phi_1)
\end{equation}

\end{itemize}

\textit{Phase Evolution towards entrainment and fixed point of Return Map for a given detuning}

Assuming a given phase response curve, one can  compute the instantaneous phase $\phi_{new}$ of a system right after a given perturbation of size $\epsilon$ (see \cite{Winfree} and Eq. \ref{PC} in Appendix)
 
 \begin{equation}
     \phi_{new}= \phi_{init} + PRC(\phi_{init},\epsilon) \quad \mathrm{mod}  2 \pi
 \end{equation}
where we use the defined the Phase Response Curve of the system in response to a perturbation of size $\epsilon$. One can then classically consider the return map defined by the periodic perturbation \cite{Glass2001}, and generalize the definition of phase response \cite{Izhikevitch2007} to get an equation relating the phase of the oscillators at the beginning of two consecutive perturbations (indexed by $n$) :

\begin{equation}
    \phi_{n+1}=\left(\phi_n+PRC(\phi_n,\epsilon)+2\pi \frac{T_{p}}{T}\right)\quad \mathrm{mod} \quad 2 \pi \label{eq:map}
\end{equation}

where $T_p$ is the period of the perturbations and $T$ the natural period of the free-running oscillator.

Entrainment is achieved when this map converges to a fixed point $\phi_{n+1}=\phi_n=\phi_*$, which gives immediately :

\begin{equation}
PRC(\phi_*,\epsilon)=-2\pi \frac{T_{p}}{T} \mathrm{mod} \quad 2 \pi = 2\pi \frac{T-T_{p}}{T} \mathrm{mod} \quad 2 \pi
\end{equation}

This simply relates the phase of entrainment $\phi_*$ at the time of perturbation to both periods of both the perturbation and the intrinsic oscillator (notice we expressed this as a function of the detuning  $T-T_{p}$ which is a standard convention of the field). This is of interest because for symmetry reasons we a priori expect that there always exists a phase $\phi_*$ such that $PRC(\phi_*,\epsilon)$ for any $\epsilon$ : this means that as soon as $T=T_p$, one should be able to entrain the system with any perturbation. The curves $T,\epsilon$ delineating entrainment regions have characteristic elongated shapes and are called `Arnold tongues' \cite{Glass2001,Granada2013}

\appsubsec{Two demonstrations of Malkin theorem}

 This theorem nicely shows that the infinitesimal phase response curve $\mathbf{Z}$ in fact is the solution of the adjoint equation on the limit cycle, i.e. :


\begin{equation} \frac{d \mathbf{Z}}{dt}= - \frac{\partial F(\mathbf{X}(t))}{\partial {\mathbf{X}}}^{\dagger}   \mathbf{Z} \label{Malkin} \end{equation}

where the derivative is computed on the limit cycle $\mathbf{X}(t)$ and $\dagger$ is the transpose operation.


This theorem is of practical importance since it gives a simple way to compute the phase response curve for any limit cycle: one just needs to integrate the equations on the limit cycle, then compute $\frac{\partial F(\mathbf{X}(t))}{\partial {\mathbf{X}}}$ and from there integrate equation \ref{Malkin} to get $ \mathbf{Z}$ (as pointed out by Izhikevich, it is useful to integrate this equation backward in time, to account for the $-$ sign).

A rigorous demonstration of the theorem can be found in \cite{Hoppensteadt2012} but is not very intuitive so here we provide two alternative demonstrations.

\textit{Intuitive demonstration}
The first demonstration is the less rigorous one but the most intuitive and follows the simpler treatment of the adjoint method in the context of Neural ODE derivations from the Appendix of \cite{Chen2018}. Let us discretize time, and consider a sequence of times $t_1, t_2, \dots, t_N$ such that $t_{i+1}=t_i+\epsilon$. We consider the phase at the end of the sequence $\theta(t_N)$ (which is possible via the definition of isochrons).One can define the sensitivity of the phase $\theta(t_N)$ to an infinitesimal perturbation made in the past  $t_i$
\begin{equation}
    \mathbf{Z}(t_i)=\frac{\partial \theta(t_N)}{\partial{\mathbf{X(t_i)}}}
\end{equation}    
which by definition corresponds to the infinitesimal phase response curve such  that  $d\theta_N= \frac{\partial \theta(t_N)}{\partial{\mathbf{X(t_i)}}}. dx(t_i)= \mathbf{Z}. dx(t_i)$. 

Notice there are many underlying assumptions here: we can properly define a phase following perturbations outside of the limit cycle (e.g. via isochrons), each of those perturbations can be then treated independently, the perturbations keep you close to the limit cycle etc... All those assumptions need to be well justified mathematically, but make perfect sense in the limit of infinitesimal perturbation.

From there, $\mathbf{Z}(t_i)$ can be connected to one another: using the chain-rule, and making explicit the partial derivatives with respect to the components $X^k(t_{i+1})$ of $\mathbf{X} (t_{i+1})$ we have

\begin{equation}
\frac{\partial \theta(t_N)}{\partial{X^k(t_{i+1})}}=\sum_l \frac{\partial X^l(t_i)}{\partial X^k(t_{i+1})} \frac{\partial \theta(t_N)}{\partial X^l(t_{i})}
\end{equation}

or in a more compact way

\begin{equation}
\mathbf{Z}(t_{i+1})=\left(\frac{\partial \mathbf{X}(t_i)}{\partial \mathbf{X}(t_{i+1})}\right)^\dagger \mathbf{Z}(t_{i}) \label{Z_step}
\end{equation}

We now have from Eq. \ref{ODE_param} at lowest order (going backward in time)
\begin{equation}
\mathbf{X}(t_i)=\mathbf{X}(t_{i+1})-\epsilon F(\mathbf{X}(t_{i+1}))
\end{equation}  

so that
\begin{equation}
\frac{\partial \mathbf{X}(t_i)}{\partial \mathbf{X}(t_{i+1})}\simeq 1-\epsilon \frac{\partial F (\mathbf{X}(t_{i+1}))}{\partial \mathbf{X}} \label{X_step}
\end{equation} 

Combining Eq. \ref{Z_step}-\ref{X_step} we thus have at lowest order in $\epsilon$:
\bwt
\begin{eqnarray}
\mathbf{Z}(t_{i+1}) &=&\left(1-\epsilon \frac{\partial F (\mathbf{X}(t_{i+1}))}{\partial \mathbf{X}}^{\dagger}\right)\mathbf{Z}(t_{i}) = \mathbf{Z}(t_{i})- \epsilon \frac{\partial F (\mathbf{X}(t_{i+1}))}{\partial \mathbf{X}}^\dagger \mathbf{Z}(t_{i})  \\
 & = & \mathbf{Z}(t_{i})+\epsilon \frac{d \mathbf{Z}(t_{i})}{dt}
\end{eqnarray} 
\ewt
Taking the limit $\epsilon=0$  gives Eq. \ref{Malkin} by direct identification.

\textit{Multiscale, algebraic demonstration}

The second demonstration is adapted from \cite{Ermentrout2019} and is a simplified, easier-to-follow version of the demonstration of \cite{Hoppensteadt2012}. 
We start again with the differential equation

\begin{equation}
\frac{d \mathbf{X}}{dt}=F(\mathbf{X}) \label{cycle_Ermentrout}
\end{equation}

and let us call $\mathbf{U}(t)$ the periodic solution (limit cycle) of this equation.

Let us consider a small perturbation $x$ of the limit cycle, its time evolution is given by the linearized equation

\begin{equation}
\frac{d \mathbf{x}}{dt}=\frac{\partial F(\mathbf{U}(t))}{\partial \mathbf{X}} \mathbf{x} \label{Floquet}
\end{equation}

We know there is one solution to this Equation: the tangent to the cycle $\mathbf{U}'(t)$. In particular, we have

\begin{equation}
L \mathbf{U}'(t)=0
\end{equation}

where we have defined the linear operator $L=\frac{d}{dt}-\frac{\partial F(\mathbf{U}(t))}{\partial \mathbf{X}}$. From Floquet theory \cite{Kuramoto}, $\mathbf{U}'$ is the unique eigenvector associated with eigenvalue $0$, and all other eigenvalues are negative.

Let us now define a dot product between two vectorial functions, calling $T$ the period of the cycle :

\begin{equation}
<u,v>=\frac{1}{T}\int_0^T u(t)^\dagger.v(t) \mathrm{d}t
\end{equation}
where $\dagger$ indicates the transpose. We define the Adjoint operator to $L^*$ such that

\begin{equation}
<u,Lv>=<L^* u,v>
\end{equation}

By integration by part, it is clear that
\begin{equation}
L^*=-\frac{d}{dt}-\frac{\partial F(\mathbf{U}(t))}{\partial  \mathbf{X}}^\dagger
\end{equation}

Since $L$ has a unique $0$ eigenvector, $L^*$ also has a unique  $0$ eigenvector, called $\mathbf{Z}$ and by definition, we thus have $L^*\mathbf{Z}=0$, i.e. 

\begin{equation}
\frac{d\mathbf{Z}}{dt}=-\frac{\partial F(\mathbf{U}(t))}{\partial \mathbf{X}}^\dagger. \mathbf{Z}
\end{equation}
so that $\mathbf{Z}$ is the solution to the adjoint equation. We normalize $\mathbf{Z}$ such that $<\mathbf{Z},\mathbf{U}'>=1$.
We now show that perturbations to the cycle can be simply computed with the help of $\mathbf{Z}$. To proceed, let us consider a perturbation of the limit cycle, in the form of

\begin{equation}
\frac{d \mathbf{X}}{dt}=F(\mathbf{X})+\epsilon \mathbf{G}(t) \label{perturbation}
\end{equation}

To study the modified behavior of $\mathbf{X}$ we proceed with a  multiscale analysis \cite{BenderOrzsag}, defining $s=t$ and a slow variable $\tau=\epsilon s$. We then expand $\mathbf{X}$ in powers of $\epsilon$ :

\begin{equation}
\mathbf{X}(t)= \mathbf{X}_0(s,\tau)+\epsilon  \mathbf{X}_1(s,\tau)+\dots \label{multiscale}
\end{equation}
Looking now for periodic solutions, we Taylor expand the RHS of Eq. \ref{perturbation} at first order in $\epsilon$

\begin{equation}
\frac{d \mathbf{X}}{dt}=F( \mathbf{X_0})+\epsilon \frac{\partial F}{\partial \mathbf{X}}  \mathbf{X_1}+\epsilon \mathbf{G}(t) \label{perturb_eps}
\end{equation}

Now we also have from the multiscale property
\begin{eqnarray}
\frac{d \mathbf{X}}{dt} & =& \frac{\partial \mathbf{X}}{\partial s}\frac{ds}{dt}+\frac{\partial \mathbf{X}}{\partial \tau}\frac{d\tau}{dt}\\
& = & \frac{\partial \mathbf{X}}{\partial s}+\epsilon \frac{\partial \mathbf{X}}{\partial \tau}\label{multiscaleODE}
\end{eqnarray} 
so that at the lowest order in $\epsilon$ we have

\begin{equation}
\frac{d \mathbf{X}}{dt}=\frac{d  \mathbf{X_0}}{dt}+\epsilon \frac{d  \mathbf{X_1}}{dt}=\frac{\partial  \mathbf{X_0}}{\partial s}+\epsilon \frac{\partial  \mathbf{X_0 }}{\partial \tau} +\epsilon \frac{d  \mathbf{X_1}}{dt}
\end{equation}
that we identify with the RHS of Eq. \ref{perturb_eps}.
 We have at order $0$
\begin{equation}
    \frac{\partial  \mathbf{X_0}(s,\tau)}{\partial s}=F( \mathbf{X_0}(s,\tau))
\end{equation}
so that $ \mathbf{X_0}$ is a solution of the unperturbed equation. In particular, this means that $ \mathbf{X_0}(s,\tau)=\mathbf{U}(s+\theta(\tau))$ where we have added a yet unknown phase shift on the cycle $\theta(\tau)$. Notice that $\theta$ is a function of the slow variable so that it will only slowly move as a function of real-time $t$.

Getting to order 1 in $\epsilon$, we have 

\begin{equation}
\frac{\partial \mathbf{U}_0}{\partial \tau}+\frac{d  \mathbf{X_1}}{dt}= \frac{\partial F}{\partial \mathbf{X}}  \mathbf{X_1}+\mathbf{G}(t)
\end{equation}

or 

\bwt
\begin{equation}
\frac{d \theta}{d \tau} \mathbf{U}'(s+\theta)+L .  \mathbf{X_1}=\mathbf{G}(t)
\implies
\frac{d \theta}{d \tau}<\mathbf{Z},\mathbf{U}'>+ <\mathbf{Z},L. \mathbf{X_1}>=<\mathbf{Z},\mathbf{G}(t)>
\end{equation}
\ewt

We now use that :
\begin{itemize}
    \item $<\mathbf{Z},\mathbf{U}'>=1$
    \item $<\mathbf{Z},L.X_1>=<L^*\mathbf{Z}, \mathbf{X_1}>=0$ by definition of $\mathbf{Z}$
\end{itemize}

to get our final result

\begin{equation}
\frac{d \theta}{d \tau}=<\mathbf{Z},\mathbf{G}(t)>=\frac{1}{T}\int_0^T \mathbf{Z}. \mathbf{G}(t) \mathrm{d}t
\end{equation}

which shows that  $\mathbf{Z}$ indeed corresponds to the instantaneous phase response.

\appsec{Complementary Discussions}

\appsubsec{Scaling Laws}
\label{app:scaling}
This section complements the discussion on scaling in the main section \myref{Species}, in light of the segmentation clock paradigm.
 
It is well known that time scales of developmental processes are very temperature sensitive, so what happens if we change temperature?  A first limit would be a situation where $v$ (speed of growth) and $T$ (period of the clock) are inversely correlated so that the somite size $S = vT $ is kept constant: this gives a simple null model for embryonic temperature compensation giving constant somite size. But this immediately suggests the presence of extra feedback in the process: if the period of the clock and the speed of the wavefront are (inversely) regulated by the same pathways, compensation to various perturbations would then naturally occur.

 There could be other compensatory mechanisms at play.  Let us assume that the total growth time of the embryo is $T_{tot}$. Let us further assume that the speed of the wavefront corresponds to a linear growth rate of the embryo so that the total length of the embryo is $L= v T_{tot}$. The somite to total length ratio then is $S/L=T/T_{tot}$: it is constant as soon as the two temporal time scales (total embryonic growth time vs period of the clock) stay proportional to one another.  One could then get another kind of compensation where the size of the embryos vary (e.g. with temperature), but still, overall embryos of different sizes would scale to one another. 
 
Some more complexity can be assumed: in the original clock and wavefront model (see section \myref{CWsection})  is assumed to be related to some \textit{temporal}  component of development. A natural model would be that the wavefront is related to the age of the cell once it exits the tail bud so that an age gradient is laid in a developing embryo. This could be translated into some signaling molecules effectively acting as a morphogen gradient, triggering somite boundary formations below some threshold. This can be done for instance through degradation of some component \cite{Dubrulle2004}, and in many modern papers the Clock and Wavefront model is presented with such underlying assumption. This graded process introduces a new temporal time scale $T_c$ (age of a cell exiting from the tail bud until it adopts the somite fate), and a new length scale $P=v T_c$, corresponding to the presomitic mesoderm length where the clock is oscillating before it defines the pattern. One could then imagine other multiple scaling laws. For instance, the ratio of somite size to oscillation zone size would then be $S/P= T/T_{c}$. Multiple situations could then occur if $T_c$ varies; in particular one can get a constant scaling of somite to embryonic size if the period of the clock is proportional to the total developmental time, but during development itself, the relative size of the differentiation zone to the total embryo size could vary. Even more complications could come if the different time scales vary during development, but those simplified situations give at least some simple intuitions on how different parameters can influence different phenotypic observables, and how even with a minimal number of observables one can have many independent variations and scaling relationships.

\appsubsec{Number of waves in a growth model}

\label{Appendix:growth}

This section complements the discussion in the main section \myref{Species}, using the Lewis Phase Mode (main section \myref{LPM}) to compute explicitly the number of waves.

It is assumed in this model that the PSM is defined between position $m=0$ and $m=L$. Importantly, the PSM is growing uniformly, with an instantaneous growth rate $\alpha$. If the PSM maintains constant length, it means that the somite size (leaving PSM) exactly is the newly created matter in one cycle ($2\pi$ in proper units), i.e. we approximately have

\begin{equation}
S=2\pi \alpha L \label{somite_growth}
\end{equation}

Now modifying equation \ref{movingframe} to account for PSM growth we have :
\begin{equation}
\frac{\partial \phi}{\partial t}= r(m)- v(m) \frac{\partial \phi}{\partial m}
\end{equation}
where $v(m)=\alpha m$ to account for expansion of the PSM. Assuming that at stationarity $\frac{\partial \phi}{\partial t}=1$, we then get a modified equation for the phase shift across the entire PSM (or similarly the number of waves).

\begin{equation}
 \Delta \psi_0 = \int_{0}^{L} \frac{1}{v(m)}\left(r(m)-1 \right) =\int_0^1 \frac{1}{\alpha x} \left(r(L,x)-1\right)dx
 \end{equation}
where we performed the change of variable $x=m/L$, and $r(L,x)=r(m)$. Experimental comparisons in \cite{Gomez2009} suggest indeed  that $r(L,x)=r(x)$, meaning that the frequency gradient is the same function of the relative position  $x=m/L$ in different species. In other words, the observed gradient of frequency scales with PSM length.  We then get immediately
\begin{equation}
 \Delta \psi_0  =\frac{1}{\alpha}\int_0^1 \frac{1}{x} \left(r(x)-1\right)dx
 \end{equation}

$\Delta \psi_0 $ is thus directly proportional to $1/\alpha$ , which from Eq.\ref{somite_growth} is proportional to $L/S$, and to some integrated quantity depending on the shape of the frequency gradient only, but lower than $1$ in absolute value since we typically observe a flat gradient close to $0$, and $r(1)=0$. For instance, taking $r(x)=1-x^2$, which is the simplest frequency profile flat in $0$ and canceling out in $1$, we recover the exact same result as in Eq. \ref{n_waves}, i.e. $N =\frac{L}{2S}$.

\appsubsec{ Doppler period shift calculations }

This section complements the main section \myref{sec:Doppler}.

The goal of this section is to describe how to compute Doppler contributions (in particular $\Lambda$) for two cases. We treat here two analytical cases: the case similar to the initial publications \cite{Soroldoni14,Jorg2015} where the frequency gradient is assumed to scale with PSM (frequency scaling), and a different case where the PSM simply shrinks without adaptation of the frequency gradient (frequency cropping), for which we do not expect much dynamical wavelength contribution. The two cases illustrate in particular how the dynamical wavelength contribution can eventually lead to an attenuation of the Doppler effect.

\textit{Shrinking and frequency scaling}

Let us assume the PSM shrinks with a constant speed $\bar{x}=\bar{x}_0-\bar{v}t$ so that, from Eq. \ref{Lambda_Doppler}
\begin{eqnarray}
\Lambda(\bar{x}(t),t) & =& \frac{\bar{v}}{v}\int_0^{\bar{x}}\frac{\partial}{\partial L}\omega\left( x',\bar{x}_0-\bar{v}\left(t-\frac{\bar{x}(t)-x'}{v}\right)\right)dx' \\
& =& \beta\int_0^{\bar{x}}\frac{\partial}{\partial L}\omega\left(x',(1+\beta)\bar{x}(t)-\beta x'\right)dx' 
\end{eqnarray}

calling $\beta=\frac{\bar{v}}{v}$. This simplification suggests to rescale $x'$ by the PSM length $\bar{x}(t)$, so that, setting $x'= \bar{x}(t) u$ we get 
\begin{equation}
\Lambda(\bar{x}(t),t) =\beta \bar{x}(t)\int_0^{1}\frac{\partial}{\partial L}\omega\left[\bar{x}(t) u,\bar{x}(t)(1+\beta (1-u))\right]\mathrm{d}u 
\end{equation}

Notice that all complicated past contributions have now been integrated into a simple coefficient $\beta$, so that $\Lambda(\bar{x}(t),t) $ now is a pure function of $\bar{x}(t)$ only (and of $\beta$).

In  \cite{Jorg2014,Jorg2015}, it is assumed that $\omega(x,L)=\omega_0 U(x/L)$, so that $\frac{\partial \omega}{\partial L}=-\omega_0 \frac{x}{L^2}U'(x/L)$. This gives immediately :

\begin{eqnarray}
\Lambda(\bar{x}(t),t)/\omega_0 &=&-\beta \int_0^{1}\frac{u}{(1+\beta(1-u))^2}U'\left(\frac{u}{1+\beta(1-u)}\right)\mathrm{d}u \\
 &=&-\beta \int_0^{1}\frac{\xi}{1+\beta \xi}U'(\xi)\mathrm{d}\xi \\
  &=&\Delta -\frac{\beta}{\beta+1} U(1)
\end{eqnarray}
After integration by parts and defining $\Delta=\beta\int_0^1 U(\xi)(1+\beta \xi)^{-2}d\xi$.
Notice this has the nice property to be completely independent from $\bar{x}$ and thus time-independent; also the introduction of quantity $\Delta$ will allow for a simplification later on. The advantage of this expression is that it allows for a more compact estimate of the anterior frequency: one gets

\begin{eqnarray}
\Omega_A &=& \omega_0-\Lambda(\bar{x},t)-\beta\left[\omega(\bar{x},\bar{x})-\omega_0+\Lambda(\bar{x},t)\right]\\
&=& \left[(1+\beta)(1-\Lambda(\bar{x},t)/\omega_0)+\beta U(1) \right]\omega_0 \\
&=& (1+\beta)(1-\Delta)\omega_0
\end{eqnarray}
(notice in particular the $U(1)$ expression cancels out).

\textit{Shrinking and frequency cropping}

Given the specific forms of frequency scaling with PSM size assumed in the previous section (a possibly rather strong assumption), it is worth examining alternative forms of Doppler contributions. 
 An alternative assumption could be that the frequency gradient does not change much with space and time: rather $\bar{x}$ defines a moving front of a literal clock stopping. Such frequency gradient would then take the  shape :
\begin{equation}
\omega(x,L)=\tilde\omega(x)\sigma(x/\bar x(t)) \label{omegacollapse}
\end{equation}
where $\sigma(y)$ is a sigmoidal function such that $\sigma \simeq 1$ for  $y<1$ and $0$ otherwise (e.g. a softmax function $\sigma(y)=\frac{1}{1+e^{\alpha(y-1)}}$ with big $\alpha$). Such a situation could happen, for instance, if the frequency gradient and the clock stopping are regulated by independent processes. 
We then get :
\begin{equation}
\Lambda(\bar{x},t) =-\beta \int_0^{1}\tilde\omega(\bar{x}u)\frac{u}{(1+\beta(1-u))^2}\sigma'\left(\frac{u}{1+\beta(1-u)}\right)\mathrm{d}u 
\end{equation}

If $\sigma$ is strongly non-linear, this integral is  dominated by the behavior close to $u\simeq 1$ and we  get :

\begin{equation}
\Lambda(\bar{x}(t),t)=\lambda \beta \tilde \omega(\bar{x}(t))
\end{equation}
where $\lambda$ depends on the non-linearity of $\sigma$ (for a softmax function in the limit $\alpha\rightarrow +\infty$ one would get $\lambda=1/2$).This is a rather intuitive result: in such a situation, an oscillator at a given position "does not know" that the PSM is shrinking, and it is only when the front is reaching it that a dynamical wavelength contribution appears. Thus the dynamical wavelength contribution depends only on the local frequency at the front $ \tilde\omega(\bar{x}(t))$.  It is worth noticing that, especially at the beginning of PSM shrinkage, we expect $\omega(\bar{x}(t))$ to be rather low to that the effect would be small.

Furthermore,  this contribution might be quite difficult to detect experimentally because in this situation it would be very localized at $\bar{x}$. To see this, one can look at what happens right before the front,  at  the position $\bar{x}(1-\epsilon)$, where one gets :

\bwt
\begin{eqnarray}
\Omega_{A-\epsilon} & =&\frac{d}{dt} \phi(\bar{x}(t)(1-\epsilon),t)=\left.\frac{\partial \phi}{\partial t}\right|_{\bar{x}(t)-\epsilon}+ \left.\frac{d \bar{x}(t)}{dt} \frac{\partial \phi}{\partial x}\right|_{\bar{x}(t)(1-\epsilon)} \\
 & = & \omega(0) - \Lambda(\bar{x}(1-\epsilon),t) \nonumber \\
 &&+\frac{\dot{\bar{x}}}{v}[\omega(\bar{x}(1-\epsilon),\bar{x})-\omega(0)+\Lambda(\bar{x}(1-\epsilon),t)]
\end{eqnarray}
\ewt
Now following the exact same reasoning one would get 

\begin{equation}
\Lambda(\bar{x}(t)(1-\epsilon),t)= -\beta \int_0^{1-\epsilon}\tilde\omega(\bar{x}u)\frac{u}{(1+\beta(1-u))^2}\sigma'\left(\frac{u}{1+\beta(1-u)}\right)\mathrm{d}u 
\end{equation}

Now this integral is $0$ if the non-linearity $\sigma$ is very strong: the reason is that right before the front $\sigma$ is essentially constant so the derivative is $0$. In such case we get the much simpler expression 

\begin{eqnarray} 
\Omega_{A-\epsilon} & =& \omega(0) -\beta[\omega(\bar{x}(1-\epsilon),\bar{x})-\omega(0)]
\end{eqnarray}

which exactly coincides with the more intuitive terms described after Eq. \ref{omegadoppler}, without any dynamical wavelength contribution. Notice that if there is a very small frequency in the anterior ($\omega(\bar{x}(1-\epsilon))\sim 0$), we get

\begin{eqnarray} 
\Omega_{A-\epsilon} & =& (1+\beta)\omega(0)
\end{eqnarray}

which coincides with the pure `Doppler' contribution of Eq. \ref{eg:doppler_delta}.
In fact, as described in  \cite{Soroldoni14}, experimentally the  $\Lambda$ term in fact goes to $0$ which is more consistent with such a model at later times.

\appsubsec{Conditions for oscillations for negative feedback oscillators with delay}

We consider the following equation
\begin{equation}
\epsilon \frac{db}{dt} = f(b(t-1))- b \label{eq:rescaled}
\end{equation}
similar to Eq. \ref{rescaled} in the main text. We consider an explicit example with a Hill-like repression where $f(b)=\frac{r}{1+b^n}$. Calling $b^*$ the fixed point ($f(b^*)=b^*$), one can consider a small perturbation close to the fixed point. Setting $b=b^*+\rho e^{(\mu+i\omega)\theta}$, injecting into equation \ref{rescaled} and Taylor expanding close to $b^*$ one gets 
\begin{equation}
 \epsilon
(\mu+i\omega)=-\frac{nrb^{*(n-1)}}{(1+b^{*n})^2}e^{-(\mu+i\omega)}-1
\end{equation}

Identifying real and imaginary parts, we get 
\begin{equation}
 \epsilon\mu=-\frac{nrb^{*(n-1)}}{(1+b^{*n})^2}e^{-\mu}\cos{\omega}-1\label{equationmu}
\end{equation}

\begin{equation}
 \epsilon\omega=\frac{nrb^{*(n-1)}}{(1+b^{*n})^2}e^{-\mu}\sin{\omega}
\end{equation}

combining into 

\begin{equation}
\tan{\omega}=-\frac{\epsilon \omega}{1+\epsilon \mu} \label{equationepsilon}
\end{equation}

Oscillations appear when $\mu>0$ (corresponding to an amplification of the perturbation close to the fixed point). At the bifurcation $\mu=0$, one gets  $\cos{\omega}=-\frac{(1+b^{*n})^2}{nrb^{*(n-1)}}$, so that $\omega=\cos^{-1}\left(-\frac{(1+b^{*n})^2}{nrb^{*(n-1)}}\right)$. Equation \ref{equationepsilon} then imposes a critical value $\epsilon_c$  

\begin{equation}
\epsilon_c=\frac{\left(\left(\frac{nrb^{*(n-1)}}{(1+b^{*n})^2}\right)^2-1\right)^{1/2}}{\cos^{-1}\left(-\frac{(1+b^{*n})^2}{nrb^{*(n-1)}}\right)} \label{epsilon_crit}
\end{equation}

Oscillation occurs for small $\epsilon$, i.e.  $\epsilon<\epsilon_c$.  For values of $\epsilon$ higher than $\epsilon_c$, the system will relax to $b^*$ and does not oscillate.

We also get along the way the necessary condition $\frac{(1+b^{*n})^2}{nrb^{*(n-1)}}<1$, and using the fact that $f(b^*)=b^*$ one gets $(n-1)r>nb^*$. This immediately shows that $n>1$ is necessary to have oscillations, which implies the existence of a cooperative \cite{}, negative feedback loop. For a given $n$, this also constrains $r$, e.g. for $n=2$, we get $r>2b^*$, which implies that $b^3-b^*>0$, or $b^*>1$, so that $r>2$.

\appsubsec{Delayed model with noise}

Here we briefly reproduce part of the calculations made in \cite{Negrete2021} to account for noise in the Delayed oscillator. Again analytical calculations are possible with minimal assumptions since the system  is linear by piece.  Equation for $b^{-}$ becomes :

\begin{equation}
    \epsilon \frac{db^{-}}{dt} = - b^{-} + b^{-}\eta
\end{equation}

which reduces to a standard Brownian motion equation $\dot x=\eta$ by the simple change of variable $x(t)=\epsilon\log b^{-}(t) +t$ so that

\begin{equation}
    b^{-}(t)=e^{-t/\epsilon+x} \label{bm}
\end{equation}

Since $x$ is a standard 1D Brownian motion, its distribution is Gaussian, and $b$ is thus given by a log normal distribution
\begin{equation}
    p(b^{-})=\frac{1}{2\pi \sigma_0^2 b^{-}}e^{-(\log b^{-}+t)^2/2\sigma_0^2}
\end{equation}

Similarly, in the production phase, we have 

\begin{equation}
    \epsilon \frac{db}{dt} = r- b + b\eta 
\end{equation}

The solution of this equation is now of the form :
\begin{equation}
  b^+(t)e^{t/\epsilon-x}=b_{min}+r\int_0^{t} e^{s/\epsilon-x(s)} ds   \label{bp}
\end{equation}

which allows for a computation of various moments (being careful for Gaussian cross-correlations).

One can also get the distributions of maximal and minimal values $b_{min}$ and $b_{max}$, which corresponds to solutions of Eqs.\ref{bm}-\ref{bp} at $t=1$ with initial conditions $b=1$, i.e.

\begin{equation}
    b_{min}=e^{-1/\epsilon+x(1)}
\end{equation}

and
\begin{equation}
  b_{max}=b_{min}+r\int_0^{1} e^{(s-1)/\epsilon+x(1)x(s)} ds  
\end{equation}

One can also use similar self-consistent approximation to get distributions of $t_{on},t_{off}$  for instance for the first part of the cycle we have

\begin{equation}
   1=b_{max}e^{-{t_{off}-1}/\epsilon+x(t_{off}-1)}
\end{equation}

\appsubsec{Practical details on the ERICA model}
For numerical integration, it is useful to formulate the ERIC model in cartesian coordinates, i.e.

\begin{eqnarray}
\dot x &=& \frac{\dot{r}}{r} x -y \\
\dot y &=& \frac{\dot{r}}{r} y + x
\end{eqnarray}
which separates the harmonic oscillator part from the radial part in $\frac{\dot{r}}{r}$
We get immediately
\bwt
\begin{eqnarray}
\dot x &=& \left(\frac{r}{r+\lambda y} - r - \frac{\lambda x}{(r+\lambda y)^2}\right).x - y =dx_\lambda(x,y) \\
\dot y &=& x + \left(\frac{r}{r+\lambda y} - r - \frac{\lambda x}{(r+\lambda y)^2}\right).y =dy_\lambda(x,y)
\end{eqnarray}
 \ewt
 Considering now a more general form

\begin{eqnarray}
\dot x &=& s(x,y)[x f(x,y) -y] \\
\dot y &=& s(x,y)[y f(x,y)+x]
\end{eqnarray}
we then get
\begin{equation}
r \dot r=x \dot x + y \dot y= s(x,y) r^2 f(x,y)
\end{equation}

so that we have 

\begin{eqnarray}
\dot x &=&  x\dot r/r-s(x,y)y \\
\dot y &=& y\dot r/r+ s(x,y)x
\end{eqnarray}

By identification in polar coordinates, we immediately get

\begin{equation}
\dot \theta = s(x,y)
\end{equation}
 
 showing that the full form of the ERIC model should be :

\begin{eqnarray}
\dot x &=& s(\theta) dx_\lambda(x,y)\\
\dot y &=& s(\theta) dy_\lambda(x,y)
\end{eqnarray}

For simplicity, and to keep the system analytical, we restricted ourselves first to $s$ functions linear by piece, i.e. $s=s_* > 1$ for one sector and $s=1$ otherwise. The sped up sector is centered at angle $\alpha$ and has width $\beta$, so that the modified period of the cycle is $T_{s_*}=\beta/s_*+(2\pi-\beta)$. It is also convenient to define the rescaled angular velocity $\omega_{s_*}=2\pi/T_{s_*}=\frac{2\pi s_*}{\beta+(2\pi-\beta)s_*}$. From there, the phase of the cycle as a function of the angle $\theta$ in the plane is a simple linear transformation (defining $\theta_0=\alpha-\beta/2$):
\begin{equation}
\phi_s(\theta) =\omega_{s_*}\theta= \theta\frac{2\pi s_*}{\beta+(2\pi-\beta)s_*} \label{1stsection}
\end{equation}
for $0<\theta<\theta_0$,
\bwt

\begin{equation}
\phi_s(\theta)  = \omega_{s_*} \left(\theta_0 + \frac{\theta-\theta_0}{s_*} \right) =\left(\theta_0 + \frac{\theta-\theta_0}{s_*} \right)\frac{2\pi s_*}{\beta+(2\pi-\beta)s_*} \label{2ndsection}
\end{equation}
for $\theta_0<\theta<\theta_0+\beta$, and 
\begin{equation}
\phi_s(\theta)  = \omega_{s_*}\left(\theta_0 + \frac{\beta}{s_*} + \theta - (\theta_0 + \beta) \right)= \left(\theta_0 + \frac{\beta}{s_*} + \theta - (\theta_0 + \beta) \right)\frac{2\pi s_*}{\beta+(2\pi-\beta)s_*} \label{3rdsection}
\end{equation}
\ewt
for $\theta_0+\beta<\theta<2\pi$. Those functions ensure that the rate of phase evolution in sector $\theta_0<\theta<\theta_0+\beta$ is $1/s_*$ times the rate in the other sectors (compare $\theta$ coefficients in  Eq. \ref{1stsection}-\ref{3rdsection}), that angle $0$ in $\theta$ is phase $\phi_s=0$, and that phase is continuous so that $\phi_s(\theta_0)=\omega_{s_*} \theta_0$ and $\phi_s(\theta_0+\beta)=(\theta_0+\beta/s_*)\omega_{s_*}$. Notice also that for $\theta=2\pi$ we get from Eq.\ref{3rdsection} $\phi_s(2\pi)=\left( \frac{\beta}{s_*} + 2\pi -  \beta \right)\frac{2\pi s_*}{\beta+(2\pi-\beta)s_*} =2\pi$ as expected after one full cycle.

In \cite{Sanchez2021}, we used Markov chains to generate distributions of parameters and took the average values to plot the model limit cycle and PRC. The optimized parameter values are: $\epsilon = 0.43$, $\lambda = 0.53$, $s_* = 5.64$, $\alpha = 1.51\pi$, $\beta = 1.15\pi$, $\phi_0 = 1.17\pi$.

\terminateappendix


\end{document}